INSTITUTO FEDERAL DO ESPIRITO SANTO
MESTRADO NACIONAL PROFISSIONAL EM ENSINO DE FÍSICA

**LUCIA HELENA HORTA OLIVEIRA**

# PROPOSTA DE SEQUÊNCIA DIDÁTICA PARA ENSINO DAS LEIS DE NEWTON, UTILIZANDO GIFs E VÍDEOS

LUCIA HELENA HORTA OLIVEIRA

# PROPOSTA DE SEQUÊNCIA DIDÁTICA PARA ENSINO DAS LEIS DE NEWTON, UTILIZANDO GIFs E VÍDEOS

Dissertação apresentada ao Programa de Pós-graduação em Ensino de Física - Mestrado Nacional Profissional em Ensino de Física, ofertado pela Sociedade Brasileira de Física em parceria com o Instituto Federal do Espírito Santo, Campus Cariacica, como requisito parcial para a obtenção do título de Mestre em Ensino de Física.

Orientador: Dr. Samir Lacerda da Silva

MNPEF
Mestrado Nacional Profissional em Ensino de Física

INSTITUTO FEDERAL
Espírito Santo
Campus Cariacica

SBF
SOCIEDADE BRASILEIRA DE FÍSICA

LÚCIA HELENA HORTA OLIVEIRA

# PROPOSTA DE SEQUÊNCIA DIDÁTICA PARA ENSINO DAS LEIS DE NEWTON UTILIZANDO GIFs E VÍDEOS

Dissertação apresentada ao Programa de Pós-graduação em Ensino de Física – Mestrado Nacional Profissional em Ensino de Física, ofertado pela Sociedade Brasileira de Física em parceria com o Instituto Federal do Espírito Santo, Campus Cariacica, como requisito parcial para a obtenção do título de Mestre em Ensino de Física.

**COMISSÃO EXAMINADORA**

Prof. Dr. Samir Lacerda da Silva
Instituto Federal do Espírito Santo
Orientador

Prof. Dr. José Bohland Filho
Instituto Federal do Espírito Santo
Membro interno

Prof. Dr. Judismar Tadeu Guaitolini Junior
Instituto Federal do Espírito Santo
Membro externo

MNPEF
Mestrado Nacional Profissional em Ensino de Física

INSTITUTO FEDERAL
Espírito Santo
Campus Cariacica

SBF
SOCIEDADE BRASILEIRA DE FÍSICA

LÚCIA HELENA HORTA OLIVEIRA

OLIVEIRA, Lúcia Helena Horta; SILVA, Samir Lacerda da. **Proposta de sequência didática para ensino das Leis de Newton utilizando Gifs e vídeos**. Cariacica: Ifes, 2020. 33 p.

Produto Educacional apresentado ao Programa de Pós-graduação em Ensino de Física – Mestrado Nacional Profissional em Ensino de Física, ofertado pela Sociedade Brasileira de Física em parceria com o Instituto Federal do Espírito Santo, Campus Cariacica, como requisito parcial para a obtenção do título de Mestre em Ensino de Física.

Aprovado em 19 de fevereiro de 2020

**COMISSÃO EXAMINADORA**

Prof. Dr. Samir Lacerda da Silva
Instituto Federal do Espírito Santo
Orientador

Prof. Dr. José Bohland Filho
Instituto Federal do Espírito Santo
Membro interno

Prof. Dr. Judismar Tadeu Guaitolini Junior
Instituto Federal do Espírito Santo
Membro externo

Dedico esse trabalho às pessoas que estiveram ao meu lado, a todo o tempo, me ajudando a seguir em frente, nos momentos de dificuldade. Minha mãe Terezinha, meu marido Eder, minha filha Aysla e principalmente meu filho Ayler, que teve a paciência de me ajudar a projetar a revista e criar um aplicativo exclusivo. Ayler, sem você esse trabalho não seria tão interessante, pois através do aplicativo conseguimos conectar a física à linguagem adolescente. Ainda teremos muitos projetos juntos, pois amo trabalhar com você e sei que tudo que você fizer será sucesso. Obrigada.

**AGRADECIMENTOS**

Agradeço ao meu orientador, Samir Lacerda da Silva, que acreditou no meu trabalho.

Ao Professor Luiz Otavio Buffon, que me incentivou e me ensinou, fazendo toda diferença para os novos caminhos que pretendo trilhar.

Ao Professor Júlio Cesar Fabres, um dos maiores cientistas que conheci, o qual me deu a honra de poder trabalhar em projetos científicos sob sua supervisão acreditando nas minhas ações.

Aos alunos do primeiro ano do ensino médio da EEEM DR. SILVA MELLO turma de 2018, que participaram ativamente na aplicação da sequência didática.

O presente trabalho foi realizado com o apoio da Coordenação de Aperfeiçoamento de Pessoal de Nível Superior – Brasil (CAPES) – Código de financiamento 001.

## RESUMO

A realidade tecnológica e digital tem se expandido amplamente, atingindo, de forma massiva, espaços que, há um século, não faziam parte desse ambiente, sendo a escola um exemplo. Por isso, esta pesquisa tem por objetivo geral desenvolver uma sequência didática, utilizando como uma de suas estratégias uma história em quadrinhos, bem como Figuras, Gifs e vídeos, programadas para serem aplicadas através de quadrinhos às Leis de Newton. A pesquisa visou a analisar de que forma essa ferramenta metodológica pode influenciar na aprendizagem do aluno. Para isso, apoitamo-nos em pesquisas como as de Carvalho e Lemos (2010), Camargo (2015), Farias (2010), Nardin (2016) e Vygostky (1930). O método utilizado para o desenvolvimento da pesquisa foi um estudo de caso, com coleta de dados de cunho qualitativo e quantitativo. O produto parece ter tido uma boa aceitação entre os adolescentes e por esse motivo pôde ajudar no aprendizado, favorecendo avanços na aprendizagem de conceitos e na aplicação de conhecimentos nos eventos do cotidiano.

Palavras-chave: Gifs no ensino de física. Realidade aumentada. Quadrinhos.

**ABSTRACT**

The technological and digital reality has broadly expanded, reaching, in a massive way, spaces, for a century, it is not part of this environment, being a school an example. Therefore, this research has as general objective to develop a didactic sequence, using as a strategy a comic book, as well as Figures, Gifs and videos, programs for activities applied to comics using Newton's Laws. A research that aims to analyze which methodological form can influence student learning. For this, it supports us in research such as Carvalho and Lemos (2010), Camargo (2015), Farias (2010), Nardin (2016) and Vygostky (1930). The method used for the development of the research was a case study, with the collection of qualitative and quantitative data. The product seems to have received a good acceptance among teenagers and for this reason it can help in learning, favoring advances in the learning of concepts and in the application of knowledge in everyday events.

Keywords: Gifs in physics education. Augmented Reality. Comics.

**LISTA DE FIGURAS**

**LISTA DE GRÁFICOS**

**SUMÁRIO**

# 1 INTRODUÇÃO

A realidade tecnológica e digital tem expandido, atingindo, de forma massiva, espaços que há um século não faziam parte desse ambiente, sendo a escola um desses exemplos. As últimas gerações de alunos têm estado potencialmente imersas na era digital, impondo às escolas novos desafios, de maneira geral, incluindo-se aí, a reformulação de disciplinas e preparo de professores. Esse painel demanda, pois, tomada de atitudes.

O trabalho de integração entre tecnologia e conteúdo disciplinar tem se tornado comum, vez que a tecnologia é uma porta de entrada para o despertar do interesse dos alunos. A partir da inclusão dessa ferramenta no ensino é possível atingir o contexto de vida dos discentes, tendo em vista que estes utilizam a tecnologia em praticamente todos os momentos de suas vidas, tais como videogames, *smartphones* e computadores.

Sendo assim, este trabalho tem como escopo o ensino de física, uma disciplina que, em geral possui um índice baixo de aceitação dos alunos por seu alto grau de dificuldades e abstração pelos próprios educandos. O ensino tradicional de física possui um grau elevado de complexidade, o que pode obstruir a ponte entre o conteúdo ministrado e o contexto social do aluno. Como consequência disso, cita-se o desinteresse dos alunos, a dificuldade de assimilação da disciplina e, paralelo a isso, a sua rejeição.

O ensino de física e as dificuldades que os educandos apresentam na compreensão de conceitos vêm sendo objeto de estudo ao longo dos anos. Pesquisas como as de Carvalho e Lemos (2010), Camargo (2015), Farias (2010) e Nardin (2016) apontam para um mesmo fator potencializador dessa dificuldade – o modo como a física é apresentada e ensinada aos alunos.

De acordo com a teoria sócio-histórica de Vygostky (1930), cuja mediação é posta como o ponto central das relações de aprendizagem, uma vez que o contexto à volta do aluno será envolvido no processo, o professor assumirá um papel importante na abordagem de conteúdos de física. Para isso, este exercerá o papel de um mediar, utilizando-se de instrumentos e signos da física, uma interação entre os alunos e os tópicos em abordagem.

São diversas as ferramentas que podem auxiliar no processo de ensino-aprendizagem do educando, como as de tecnologia multimídia, por exemplo, cuja utilização cresce significativamente no ambiente escolar, como afirmam Cintra e Solttau (2018). A inclusão de recursos digitais e tecnológicos é garantida também pelos Parâmetros Curriculares Nacionais (PCN's) dando amparo ao professor para trabalhar tais recursos de maneira livre.

O enfoque desta pesquisa aponta para o uso da realidade aumentada, por meio de Gifs e vídeos, no ensino de física, algo que vem sendo discutido e amplamente utilizado nas aulas desta disciplina. Tal fato nos leva a destacar a importância de se discutir o assunto, pois, apesar do que se tem feito, torna-se necessário delinear os passos e instruções para outros professores, ofertando-lhes o auxílio acadêmico e o suporte teórico.

# 2 OBJETIVOS

## 2.1 OBJETIVO GERAL

Esta pesquisa tem como objetivo geral desenvolver uma sequência didática, utilizando-se de quadrinhos, Gifs e 3D com programação de vídeos, e realidade aumentada, de modo a abordar as Leis de Newton.

## 2.2 OBJETIVOS ESPECÍFICOS

Como objetivos específicos, esta pesquisa pretende:

Analisar o nível de adaptação do professor ao uso do material escolhido para as aulas ministradas a partir da realidade aumentada.

Utilizar de pequenos vídeos (Gifs) e Figuras 3D para demonstrações virtuais, representando situações de acordo com as leis e principais forças estudadas, utilizando a mediação de Vygotsky.

Observar o nível de motivação do aluno no aprendizado da física, utilizando como signo a tecnologia para complementar o aprendizado.

## 2.3 HIPÓTESE

Esta pesquisa decorre da seguinte hipótese: o uso de vídeos e Gifs, como mecanismo de mediação no ensino das leis de Newton pode auxiliar na socialização e na compreensão do conteúdo de Física?

Considera-se para esse trabalho que o uso da realidade aumentada, junto ao material didático de Física, pode possibilitar uma melhor compreensão de modelos científicos, melhorando a interpretação de textos.

# 3 REFERENCIAL TEÓRICO

As bases pedagógicas do ensino fundam-se em teorias propostas por especialistas psicólogos, que desenvolveram suas pesquisas ao longo de décadas, entre os quais estão Jean Piaget e Lev Vygotsky. Pensando de forma específica na teoria de Vygotsky, este propunha o processo de desenvolvimento do indivíduo por meio de suas respectivas relações sociais, isto é, seu conhecimento se constrói a partir das relações construídas com o ambiente externo.

Lev Vygotsky propõe, ainda, a existência de uma zona específica onde ocorre o aprendizado, conhecida como Zona de Desenvolvimento Proximal. Entende-se que essa área, em específico, deva ser trabalhada pelo professor no aluno, de forma que este possa avançar no aprendizado (FIORAVANTE-TRISTÃO, 2010). A ligação direta entre o nível de desenvolvimento real e o potencial chama-se Zona de Desenvolvimento Proximal, a qual demanda a mediação de algo ou alguém para que a criança alcance o desenvolvimento para a tarefa proposta.

A mediação, para Vygotsky, é algo de suma importância, sendo considerada o fator central de sua psicologia (VYGOTSKY, 1930). Várias são as discussões acerca das formas de mediação e, em algumas abordagens, esse processo pode se aproximar do foco na atividade. No entanto, *grosso modo* pode-se entender a mediação como um processo de intervenção, em que se insere um elemento em uma relação, a qual deixa de ser direta, passando a ser mediada pelo elemento.

De acordo com Lev Vygotsky, instrumentos como ferramentas materiais, signos, ferramentas psicológicas ou seres humanos podem ser os meios pelos quais ocorre a mediação. Tendo como ponto de partida o princípio marxista de que o trabalho é fruto mediador das relações dos seres humanos entre si e com a natureza, Vygotsky (1930) deu enfoque à função mediadora de instrumentos elaborados pelos homens como estimulantes de mudanças externas, pois estes tornam possível a intervenção na natureza por meio da ampliação da capacidade produtiva do indivíduo.

Corroborando com os estudos de Vygotsky, Rego (2001) afirma que os seres humanos têm a capacidade de pensamento, elaboração e produção de seus próprios instrumentos, que resultarão na realização de tarefas específicas e na conservação destes instrumentos para que possam ser utilizados posteriormente. Isso significa que os indivíduos podem transmitir a função destes para outros membros de suas comunidades, recriando e aprimorando tais instrumentos.

No que se refere à mediação pedagógica, o processo de intervenção – que se dá por meio de elementos externos – é considerado intencional e sistematizado, uma vez que ocupa um papel de importância dentro do processo de aprendizagem do aluno. O modelo de aprendizagem que fundamenta as necessidades da sociedade aponta para o dinamismo, tendo em vista os conhecimentos e as interconexões mentais do discente (SANTOS, 2008). Por isso, a mediação pedagógica assume grande valor, uma vez que se faz necessária para que o aluno se torne capaz de construir seus conhecimentos, formar novos conceitos a respeito do mundo e, com isso, sobre ele intervir.

A mediação pedagógica trata-se de um processo de comunicação e construção de significados, tendo por objetivo a ampliação das possibilidades de argumento e diálogo, de modo a desenvolver criticamente os processos e conteúdos explorados nos espaços educacionais (SHECTMAN, 2009). Busca, também, incentivar a construção do pensamento crítico e contextualizado, possível a partir da relação entre professor e aluno (VARELA *et al*, 2014), intermediado por ferramentas culturais e signos.

A sala de aula pode ser vista como um local de incentivo à interação entre sujeitos e a produção de conhecimento, porém, tal interação será efetivada somente a partir da qualidade da mediação pedagógica que ocorre neste contexto. Sobre isso, Romanovsky e Martins (2008) afirmam que a práxis social é o elemento básico da aula, a qual se materializará nas relações sociais coletivas, compartilhadas e interessadas. Moraes (2008) contribui para o conceito de mediação pedagógica, defendendo que esta acontece por meio da comunicação, isto é, da ação sobre o outro.

Este tipo de mediação tem por enfoque a atitude e o comportamento do professor, uma vez que este é o promotor de aprendizagem, atuando como o elo entre o aprendiz e o conhecimento a partir do diálogo, das experiências compartilhadas, da argumentação e da resolução de problemas (MASETTO, 2000). Ao tratar sobre o ensino, Paulo Freire (2002, p. 134) afirma que,

> [...] ensinar não é transferir conteúdo a ninguém, assim como aprender não é memorizar o perfil do conteúdo transferido no discurso vertical do professor. Ensinar e aprender têm que ser com o esforço metodicamente crítico do professor de desvelar a compreensão de algo e com o empenho igualmente crítico do aluno de ir entrando, como sujeito de aprendizagem, no processo de desvelamento que o professor ou professora deve deflagrar (FREIRE, 2002, p. 134).

Todos nós trazemos os conceitos citados, os quais usamos no processo de mediação para uma aplicação prática, utilizamos, também, os instrumentos que tornam possível a mediação entre professor, conteúdo e aluno, entre os quais estão as tecnologias da informação e da comunicação – TIC.

As inovações em tecnologia podem ser consideradas como agentes de mudança, vez que propiciam formas de acesso à informação de diferentes e novos modos de raciocínio e conhecimento, ampliando determinadas capacidades cognitivas humanas, quais sejam, a memória, a imaginação e a percepção (LÉVY, 1999). Usando um hipertexto por exemplo, pode haver a quebra da linearidade da leitura que um sujeito faz de um texto escrito, fazendo com que surjam novos estilos de aprendizagem.

A partir do compartilhamento de tecnologias entre os indivíduos, torna-se possível, também, o aumento do potencial de inteligência coletiva, uma vez que todos os envolvidos são postos a pensar e a desenvolver algo. Para Vygotsky (1981, p. 137),

> A introdução de uma nova ferramenta cultural num processo ativo, inevitavelmente o transforma. Nessa visão, recursos mediadores como a linguagem e as ferramentas técnicas não facilitam simplesmente as formas de ação que irão ocorrer, mas altera completamente a estrutura dos processos mentais (VYGOTSKY, 1981, p. 137).

Dessa forma, se torna possível variar as práticas pedagógicas na sociedade da informação, uma vez que o professor se vê diante de um novo contexto de conhecimento, a era digital. Segundo Behrens (2000), reconhecer a era digital como uma nova maneira de categorização do conhecimento não abre precedentes para o descarte do caminho trilhado até então pelo modelo tradicional de linguagem oral e escrita, nem para a mitificação do uso exacerbado de computadores e afins no processo de ensino. Por outro lado, implica adequar o uso de ferramentas eletrônicas e/ou tecnológicas para uma construção dos processos metodológicos mais significativos.

Para que o processo de ensino aprendizagem e a mediação pedagógica, por meio dessas ferramentas, sejam harmoniosos, é necessário que o professor entenda a possibilidade de ir além, ainda que permaneça dentro da sala de aula. Dessa forma, permitirá a criação de encontros presenciais e de encontros virtuais que façam com que o aluno acesse informações disponíveis dentro da sociedade do conhecimento. Recursos tecnológicos não caracterizam o encerramento da aprendizagem, mas assinalam possíveis meios de se instigarem novas metodologias que encaminhem o aluno para a aprendizagem com interesse, criatividade e autonomia.

Para Behrens (2000), é importante que o professor não se furte da articulação de projetos de aprendizagem que envolvam, ou tenham como base a tecnologia, mesmo porquê, na maior parte das vezes, esta já faz parte do contexto das instituições de ensino. Portanto, é a partir da perspectiva epistemológica histórico-cultural que compreendemos o uso da tecnologia, outros recursos e de linguagens digitais como colaboradores significativos para a eficácia da educação. O uso das TICs pode proporcionar mudanças importantes no modo pelo qual os materiais educacionais serão projetados e utilizados por professores e alunos (GARCIA, 2016).

Pesquisas como as de Tijiboy *et al* (2009), Garcia (2016), Costa *et al* (2015), Gehlen e Delizoicov (2012), Perrone (2018) e Oliveira (2016) atestam que o uso de recursos tecnológicos agrega resultados positivos no que tange ao alcance de novos conhecimentos e apropriação da aprendizagem por parte do aluno. Em todas as pesquisas citadas, ocorreu a mediação por parte de um recurso que se difere da linguagem oral e escrita, demarcando como a mediação pedagógica, por meio de ferramentas externas, ocorre em conjunto com o trabalho do professor.

As pesquisas de Perrone (2018) e Oliveira (2016), especificamente, tratam do ensino de física e matemática, respectivamente, a partir da utilização da realidade aumentada. Para Perrone (2018), a realidade aumentada configura-se como uma ferramenta tecnológica contemporânea que se integra ao contexto histórico dos alunos da sociedade atual, sendo esta uma ferramenta de valor, uma vez que desperta o interesse do aluno pelo conteúdo.

Uma comprovação disto é o resultado da pesquisa elaborada por Forte (2009), em que, a partir da aplicação de um questionário, observou que a utilização deste recurso pode gerar motivação em aprender e curiosidade sobre o conteúdo por parte do aluno. O autor destaca, ainda, que o uso da realidade aumentada proporciona ao aluno a oportunidade de observação dos fenômenos de forma repetida e em diferentes situações.

No entanto, é necessário compreender que toda metodologia pedagógica utilizada em sala de aula deve obrigatoriamente ser planejada, pois a sistematização do procedimento é um fator de excelência no processo de aprendizagem. O uso da realidade aumentada, bem como de qualquer recurso tecnológico, sem a orientação do professor, não produz a aprendizagem significativa. Para Forte (2009), a realidade aumentada, quando utilizada para mera visualização, com ausência da intermediação do professor, não agrega valor ao processo de aprendizagem do aluno.

Portanto, o processo de mediação pedagógica deve ser intencional e sistematizado, não descartando a ação do professor e o compartilhamento de ideias, do diálogo e da resolução de problemas por meio das relações sociais e tecnológicas, trabalhadas em conjunto.

# 4 FÍSICA TEÓRICA ENVOLVIDA NA APLICAÇÃO DA SEQUÊNCIA DIDÁTICA

A dinâmica, conhecida como mecânica clássica, foi desenvolvida entre os séculos XVII e XVIII. Algumas teorias mais recentes como a relatividade especial, a relatividade geral e até mesmo a mecânica quântica trazem novos estudos, mas para objetos um pouco mais distantes de nossas realidades. No entanto, teorias como a mecânica quântica conseguem apresentar resultados comprováveis apenas em ambientes específicos no nosso cotidiano, já a mecânica clássica consegue explicar os fenômenos com maior visibilidade, com maior tranquilidade. Desde a antiguidade, os filósofos discutiam sobre questões que envolviam movimentos e sua causa.

Questões envolvendo movimentos e suas causas persistiram até o século XVII em que Galileu e Isaac Newton conseguiram desenvolver formas diferentes de estudar o movimento. Essas formas ficaram conhecidas como a mecânica clássica. Newton formulou as três leis do movimento, por volta do século XVII no seu livro *Filosofia Naturalis Principia Mathematica,* conhecido como Princípia.

As leis de Newton constituem a base do que entendemos hoje sobre o movimento e suas causas e até mesmo as suas limitações. No século XX, através da descoberta da física quântica e da relatividade geral e específica, foram reveladas as limitações que existiam na lei do movimento de Newton. O principal interesse da mecânica clássica está no movimento do objeto particular que, ao interagir com os objetos que estão na sua vizinhança, tem sua velocidade acelerada.

Essa interação do corpo com a sua vizinhança, que colocamos como um referencial, é descrita como uma força, a qual determina uma ação que pode ser empurrar ou puxar, em uma determinada direção. Essas forças são representadas por vetores. As forças serão combinadas utilizando as regras de adição de vetores. Nessas condições, conseguimos calcular a força resultante, sempre, no movimento unidimensional.

As forças devem ser analisadas e identificadas no corpo onde estão atuando, desta forma calculamos os movimentos desejados. Então, a mecânica clássica estabelece as leis de força como básicas nas propriedades de um corpo e sua vizinhança. Desta forma, o sistema fornecerá resultados que estarão de acordo com os experimentos, as leis de força são simples e razoáveis, para o entendimento do nosso cotidiano.

## 4.1 A PRIMEIRA LEI DE NEWTON

Como a força afeta o movimento? Podemos responder a essa pergunta observando o que acontece quando a força resultante em um corpo é igual a zero. Um corpo está parado e se não existe nenhuma força puxando-o ou o empurrando (força resultante), ele deve permanecer em repouso. Mas se o corpo estiver se movendo e a força resultante existente também for igual a zero?

Usando um pequeno disco usado para hóquei, como exemplo, a princípio, empurramos o disco numa mesa de madeira. No momento em que você retirar a mão e parar de empurrar o disco ele andará apenas alguns centímetros e irá parar, não se movendo mais. Se quiséssemos que o disco continuasse a se mover, teríamos que continuar exercendo força sobre nele, ou seja, seria necessária uma força para manter o movimento do disco (Figura 1-a).

Se colocarmos um pequeno disco tipo, o usado para hóquei, sobre uma superfície plana de gelo (Figura 1-b) e dermos um empurrão no disco, ele percorrerá um espaço um pouco maior, depois que pararmos de empurrá-lo, até parar. Podemos continuar com a experiência colocando o disco usado para hóquei sobre uma mesa com um colchão de ar, onde supostamente ele possa flutuar. Nesse caso, ele irá percorrer um espaço bem maior que sobre a superfície de gelo (Figura 1-c).

Figura 1 – Movimento do disco

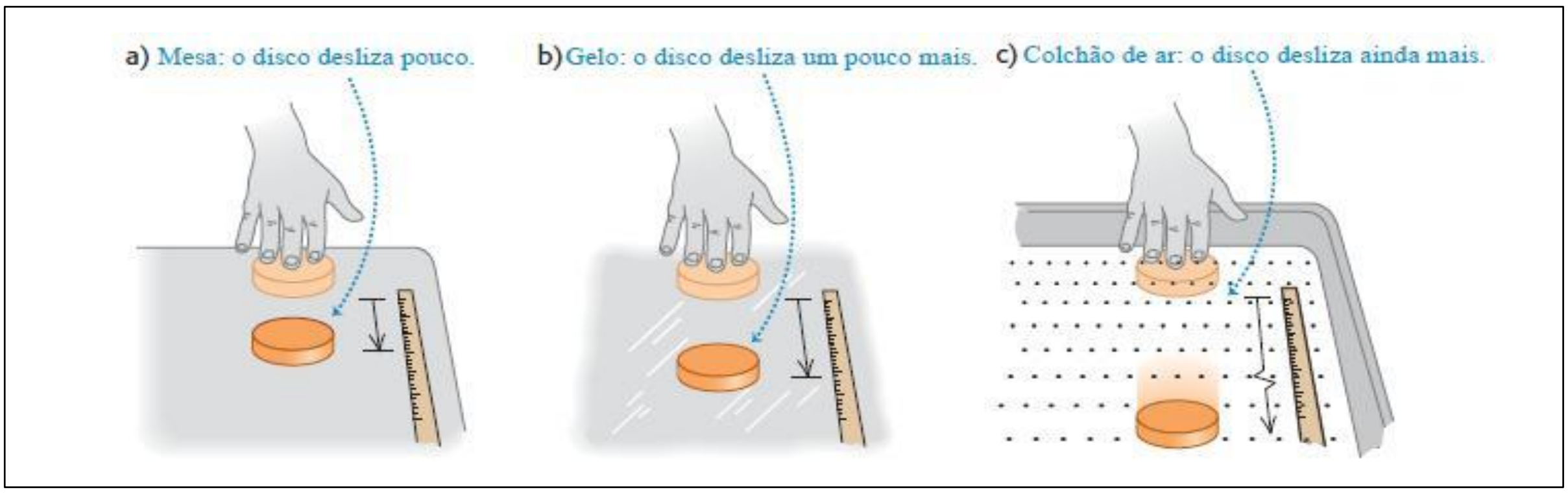

Fonte: YOUNG; FREEDMAN (2009)

A primeira lei de Newton diz que não precisamos necessariamente de uma causa para o objeto se movimentar, ou seja, que é uma força que faz o objeto variar a sua velocidade. O fato de o corpo continuar em movimento depois de iniciado, decorre da propriedade chamada inércia e podemos nos referir à primeira Lei de Newton como a lei da inércia: “Se um Objeto não estiver em movimento, a propensão é de continuar em repouso e se está se movimentando vai continuar no mesmo movimento com a mesma velocidade se a força resultante for nula”.

O importante, na primeira lei, é conhecer a força resultante. Quando a força sobre o objeto é nula, o objeto está em equilíbrio mecânico. Para a primeira lei de Newton, temos duas formas diferentes, se o objeto estiver parado, teremos o equilíbrio estático e se o objeto se movimenta em linha reta e com a mesma velocidade, teremos o equilíbrio dinâmico.

Para que um corpo estar em equilíbrio teremos: $\sum \vec{F} = \vec{0}$ e, para que isso aconteça, cada um dos componentes que tivemos da força ($\vec{Fx}$ $e$ $\vec{Fy}$ ) deve ser igual a zero ou seja $\sum \vec{Fx} = \vec{0}$ $e$ $\sum \vec{Fy} = \vec{0}$

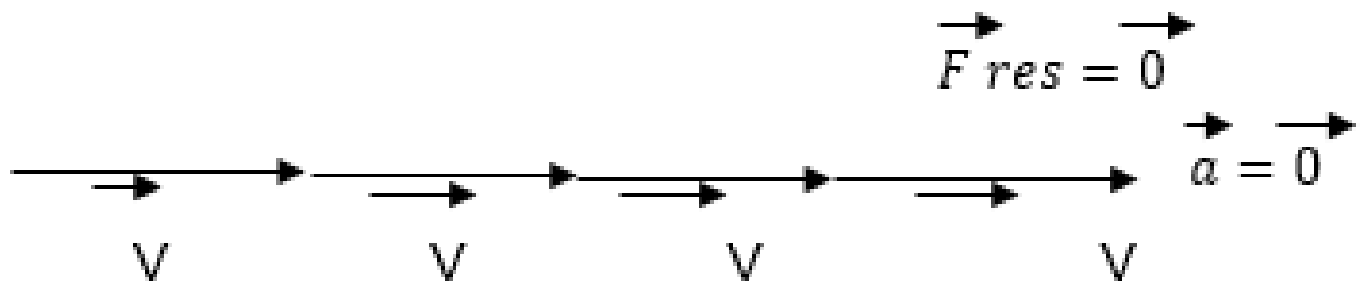

Pela primeira lei de Newton não é necessária nenhuma “ causa “ para que haja movimento. O movimento constante uniforme é considerado “estado natural do corpo”. Outra coisa muito importante da primeira lei de Newton é que ela só se aplica a referenciais inerciais.

Para testarmos a primeira lei de Newton para referenciais inerciais ou não inerciais usamos $\vec{a} = \vec{0}$ (objeto parado ou movendo-se à velocidade constante), $\vec{F} res = \vec{0}$ o sistema referencial em relação a $\vec{a}$ é inercial.

Não são todos os sistemas de referência que são inerciais e, para distinguirmos, usamos, como exemplo, um avião, no qual colocamos uma bola no chão bem no corretor. Se esse avião viaja com velocidade constante, a bola permanecerá parada pois não existem forças horizontais agindo, por isso a bola fica em repouso. Temos nesse caso $\vec{a} = \vec{0}$ quando $\vec{Fres} = \vec{0}$.

Se colocarmos a bola no chão do avião e o avião decolar, a bola irá rolar para trás, pois, apesar de não ter uma força de contato horizontal agindo sobre a bola, ela está acelerando, se comparado ao sistema referencial do avião, o que viola o princípio da primeira lei e torna o avião como referencial não inercial, quando ele decola, ou seja $\vec{Fres} \neq \vec{0}$ e $\vec{a} \neq \vec{0}$.

Podemos definir referenciais inerciais como referenciais que se movem com a mesma velocidade ou seja com $\vec{a} = \vec{0}$.

### 4.1.1 Força

Se de acordo com a primeira Lei de Newton a ausência de força leva à ausência de aceleração, levanta-se o questionamento sobre o que acontece quando há presença de força. Um corpo acelera quando uma força é aplicada sobre ele.

Para verificarmos uma medição, precisamos buscar a primeira Lei de Newton e verificar se o sistema de referência usado é um sistema inercial, ou seja: se o corpo está em repouso, permanecerá em repouso. Mas, e se for colocado em movimento, com velocidade constante? Será que ele permanecerá nesse mesmo movimento com velocidade constante? Provavelmente se ele estiver em velocidade constante, poderá ter sua velocidade reduzida lentamente, devido ao atrito. Para isso, teremos que construir um aparato de prova em um ambiente em que deverá existir o mínimo de atrito possível, fazendo o objeto praticamente flutuar. O objetivo seria aplicar uma força no objeto e verificar a sua aceleração em ambiente sem atrito.

### 4.1.2 Massa

Qual efeito obtemos quando aplicamos a mesma força a corpos diferentes? Podemos notar em experiências do nosso dia a dia que é muito mais fácil acelerar uma bicicleta do que empurrar um carro. Nesse caso, vemos que a força produz acelerações diferentes, quando aplicada a diferentes corpos.

A massa é a propriedade que um corpo tem e determina a sua resistência à mudança de movimento. A relação entre força e massa é feita pela aceleração do corpo, com massas diferentes usando um conjunto de molas calibradas, por exemplo. Observando-se diversos objetos parecidos, podemos concluir que quanto maior for a massa de um objeto, menor aceleração produzida por determinada força. A aceleração gerada é dada pela força que é inversamente proporcional à massa, então a massa do corpo pode ser vista como uma medida quantitativa da resistência inercial de um corpo.

A razão de massas dos dois corpos será igual à razão inversa das acelerações. Em experiências, podemos observar, também, que, quando duas massas são acopladas, elas agem mecanicamente como se fossem uma massa única, demonstrando que as massas são acionadas como uma grandeza escalar.

## 4.2 SEGUNDA LEI DE NEWTON

Podemos resumir a segunda Lei de Newton na única equação fundamental da mecânica clássica, a qual representa a somatória da forças ( $\sum \vec{F}$), que é a soma vetorial das forças que agem em um corpo. Usamos M para nos referir à massa do corpo e $\vec{a}$, para aceleração. Usualmente $\sum \vec{F}$ é denominado força resultante pela segunda Lei de Newton.

Um objeto cuja massa $m$ sofre aceleração. Indicamos a fórmula:

$$a = \frac{1}{\mathrm{m}} F\,res$$

Onde: $F\,res = F1 + F2 + f3$ ... esse será o vetor soma para todas as forças individuais que forem executadas no objeto.

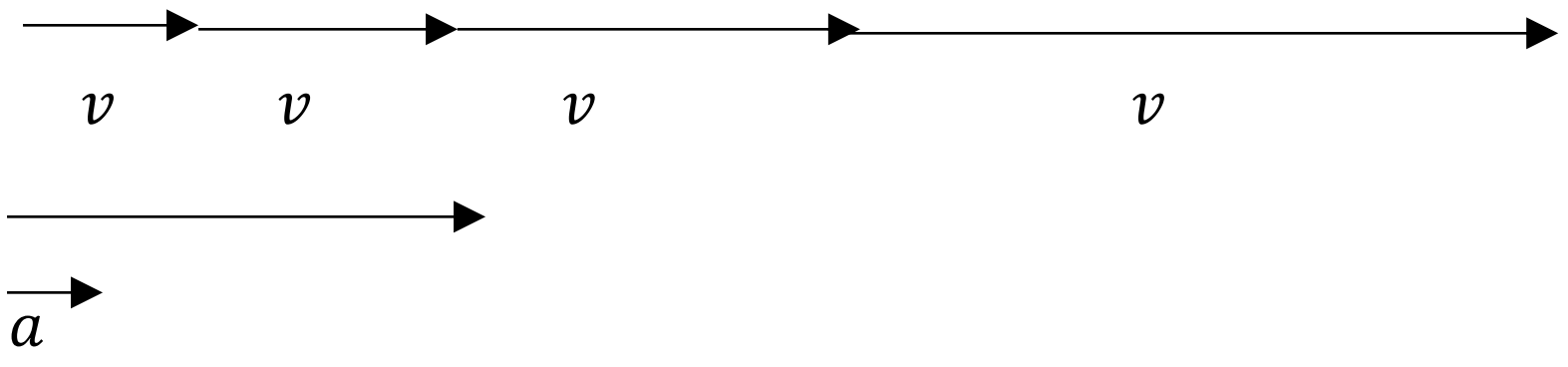

A segunda Lei de Newton relaciona matematicamente a quantidade de movimento de um corpo e sua variação com a ação de uma força externa. Caracterizamos o momento por P, e temos, então, a equação 1:

$$\vec{F} = \frac{\overrightarrow{dp}}{dt}$$

Na equação 2, a representação onde a ação da força externa causa a variação da quantidade de movimento, pode ser assim representada:

$$m.\vec{v}$$

Para massa constante, usamos a equação 3:

$$\vec{F} = m\frac{\overrightarrow{dv}}{dt}$$

Da equação 3, teremos a expressão 4:

$$\sum \vec{F} = m\,.\vec{a}.$$

Se a massa variar, temos que observar a variação do movimento total. Podemos usar, então, uma fórmula geral, a fórmula 5, que abrange todos os casos:

$$\vec{F} = \frac{d\overrightarrow{mv}}{dt} = \frac{\mathrm{md}\vec{\mathrm{v}}}{dt} + \frac{\vec{\mathrm{v}}\mathrm{dm}}{dt}$$

A Figura 2 apresenta um diagrama de forças de algumas situações onde aplicamos a mesma força, mas obtemos acelerações diferentes, por causa da diferença de massa dos corpos.

Figura 2 – Diagrama de forças

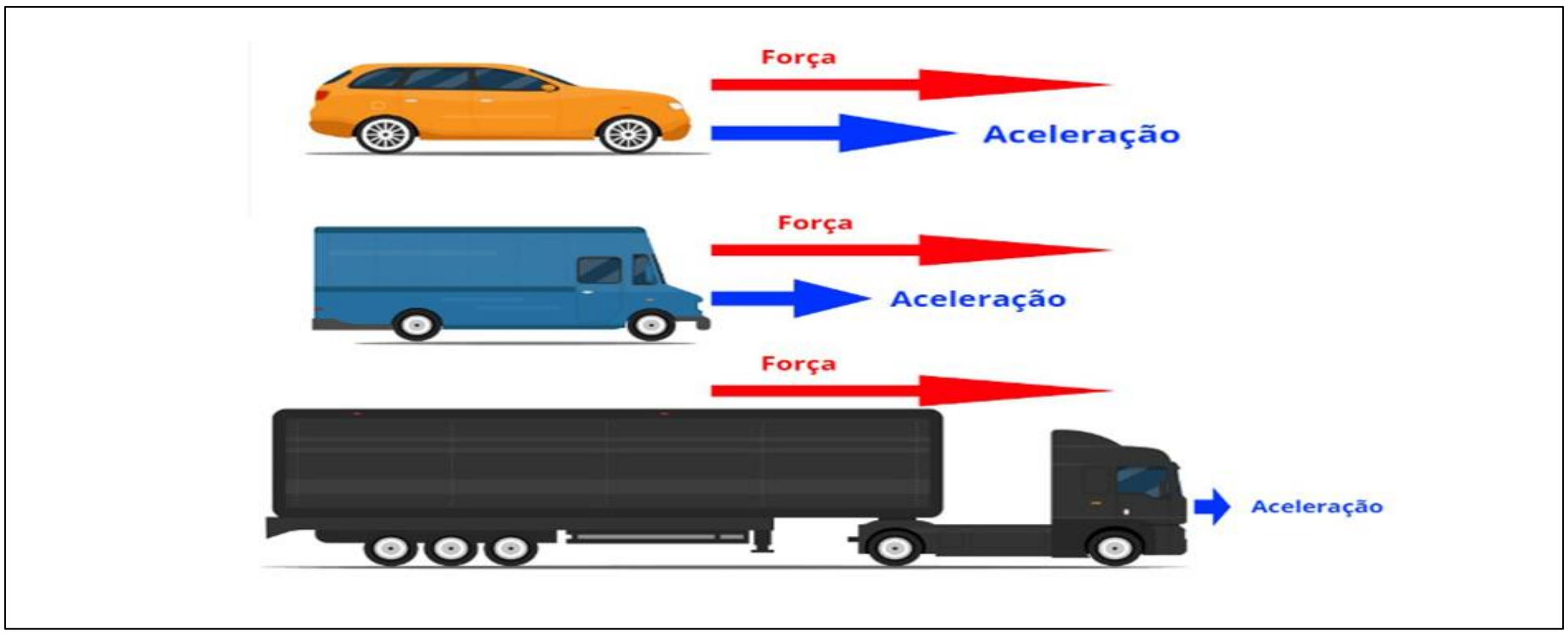

Fonte: Rafael Helerbrock (0000)

Percebemos que uma força de aceleração será inversamente proporcional à massa do corpo, proporcional diretamente à força resultante que está atuando nele e também paralela à direção dessa força. Percebemos que uma força de aceleração será inversamente proporcional à massa do corpo. Podemos observar, também, que a primeira lei está contida na segunda lei, como um caso especial, onde a somatória de forças é igual a zero.

A aceleração será zero se a força resultante em um corpo for nula. O corpo estará em repouso ou em movimento com velocidade constante, conforme estabelecido pela primeira lei. Não podemos esquecer que a primeira lei possui um papel independente importante na definição do sistema de referência inercial, se não houvesse essa definição, não se poderia escolher sistemas de referência para aplicar à segunda lei, então ambas as leis são necessárias para o sistema mecânico completo.

A representação de uma soma vetorial assim como qualquer equação vetorial pode ser escrita através de três equações dimensionais, as quais relacionam componentes x, y e z, da força resultante com os componentes da $\vec{ax}$, $\vec{ay}$ e $\vec{az}$. Os sinais das componentes e os sentidos relativos das forças devem ser levados em conta ao efetuar a soma algébrica.

### 4.2.1 Força Gravitacional ($g$)

A força gravitacional é uma força especial de atração que é executada em um primeiro corpo e atua no segundo corpo, atraindo-o. Geralmente usamos a seguinte equação:

$$Fg = mg$$

Newton observou a atração reciproca entre os corpos. Seu estudo sobre essa ação reciproca, agindo sobre os planetas, o fez propor a lei da gravitação universal que também apresenta explicações sobre a altura dos mares e a vida através do ciclo das estrelas.

Para a gravitação universal temos a seguinte equação:

$$\mathrm{F} = G\frac{Mm}{d^{\ 2}}$$

### 4.2.2 Força Peso (P)

Peso é o módulo da força exercido sobre um corpo em direção vertical e sentido para cima que equilibra a força gravitacional.

O peso de um objeto relaciona-se à sua massa pela equação $P = mg$. Podemos apresentar, também, que o peso de um objeto é associado à força gravitacional que esse objeto recebe com a ação da atração gravitacional a que ela é submetida por outro corpo com massa. Associamos a massa do objeto à força, observando a atração gravitacional exercida pelo outro corpo.

Podemos considerar que existe a seguinte simetria: O corpo 1 ficará propenso à ação do corpo 2 e o corpo 2 ficará propenso à ação da gravidade do corpo 1 e, nesse contexto, esse peso será igualmente em modulo ao peso do mesmo objeto pela terceira lei de Newton. Temos na Figura 3 a seguir um diagrama de representação da força peso.

Figura 3 – Representação da força peso

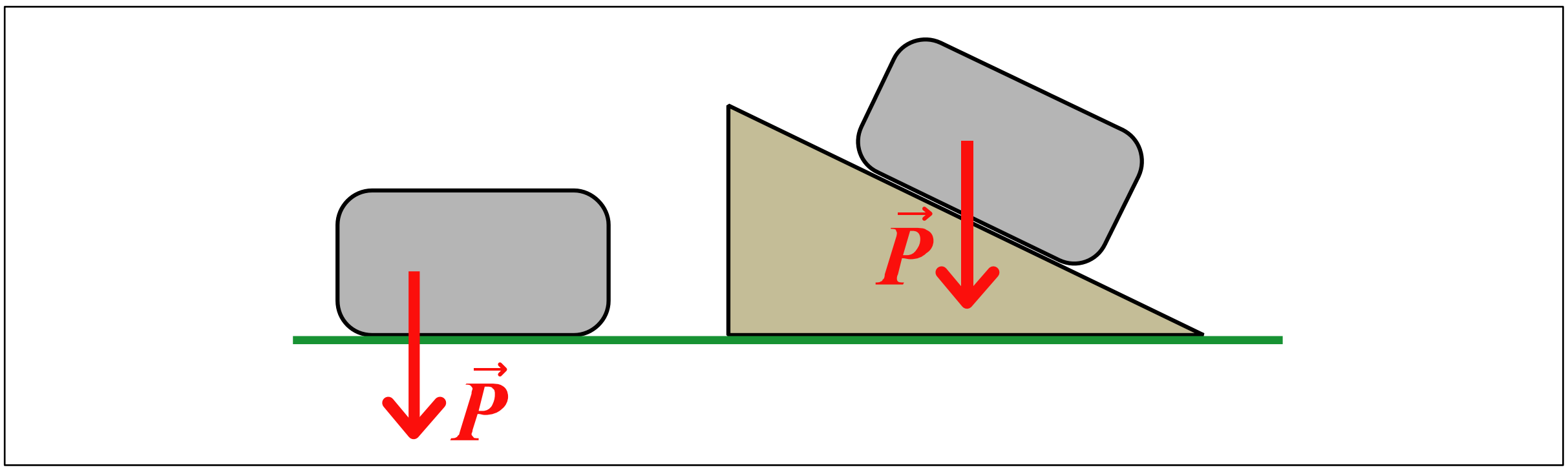

Fonte: Elaborado pela autora (2019)

### 4.2.3 Força Normal ($\vec{FN}$)

Podemos considerar força normal aquela executada em um corpo no encontro de apoio desse corpo com a superfície.

Usaremos como exemplo para a força normal um bloco sobre uma superfície de mármore, o qual será deformado pelo bloco (mesmo que de forma desprezível) e o empurrará para cima. Esse empurrão, que acontece no contato entre o bloco e a superfície de mármore, é chamado de força normal. O nome normal significa perpendicular, então a força que o mármore exerce no bloco é perpendicular à superfície de mármore. Podemos observar o diagrama de força normal na Figura 4 a seguir

Figura 4 – Diagrama de força normal

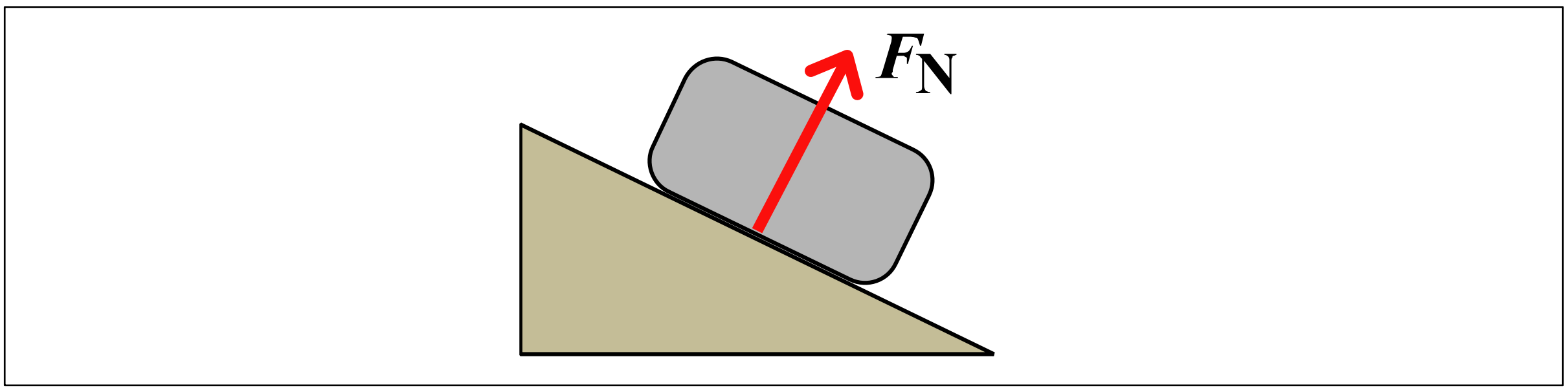

Fonte: Elaborado pela autora (2019)

### 4.2.4 Força de atrito

Quando estamos empurrando um móvel sobre o chão, os átomos do móvel interagem com os átomos do chão e, nessa interação, geralmente, ocorre uma resistência para efetuarmos o movimento. Essa resistência é considerada como sendo uma única força $\vec{f}$, a qual chamamos de força de atrito, ou apenas atrito. Essa força age paralela à superfície, mas em sentido contrário ao movimento executado. Apresentamos a Figura 5 que demonstra a força de atrito agindo.

Figura 5 – Força de atrito

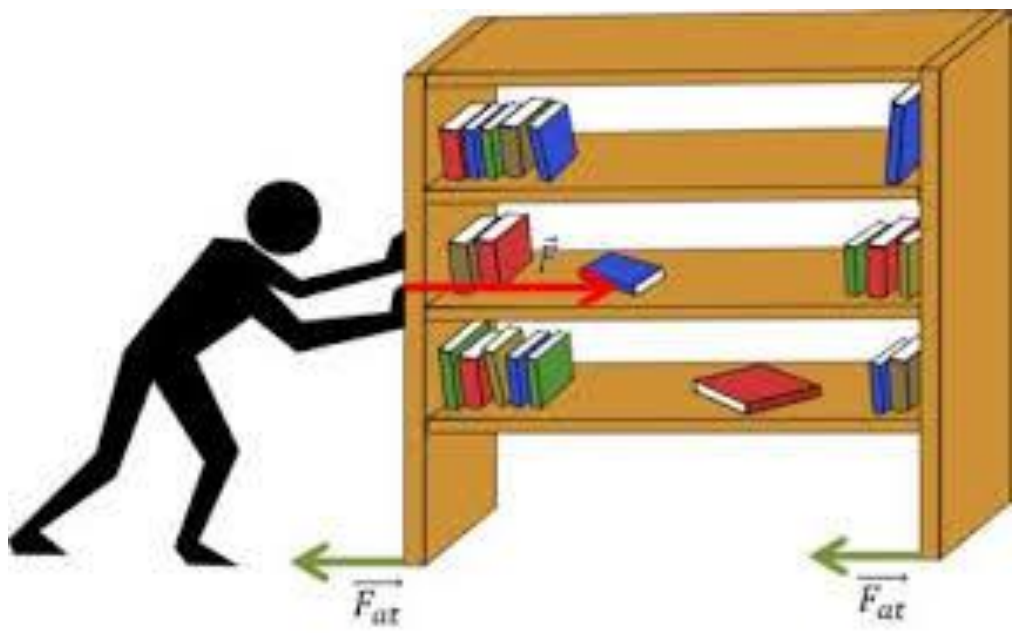

Fonte: RESPONDEAÍ (2018)

## 4.3 TERCEIRA LEI DE NEWTON

Sabemos que quando um corpo 1 executa uma força sobre o corpo 2, logo o corpo 2 executará uma força sobre o corpo 1. Neste caso, as forças sempre são iguais em intensidades e opostas em sentido. Podemos considerar a Terra e a Lua. A Terra executa uma força gravitacional sobre a Lua e a Lua executa uma força gravitacional sobre a Terra. Nesse sistema de dois corpos (poderíamos ter mais), teremos todas as forças partindo da interação entre os dois corpos.

De acordo com a terceira Lei de Newton, não é possível uma situação em que apenas uma única força atue isoladamente. No sistema Terra-Lua, a intensidade da força executada pela Terra sobre a Lua é igual à intensidade da força executada pela Lua sobre a Terra. As forças serão opostas em sentido e se imaginarmos a existência de uma linha ligando a Terra à Lua, então a força exercida pela Terra sobre a Lua vai agir ao longo dessa linha, que irá em direção à Terra e a força executada pela Lua sobre a Terra agirá também na mesma direção, porém, em sentido contrário em direção à Lua.

Observamos, então, que a terceira Lei de Newton se resume a quando um corpo 1 exercer uma força sobre o corpo 2, o corpo 2 exerce uma força sobre o corpo 1, essas duas forças são sempre iguais em intensidade e opostas em sentido. Podemos indicar sinais diferentes para representação de sentidos opostos. Podemos rotular as forças como forças de ação e reação por causa da interação mútua dos corpos.

Qualquer das duas forças poderá ser chamada de ação ou de reação, então esses rótulos devem ser arbitrários. As duas forças existem por causa da interação mútua, optando-se, assim, onde, se uma força é ação, automaticamente a outra passa a ser a reação. A regra é que toda vez que tivermos uma ação, teremos uma reação igual, mas com sentido oposto.

As forças da terceira Lei de Newton, ação e reação, sempre estarão trabalhando em diferentes corpos. Quando existem situações em que forças serão iguais em intensidade e com sentido oposto atuando no mesmo corpo, não podem ser considerados um par de forças de ação e reação, porque agem sobre o mesmo corpo. No par ação e reação verdadeiro uma força atua em um corpo A e outra no corpo B (Quadro 1).

Quadro 1 – Relação entre ação x reação

| ACÃO | REAÇÃO |
|---|---|
| Força sobre o livro exercida pela mesa | Força sobre a mesa exercida pelo livro |
| Força da Lua exercida pela Terra | Força sobre a Terra exercida pela Lua |
| Força sobre o elétron exercida pelo núcleo | Força sobre o núcleo exercida pelo elétron |

Fonte: Elaborado pela autora (2018)

Na representação do diagrama de forças na Figura 6 temos duas bolas diferentes exemplificando o par ação e reação. Elas se chocam exercendo a força de uma (A) sobre a outra (B). Após se chocarem, seguem caminhos diferentes. Essas forças são de grandeza vetorial, então, têm sentido, direção e módulo distintos.

Figura 6 – Diagrama de forças ação–reação

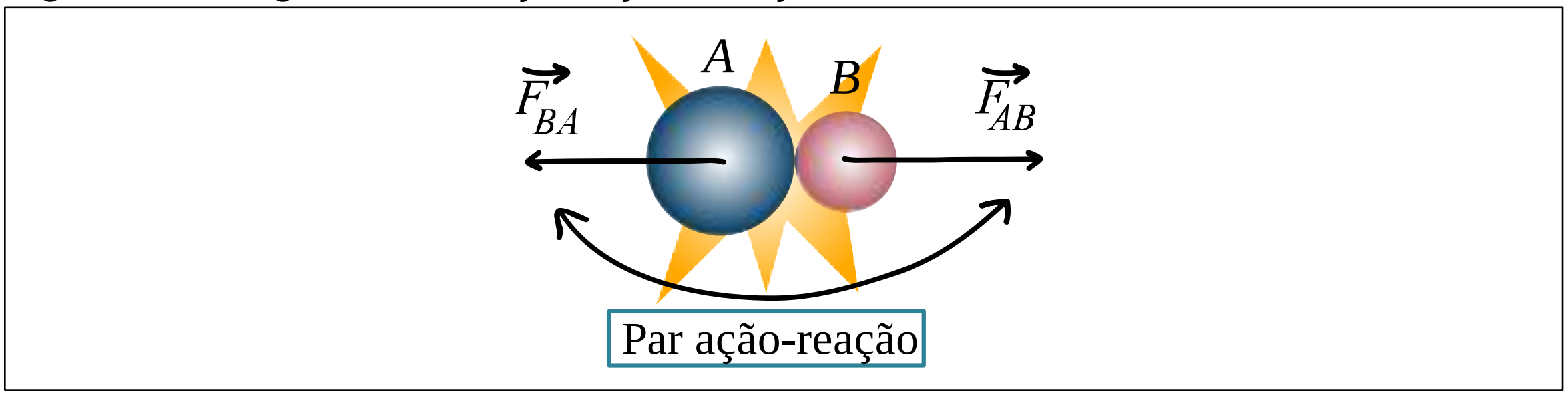

Fonte: Elaborado pela autora (2019)

No estudo da dinâmica de um corpo, consideramos apenas um elemento do par ação e reação, pois o outro elemento só é considerado se também estiver sendo estudada a direção do segundo corpo.

Toda força existe como parte de um sistema par de forças ação/reação. Esses dois membros pares ação e reação são executados em objetos diferentes e em sentidos opostos, mas tem módulos iguais.

$$\vec{F}_{A\ sobre\ B} = -\vec{F}_{B\ sobre\ A}$$

## 4.4 PESO E MASSA

Medimos o peso em unidades de força que denominamos de newtons (N). Para medirmos diretamente o peso, colocamos um objeto sobre uma balança de banheiro, medimos a quantidade de força que a plataforma da balança exerce no corpo. O corpo estando parado, a força vertical resultante será zero. Por esse motivo a força para cima no corpo exercido pela plataforma da balança será igual à força para baixo, então a força vertical resultante será zero.

Dessa forma, a força para cima, sobre o corpo exercido pela plataforma, deve ser igual à força para baixo, no corpo que a terra exerce, essa força também podemos medir em uma balança de molas ou dinamômetro. A força resultante no corpo deverá ser nula e será lida na balança a força da mola que é exercida para cima, com intensidade igual à força para baixo, medida pela massa multiplicada pela gravidade ($mg$). Toda vez que desenharmos um diagrama de corpo livre próximo à superfície da Terra, devemos incluir uma força $mg$ voltada para o centro da Terra, a qual irá representar o peso, ou seja a força gravitacional no corpo feita pela Terra.

### 4.4.1 Diferença entre peso e massa

Observamos que o peso depende da massa: quanto maior for a massa maior será o peso. Um segundo corpo com o dobro de massa do primeiro terá duas vezes o peso no mesmo local.

A massa de um corpo possui o mesmo valor em qualquer local, mas o seu peso varia com a posição na superfície onde a aceleração do corpo livre varia com a localização. Mesmo na superfície da Terra, existem determinadas localizações em que a gravidade é diferente e o peso passa a ser diferente também, mas a massa continuará a mesma. Temos como exemplo a linha do Equador, onde uma massa de 1 kg possui um peso de 9,78 N, enquanto próximo aos polos a gravidade muda de 9,78 m/s$^2$ para 9,83 m/s$^2$. Isso ocorre porque os polos estão mais próximos do centro da terra do que o Equador.

# 5 TECNOLOGIA APLICADA AO ENSINO

As Tecnologias de Informação e Comunicação (TICs) referem-se aos produtos utilizados para ensino que incluem ideias inovadoras que facilitam os métodos de ensino e a aprendizagem. Essas tecnologias podem ser usadas com o intuito de funcionar como ferramentas auxiliares no aprendizado de Física. Trazer aos alunos de Física a teoria por meio de tecnologia, como o uso de Gifs (Graphics Interchange Format) e de Realidade Aumentada (RA) pode ajudar a transformar a falta de interesse em motivo para aprender.

## 5.1 O QUE SÃO GIFS

Gif é a sigla para o termo inglês Graphics Interchange Format que significa basicamente formatos para intercambio gráfico, ou seja, é um formato de imagem que serve tanto para animações, quanto para imagens fixas. Quase sempre são atrelados ao humor e estão fortemente presentes em nosso cotidiano seja no *twiter*, no *facebook*, no e-mail ou em páginas da *web*. Geralmente são utilidades na internet para comunicação rápida. Os Gifs são universais por isso, quando escolhidos da maneira correta, conseguem transmitir mensagens a qualquer pessoa, em qualquer lugar. Eles geralmente são formulados para durar no máximo 8 segundos para captar a atenção do observador.

## 5.2 O QUE É REALIDADE AUMENTADA?

A realidade aumentada configura-se como a interação entre ambientes virtuais e o mundo físico. Ela faz a interação entre elementos ou informações virtuais e a visualização do mundo real por meio de um aplicativo e uma câmera que proporciona uma experiência interativa com o mundo real, na qual os objetos que existem no mundo real são “acentuados” por informações criadas por captadores virtuais. A realidade aumentada construtiva agrega o virtual ao ambiente natural

Como o uso da RA, através de Gifs no ensino das leis de Newton, pode auxiliar o professor na construção de conhecimento a qual, ao ser apresentada por meio de símbolos, busca atrair a atenção do aluno? Essa forma de apresentação de objetos e fenômenos traz a tecnologia para realidade, possibilitando-nos viajar ao espaço em nossas próprias salas de aula e observar os planetas girando com a atração da gravidade em seu redor. Essa tecnologia permite-nos visualizar de forma virtual pequenos vídeos (Gifs) demonstrando fenômenos de nosso cotidiano, sendo possível modificar a visão dos nossos livros didáticos e adaptá-los com a proposta de utilizar a RA para ajudar o professor na inclusão da tecnologia e facilitar a comunicação. Nosso objetivo é implementar TICs, através de Gifs e RA, em uma revista em quadrinhos, onde os personagens serão baseados em *animes*, devido à grande aceitação dos jovens a esse modelo de revista.

Nessa revista que propomos, os personagens dialogam com linguagem simples explicando as leis de Newton. A cada exemplo, teremos figuras em Gifs ou 3D, as quais serão engatilhadas por um aplicativo de celular para leitura dos símbolos criados como transmissores da linguagem para que alunos observem e sejam capazes de interpretar os fenômenos físicos das leis de Newton.

Esse aplicativo guardará dados de leitura dos símbolos, que serão programados de acordo com o assunto abordado e visualizado pelo programa. Através da tela do celular, o aluno poderá observar os exemplos em movimento, projetados na folha da revista Figuras em Gifs ou em 3D, direcionados ao contexto estudado. Nossa intenção é que esse aplicativo seja colocado à disposição para ser baixado no Google Play e possa ser adquirido por qualquer pessoa em qualquer lugar. Ele pode ser acionado através de marcações existentes na revista. A inovação poderá ser utilizada por todo país sem utilização da internet, já que o aplicativo guardará as informações (Gifs e 3D), utilizados na revista. O aplicativo também pode ser baixado no blog do autor, *https://physics7.school.blog*, onde já está disponível.

A busca por novas tecnologias que possam atrair nossos alunos também está expressa nos Parâmetros Curriculares Nacionais (PCN), que defendem uma alfabetização científica, a qual pode ocorrer por meio de:

> [...] novas e diferentes formas de expressão do saber da Física, desde a escrita, com elaboração de textos ou jornais, ao uso de esquemas, fotos, recortes ou vídeos, até a linguagem corporal e artística. Também deve ser estimulado o uso adequado dos meios tecnológicos, como máquinas de

> celular, ou das diversas ferramentas propicias pelos microcomputadores (BRASIL, 2002).

O uso de Tecnologia de informação e comunicação (TICs) aponta para a tentativa de incentivar o aprendizado participativo, tornando o aluno um sujeito ativo desse processo. O desinteresse hoje apresentado por muitos alunos nos leva a buscar formas diferentes de ensinar o conteúdo pois trazer o ensino clássico ao século XXI não é tarefa fácil.

Todavia, a maioria dos professores se recusa a adaptar-se a novas TICs. Existem resistências principalmente dos professores mais velhos pelo uso de tecnologia. A boa e velha aula expositiva que já trouxe muito aprendizado não tem dado tanto resultado, pois nossos alunos mudaram com o uso de tecnologias como computador, celular, internet.

Nossos alunos revelam-se dedicados e envolvidos quando estão comprometidos com seus jogos eletrônicos, nas redes sociais, nos programas do *youtube* porque essas tecnologias os atraem. Engajar a tecnologia a materiais de ensino é mostrar que o estudo não está fora de moda, razão pela qual temos que trazer o ensino para nossa realidade, a qual se destaca pelo uso de tecnologia diariamente. Nossos alunos viraram reféns dessas tecnologias e a que mais é utilizada é o celular e seus aplicativos, por serem mais acessíveis.

É importante tentar a mudança esperando, com isso, que eles aprendam nossos conteúdos por meio dessas tecnologias. Cabe às instituições de ensino fornecer capacitação para os professores e a modernização dos equipamentos disponíveis. A transformação da prática pedagógica depende de esforços individuais e coletivos. As TICs devem estar sempre presentes nas formações. A Conferência Nacional de Educação (CONAE, 2010), ocorrida em 2010, prevê no plano de educação 2011-2020:

> Garantia do desenvolvimento de competências e habilidades para o uso das tecnologias de informação e comunicação (TIC) na formação inicial e continuada do/das profissionais da educação, na perspectiva de transformação da prática pedagógica e da ampliação do capital cultural dos /das professores/as e estudantes (CONAE, 2010, p. 81).

Criar materiais didáticos diversificados com uso da tecnologia pode auxiliar na concentração e na aprendizagem do aluno. Alguns aplicativos têm feito esse papel de mediação ao ensino. Essa nova TIC é muito pouco utilizada no ensino, apesar de permitir ao aluno observar o mundo virtual se transportando da folha para o aplicativo. Ela surge em um momento de necessidade de modernização, para complementar e tentar potencializar a compreensão do conteúdo ministrado pelo professor. Azuma (1997) define que a realidade aumentada utilizada no ensino deve possuir as propriedades de combinar o real e o virtual e permitir que haja interação em tempo real e adaptar os objetos virtuais ao mundo real. A Figura 7 representa um exemplo dessa tecnologia intitulada realidade aumentada.

Figura 7 – Planeta Terra em RA

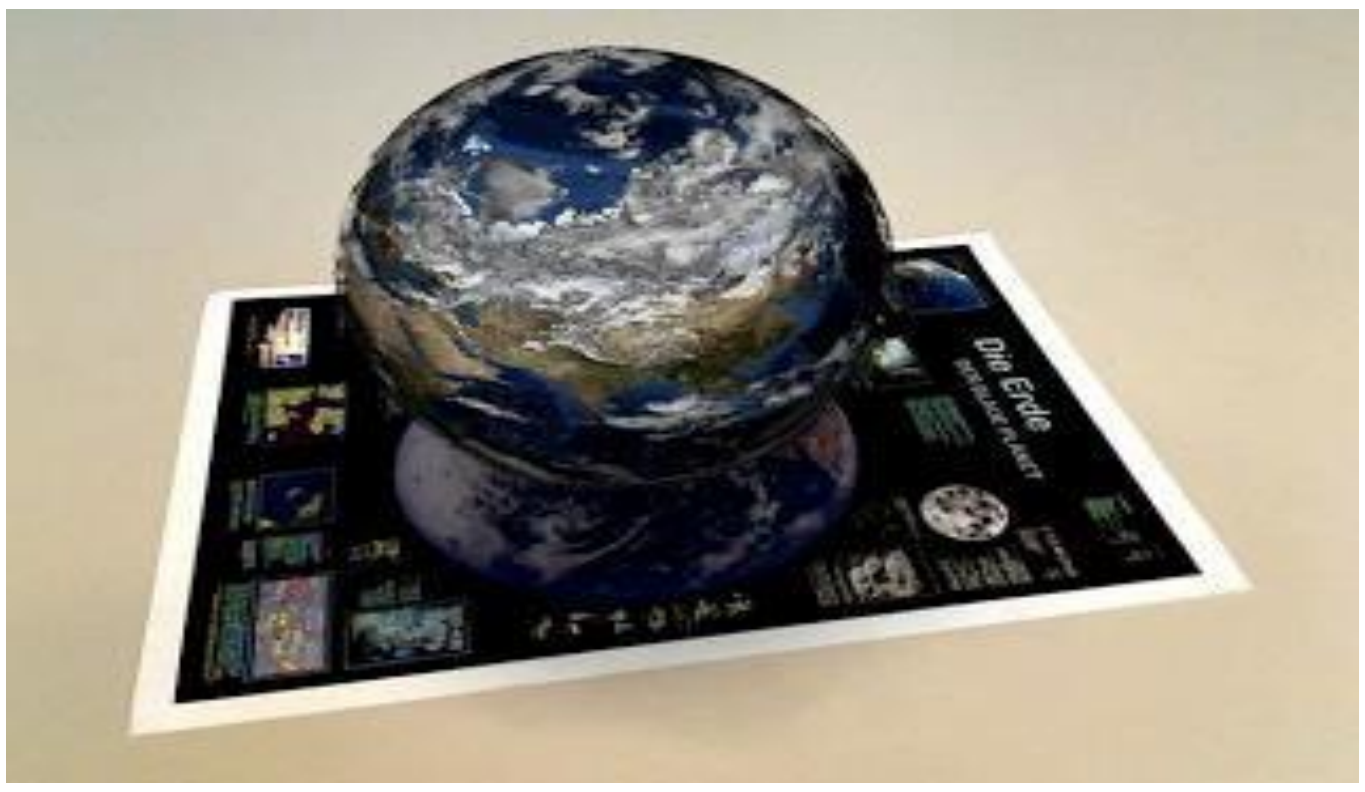

Fonte: LEONE (2009)

O fato de estar prevista a utilização das TICs na formação inicial dos professores favorece sua utilização e gera incentivos para criação de diversos materiais, de modo a deixar as aulas mais atrativas e interessantes para os alunos. Conforme documentos apresentados (PCN 2002, CONAE, 2010), incentivar a utilização dessas tecnologias se tornará uma tendência na área de ensino.

No nosso dia a dia, estamos tão conectados com certas tecnologias digitais como computadores, notebooks, *tablets, smartphones*, videogames, *smart* TVs, aplicativos diversos, redes sociais, entre outros, que se torna necessária a inserção de materiais didáticos com aplicações de TIC's na educação, pois essas tecnologias são presentes direta ou indiretamente em sala de aula. Usar com inteligência pode trazer um diferencial na sala de aula que motivará o aprendizado.

Existem diversos TIC's que normalmente são utilizados em sala de aula. Alguns mais comuns como editores de textos, apresentações, programas de busca e outros usados em disciplinas como ciências, matemática, como os simuladores e etc. Esse trabalho propõe a utilização de uma nova TIC. Apesar de ser pouco explorada, pode ser um diferencial, tornando as aulas mais atrativas, podendo auxiliar os alunos a observar as leis da Física em nosso dia a dia.

A RA pode modificar a visão do mundo através de objetos virtuais sobrepostos dentro do mundo real. Na verdade, ele complementa a realidade, não a substitui completamente. Azuma (1997) define que os sistemas de RA devem combinar o real e o virtual, permitir interação em tempo real e alinhar os objetos virtuais no mundo real.

Os recursos da RA têm sido usados com uma grande frequência. Uma das maiores utilizações está no *Marketing,* que chama a atenção a produtos sendo muito difundidos como forma diferencial de propaganda. A utilização no ensino de diversos materiais tem ocorrido de maneira tímida, mas que vem ganhando adeptos de todas as áreas de ensino. Como proporciona uma viagem virtual interativa, manuais instrutivos, mesmo da área médica, têm apoiado a ideia e aderido à utilização.

Em capítulo adiante trataremos com mais detalhes de algumas utilizações tanto no ensino quanto em outras áreas. O fato é que a realidade aumentada é associada a uma simulação e não substitui o raciocínio mas prolonga e transforma a capacidade de imaginação e pensamento (LEVY, 1999). O professor terá o papel de mediar e desenvolver a criatividade, ilustrando com vídeos de RA que têm movimento e, por si, transmitem formas de melhorar o diálogo entre aluno x professor e aguçam a investigação do problema na situação apresentada Buscamos em Levy (1999, p. 171), esse amparo para a relevância da mediação do professor:

> O professor torna-se um animador da inteligência coletiva dos grupos que estão ao seu encargo. Sua atividade será centrada no acompanhamento e na gestão das aprendizagens: o incitamento à troca de saberes, a mediação relacional e simbólica, a pilotagem personalizada dos recursos de aprendizagem, etc (LEVY, 1999, p. 171).

O desafio de trabalhar com ensino de matemática lógica e desembaraçada aos nossos alunos não é tarefa fácil. O uso da tecnologia para que tenhamos condições de ajudar nessa compreensão torna-se um facilitador no processo.

Ao tratarmos sobre o ensino de Física, vem, quase sempre, à mente do aluno a dificuldade matemática, vez que a disciplina mistura a teoria com expressões algébricas interpretativas, diferentes da matemática usual, que, muitas vezes, até problemas fáceis se tornam complicados pela carência da habilidade de interpretação do enunciado do exercício. Muitos alunos não conseguem expressar essa interpretação corretamente nas expressões algébricas.

As tecnologias não são o remédio para essa dificuldade mas podem ajudar na interpretação do conteúdo exposto e organizar a estruturação dos cálculos necessários. Trabalhar com duas disciplinas pode se tornar uma ferramenta poderosa quando colaborada pela tecnologia.

Como exemplo citamos o artigo usando realidade aumentada para aprendizagem de física e matemática da universidade Federal de Itajubá (Usando Realidade Aumentada no Desenvolvimento de Ferramenta para Aprendizagem de Física e Matemática), o qual apresenta uma discussão sobre o uso tecnologia com ênfase nas disciplinas de Matemática e Física com aplicações baseadas em realidade aumentada.

Pensando na adoção de recursos tecnológicos como ferramenta facilitadora do processo aprendizagem, a escola pode acolher tais recursos por eles se apresentarem potencialmente capazes de promover o desenvolvimento do ensino e da aprendizagem. A utilização de recursos ricos se torna facilitador no auxílio da aula expositiva. Podemos citar o uso de fotos, desenhos, sons, vídeos, figuras 3D (todos podem ser considerados realidade aumentada desde que sejam engatilhados por um código ou QR), os quais dão ao educando a possibilidade de melhor entendimento de conceitos expostos pelo professor.  A utilização dos recursos da realidade aumentada tem despertado grande interesse da comunidade educacional como facilitador no aprendizado, pois amplia o interesse e incrementa a motivação da descoberta científica.

# 6 APRENDENDO FÍSICA COM QUADRINHOS E TICs

As histórias em quadrinho têm tido grande sucesso há muito tempo, divertindo crianças e adultos, atraídos por sua formatação própria e linguagem universal. Esse material costuma ser bem conhecido entre os jovens por suas formas simples e acessíveis de linguagem popular e padrões linguísticos que visam à catarse (queda do estresse por parte do leitor) e uma forte ligação com o cognitivo do indivíduo que mergulha na sua narrativa.

O ensino de Física geralmente é baseado em metodologias expositivas, em que o professor cumpre uma grande quantidade de conteúdo. Essa abordagem costuma ser voltada para resolução de exercícios que comprovam a competência, ou não, do aluno em substituir números em fórmulas, deixando a desejar, na maioria dos casos, na compreensão dos fenômenos apresentados.

Abi (1988) critica a concepção tradicional de ensino e conclui que ela se baseia na falsa hipótese de que " basta que o professor explique bem para que a transmissão do conhecimento ocorra", ficando clara a posição receptiva/passiva do discente, sendo que o “conhecimento” estaria centralizado na figura do professor. O resultado desta metodologia é a simples memorização de fórmulas, as quais, geralmente, não fazem nenhum significado para o aluno.

Devido a esse abismo entre a linguagem do professor e a compreensão do aluno a utilização das revistas em quadrinho torna-se uma ferramenta que pode diminuir essa diferença, inclusive pelo mecanismo psicológico da catarse, ou seja, retirando a tensão da aprendizagem da física propondo uma atividade de ensino lúdica.

## 6.1 LÚDICO

Podemos considerar a revista como um jogo lúdico (Ramos, 1990; Huizinga, 2001). A maioria das revistas em quadrinhos tem humor e regras para leitura. O acompanhamento do conteúdo programático asserido à forma lúdica de compreensão, segundo Quella-Guyot (1994), ao explorar os ideogramas (que nesse caso podem ser ligados aos símbolos apresentados pela realidade aumentada), utilizado sempre nos quadrinhos, trará um equilíbrio entre arte e ludicidade.

Para que a atividade possua efeito, o professor terá que despertar o interesse do aluno desde o início. Buscar através da catarse a libertação dos alunos das aulas tradicionais, retirando a tensão da obrigação de decorar o material e simplesmente ler, observar, buscar interpretação, pesquisar e aplicar em seu cotidiano, pode fazer com que o jogo/brinquedo, realizado/utilizado, fique gravado no subconsciente, fazendo com que ocorra aprendizagem, quase que inconsciente, por parte do discente (RAMOS, 1990).

A atividade lúdica poderia vir a ser tratada como não-séria, mas segundo Huizinga (2001), o caráter "não-sério" (o cômico, os risos que acompanham o ato lúdico, a realidade aumentada) não significa que a brincadeira deixe de ser séria. Quando o jogador brinca, ele o faz de modo comprometido. Esta objetividade inerente à atividade lúdica faz sentido, pois ela consegue propor situações que despertam o real interesse do participante – o desafio lúdico.

A atividade lúdica está inteiramente relacionada ao desafio. Desafiar o aluno a interpretar e mesmo reproduzir os vídeos implica trazer situações e obstáculos comuns do seu cotidiano para resolvê-los cientificamente. Discutir e entender como ocorrem esses fenômenos, em nosso cotidiano, se torna leve e concreto, contribuindo para uma educação preparatória para atuação do sujeito na sua realidade.

## 6.2 LINGUAGEM

Usar como forma de demonstração a apresentação de signos é uma estratégia que tem sido bem utilizada. Machado (1999) define os processos em que, combinados à escrita formam a estrutura como única e imutável. Tais formações são definidas por ela como gêneros dos discursos:

> Os códigos estão cada vez mais diferentes. Logo, as possibilidades de variedades distintas integrarem o mesmo circuito comunicativo aumentaram muito. (...) na era da informação, em que o hibridismo salta à vista, nem o mais simples exemplar daquilo que se chama texto pode prescindir da combinatória de elementos que se reportem à diversidade de códigos. Nada

> impede que diferentes códigos estejam relacionados numa mesma combinação textual capaz de exprimir uma unidade significativa na cultura (MACHADO, 1999, p. 46).

O texto incorporado à figura explica, de forma resumida, o conteúdo explorado e as imagens geradas pelos códigos expostos na revista e lidos através do aplicativo AR Physics. Esse aplicativo é de confecção própria da autora e que tem o objetivo de acionar as marcações da revista para a visualização dos Gifs e RA. O Ar Physics está disponível para ser baixado no site *https://physics7.school.blog/*. Os Gifs e RA acionados explicam os fenômenos das leis de Newton, através de representações animadas, posicionadas de acordo com assunto abordado, instigando o aluno a projetar a situação no seu cotidiano. Essa divisão conteúdo x animação transforma o texto em um sistema dinâmico. A utilização da revista em sala de aula implica uma montagem cuidadosa para que se cumpra o desejado, tornando as histórias em quadrinho (HQs) uma rica estratégia de ensino de física.

História em quadrinhos - ou HQ - é o nome dado à arte de narrar histórias por meio de desenhos e textos dispostos em sequência. As histórias em quadrinhos (HQs) são um tipo de arte que mistura texto e desenho de maneira única.

## 6.3 COGNITIVO

Trabalhar a imaginação, interpretar os sinais, as imagens e se divertir com isso, tudo isso é possível nas revistas em quadrinho. Mesmo nas animações mais irreais a imaginação trabalha transformando o leitor e transportando-o até o cenário. Levar o aluno à capacidade de estabelecer relação com os personagens e as animações existentes, explorando uma série de ações cognitivas a serem observadas e trabalhadas, permite a aprendizagem do conteúdo específico a ser ensinado. A relação de identificação e participação do estudante no enredo proposto pela HQ será parte da estratégia de envolvimento do aluno com a aprendizagem do conteúdo. Sendo assim, propomos a utilização da revista como desencadeadora do conflito cognitivo, usando a mediação como base para a construção do conhecimento.

# 7 MÉTODOS E RESULTADOS

Essa pesquisa visa a analisar a utilização do material didático criado em sala de aula, avaliando o interesse dos alunos pelo material, a aprendizagem do conteúdo programático (ensino das leis de Newton), a compreensão dos fenômenos estudados e descritos e a interpretação da realidade aumentada como estratégia de ensino, com um intuito de investigar se o uso da RA no ensino de Física auxilia na compreensão dos conceitos utilizados em modelos científicos. Foi desenvolvida uma pesquisa qualitativa e quantitativa para coleta de dados.

## 7.1 CRIAÇÃO DOS OBJETOS VIRTUAIS ADAPTADOS À REVISTA PARA APLICAÇÃO DA SEQUÊNCIA DIDÁTICA

O processo de criação e customização dos códigos inseridos na revista traz uma forma diferenciada para pesquisa, pois assume um material de poucas referências iguais existentes no mercado. O desafio era confeccionar um material com linguagem jovem e o mais simples possível, resguardando conteúdo original para o ensino das Leis de Newton, usando, como atrativo, os códigos que ativam pequenos Gifs (série de imagens reproduzidos em um loop de tempo x, de uma determinada situação) e figuras de realidade aumentada com modelagem 3D.

Em primeira ordem, o protótipo da revista foi criado em PowerPoint e tem a sequência da história contada por figuras representativas retiradas de páginas na internet. Após esta parte da formulação, pesquisamos se junto com o responsável pela programação do aplicativo (Pap.), os Gifs que poderiam ser utilizados em cada situação e onde seriam inseridos na revista. Foi descrito ao desenvolvedor do sistema como deveria ser direcionado o aplicativo que leria os códigos e ativaria os signos existentes. O desenvolvedor, através da plataforma Unity, em conjunto com *plug-ins* (Vuforia) e algumas opções de configuração para realidade aumentada, adicionou às cenas uma série de planos e, nos mesmos, através de um código, utilizando a linguagem de programação, java scripit e a orientação a objetos inseriu os Gifs.

Os Targets serviram de gatilho para ativação dos códigos que exibem os Gifs na tela do dispositivo. Para criação e engatilhamento de cada imagem, foi gasto uma média de 4 dias de programação. A programação e criação do aplicativo foi feita por um estudante de sistema de informação da Universidade Federal do Espírito Santo.

## 7.2 OBJETOS VIRTUAIS DESENVOLVIDOS E ACIONADOS NA REVISTA

O aplicativo Ar Phisycs - Leis de Newton foi desenvolvido exclusivamente para esse projeto e está em processo de publicação para que fique disponível na Google Play Store, onde poderá ser baixado na versão gratuita.

Na primeira Lei de Newton, foi colocado um Gif onde uma pessoa é imobilizada em uma maca, colocada em uma caminhonete aberta, sem nenhum tipo de objeto prendendo a maca (p. 6). Quando ocorre o movimento (aceleração e a curva), a tendência do corpo é permanecer parado ou em MRU, por isso ele cai do veículo.

Na segunda lei, são apresentados dois Gifs e um deles apresenta um pequeno cão de raça Pug, que empurra um carrinho com um cãozinho de pelúcia dentro (p. 7). O segundo Gif apresenta um senhor de idade mais avançada que estaciona o carro apenas o empurrando, esse Gif foi escolhido buscando a investigação do fenômeno apresentado (p. 8).

Na terceira lei de Newton, temos três Gifs, o primeiro descreve a decolagem de um foguete saindo do solo terrestre (p. 9), o segundo dispõe de uma bala de revólver que estilhaça na parede (p. 10) e o terceiro contém o encontro de ação e reação do rosto de um homem com um balão cheio de agua com tinta (p. 11).

A revista ainda demonstra as grandezas: Força - com o Capitão América segurando um helicóptero (p. 12), Força resultante - figura animada de um homem que demonstra vetores da força normal, força de atrito, força peso (p. 13), Força peso - com a demonstração de um peixe bem grande pesado por um dinamômetro (p. 14).

Temos a representação das forças de interação à distância representada por uma Figura 3D demonstrando a interação Lua x Terra, (p. 15) e a interação por contato apresentado por um carro de fórmula 1 atritando o solo (p.15).

A força normal foi dividida em três Figuras montadas juntas, em blocos com posições diferentes, visando a apresentar, de forma simples, que a normal aparece no momento de encontro dos objetos (p. 16). Foi acrescentada na página 17 uma curiosidade que demonstra a força elástica com um Gif demonstrando a queda de uma mola e, na capa, encontra-se um Gif demonstrando o movimento de um pêndulo.

## 7.3 CONTEXTO DE APLICAÇÃO DA SEQUÊNCIA DIDÁTICA

A aplicação foi feita na escola de Ensino médio Dr. Silva Mello na cidade de Guarapari, no estado do Espirito Santo, em uma turma de primeiro ano do ensino médio, escolhida aleatoriamente, no transcurso dos meses de outubro e novembro de 2018.

A aplicação foi feita no final do ano devido à grande dificuldade da turma de primeiro ano com matérias básicas, sendo necessária uma ação de reforço matemático e interpretativo no primeiro trimestre do ano, atrasando todo o conteúdo de física que deveria ser estudado.

Foi necessário introduzir, no primeiro trimestre, os primeiros conceitos de soma, subtração, divisão e multiplicação de potencias, frações, notação científica, transformações de unidades não somente como revisão de conteúdo, mas como matéria nova. Também foi trabalhada a história da Física, com interpretação de textos.

A pesquisadora percebeu que alguns alunos, vindos do sistema municipal, foram promovidos para anos à frente sem o conhecimento necessário, para não ficarem defasados pela idade, chegando ao ensino médio sem base em matemática e português.

# 8 APLICAÇÃO DA SEQUENCIA DIDÁTICA

## 8.1 QUESTIONÁRIO DE CONHECIMENTOS PRÉVIOS (1ª APLICAÇÃO)

Antes de iniciar a aplicação da sequência didática, foi apresentado aos alunos um questionário para verificação do conhecimento prévio (Apêndice 1), sobre o tema: As leis de Newton, conhecendo o que foi aprendido no ensino fundamental para realização das atividades.

O questionário foi aplicado para uma classe de 1º ano de ensino médio com 35 alunos. A escolha da turma foi totalmente aleatória. Logo na primeira pergunta: "Como você percebe se o corpo está parado ou em movimento? " 78% dos alunos responderam de forma errada,10% de forma incompleta, apenas 9% de forma certa e 1% não soube responder.

Nas demais perguntas, os alunos seguiram o mesmo padrão, no total foram 4% de acertos, 12% de respostas incompletas, 55% respostas erradas e 29% dos alunos deixaram em branco ou não souberam responder.

Com base nesses dados, uma média de 67% não demonstrou conhecimento prévio sólido para responder às perguntas do questionário inicial.

Observando esse fato, julgamos necessário, antes da aplicação da sequência didática, ministrar uma aula introdutória para organização dos conhecimentos prévios e uma aula mediada com recurso tecnológico através do *software* que trabalha com simulações com Tecnologia Educacional em Física (PHET), abordando conceitos básicos que deveriam ter sido aprendidos no 9° ano, mas que parecem distantes, conforme as respostas apresentadas no primeiro questionário (Gráfico 1).

Os conceitos das leis de Newton não foram tratados nessa intervenção, apenas conceitos básicos sobre massa e peso, a qual se mostrou necessária porque o número de respostas erradas com relação à Segunda Lei de Newton foi grande.

Gráfico 1 – Questionário de conhecimentos prévios

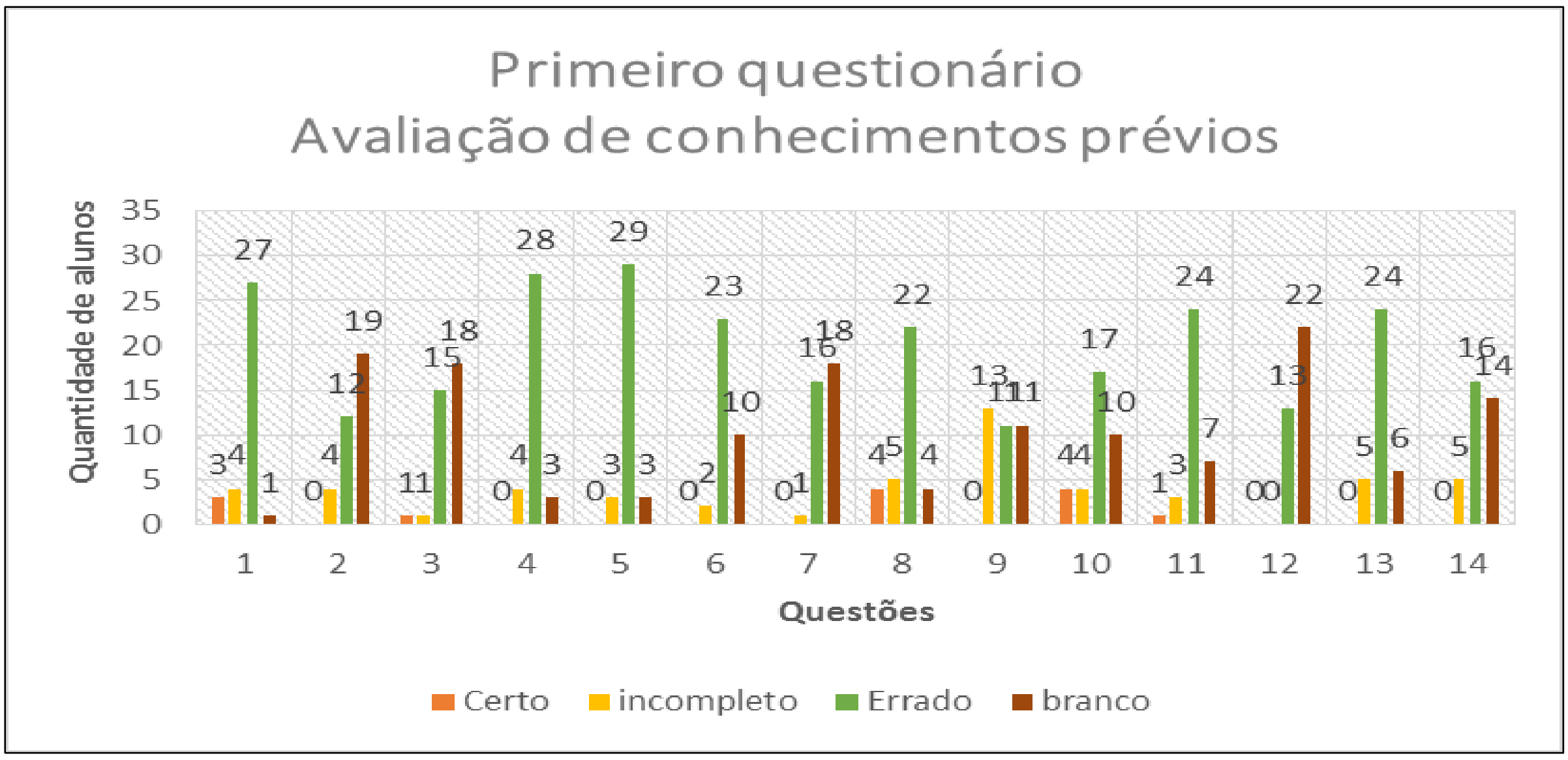

Fonte: Elaborado pela autora (2019)

O Gráfico 1 mostra que todas as perguntas tiveram respostas erradas e em branco. Deste modo, concluímos que os alunos não tinham conhecimentos prévios básicos necessários para compreensão das leis de Newton e suas aplicações.

Algumas questões ficaram mais evidentes que outras, conforme análise do Gráfico 1. A questão 01 pergunta como o aluno percebe se o corpo está parado ou em movimento? E pede que ele justifique sua resposta, na qual tivemos apenas um aluno que deixou a resposta em branco, mas tivemos 27 respostas erradas. Outra questão que demonstra esse tipo de índice é a questão 04 que questiona sobre um automóvel que trafega a uma certa velocidade em movimento retilíneo uniforme (MRU) e leva 1 segundo para parar em uma freada de emergência. A questão pergunta o que aconteceu com a velocidade do carro e se houve aceleração. Temos 3 respostas em branco, 4 incompletas e 28 erradas. A questão 5 refere-se a um corpo de massa que está sujeito à ação de uma força F que desloca segundo um eixo vertical em sentido contrário ao da gravidade. Para essa questão que pergunta porque esse corpo se move com velocidade constante, tivemos 3 respostas erradas, 3 incompletas e 29 erradas. Essas três questões obtiveram menores índices de respostas em branco, mas também apresentaram grande número de questões erradas, o que pode ser um indicativo que foi feita uma ligação entre a pergunta e o cotidiano dos alunos.

Apresentamos a Figura 8 que mostra uma resposta para a questão 01, na qual os alunos, ao serem questionados sobre quando o corpo está parado ou em movimento, respondem como entendem o movimento no seu dia a dia, ou seja, estar em movimento é ir de um lado para outro, ou não sair do lugar significando estar parado. Ocorre que não se trata de um conceito científico pois ele não justifica a resposta e não liga a pergunta a um ponto de referência.

Figura 8 – Resposta 1º questionário de avaliação de conhecimentos prévios: individual, aluno 01

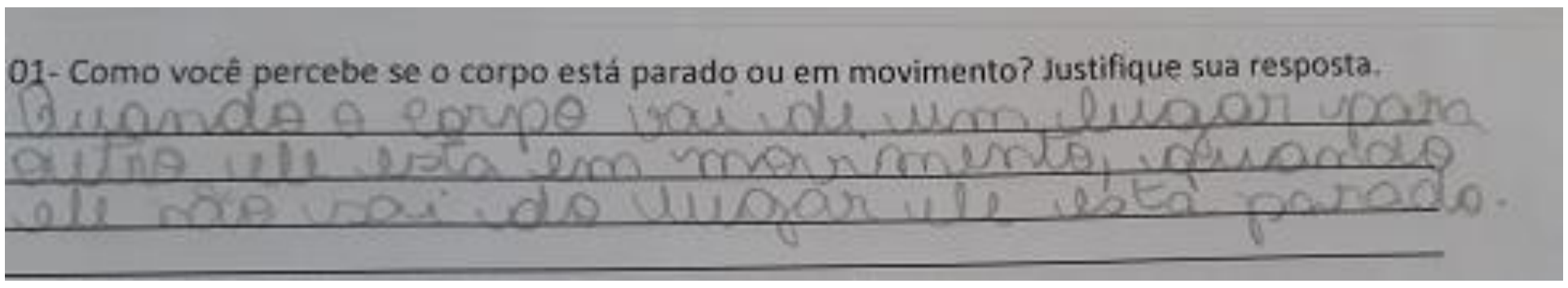

Transcrição: “Quando o corpo vai de um lugar para outro ele está em movimento, quando ele não sai do lugar ele esta parado”.

Fonte: Elaborado pela autora (2019)

O questionário inicial foi entregue aos alunos para ser respondido individualmente e as respostas dos alunos demonstraram, para a mesma pergunta, percepções bem diferentes tanto do conceito interpretativo de seu cotidiano, quanto do ponto de vista cientifico. Temos as respostas a seguir (Figuras 9 e 10) que revelam essas diferenças, sendo que a mesma pergunta recebeu respostas totalmente diferentes.

Figura 9 – Resposta 1º questionário de avaliação de conhecimentos prévios: individual, aluno 03

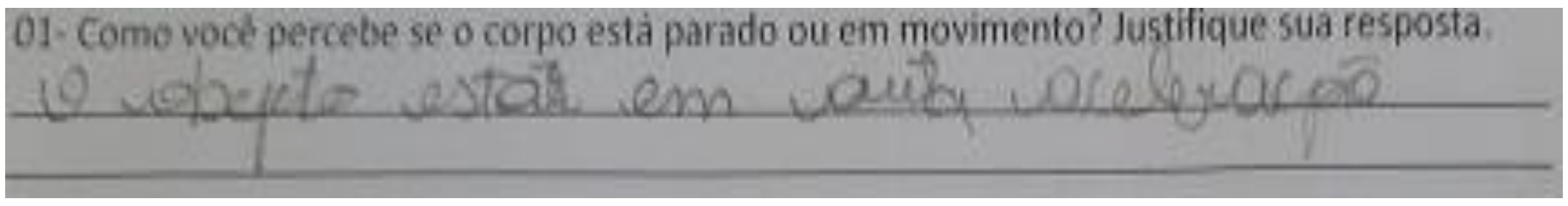

Transcrição: “O objeto está em auta aceleração”.

Fonte: Elaborado pela autora (2018)

Figura 10 – Resposta 1º questionário de avaliação de conhecimentos prévios: individual, aluno 04

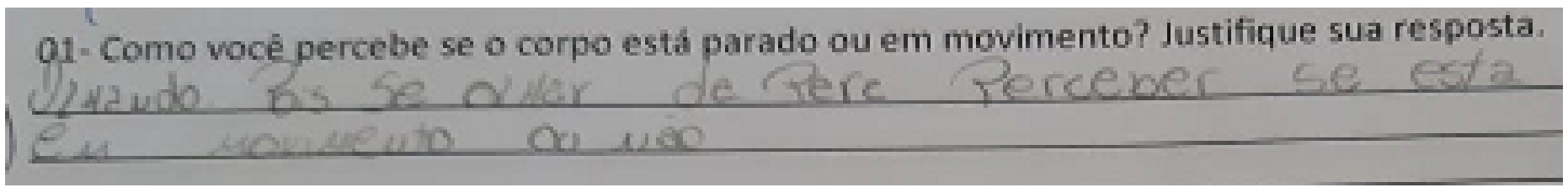

Transcrição: “Quando, pois, se olhar da para perceber se esta em movimento ou não”.

Fonte: Elaborado pela autora (2018)

Apesar de as respostas estarem desconectadas do fato de o movimento depender do referencial, elas apresentam um conhecimento do senso comum que as pessoas possuem sobre o movimento dos corpos. Na Figura 9 temos a resposta de que a percepção do movimento está ligada à aceleração e na Figura 10 de que o movimento está ligado ao olhar. A resposta do aluno permite constatar isso.

A resposta “não sei” também aparece muitas vezes, pois os alunos apresentam muita deficiência de conceitos. Essa resposta foi classificada no Gráfico 1 como resposta em branco e está diretamente ligada ao fato de não apresentar indícios de conhecimentos prévios sobre as leis de Newton.

Respostas analisadas como erradas por conterem elementos não científicos também se mostraram presentes. Temos, como exemplo, a pergunta sobre força (apesar de não ter sido feita de forma completa, dando margem a diversas interpretações) que foi uma das que obteve muitas respostas erradas. A intenção dessa pergunta era saber qual ideia o aluno tinha do conceito de força, sendo importante para analisar se havia conhecimento científico ou somente conhecimento vindo da experiência adquirida durante a vida. As Figuras 11, 12, 13 e 14, a seguir, contêm interpretações do que o aluno liga ao termo força. O termo “levantamento de peso” (Figura 11), demonstrando conhecimento de que se precisa de força para que o peso seja deslocado, a força de vontade, que não é força física mas indica capacidade de esforço, em consequência de um objetivo (Figura 13), ou em “força todos temos um pouco” (Figura 12) e “ força espiritual “ (Figura 14), a força que complementa o dia (para o aluno com pensamento religioso).

Figura 11 – Pergunta 08 (aluno 08)

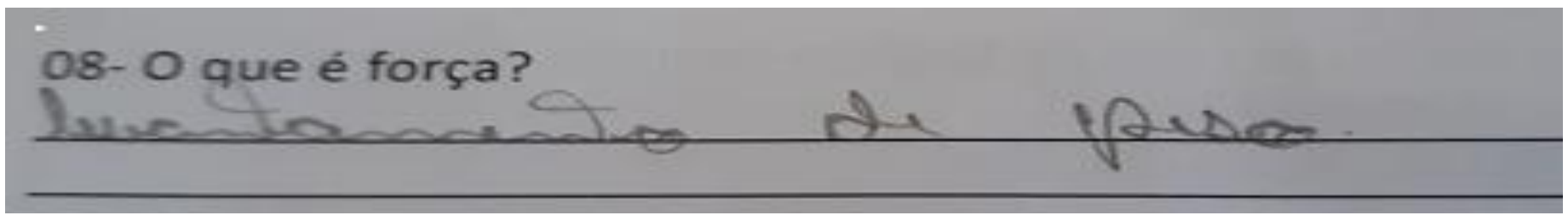

Transcrição: “levantamento de peso”

Fonte: Elaborado pela autora (2018)

Figura 12 – Pergunta 08 (aluno 09)

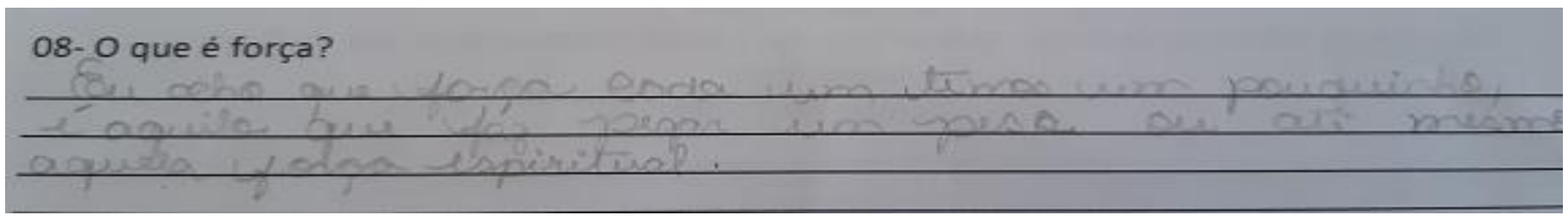

Transcrição: ”Eu acho que força cada um tem um pouquinho, é aquilo que faz pegar um peso ou ate mesmo aquela força espiritual”.

Fonte: Elaborado pela autora (2018)

Figura 13 – Pergunta 08 (aluno 10)

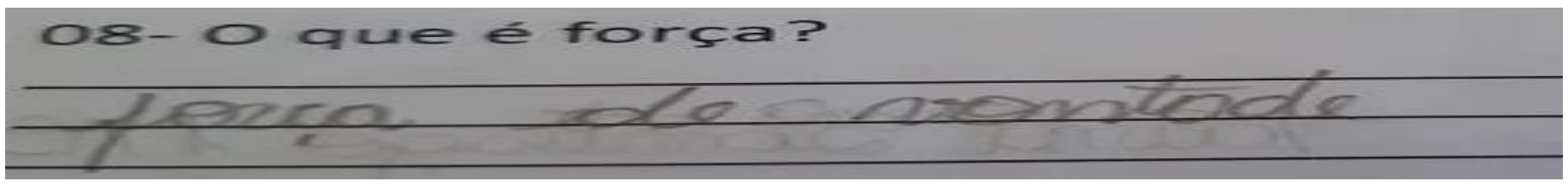

Transcrição: “força de vontade”

Fonte: Elaborado pela autora (2018)

Figura 14 – Pergunta respondida pelo aluno 11

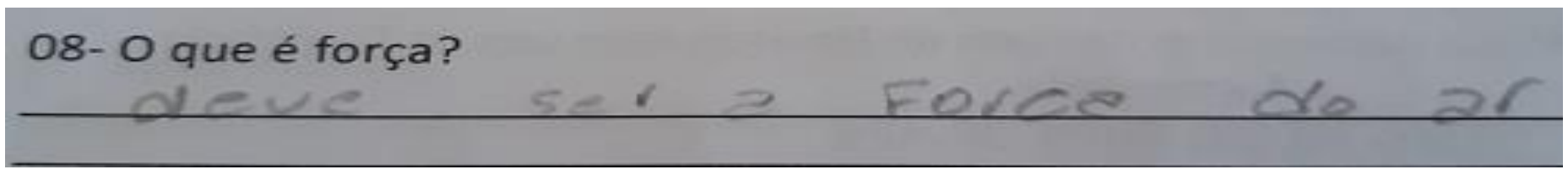

Transcrição: “deve ser a força do ar”

Fonte: Elaborado pela autora (2018)

Diferentes leituras da pergunta com diferentes interpretações foram apresentando caminhos que nos ajudaram a montar o material, para introduzir o conhecimento sobre as leis de Newton. Em média 4 alunos mostraram ter tido contato, em algum momento, com a parte teórica, apresentando respostas que puderam ser avaliadas como corretas, dentro do contexto.

Apresentamos algumas respostas dadas por esses alunos nas Figuras 15,16,17 e 18.

Figura 15 – Perguntas variadas: primeiro questionário (Aluno 12)

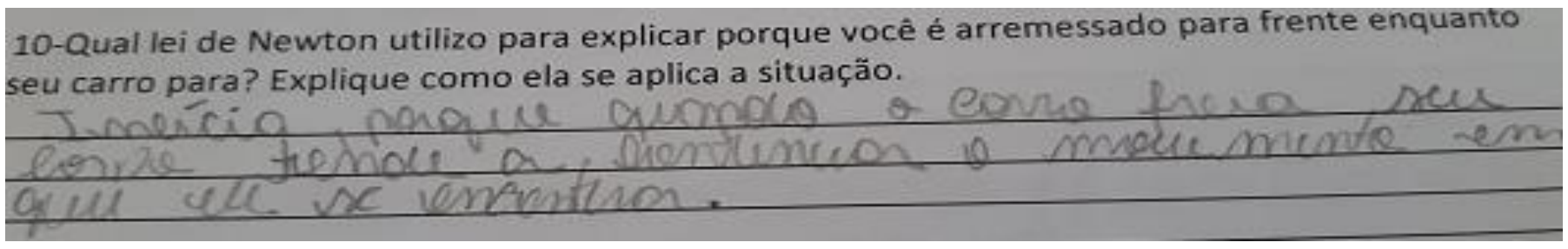
10-Qual lei de Newton utilizo para explicar porque você é arremessado para frente enquanto seu carro para? Explique como ela se aplica a situação.

Transcrição: “Inércia, porque quando o carro freia seu corpo tende a continuar o movimento em que ele se encontra”.

Fonte: Elaborado pela autora (2018)

Figura 16 – Perguntas variadas: primeiro questionário (Aluno 13)

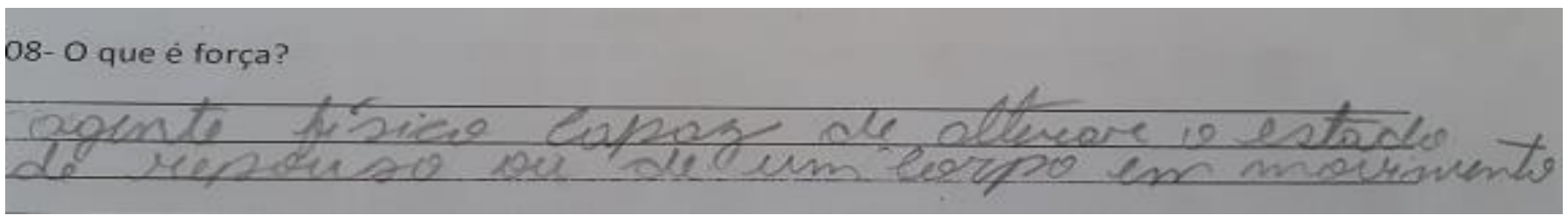
08- O que é força?

Transcrição: “agente físico capaz de alterar o estado de repouso de um corpo em movimento”

Fonte: Elaborado pela autora (2018)

Figura 17 – Perguntas variadas: primeiro questionário (Aluno 14)

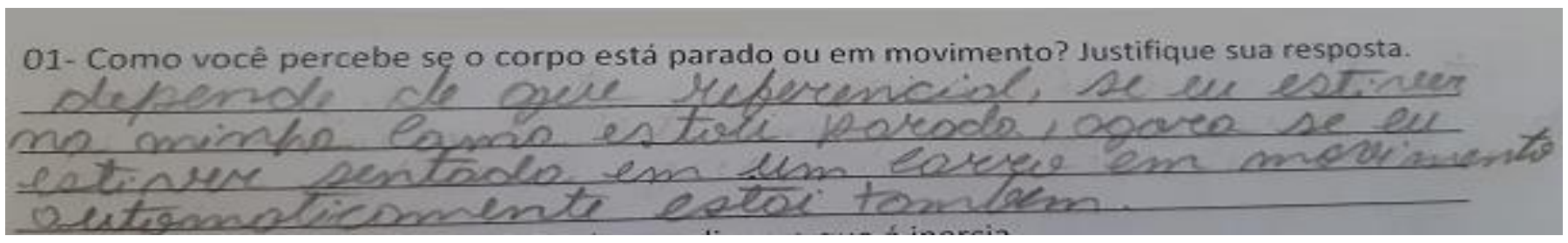
01- Como você percebe se o corpo está parado ou em movimento? Justifique sua resposta.

Transcrição: “depende de que referencial, se eu estiver na minha cama estou parada, agora se eu estiver sentada em um carro em movimento automaticamente estou também”.

Fonte: Elaborado pela autora (2018)

Figura 18 – Perguntas variadas: primeiro questionário (Aluno 15)

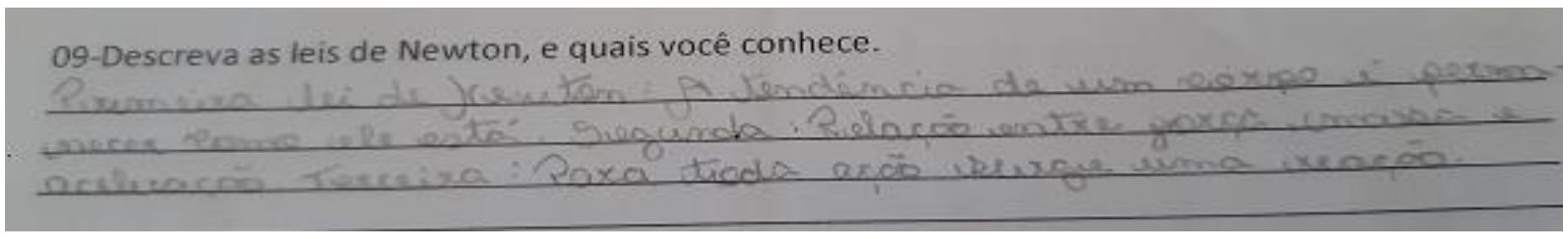
09-Descreva as leis de Newton, e quais você conhece.

Transcrição: “Primeira lei de Newton: A tendência de um corpo é permanecer como ele está. Segunda relação entre força, massa e aceleração. Terceira: para toda ação surge uma reação”.

Fonte: Elaborado pela autora (2018)

## 8.2 APRESENTAÇÃO DE CONHECIMENTOS PRÉVIOS. MASSA X PESO (2ª APLICAÇÃO)

Com a análise do Gráfico 1, cujos dados decorreram das respostas apresentadas no questionário de conhecimentos prévios (Apêndice 1), uma aula introdutória foi necessária para organização de conhecimentos prévios que não foram aprendidos anteriormente.

Os conteúdos abordados no primeiro ano do início de 2018, até o mês de aplicação da sequência didática, foram desenvolvidos de forma dialogal, abordando-os de modo preliminar devido à grande defasagem de ensino em disciplinas básicas como Matemática e Português, que precisaram de reforço contínuo, demandando mais tempo para a resolução das questões. Os conteúdos revisados antes da aplicação foram: Física básica, introdução à Física, História da Física (interpretação e leitura), Notação científica (matemática), Transformações, grandezas e medidas, representações vetoriais, operações com vetores, estudo do movimento, M.R.U (movimento retilíneo uniforme) e M.R.U.V (Movimento uniformemente variado). Não houve tempo de abordar os movimentos sob a ação da gravidade. Por entender que os alunos precisariam desse conteúdo, optamos por essa aula expositiva dialogada introdutória.

Foi preparado um material para essa aula, com exposição de slides e discussões, sem material impresso. Os alunos foram encaminhados para a sala de vídeo, onde a aula foi projetada em PowerPoint (Apêndice 2, slide de 1 a 16). Introduzimos conceitos demonstrando as diferenças entre massa e peso, depois focamos na aceleração da gravidade como estratégia de aprendizagem, por ser um tema que desperta curiosidade nos alunos. Abordamos o tema salientando a proporcionalidade entre massa e peso. A diferença entre massa e peso foi explicada e apresentada de forma lúdica. Os alunos, ao subirem na balança da escola, puderam calcular o seu peso em diversos planetas com diferentes acelerações da gravidade e entender que o conceito de massa não se modifica. Mostramos aos alunos a associação entre força gravitacional e peso. Desta forma, pudemos trabalhar a diferença de significados entre massa e peso e mostrar os aparelhos para a medida de cada um deles. Além disso, discutimos variação do peso, com a localização no globo terrestre, devido à variação do campo gravitacional.

Explicamos a definição de grandeza e suas características. Após a aula expositiva, que consistiu de discussões e experiências na balança da escola, foi feita uma dinâmica de perguntas e respostas sobre peso e aceleração da gravidade em diversos planetas, demonstrando com uma bola de papel amarrado em um barbante o movimento Terra contra Lua. Essa apresentação teve a intenção de mostrar que a Terra (mão), puxa a Lua (bola de papel), para si o tempo todo, e a Lua não cai na Terra porque tem um movimento curvilíneo que a impede.

Figura 19 – Slide 11 (Apêndice 2) da sequência de aplicação da primeira aula: interação peso, massa e aceleração da gravidade

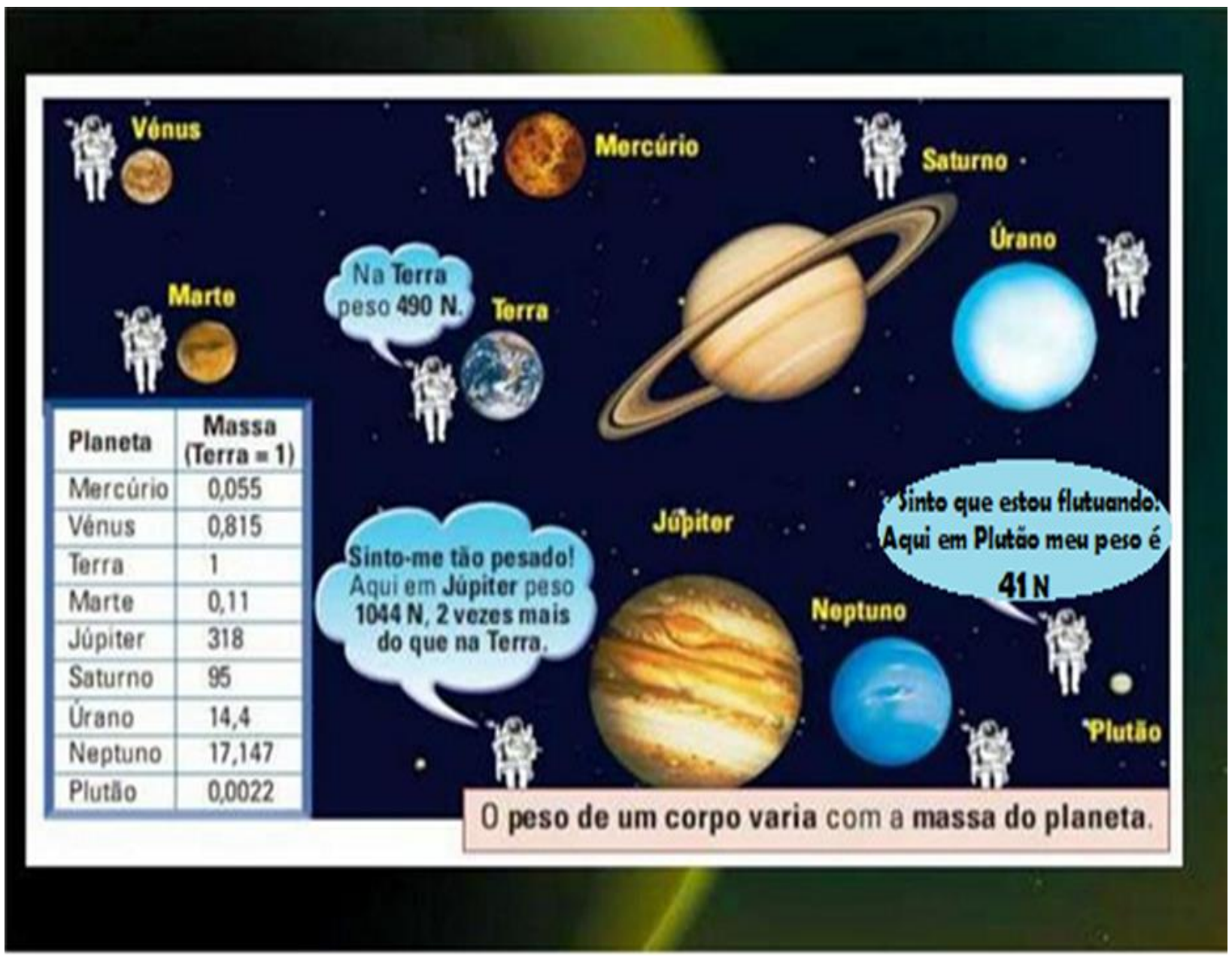

Fonte: NUNES (2012)

A Figura 19 foi apresentada na primeira aula expositiva, no slide 11, para introdução de conceitos como massa, peso, força e aceleração da gravidade, com o objetivo de ajudar na assimilação de conhecimentos básicos para o melhor aproveitamento da proposta didática.

Mostramos a diferença entre aceleração da gravidade nos planetas do sistema solar e como nosso peso será modificado em cada um deles, o que ocorre porque as massas dos planetas do sistema solar são diferentes.  Nessa aula, o aluno pôde evidenciar que a massa não se modifica independentemente da aceleração da gravidade existente e que o peso é diferente e varia de acordo com a aceleração da gravidade a que ele está submetido.

Ao final da sequência, foi distribuído um questionário com três perguntas para ser respondido individualmente pelos alunos. Nesse dia contamos com a presença de 35 alunos na sala.

Os alunos, durante a explicação, perguntaram muito e foram bastante participativos. As perguntas escritas entregues aos alunos após a aula foram respondidas individualmente, sem consulta ao colega nem ao material exposto. Também não foi permitido uso de calculadoras ou celular. As respostas foram rápidas pois não apresentando grau de dificuldade.

Gráfico 2 – Resultado do questionário referente à aula expositiva (slide 14)

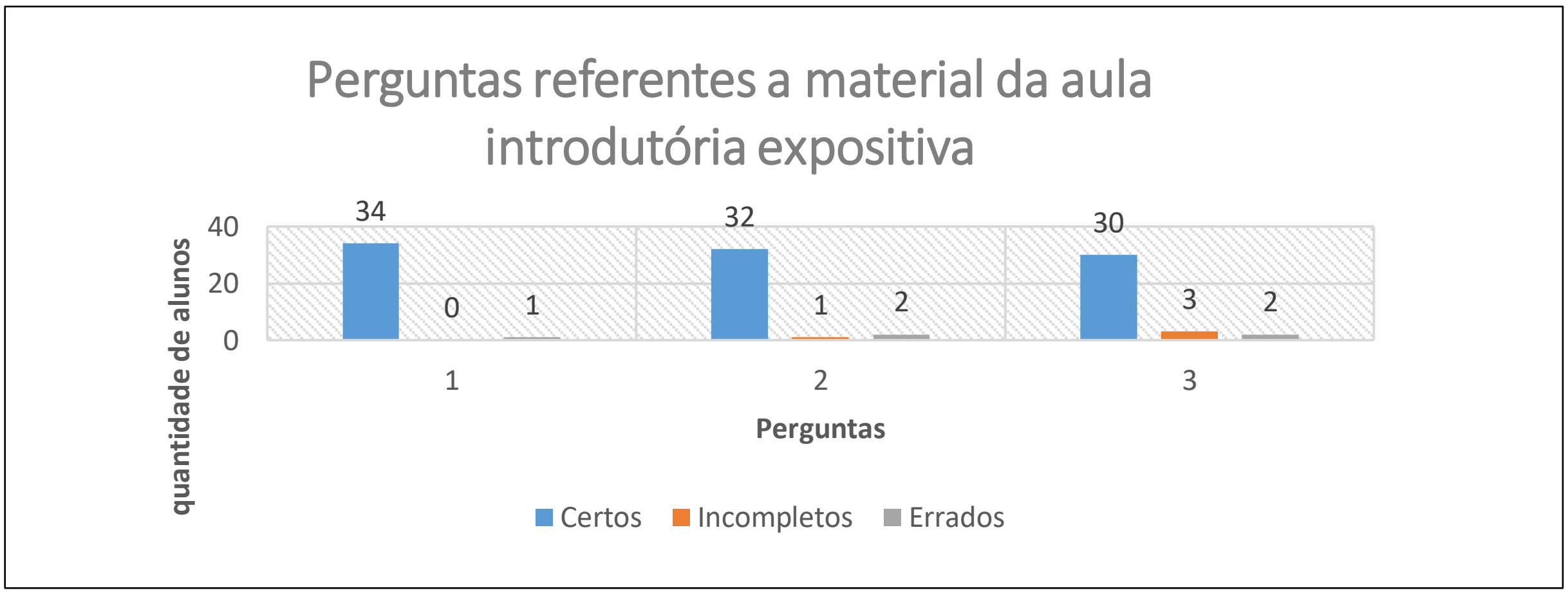

Fonte: Elaborado pela autora (2019)

O Gráfico 2 mostra que houve interesse no assunto abordado já que tivemos participação e engajamento de todos os estudantes. Entre os 35 alunos, o índice de acertos foi de 31 alunos, 2 responderam com respostas incompletas e 2 responderam errado.

Do total, 91% dos alunos realizaram as atividades de forma simples, rápida e participativa. Mesmo com a possibilidade da dificuldade normal de prender a atenção dos alunos nas aulas expositivas, o assunto foi interessante para a maioria deles. Dadas algumas exceções pontuais, eles demonstraram ter aprendido algo a mais do que revelaram no questionário inicial, possibilitando uma abertura para concretização de conhecimentos prévios antes não existentes.

Antes de iniciar a sequência didática, utilizamos o simulador PHET para que os alunos experimentassem diferentes situações aplicadas aos conhecimentos adquiridos na aula anterior. Essa prática foi apresentada devido às dúvidas e questionamentos apresentados relativos ao conteúdo da aula expositiva.

Com a utilização do PHET, a teoria foi aplicada utilizando as funcionalidades do simulador.

## 8.3 AULA MEDIADA COM RECURSOS TECNOLÓGICOS – SIMULADOR PHET (3ª APLICAÇÃO)

Na sequência usamos o simulador PHET para estudar e entender a aceleração da gravidade. Projetamos a simulação, órbitas e gravidade no quadro (Figura 20).

Figura 20 – Simulações PHET

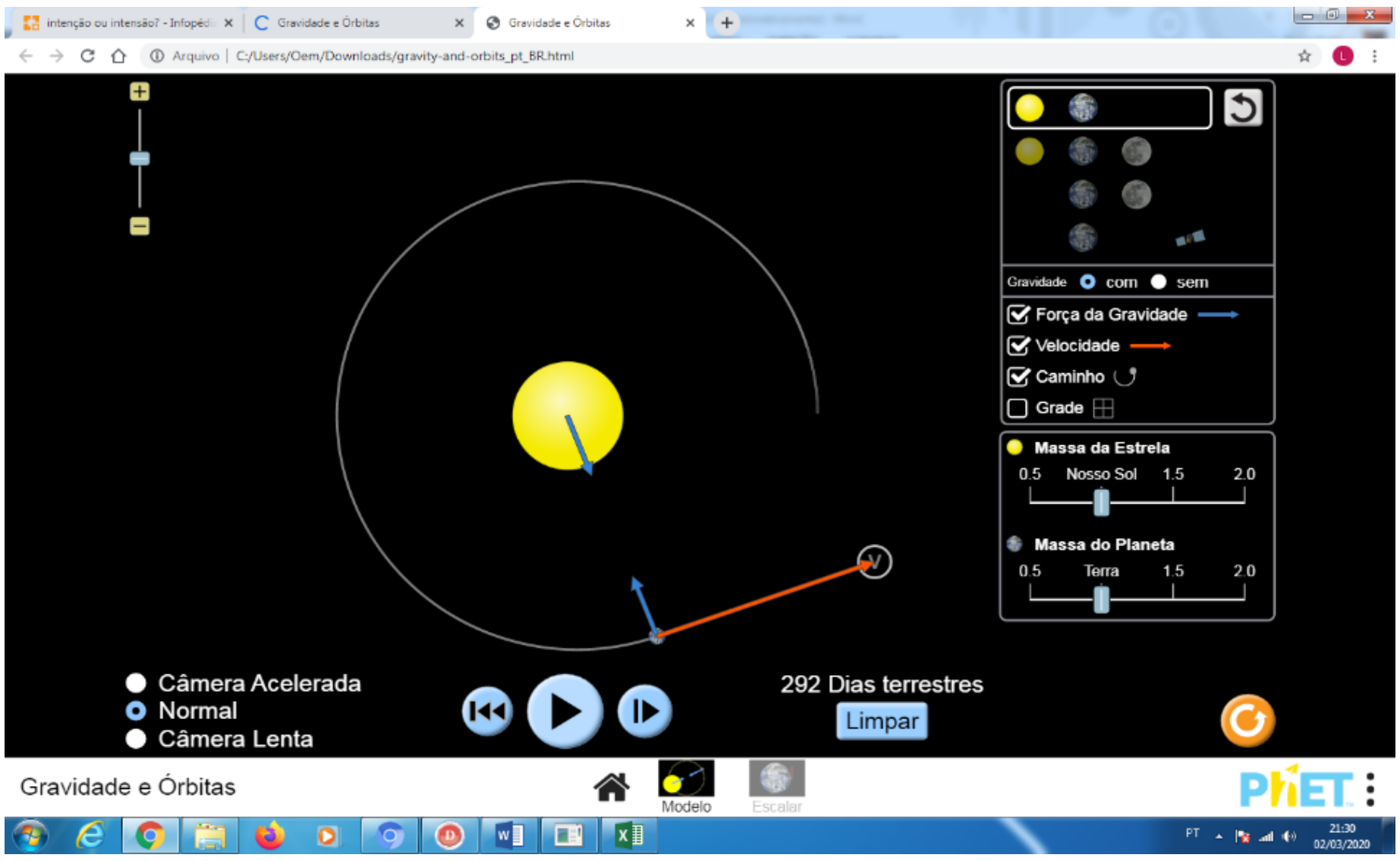

Fonte: PHET (2020)

Seguindo a sequência de explicações, foi feito um pequeno resumo do que foi explicado na aula anterior. Após o resumo apresentado, foi proposta a seguinte questão:

Na superfície da Lua, o peso de um astronauta – Lua x astronauta – é aproximadamente seis vezes menor que na superfície terrestre.

Em Júpiter seria 23 vezes maior. Não é possível ter “ Peso” ou ter “Força”. Peso é a ação que a terra exerce sobre alguma coisa ou pessoa. Ele não é característica da pessoa como a massa. E: A massa de um astronauta é 80 Kg.

Determine o módulo do peso desse astronauta:

a) Na Terra com g = 9,8m/s$^2$
b) Em órbita a 1000Km de altitude, onde o módulo de g= 7,4m/s$^2$
c) Na superfície da Lua em módulo g=1,7m/s$^2$

Os alunos copiaram o problema e responderam com 97% de acertos, mas 3% deles não responderam ou responderam de forma incompleta.

A simulação com o PHET teve o objetivo de apresentar aos alunos como simular diferentes velocidades, caminhos ou rotas a serem seguida pelo planeta Terra, Sol, Lua e satélites artificiais; utilizar diferentes velocidades; visualizar vetores direcionais e de atração, interação gravitacional, diferentes massas e acelerações da gravidade e muitas outras possibilidades apresentadas pelo programa, demonstrando como a gravidade pode agir em um corpo e a relação peso, massa, aceleração da gravidade.

Os alunos eram convidados a utilizar o programa que foi projetado no quadro, momento em que eles aproveitaram para simular diversas situações e esclarecer suas dúvidas teóricas, trocando informações com toda turma.

## 8.4 UM BREVE RELATO: HISTÓRIA DA FÍSICA (4ª APLICAÇÃO)

Para analisar se os alunos possuíam conhecimento prévio para trabalharmos com os conceitos que serão introduzidos no material na sequência didática, baseado na observação dos índices do primeiro Gráfico 1 e nas dificuldades observadas na aula expositiva, uma pequena introdução, utilizando a história da Física foi necessária. Essa introdução foi apresentada através de slides (Apêndice 2, slides 13 e 14) e foi utilizada em função de demonstrar aos alunos a necessidade e a vontade que o homem sempre teve de conhecer o mundo natural.

A partir desse pensamento, discutimos sobre como Galileu iniciou o estudo do movimento observando e experimentando o movimento de bolas rolando em planos inclinados (Apêndice 2, slide 14). Também foi citado o fato que já havia um conceito de inercia ligada à massa do corpo, vinda dos estudos de Descartes.

A história de Isaac Newton (Apêndice 2, slide 15), foi resumida para que se entendêssemos como surgiu o estudo das leis de Newton. A inércia foi apresentada como aprimorada de Joannes Buridanus (1300 — 1358), filósofo e religioso francês que deu a base para as conclusões de Galileu, que partia do fato de que um ponto material isolado está em repouso ou em movimento retilíneo uniforme e que isso significa que um ponto material isolado possui velocidade vetorial constante.

## 8.5 APLICAÇÃO MEDIADA DA PRIMEIRA LEI DE NEWTON COM A REVISTA (5ª APLICAÇÃO)

Nessa etapa, os alunos foram separados em grupos de cinco componentes e foi distribuída a revista, para que eles fizessem uma leitura e representassem cada personagem da revista (Figura 21).

Foram distribuídos os quatro personagens para leitura da revista para quatro alunos. Cada um foi lendo a sequência de frases representando seus personagens. Na introdução da revista (p. 1 e 2), temos a história de um grupo de alunos que não gostam de física e vão fazer uma prova sobre as três leis de Newton. Os personagens se parecem com típicos estudantes de ensino médio que estão perdidos na matéria e acreditam que será muito difícil entender os conceitos apresentados. O texto é curto e tem linguagem simples, voltada para o adolescente do século XXI. No entendimento da pesquisadora, fez-se necessário mudar um pouco a linguagem científica substituindo-a por uma linguagem utilizada por adolescentes em grupos de WhatsApp.

Figura 21 – Página 1 da revista

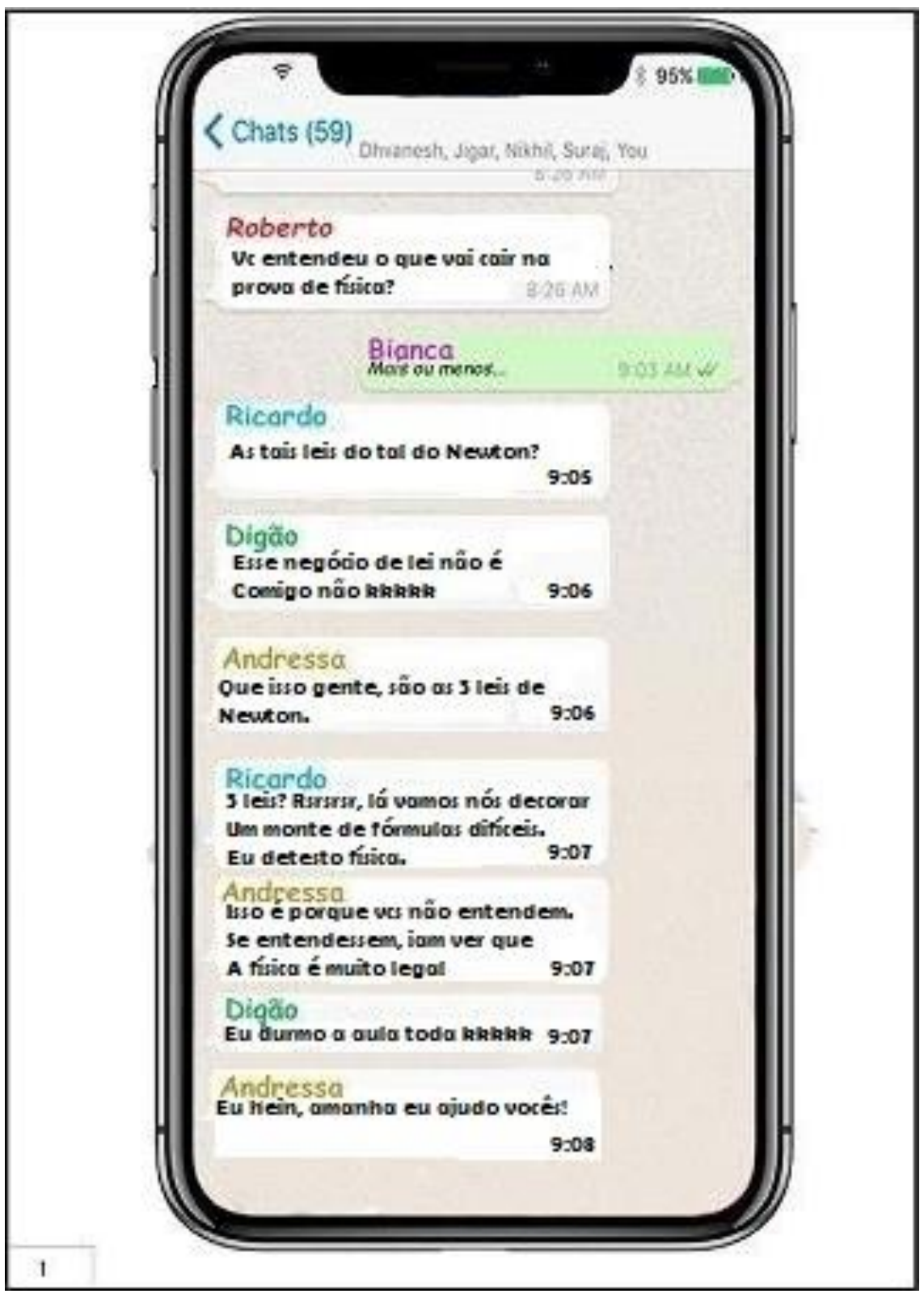

Fonte: Elaborado pela autora (2018)

A Figura 21 apresenta uma discussão em um grupo de aplicativo de mensagem onde os alunos elegem suas dificuldades com relação ao entendimento das três leis de Newton, conteúdo que será cobrado na prova de Física. Observando o teor das das mensagens, uma colega se propõe a ajudar os amigos, uma vez que aprendeu o conteúdo com maior facilidade. Eles marcam um encontro para que ela os ajude a entender o conteúdo da prova (Figura 22).

Figura 22 – Página 2 da revista

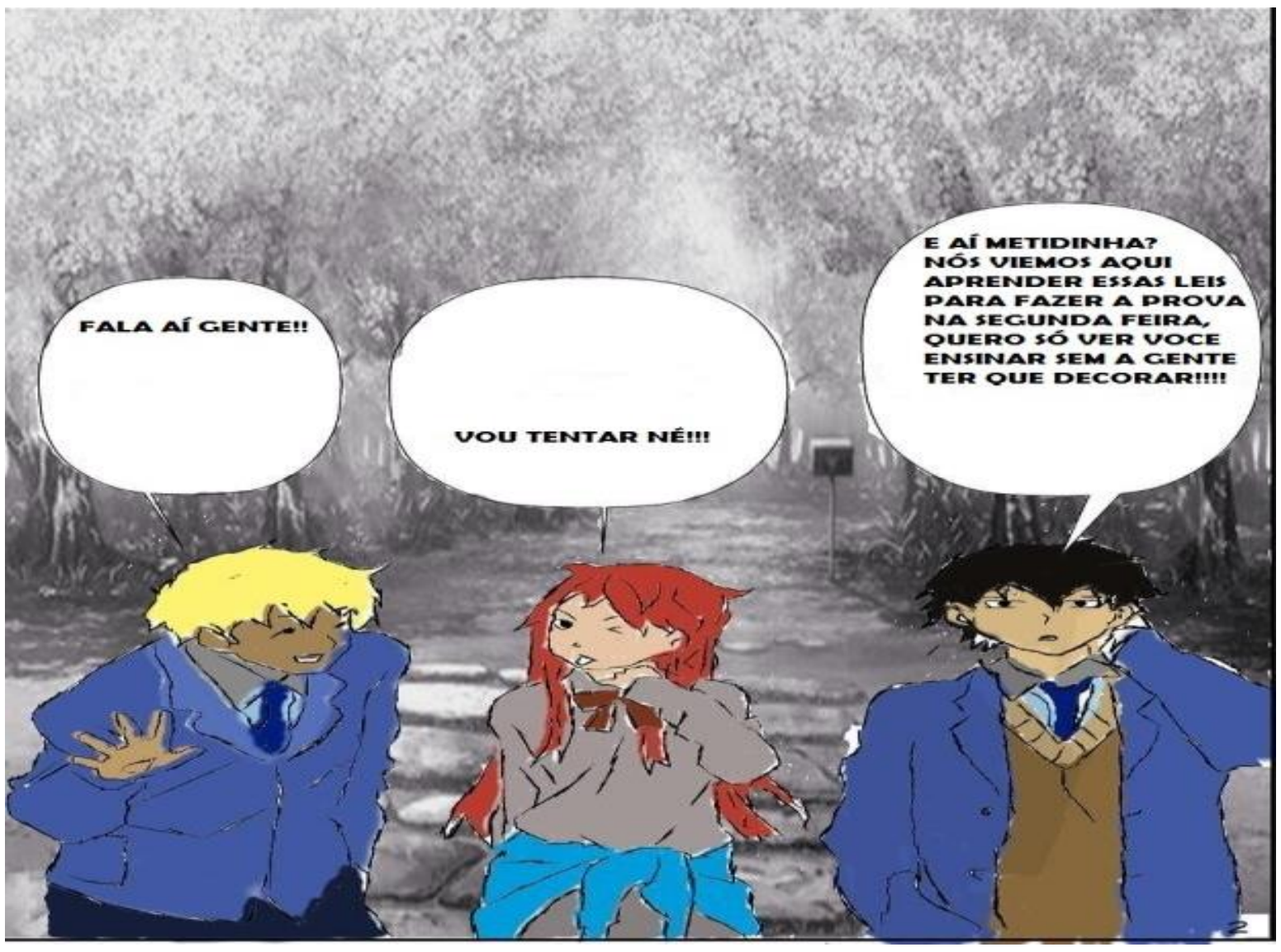

Fonte: Elaborado pela autora (2018)

Nas páginas sequentes (páginas 3,4 e 5), representadas na Figura 23, ela aborda um pouco sobre a história de Isaac Newton e inicia a introdução à primeira Lei de Newton.

Figura 23 – Página 3 da revista

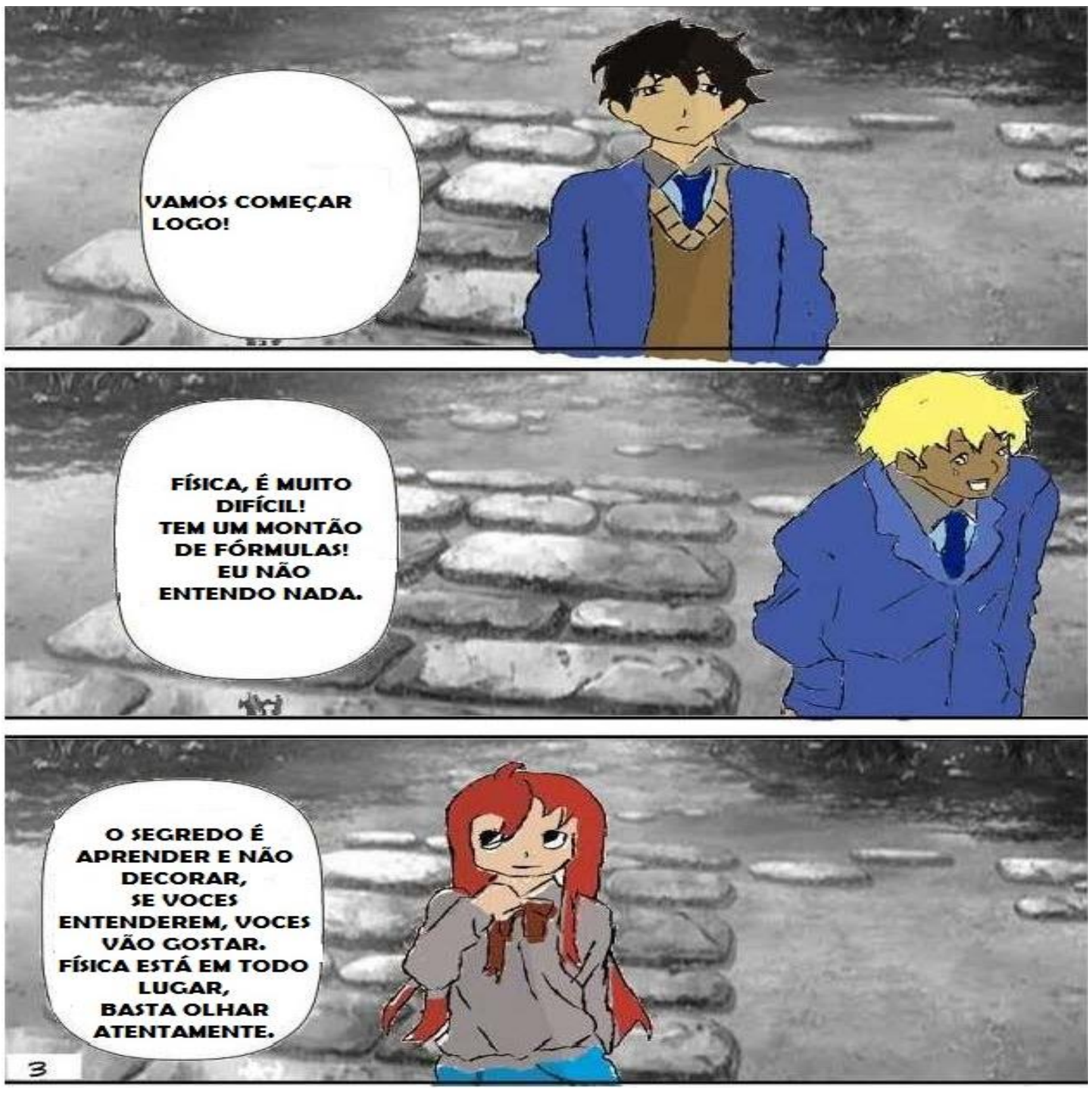

Fonte: Elaborado pela autora (2018)

Figura 24 – Página 4 da revista

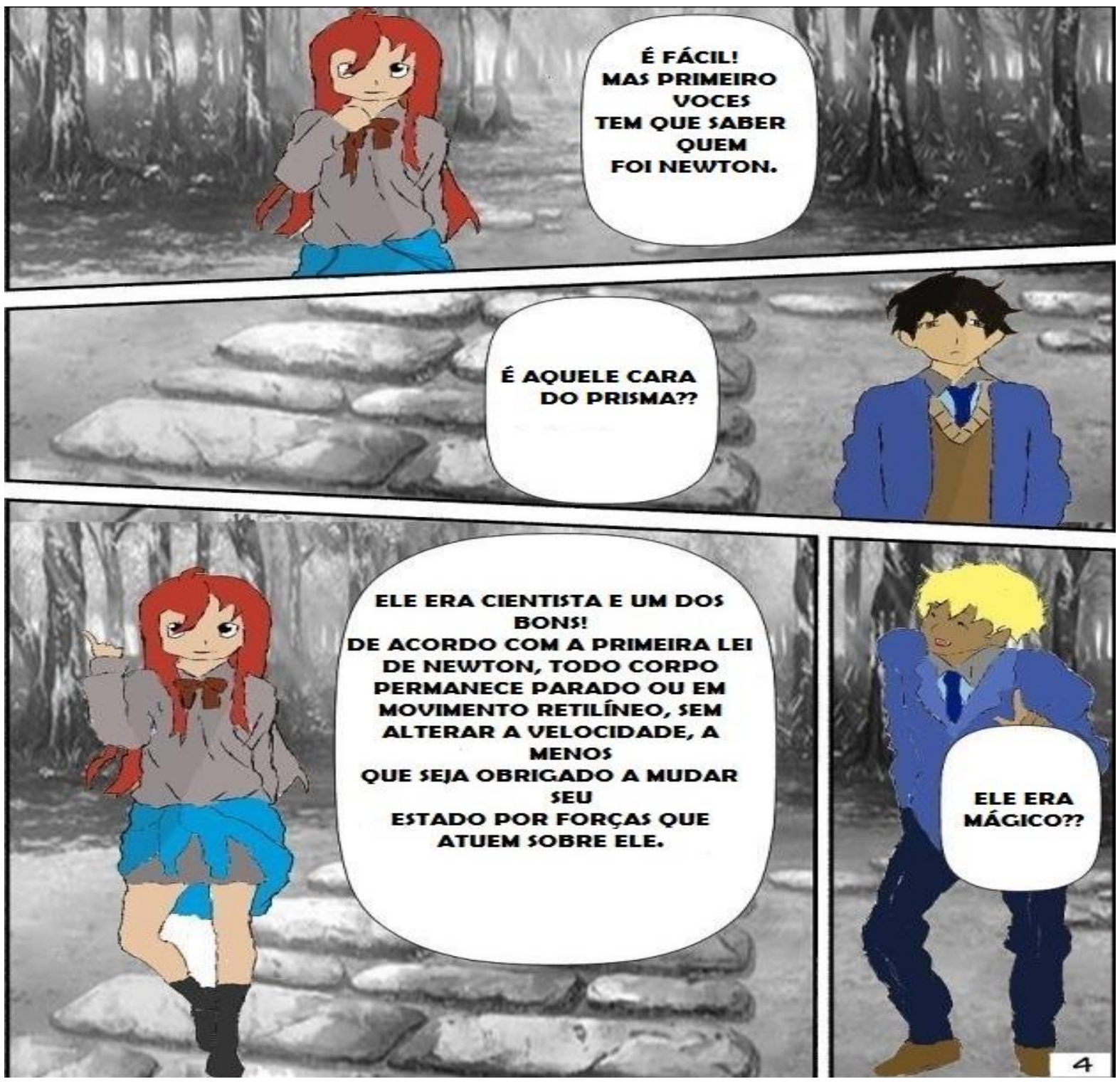

Fonte: Elaborado pela autora (2018)

Figura 25 – Página 5 da revista

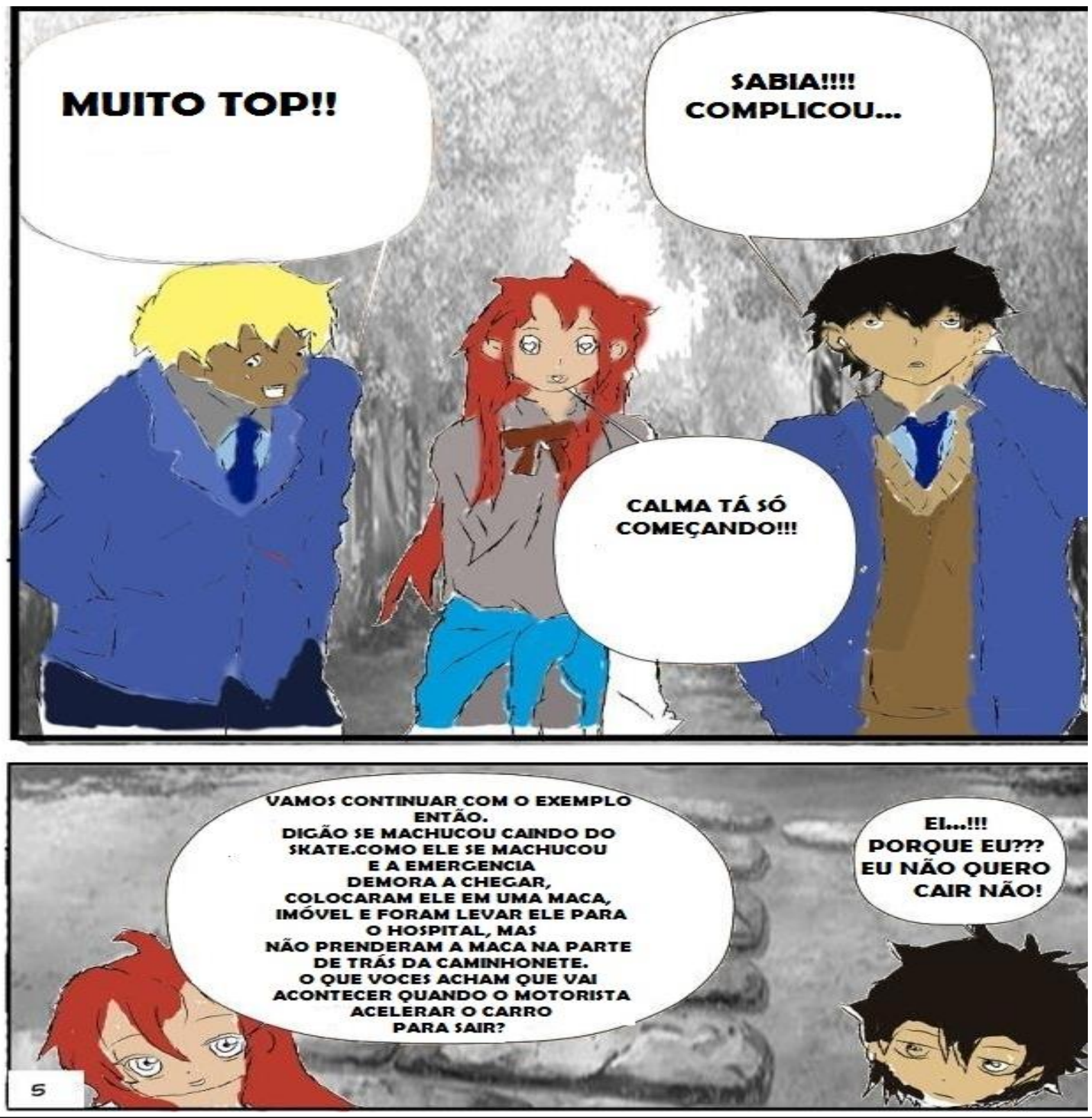

Fonte: Elaborado pela autora (2018)

Na Figura 23, o aluno demonstra estar perdido, sem entender nada da matéria e a colega (Andressa) começa a dialogar e expõe o fato de estarmos constantemente em contato com os fenômenos físicos e, muitas vezes, não percebermos, colocando a física como uma teoria difícil e representada por grandes fórmulas matemáticas.

Na Figura 24, Andressa trata um pouco sobre Isaac Newton e começa a descrever a lei da inércia, simplificando o conceito.

A explicação do fenômeno começa com um exemplo que a personagem (Andressa) descreve, utilizando-se de uma pequena experiência mental, e busca a curiosidade para que se descubra o resultado da cena criada (Figura 25)

A partir da introdução, os alunos ficaram muito motivados e demonstraram uma interação com a leitura da revista, se mostrando envolvidos com a aprendizagem. A partir da página 6, ao final da leitura, a professora instruiu aos alunos, que haveria um Target na página, descrevendo a experiência mental mencionada na página 5 (Figura 25). Eles deveriam acionar o Target com o aplicativo para celular AR Phisycs instalado no telefone da professora e analisar o Gif que abriria. Esse Gif (Figura 22) demonstra uma situação real onde o fenômeno aparece e cada grupo deveria analisar a cena e responder à pergunta entregue, referente ao tema primeira Lei de Newton (Figura 26). O aplicativo foi passado para os grupos sendo um de cada vez e os grupos não deveriam se comunicar uns com os outros. Foi entregue uma pergunta diferente à cada grupo, sorteada aleatoriamente, à qual eles deveriam responder, baseados apenas nas informações contidas na revista e na representação visual (Apêndice 3). Eles observaram e analisavam a situação e, após discussão sobre o Gif (Figura 27), pelos grupos, eles responderam à pergunta entregue pela professora.

Figura 26 – Página 6 da revista

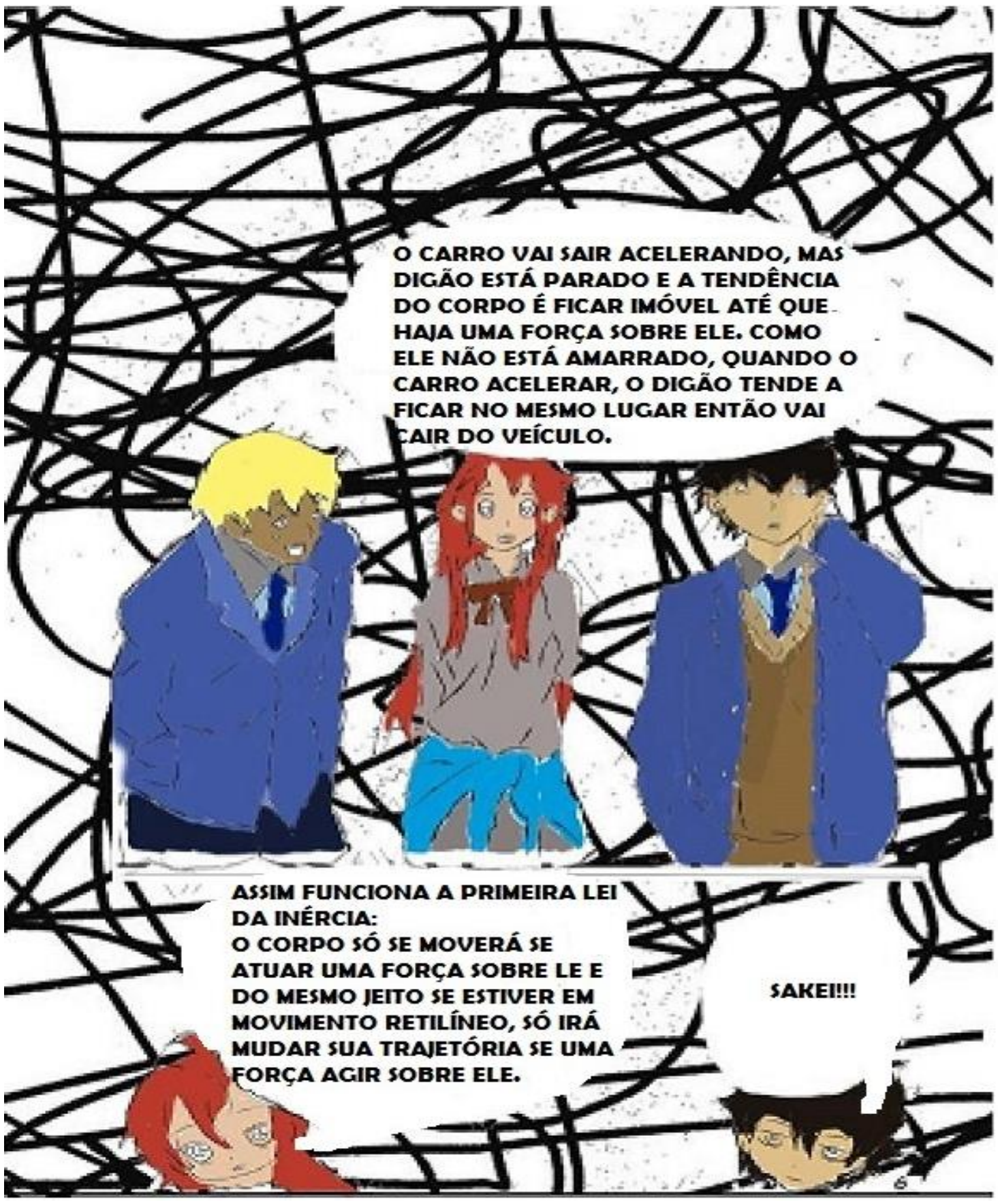

Fonte: Elaborado pela autora (2018)

Figura 27 – Acionamento do aplicativo AR- Physics: página 6, 1º Gif, Primeira Lei de Newton, homem na maca

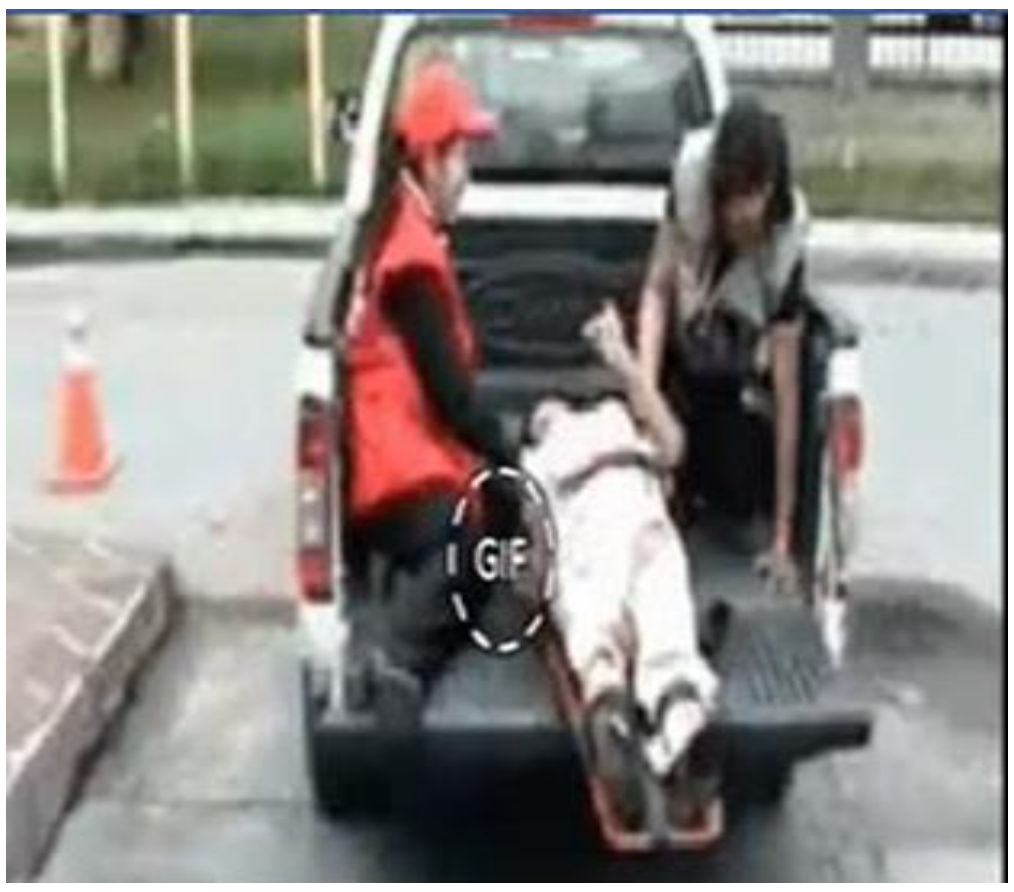

Fonte: SCHINATO (2009)

Além da pergunta que deveria ser respondida, os grupos foram instruídos a observar o Gif da primeira lei (Figura 27) e interpretar escrevendo uma pequena resenha para entregar à professora. Essa interpretação foi pedida a todos os grupos, que não podiam pedir ou trocar informações com os outros grupos.

Além da resenha do Gif, foi entregue uma pergunta para cada grupo. As sete perguntas foram sorteadas e respondidas, cada uma, por um grupo diferente. Algumas perguntas pediam que o grupo respondesse e explicasse para toda a sala. Dos sete grupos, 6 chegaram a respostas satisfatórias e 1 respondeu de forma incompleta ou inconclusiva. Podemos constatar que as respostas foram espontâneas indicando um ganho de conhecimento quando comparados às perguntas do questionário inicial (Apêndice 1).

O grupo 4 (Figura 28) fez a seguinte descrição do Gif:

Figura 28 – Interpretação do primeiro Gif (Figura 27) feita pelo grupo 4

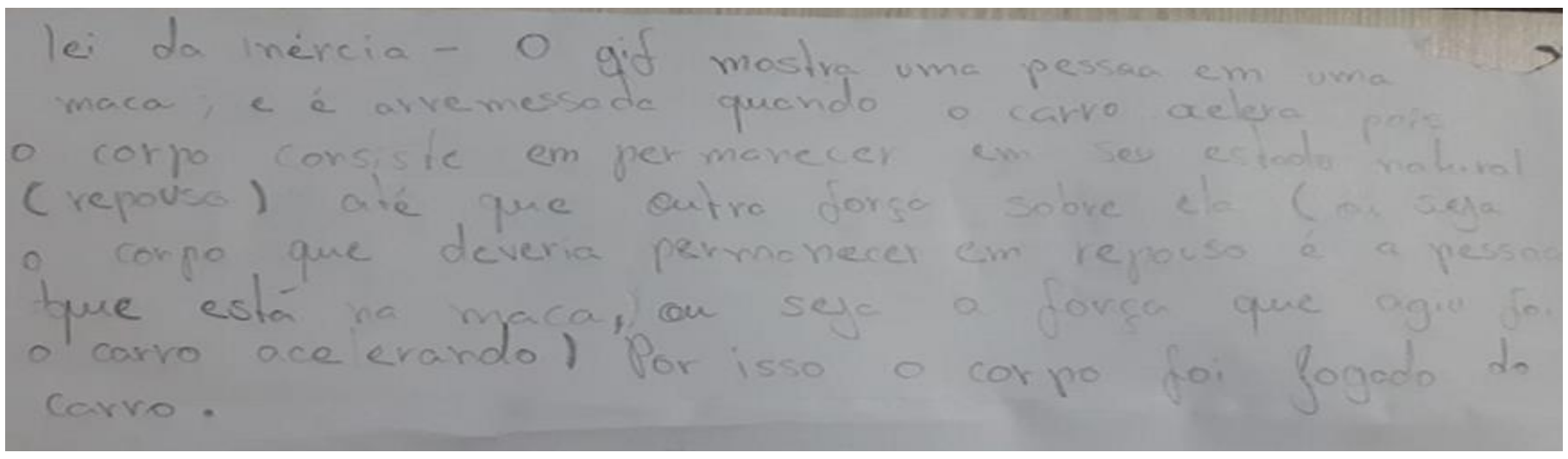

lei da inércia – O gif mostra uma pessoa em uma maca, e é arremessada quando o carro acelera pois o corpo consiste em permanecer em seu estado natural (repouso) até que outra força sobre ele (ou seja o corpo que deveria permanecer em repouso é a pessoa que está na maca, ou seja a força que agiu foi o carro acelerando) Por isso o corpo foi jogado do carro.

Transcrição: “Lei da inércia – O Gif mostra uma pessoa em uma maca e é arremessada quando o carro acelera, pois, o corpo consiste em permanecer em seu estado natural (repouso) até que outra força sobre ele (ou seja o corpo que deverá permanecer em repouso e a pessoa que está na maca, ou seja a força que age foi o carro acelerando). Por isso o corpo foi jogado do carro”.

Fonte: Elaborado pela autora (2018)

Após a análise e entrega do material, verificamos que todos os grupos descreveram o fenômeno de maneira próxima à científica. Não houve nenhum grupo que interpretou o Gif de maneira incoerente de acordo com o assunto abordado.

Colocamos mais uma interpretação, agora do grupo 7, para análise do primeiro momento apresentado pela revista (Figura 29). Além das interpretações tivemos as sete perguntas sorteadas e respondidas cada uma por um grupo diferente. Dos sete grupos 6 chegaram a respostas satisfatória

Figura 29 – Análise do Gif feita pelo grupo 7

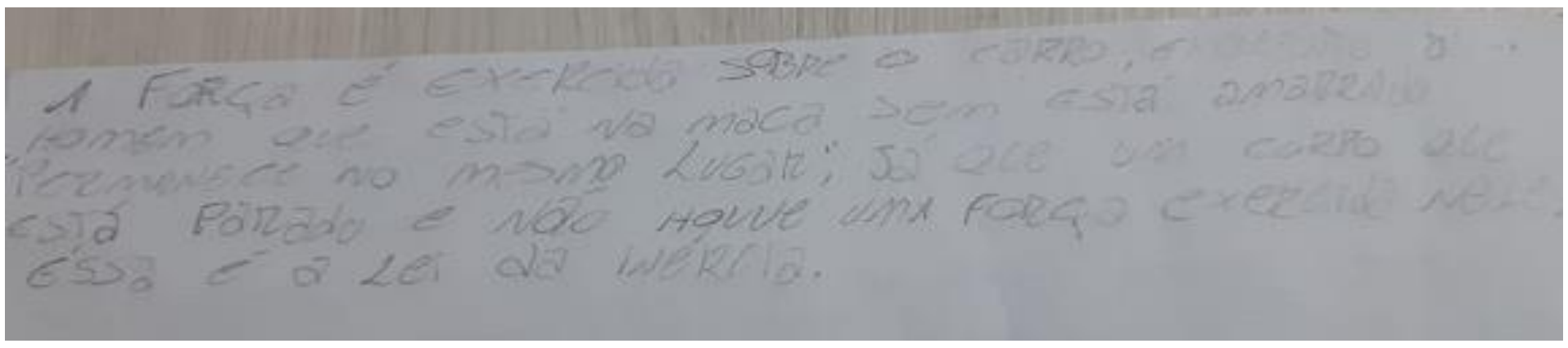

Transcrição: "A força é exercida sobre o carro envolvendo o homem que esta na maca sem esta amarrado " permanece no mesmo lugar" já que um corpo ate esta parado e não houve uma força exercida nele essa é a lei da inércia".

Fonte: Elaborado pela autora (2018)

O grupo 7 também demonstra uma interpretação favorável à aprendizagem, o conceito está correto, apesar de mal redigido. No entanto, não pode ser considerado errado, pois podemos observar que, mesmo com dificuldade, o grupo consegue socializar a ideia do fenômeno analisado no Gif, referente à primeira Lei de Newton.

Como última interpretação da primeira Lei de Newton, descrevemos a resposta do grupo 3 (Figura 30).

Figura 30 – Interpretação do primeiro Gif, feita pelo grupo 3

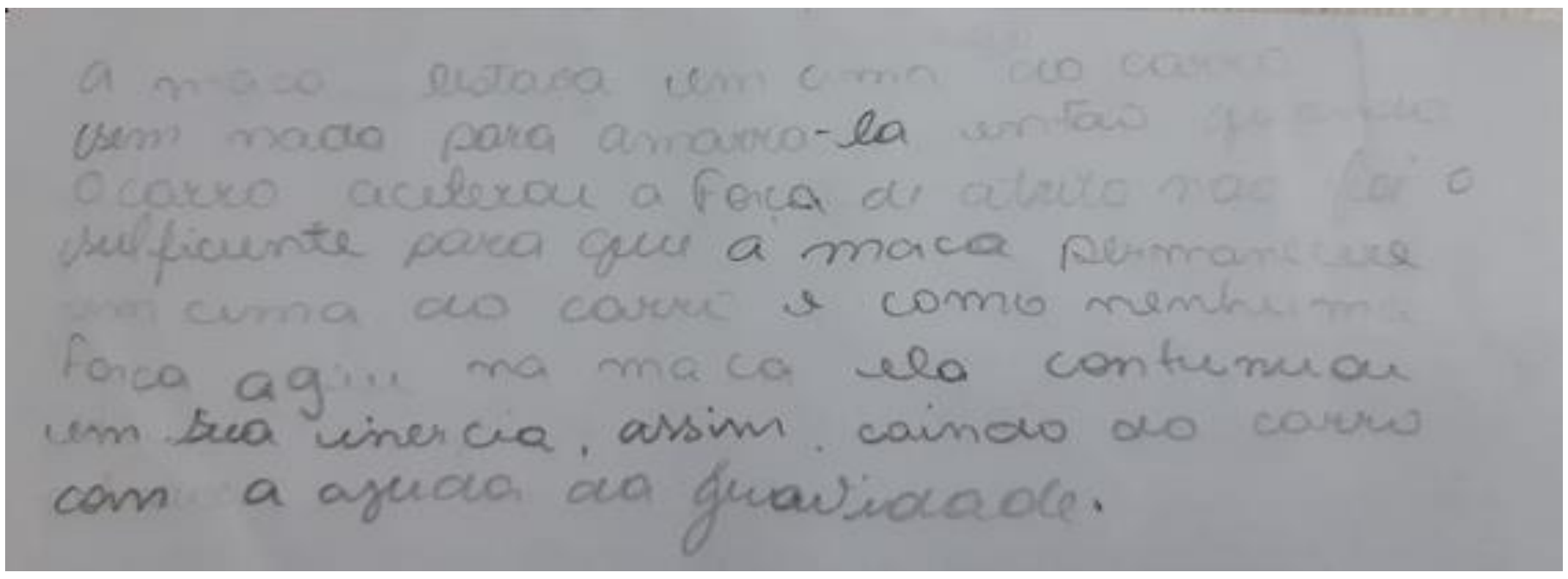

Transcrição: "A maca estava em cima do carro sem nada para amarra-la então quando o carro acelerou a força de atrito não foi o suficiente para que a maca permanecesse em cima do carro e nenhuma força agiu na maca ela continuou em sua inercia, assim caindo do carro com a ajuda da gravidade".

Fonte: Elaborado pela autora (2018)

Como podemos verificar nas Figuras 28, 29 e 30, as interpretações apresentaram linguagem científica e o direcionamento mostra que foram adquiridos conceitos que antes não existiam.

Além das interpretações, tivemos as sete perguntas sorteadas, respondidas, cada uma por um grupo diferente. Dos sete grupos, 6 chegaram a respostas satisfatórias e 1 respondeu de forma incompleta ou inconclusiva. Podemos constatar que as respostas foram espontâneas, indicando um ganho de conhecimento quando comparados às respostas ao questionário inicial. Essas perguntas respondidas foram escolhidas do material de pesquisa aleatoriamente. Temos a resposta do grupo 2 a seguir (Figura 31).

Figura 31 – Resposta do grupo 2 –1ª lei de Newton

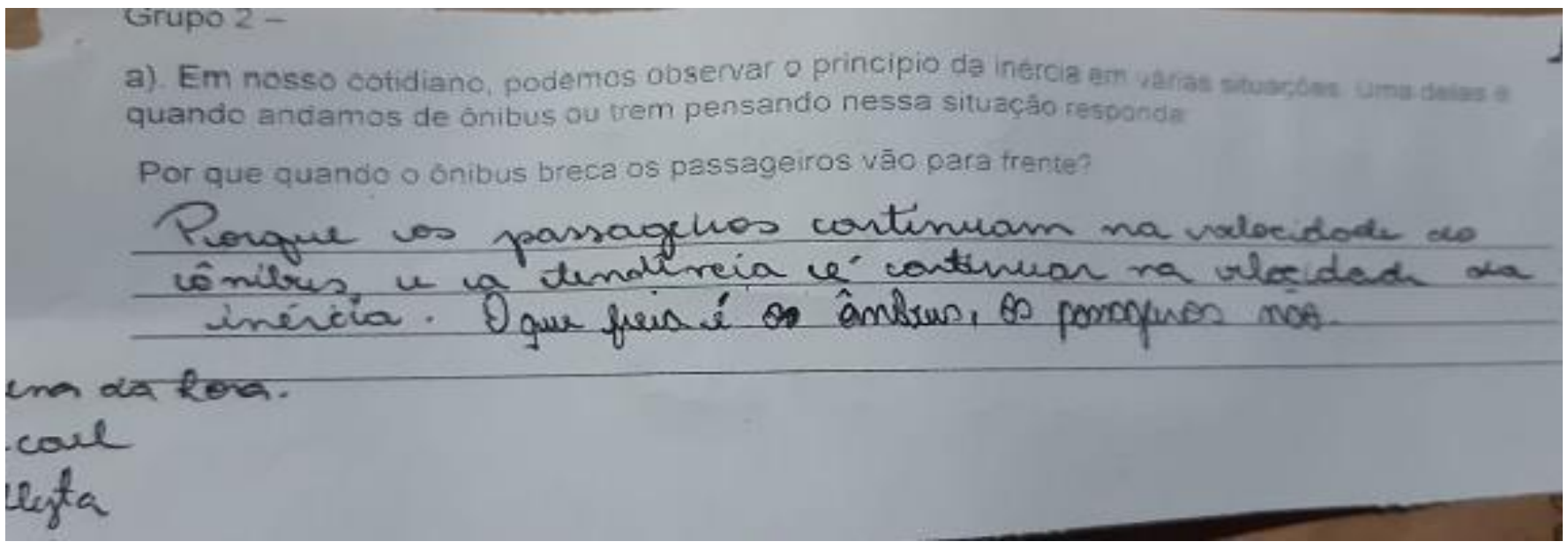
Grupo 2 –
a). Em nosso cotidiano, podemos observar o princípio da inércia em várias situações. Uma delas é quando andamos de ônibus ou trem pensando nessa situação responda:
Por que quando o ônibus breca os passageiros vão para frente?

Transcrição: “Porque os passageiros continuam na velocidade do ônibus, e a tendência é continuar na velocidade da inércia. O que freia é o ônibus, os passageiros não”.

Fonte: Elaborado pela autora (2018)

A resposta do grupo 2 (Figura 31) apresenta a teoria de forma parcialmente correta já que na hora de descrever o conceito, o aluno ligou a inercia à velocidade e não à massa. Apesar disso, o restante da resposta demonstra que houve um avanço em relação ao questionário inicial.

Analisando a resposta do grupo 5 (Figura 32), observamos que as respostas começam a ganhar uma linguagem científica, apresentando evidências de aprendizagem.

Figura 32 – Resposta do grupo 5

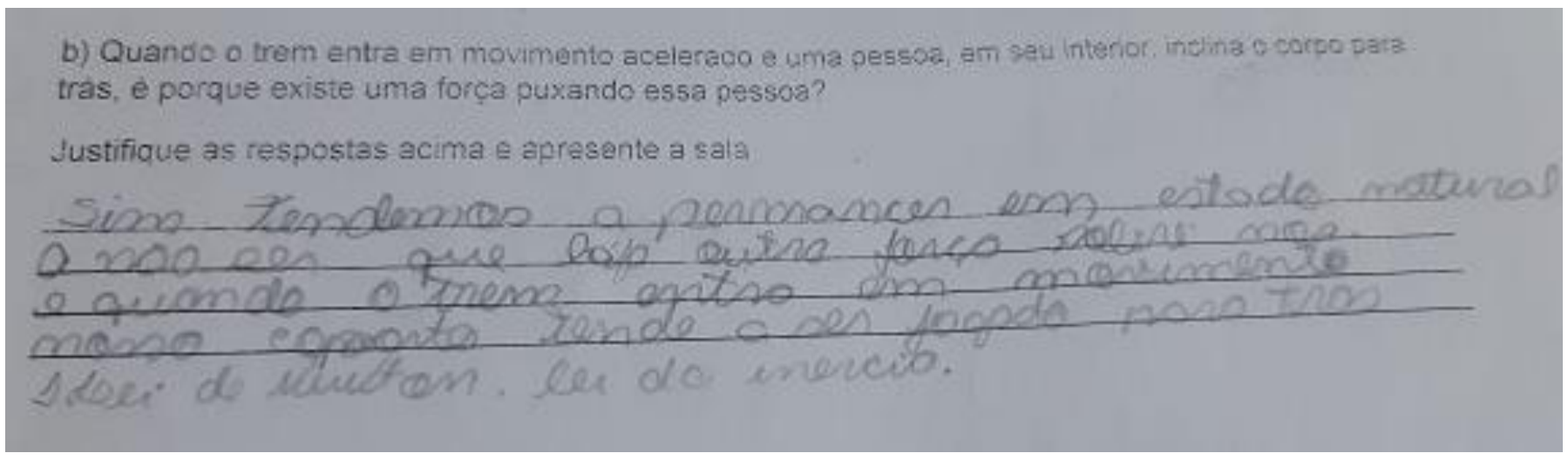
b) Quando o trem entra em movimento acelerado e uma pessoa, em seu interior, inclina o corpo para tras, é porque existe uma força puxando essa pessoa?

Justifique as respostas acima e apresente a sala

Transcrição: ”Sim tendemos a permanecer em estado natural a não ser que haja outra força sobre nos e quando o trem entra em movimento nosso corpo tende a ser jogado para tras 1 Lei de Newton. Lei da inercia”.

Fonte: Elaborado pela autora (2018)

Nessa pergunta (Figura 32), o grupo 5 apresenta uma coerência na resposta apresentada. Além disso o desenvolvimento do texto melhorou, descrevendo o fenômeno de forma simples, resumida, mas satisfatória. Apesar de ter iniciado a resposta com uma afirmativa que uma pessoa, no interior do trem, é arremessada para frente quando o trem anda demonstrando força na inércia, logo após a afirmativa, descreve corretamente a inercia e os demais conceitos ligados a ela.

Esse grupo descreveu o conceito cientifico subjacente à resposta, concernente à pergunta, para toda a turma, respondendo às dúvidas quando questionados. A professora, após o relato do grupo, fez os esclarecimentos necessários referente ao equivoco descrito pelo grupo 5 na resposta (Figura 32), para toda turma.

Além da pergunta, também foi pedido que eles apresentassem uma interpretação do Gif demonstrado.

O grupo 3 e o grupo 1 produziram as melhores respostas mostrando mais avanços na aprendizagem, com a aplicação do produto. Também verificamos que os grupos 6 e 7 foram os que tiveram maiores dificuldades, apresentando respostas incompletas nas atividades propostas.

A primeira apresentação da sequência didática utilizando a revista suscitou um grande empenho por parte dos estudantes, fazendo com que eles se envolvessem no processo de aprendizagem de forma participativa (Figura 33).

Figura 33 – Grupo 1 reunido para a dinâmica

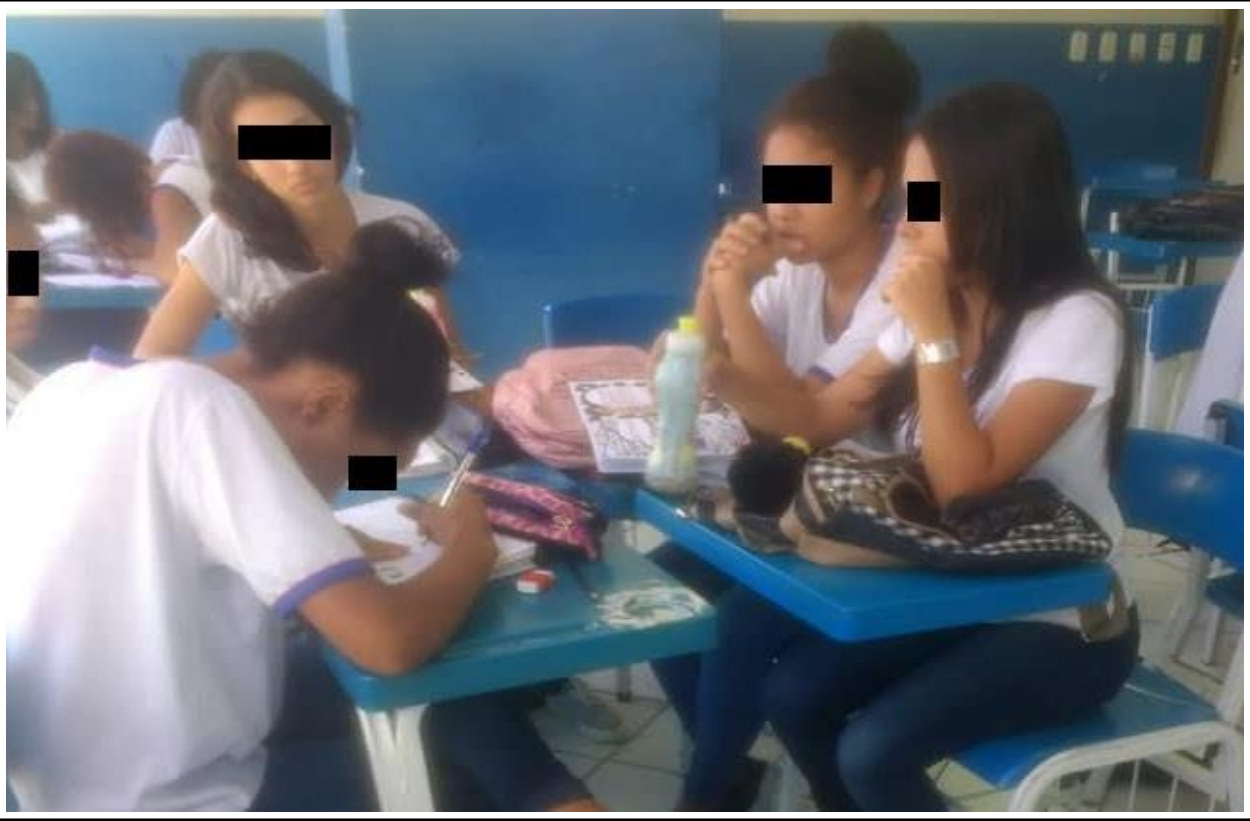

Fonte: Elaborado pela autora (2018)

Encerrando a primeira Lei de Newton foi iniciada a aplicação seguinte que foi a introdução da segunda Lei de Newton.

## 8.6 SEGUNDA LEI DE NEWTON (6ª e 7ª APLICAÇÃO)

Para a introdução da segunda Lei de Newton, seguimos com a leitura das páginas 7 (Figura 34) e 8 (Figura 39). Os alunos mantiveram o mesmo processo de leitura, acompanhando, passo a passo, os quadrinhos em voz alta. Ao final da leitura da página 7 (Figura 34), a estratégia de ensino se repetiu com a formação dos grupos e a entrega das perguntas (Apêndice 4) que foram sorteadas, uma para cada grupo. Ao final da entrega do questionário da segunda lei de Newton (Apêndice 4), a professora novamente acionou o aplicativo AR Physics em seu celular. O celular foi passado de grupo em grupo com a instrução para que os alunos observassem as páginas 7 (Figura 34) e 8 (Figura 39).

De acordo com o texto e a interpretação dos Gifs do cãozinho empurrando o carrinho e o homem estacionando o carro, os alunos deveriam responder às perguntas e interpretar o fenômeno que cada um dos Gifs demonstrava, explicando-os de acordo com sua compreensão. Nessa dinâmica, foram utilizadas duas aulas geminadas de 55 minutos cada.

Figura 34 – Página 7 da revista

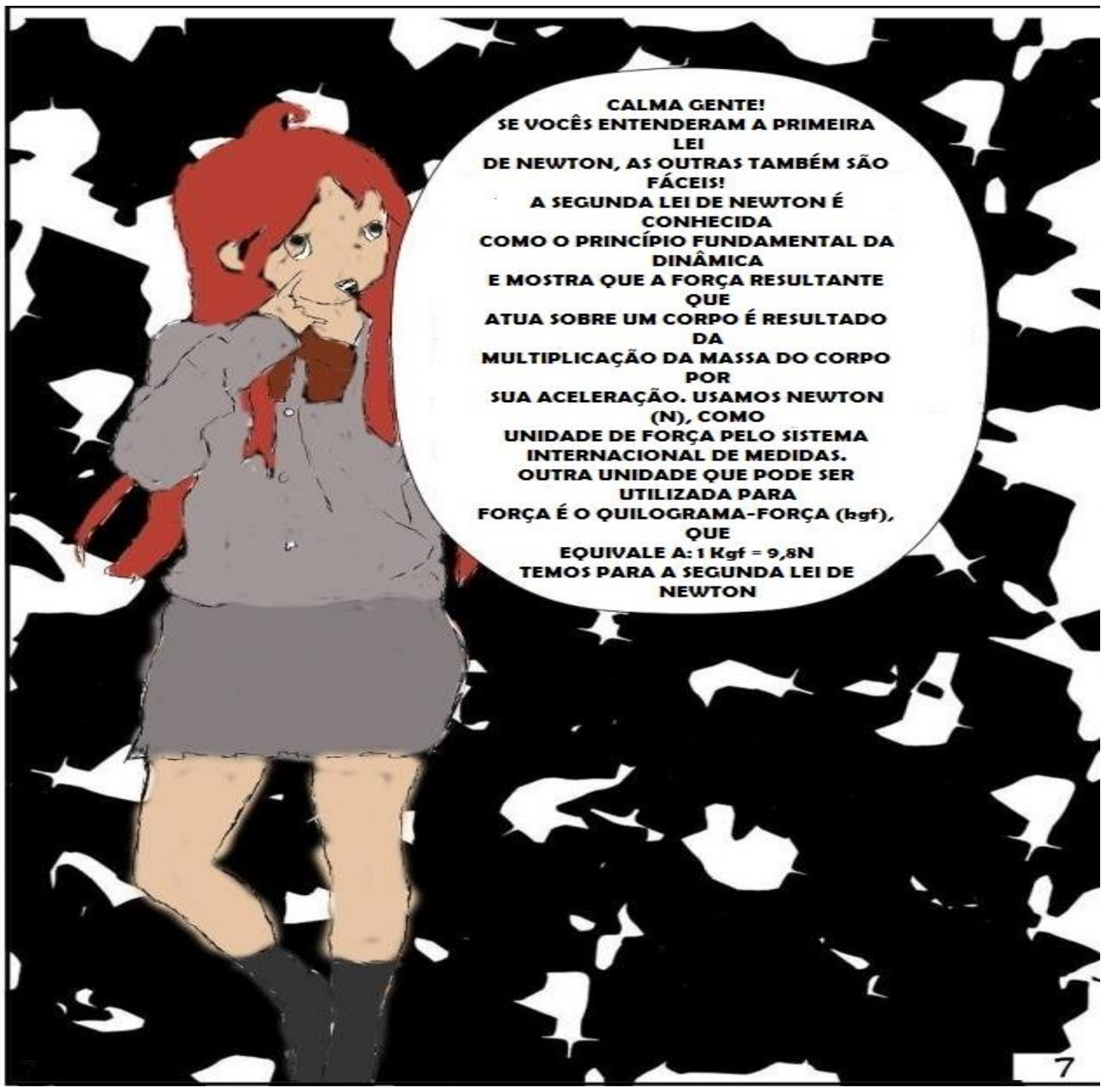

Fonte: Elaborado pela autora (2018)

Figura 35 – Acionamento do AR-Physics: página 7, 2º Gif, cão exercendo força sobre uma massa

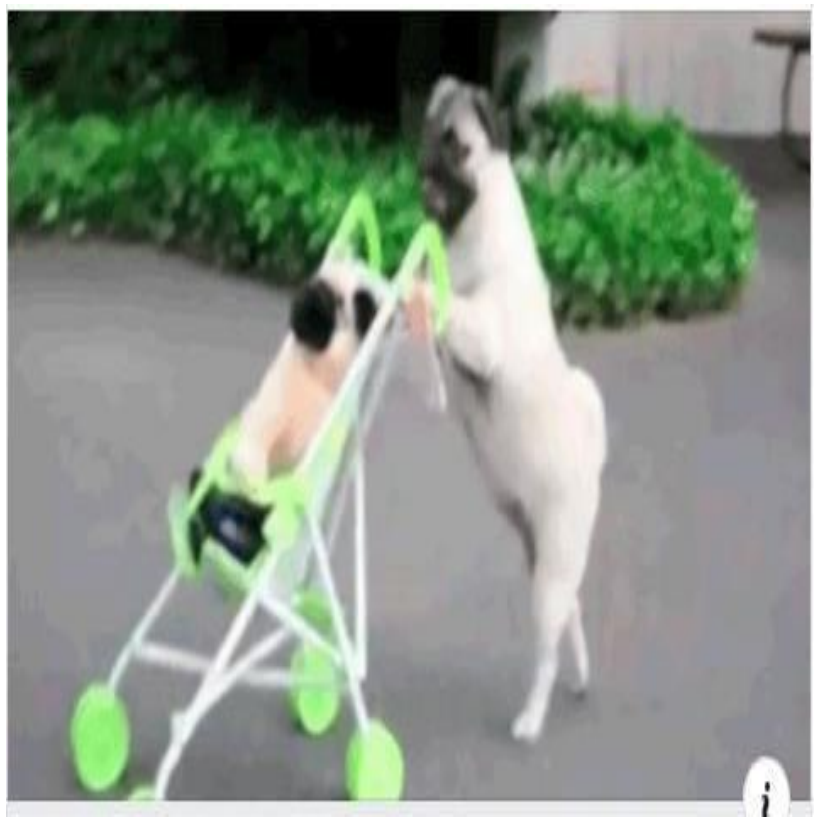

Fonte: YOUTUBE (2012)

A página 7 apresentou um Gif (Figura 35) em que um cãozinho empurra um carrinho de bebê. Neste Gif, o cãozinho exerce uma força no carrinho, mantendo-o em movimento retilíneo e uniforme. A imagem foi escolhida para conduzir o olhar do aluno a perceber que a força de atrito contrabalanceia a força exercida pelo cãozinho, demonstrando a segunda Lei de Newton.

Figura 36 – Resposta grupo 7: 2ª lei de Newton

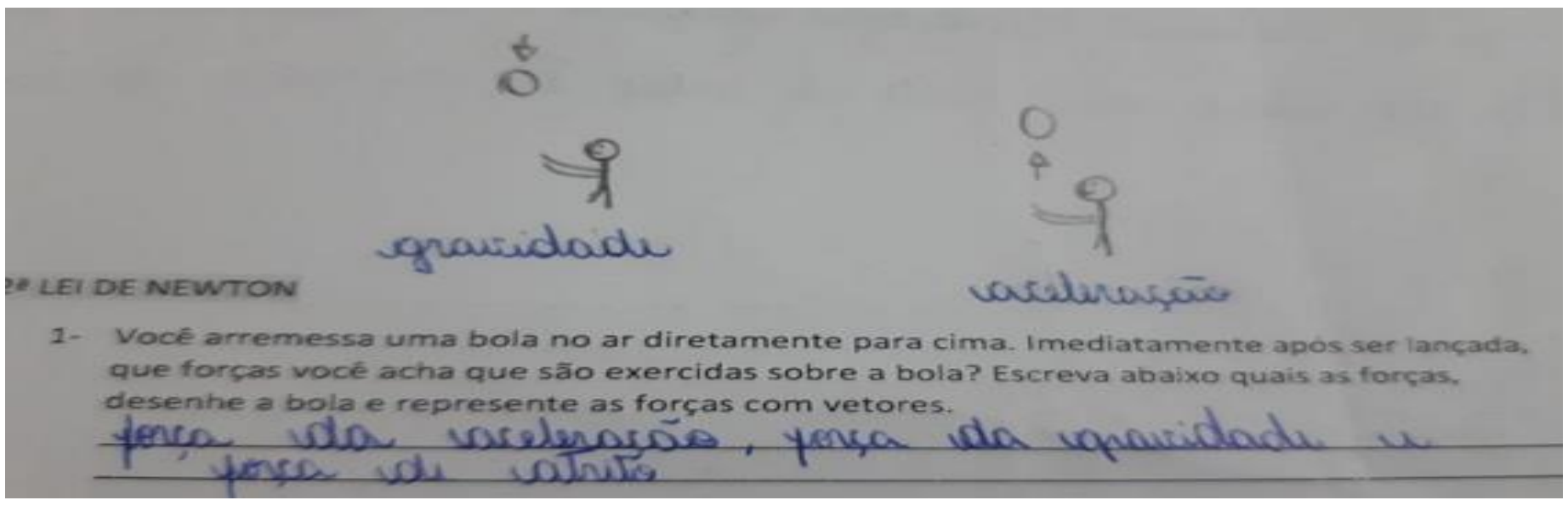

Transcrição: “força da aceleração, força da gravidade e força de atrito”.

Fonte: Elaborado pela autora (2018)

Na Figura 36, observamos que o grupo 7 não conseguiu ligar a aceleração à força da gravidade, pois o desenho está com a aceleração para cima e a gravidade para baixo. A construção da resposta está confusa, pois, pelas discussões feitas entre o grupo, escutavam-se certas dúvidas sobre o assunto. Houve dificuldades na interpretação teórica da pergunta, a representação do desenho mostra uma construção incompleta e, até mesmo, equivocada, pois o educando não conseguiu compreender que a aceleração aponta na mesma direção, e sentido, da força resultante. A confirmação da falta de conhecimentos prévios demonstra que a revista, ainda, não ofereceu informações suficientes para ajudar a responder corretamente à pergunta ou a pergunta poderia ter sido melhor escolhida para essa interpretação para o grupo 7.

Temos a resposta à outra pergunta, agora pelo grupo 2, que demonstra ganho de conhecimento.

Figura 37 – Resposta do grupo 2: 2ª lei de Newton

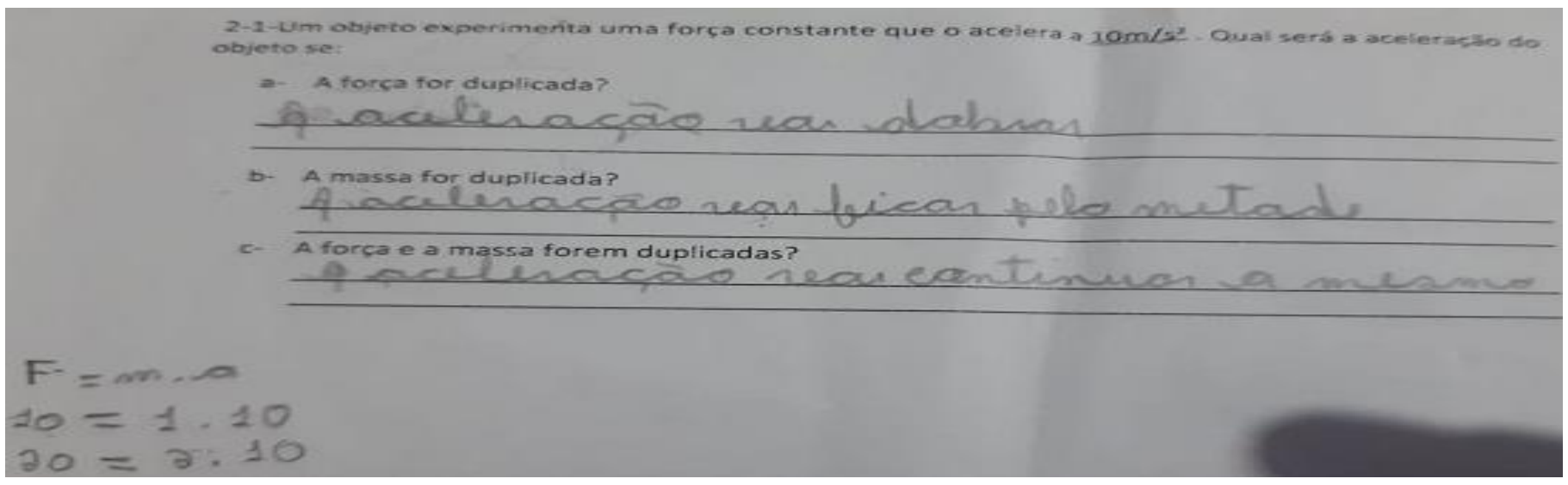

Transcrição:

a – “A aceleração vai dobrar”

b – “A aceleração vai ficar pela metade

c – “A aceleração cai continuar a mesma

F=m.a

10= 1.10

20=2.10”

Fonte: Elaborado pela autora (2018)

Na resposta (Figura 37), o grupo 2 conseguiu interpretar a pergunta de uma forma mais concisa, observamos um entendimento científico na relação entre força, massa e aceleração. O grupo, porém, não conseguiu associar a direção da aceleração com a gravidade.

Temos a resposta do grupo 2 (Figura 38), que também apresenta avanços no conhecimento, tendo em vista a interpretação do Gif feita pelo grupo.

Figura 38 – Interpretação do grupo 2, segundo Gif (Figura 35), 2ª Lei de Newton

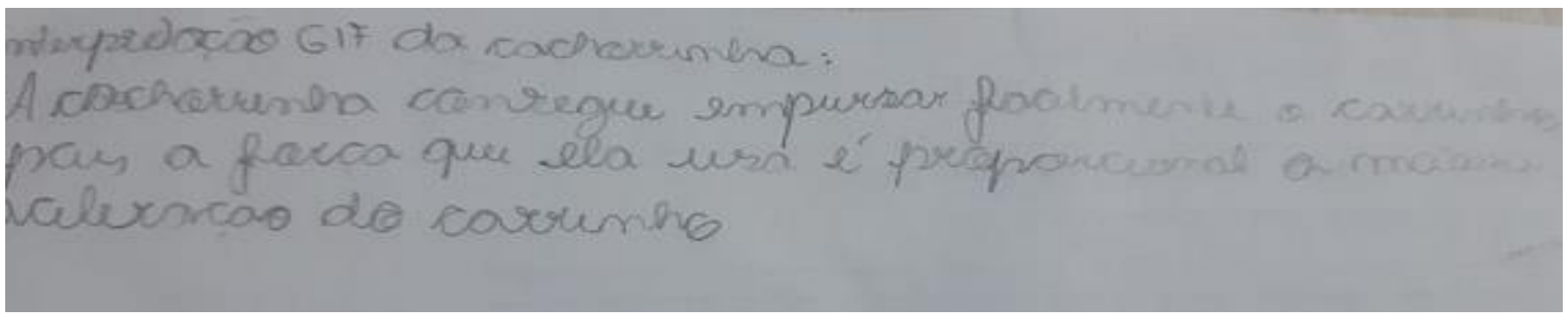

Transcrição: “Interpretação GIF da cachorrinha: A cachorrinha consegue empurrar facilmente o carrinho mais a força que ela usa é proporcional a massa e aceleração do carrinho”.

Fonte: Elaborado pela autora (2018)

Analisamos a página 8 (Figura 39) e o segundo Gif apresentado na 2ª lei (Figura 40), no qual um senhor aplica uma força sobre um veículo, mas sua força não é suficiente para mover o veículo devido ao atrito. Como a massa é muito grande, ele utiliza patins hidráulicos usados em estacionamentos, não sendo necessário ligar o motor, podendo, assim, estacionar veículos em locais com pouco espaço, reduzindo a força de atrito, vencendo a inércia, colocando o veículo em movimento.

Figura 39 – Página 8 da revista: exemplo dois da 2ª lei de Newton

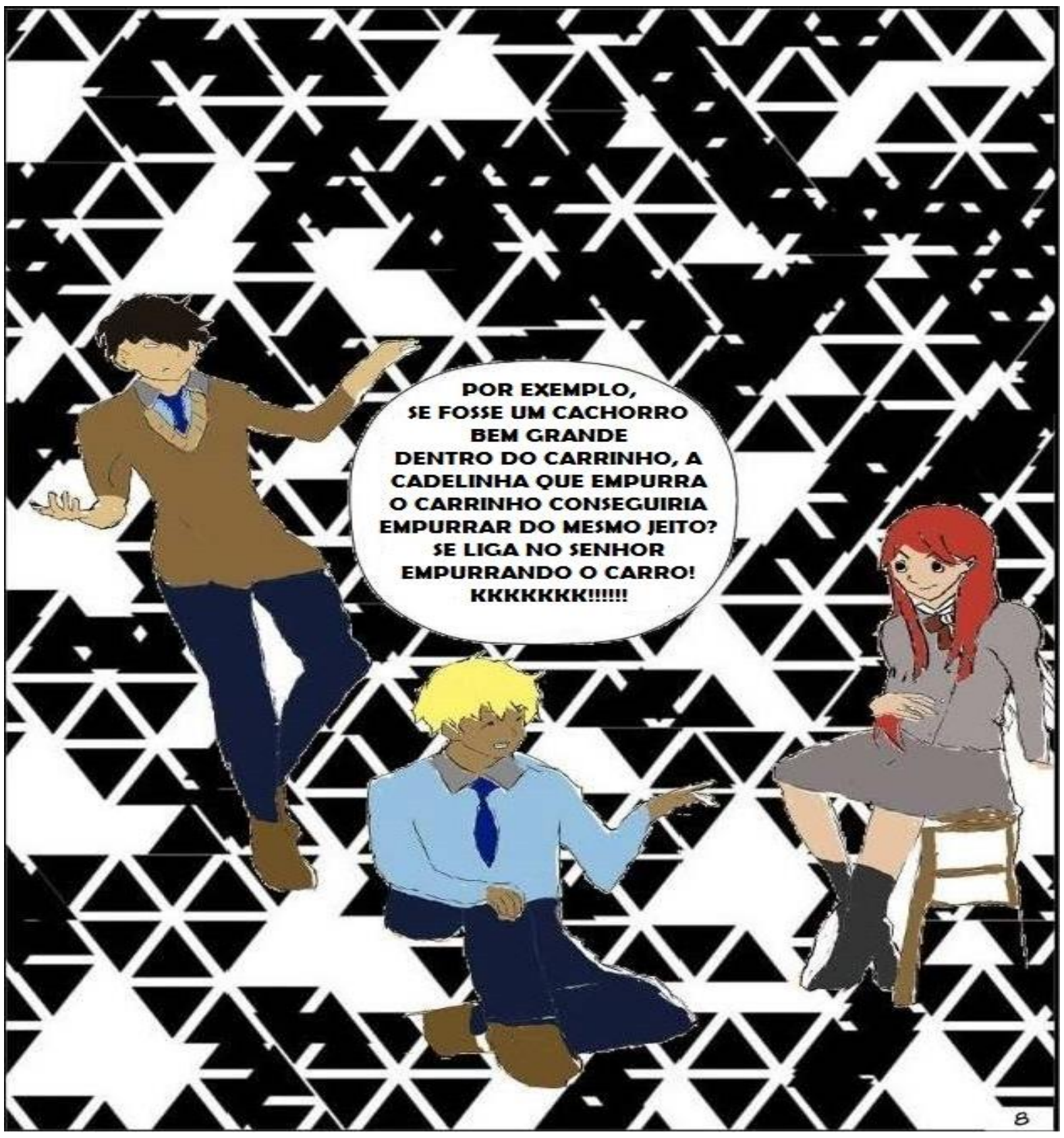

Fonte: Elaborado pela autora (2018)

Figura 40 – Acionamento do AR-Physics na página 8 da revista: 3º Gif, 2ª lei de Newton

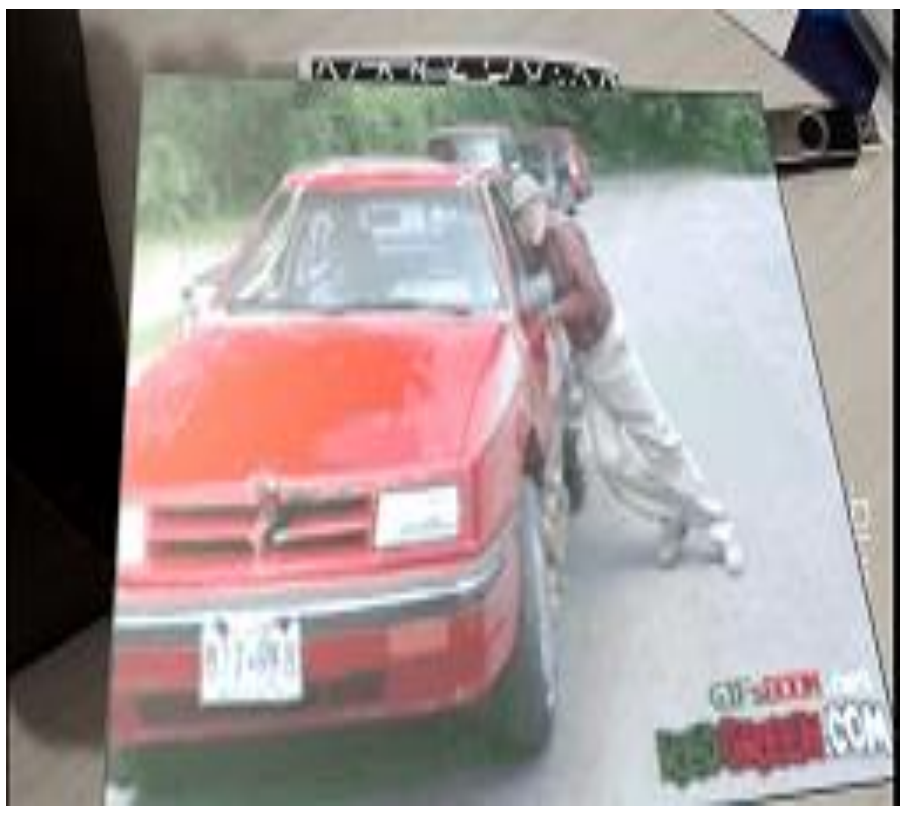

Fonte: GIPHY (2018)

Com relação ao terceiro Gif (Figura 40), temos a resposta do grupo 4, que interpreta a cena representada.

Figura 41 – Interpretação do grupo 4 para o terceiro Gif: 2ª lei de Newton

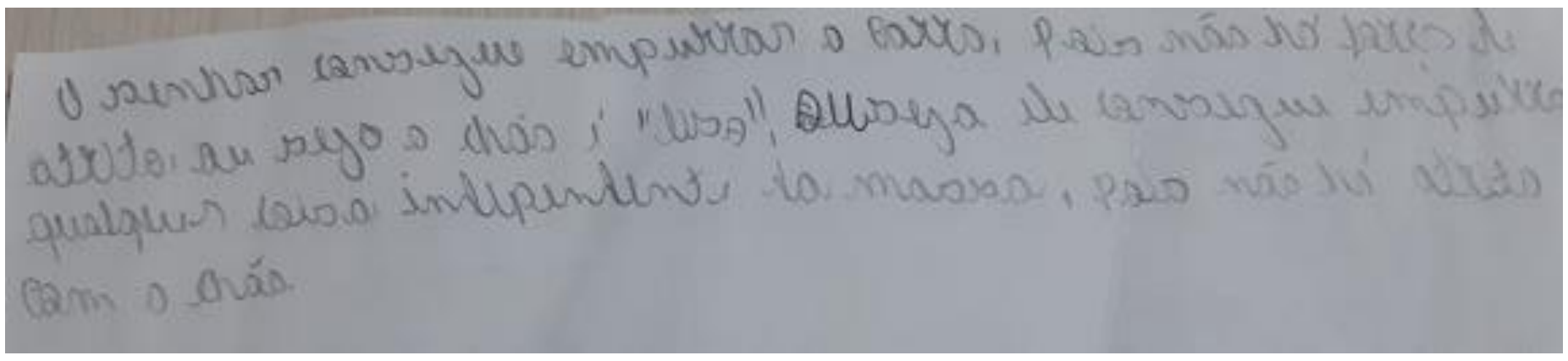

Transcrição: “O senhor consegue empurrar o carro, pois não há força de atrito, ou seja, o chão é “liso”, ou seja, ele consegue empurrar qualquer coisa independente da massa, pois não é atrito com o chão”.

Fonte: Elaborado pela autora (2018)

Na resposta do grupo 4, os alunos associaram um dispositivo à redução da força de atrito, porém a interpretação foi equivocada pois eles associaram a independência da massa à força de atrito estática, situação em que o senhor só precisa vencer a inércia. A resposta apresenta uma interpretação com ganho de conhecimento científico se comparada ao questionário de conhecimentos prévios.

Na segunda discussão da segunda lei de Newton, foram ligados dois Gifs diferentes para serem analisados, com a intenção que os alunos observassem os fenômenos e forças representados. As perguntas foram entregues aleatoriamente aos grupos. Além de responder à pergunta, os alunos também deveriam interpretar os dois vídeos (Figuras 35 e 40) explicando como a 2ª lei de Newton se aplica à cada situação.

Ao final da segunda aula (6ª aplicação), o grupo 7 apresentou a interpretação do Gif escolhido que foi o do cãozinho empurrando o carrinho. Interpretaram o Gif de maneira correta e discutiram essa interpretação com os demais colegas de sala. A professora interveio quando necessário para manter a exposição de forma correta na apresentação do grupo.

Foram utilizadas, para essa atividade, 7 perguntas sorteadas para os 7 grupos. O grupo 1 respondeu à questão 3.2 do questionário da segunda Lei de Newton que apresentava gráficos para cálculo de força x aceleração. Optou por apresentar nesta dissertação apenas 2 perguntas com respostas conceituais pois a pesquisa teve a intenção de apresentar avanços no conhecimento do conceito ligado à observação dos fenômenos estudados.

## 8.7 TERCEIRA LEI DE NEWTON (8ª E 9ª APLICAÇÕES)

Na sequência, foi apresentada a terceira Lei de Newton a partir da revista e do acionamento de três Gifs. Os alunos fizeram a leitura das páginas e os grupos voltaram a se reunir com os mesmos componentes com que trabalharam nas aulas anteriores. As perguntas sobre a terceira lei de Newton (Apêndice 5) foram cortadas em tiras e cada grupo pegou uma para responder. A professora instruiu que eles deveriam discutir essas perguntas entre o grupo para formularem as respostas de acordo com a interpretação e aprendizagem das referidas páginas estudadas para essa atividade. A sequência de leitura fora nas páginas 9, 10 e 11 do livro.

Na página 9 (Figura 42) temos a apresentação da Terceira Lei de Newton com um exemplo sobre lançamento de foguetes. Após a leitura, temos a apresentação do Gif, onde constatamos o lançamento de um foguete usando o princípio da ação e reação. Nesse momento, os alunos deveriam observar o lançamento e ler a explicação na revista.

Figura 42 – Página 9 da revista

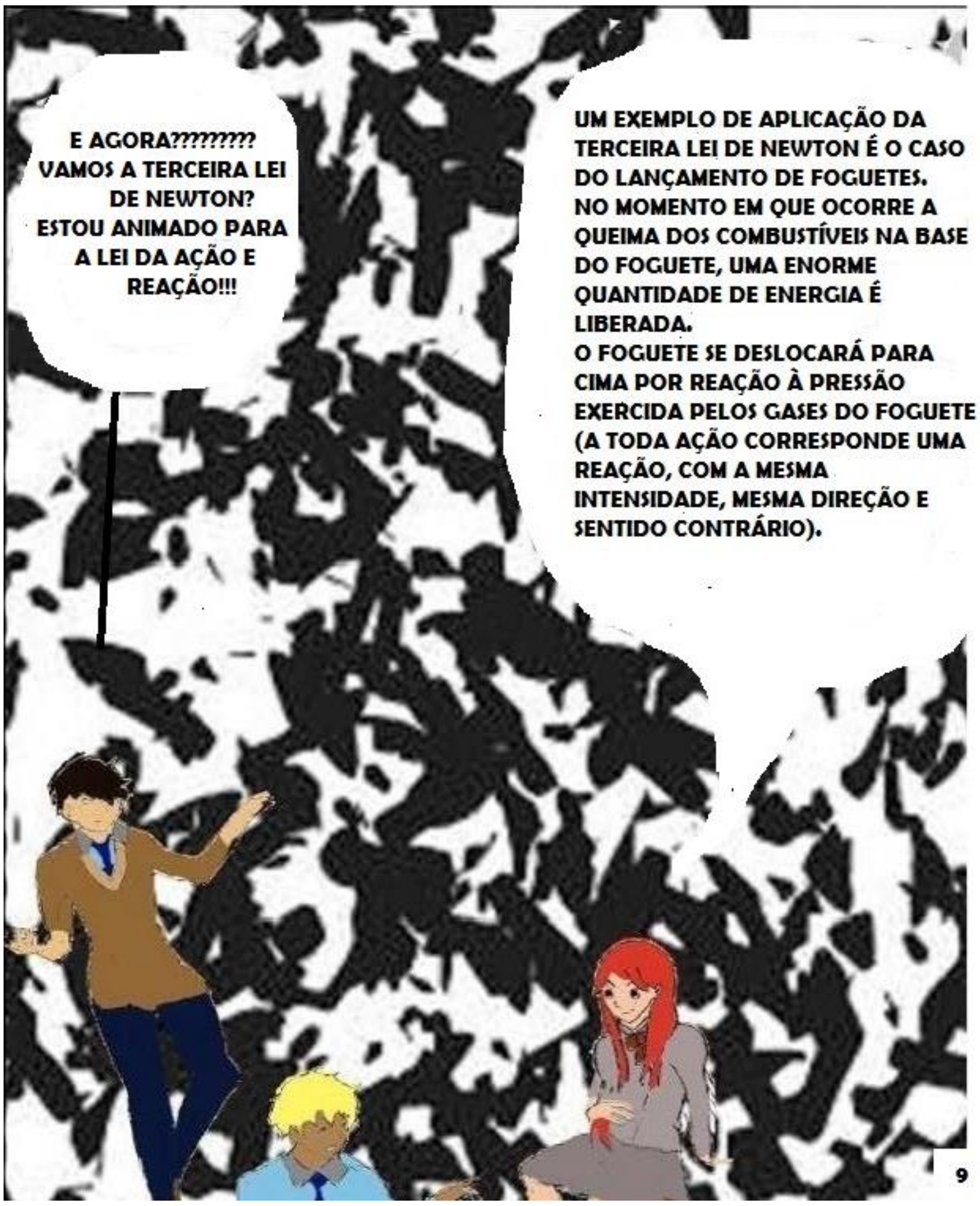

Fonte: Elaborado pela autora (2018)

Figura 43 – Acionamento do AR-Physics página 8 da revista, 4º Gif, 2ª lei de Newton, foguete decolando

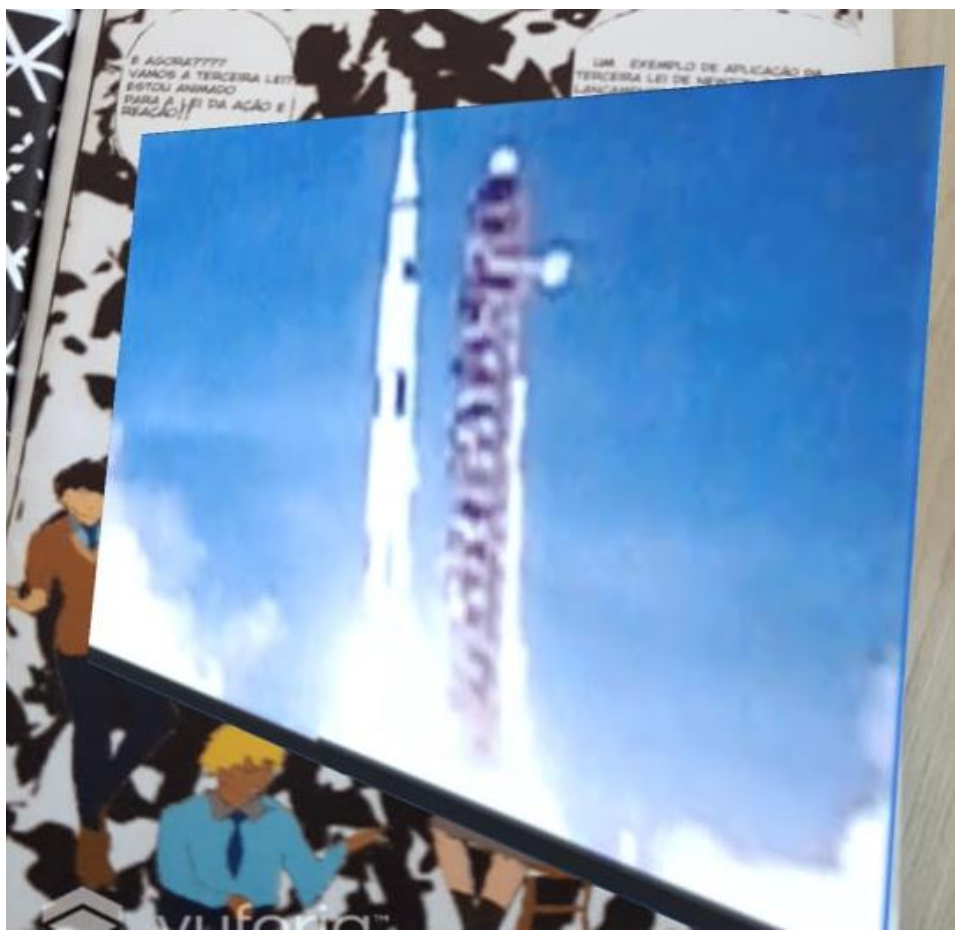

Fonte: HECRIPTUS (2018)

Após a apresentação do Gif (Figura 43), a professora pediu que os alunos produzissem um texto para relacionar o conteúdo da 3ª de Newton com os vídeos.

Temos, na Figura 44, o texto com registro do grupo 6, cuja interpretação demonstra avanços no raciocínio, quando comparado ao questionário de conhecimentos prévios. Ele apresenta uma resposta bem elaborada, o que pode ser analisado como um ganho de aprendizagem, já que nas perguntas da aplicação da 1º lei de Newton os grupos tinham dificuldades para interpretar e para organizar respostas que deveriam ser formuladas.

Figura 44 – Resposta do grupo 6, 4º Gif (Figura 43)

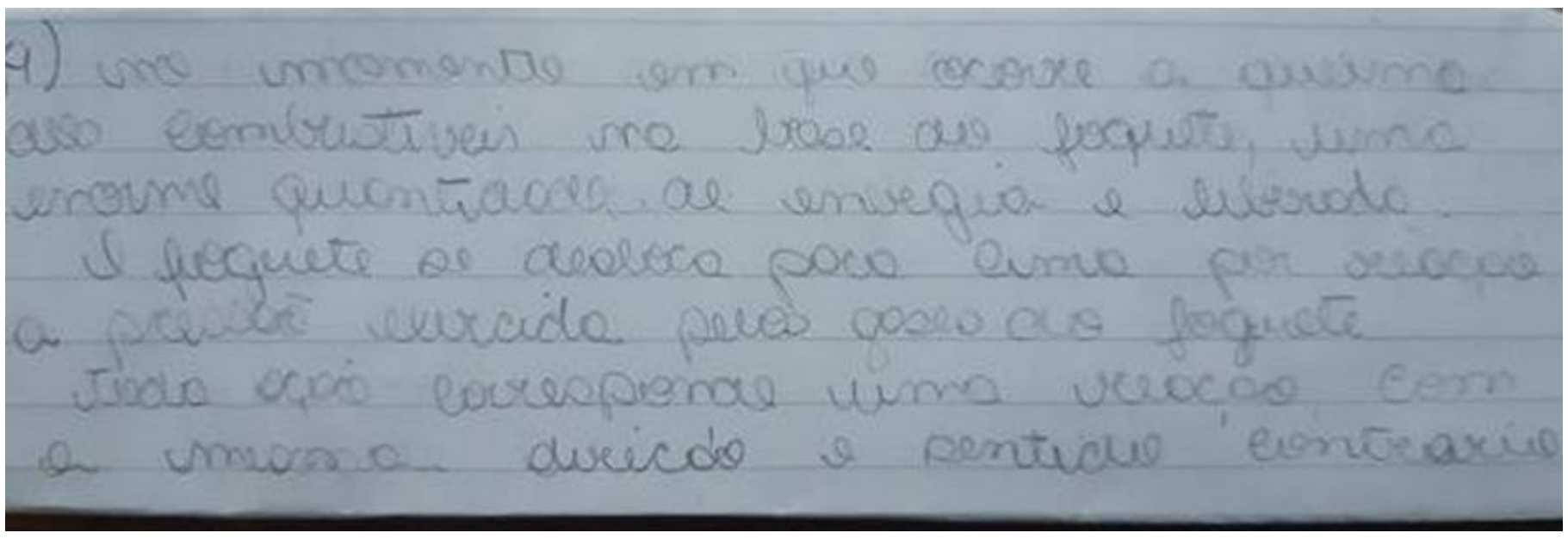

Transcrição: "no momento em que ocorre a queima dos combustíveis na base do foguete, uma enorme quantidade de energia e liberada. O foguete se desloca para cima por reação a pressão exercida pelos gases do foguete. Toda ação corresponde uma reação com a mesma direção e sentido contrário".

Fonte: Elaborado pela autora (2018)

Nessa exposição (Figura 44) os alunos do grupo 6 utilizam dados da revista e a observação do Gif (Figura 43), para redigir o texto pedido.

Seguindo a sequência didática, a professora pediu para que os alunos lessem a página 10 da revista onde temos outro exemplo da terceira lei de Newton exemplificada por uma bala explodindo na parede de concreto (Figura 46).

A personagem da revista (Andressa), explica a ação da bala e demonstra como a terceira lei de Newton se aplica. Desta maneira os alunos podem observar no Gif (Figura 45), o exemplo apresentado na prática.

Figura 45 – Página 10 da revista

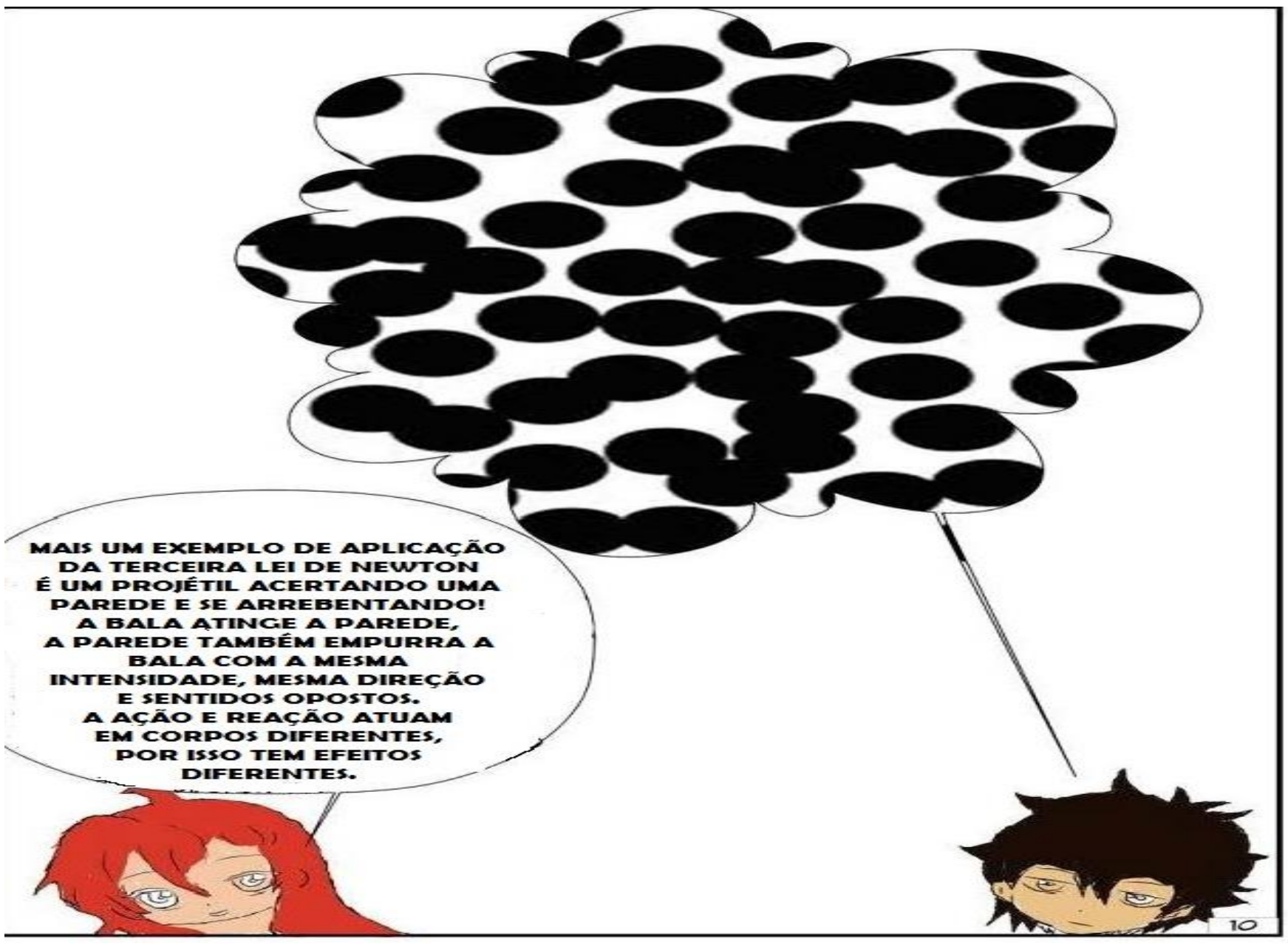

Fonte: Elaborado pela Autora (2018)

Figura 46 – Acionamento do aplicativo AR-Physics página 10, 5º Gif “bala explodindo na parede”

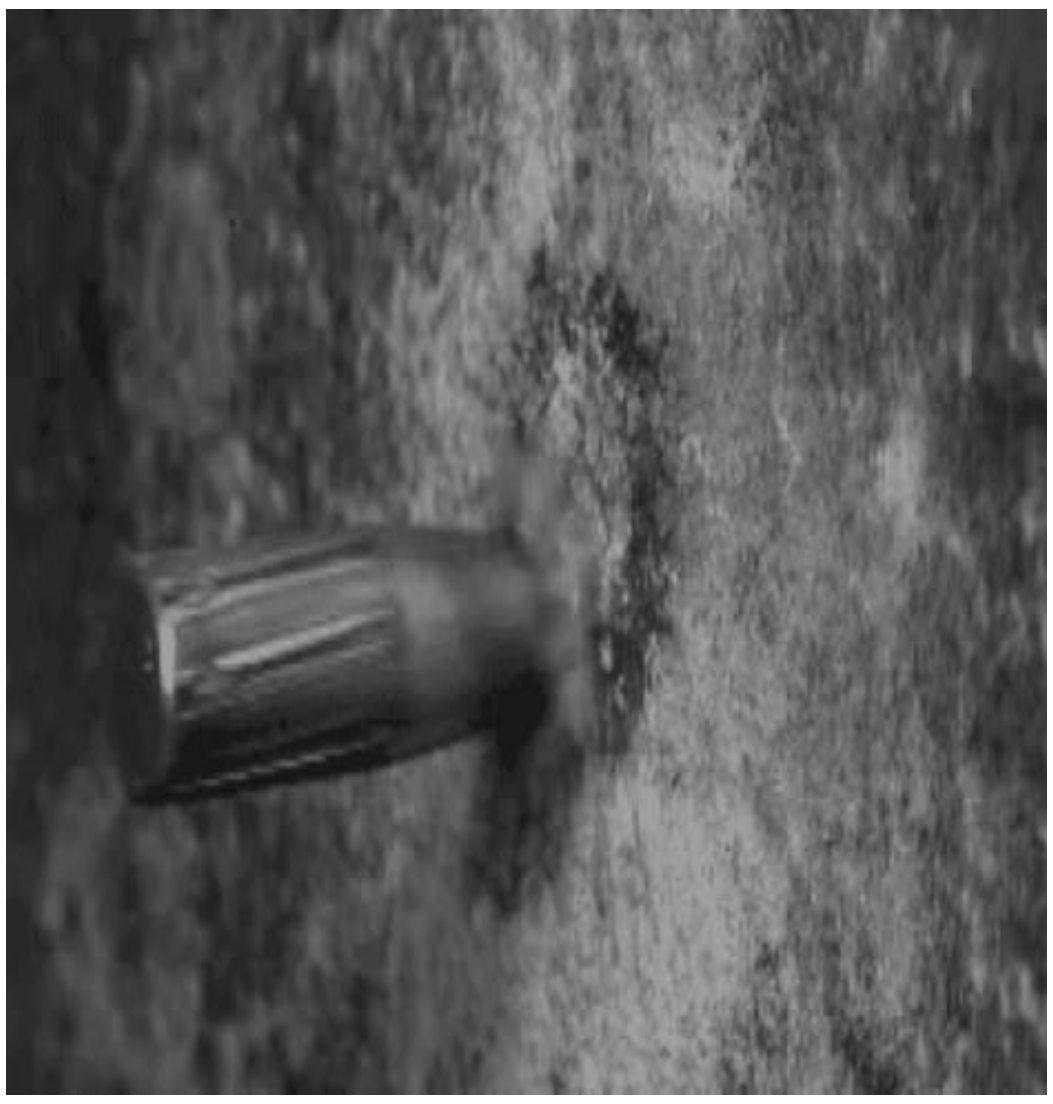

Fonte: MEDEIROS (2018)

A Figura 47 representa a exposição do grupo 1, que descreve de maneira correta o impacto da bala na parede, conforme a Terceira Lei de Newton, delineando as forças de ação e reação, demonstrando conhecimento adquirido.

Figura 47 – Apresentação do grupo 1: descrição do 5º Gif (Figura 46)

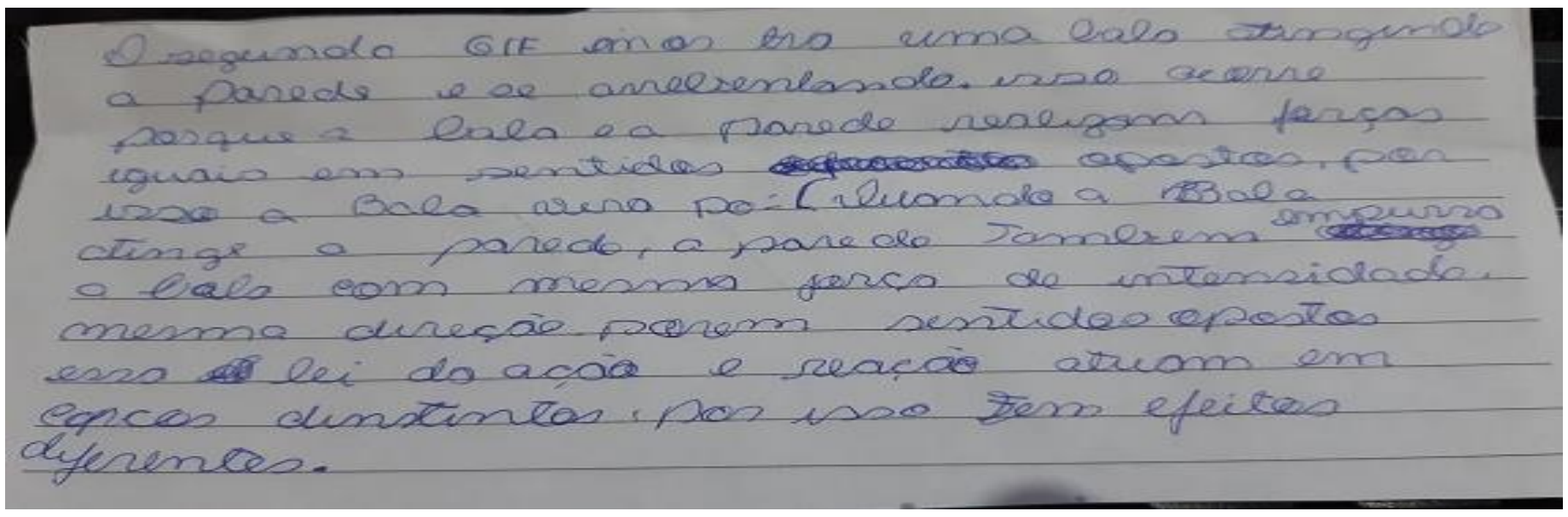
O segundo GIF mostra uma bala atingindo
a parede e se arrebentando. isso ocorre
porque a bala e a parede realizam forças
iguais em sentidos opostos, por
isso a Bala vira pó. (quando a Bala
atinge a parede, a parede Tambem empurra
a bala com mesma força de intensidade.
mesma direção porem sentidos opostos
essa lei da ação e reação atuam em
corpos distintos. por isso tem efeitos
diferentes.

Transcrição: ”O segundo Gif mostra uma bala atingindo a parede e se arrebentando. Isso ocorre porque a bala e a parede realizam forças iguais em sentidos opostos, por isso a bala vira pó (quando a bala atinge a parede também empurra a bala com a mesma força de intensidade mesma direção, porém sentidos opostos essa lei da ação e reação atuam em corpos distintos, por isso tem efeitos diferentes”.

Fonte: Elaborado pela autora (2018)

O grupo 1 descreve a 3ª lei de Newton interpretando o Gif da bala (Figura 46) de maneira correta, descrevendo as forças de mesma intensidade e mesmo sentido, em corpos distintos. A resposta é mais científica o que demonstra uma diferença do entendimento das forças de ação e reação apresentadas no primeiro questionário de conhecimentos prévios.

Na página 11 da revista (Figuras 48 e 49), temos a demonstração de mais um exemplo de aplicação prática da terceira Lei de Newton. No Gif, o homem recebe a ação do balão cheio de água em seu rosto (Figura 49). Os alunos deveriam identificar a 3ª lei de Newton nessa imagem.

Figura 48 – Página 11 da revista

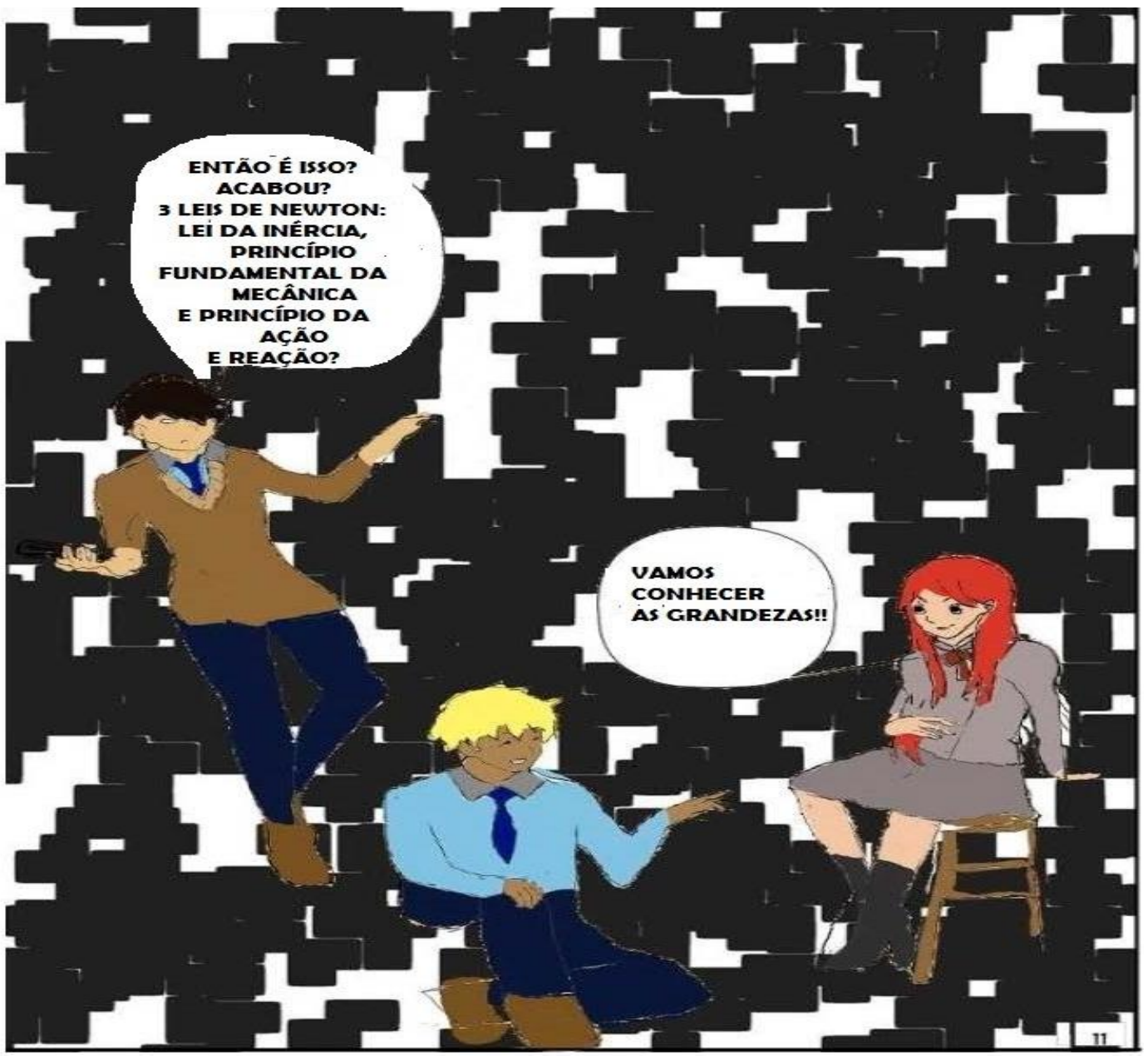

Fonte: Elaborado pela autora (2018)

Figura 49 – Acionamento do AR-Physics página 11 da revista, 3ª lei de Newton, 6º Gif “balão na cara”

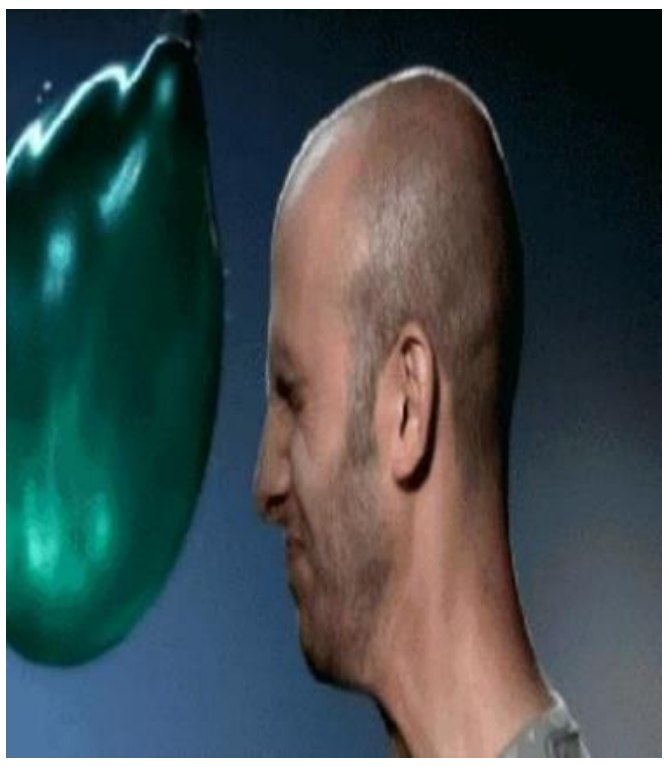

Fonte: MENDE (2019)

A resenha do grupo 3 apresenta o princípio de ação e reação de acordo com o relato (Figura 50). O grupo entende o contexto da ação e da reação, explicitando que as forças são exercidas em corpos diferentes. A ideia é mais científica, o que demonstra diferença comparado ao questionário de conhecimentos prévios (Apêndice 1). Nessa fase, os grupos já conseguem observar a ação e a reação em corpos diferentes, conforme consta do relato escrito a seguir.

Figura 50 – Relato grupo 3: descrição do 6º Gif (Figura 49)

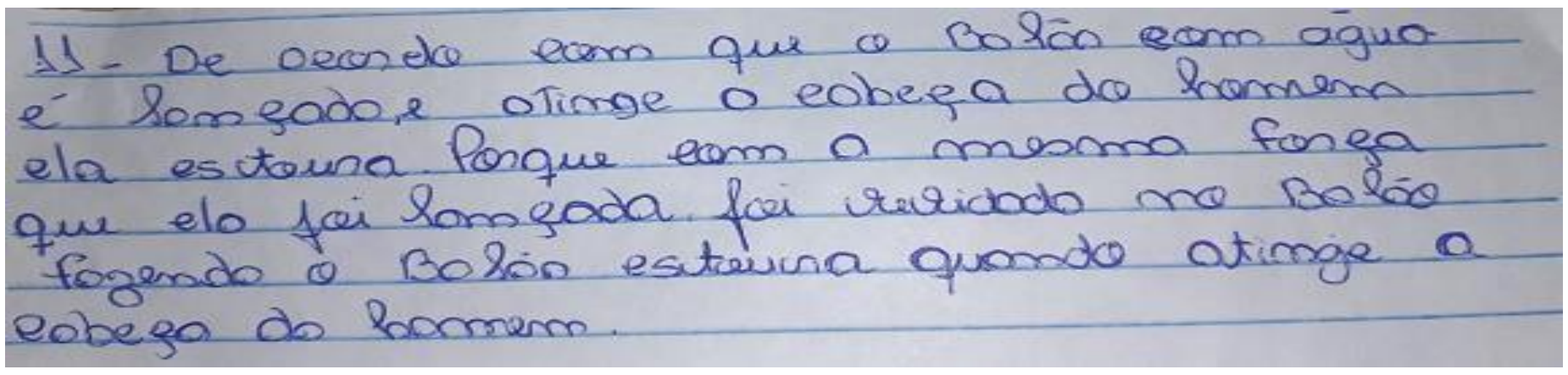

Transcrição: “De acordo com que o balão com água é lançado e atinge o cabeça do homem ela estoura. Porque com a mesma força que ela foi lançada foi revidado no balão fazendo o balão estoura quando atinge a cabeça do homem”.

Fonte: Elaborado pela autora (2018)

Como na interpretação dos outros dois Gifs referentes à terceira lei de Newton (Figuras 43 e 46), a exposição é mais clara, o que demonstra ganho de aprendizado.

Após a leitura das páginas, além de procederem às interpretações dos Gifs apresentados, a professora pediu para que os alunos respondessem às perguntas sorteadas. Temos na Figura 51 uma pergunta respondida pelo grupo 6.

Figura 51 – Pergunta respondida pelo grupo 6

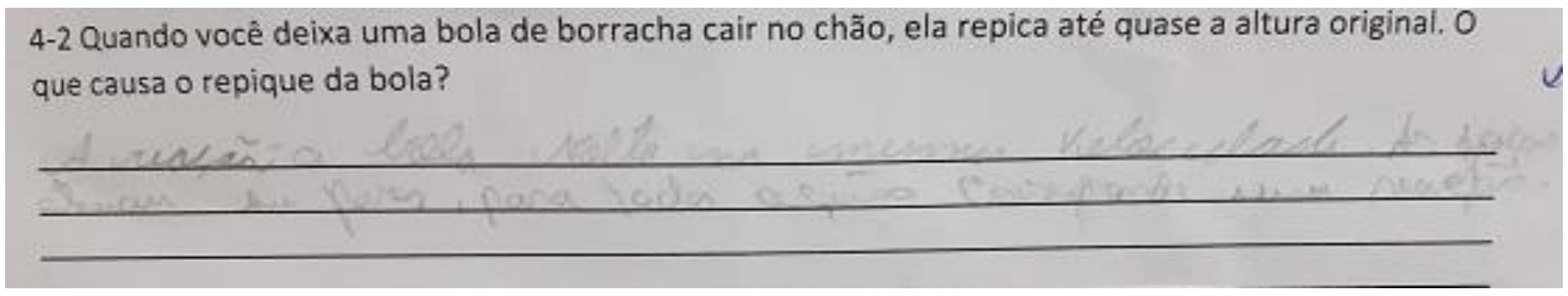

Transcrição “A reação a bola volta na mesma velocidade. As forças atuam em pares, para toda ação corresponde uma reação”.

Fonte: Elaborado pela autora (2018)

A resposta do grupo 6 apresenta o conceito de ação e reação e explicita a atuação em pares. Isso demonstra um crescimento no conhecimento se comparado às respostas do questionário inicial (Apêndice 1).

Se observarmos as respostas apresentadas, verificaremos que o conhecimento científico, embora incompleto, aparece sempre nas respostas, sinalizando avanços no conhecimento científico, comparando-se com o primeiro questionário (Apêndice 1).

Os alunos relacionaram as respostas do questionário aos Gifs observados. Essa habillidade cognitiva foi percebida em algumas respostas dos grupos como na Figura 52. A resposta da pergunta mostra uma certa ligação com o Gif.

Figura 52 – Resposta do grupo 6: 3ª lei de Newton

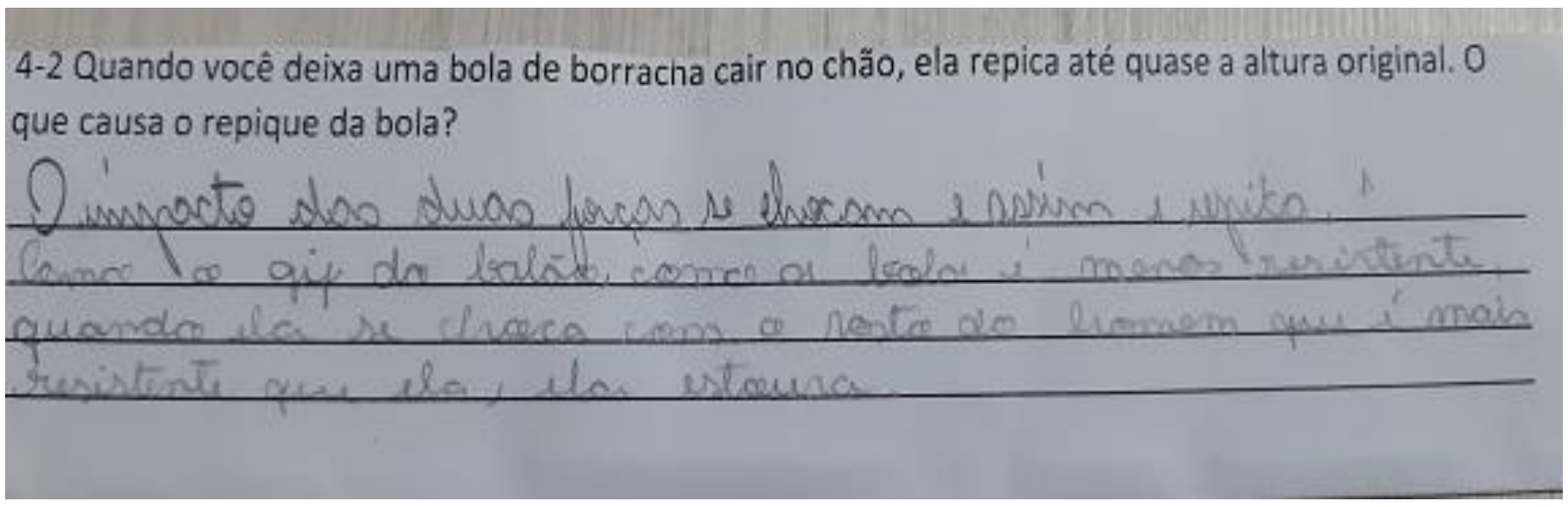
4-2 Quando você deixa uma bola de borracha cair no chão, ela repica até quase a altura original. O que causa o repique da bola?

O impacto das duas forças se chocam e assim e repica. Como o gif do balão, como a bola é menos resistente, quando ela se choca com o rosto do homem que é mais resistente que ela, ela estoura

Transcrição: “O impacto das duas forças se chocam e assim e repica. Como o Gif do balão, a bola é menos resistente, quando ela choca com o rosto do homem que é mais resistente que ela, ela estoura”.

Fonte: Elaborado pela autora (2018)

A aplicação da dinâmica da 3ª lei de Newton também durou 2 aulas de 55 minutos cada (8ª e 9ª aplicações). Ao final da segunda aula (9ª aplicação), foi escolhido um grupo para falar sobre a pergunta sorteada e sobre a interpretação do Gif à escolha do grupo.

Na 3ª lei de Newton, foi escolhido o Gif do balão na cara (Figura 49) e o grupo que apresentou seu resultado foi o grupo 02 que descreveu para toda turma a aplicação da 3ª lei de Newton na situação apresentada, demonstrando que a ação e a reação são em corpos diferentes. O grupo, durante a explicação, tentou estabelecer a ligação entre a cena e o conteúdo que estava sendo estudado. A pesquisadora interveio quando necessário, para que o conceito fosse socializado de maneira correta.

## 8.8 CONHECENDO AS FORÇAS (9ª MOMENTO)

Partindo do pressuposto que, ao apresentarmos as leis de Newton, iniciamos o estudo do movimento e suas causas, passamos a ajudar na observação do movimento dos corpos, buscando as causas e relacionando-as com suas consequências, para reconhecimento das aplicações no cotidiano do aluno. Nessa etapa, os grupos analisaram e discutiram em partes, respondendo a perguntas e fazendo relatórios sobre interpretações desenvolvidas por eles, aos novos Gifs visualizados.

Iniciamos com o conceito de força e, para sua representação, foi demonstrado um Gif do Capitão América segurando um Helicóptero (Figura 54). As explicações contidas no diálogo explicam o conceito que o aluno deve reconhecer no Gif e descrever interpretando o fenômeno de acordo com a teoria.

Figura 53 – Página 12 da revista

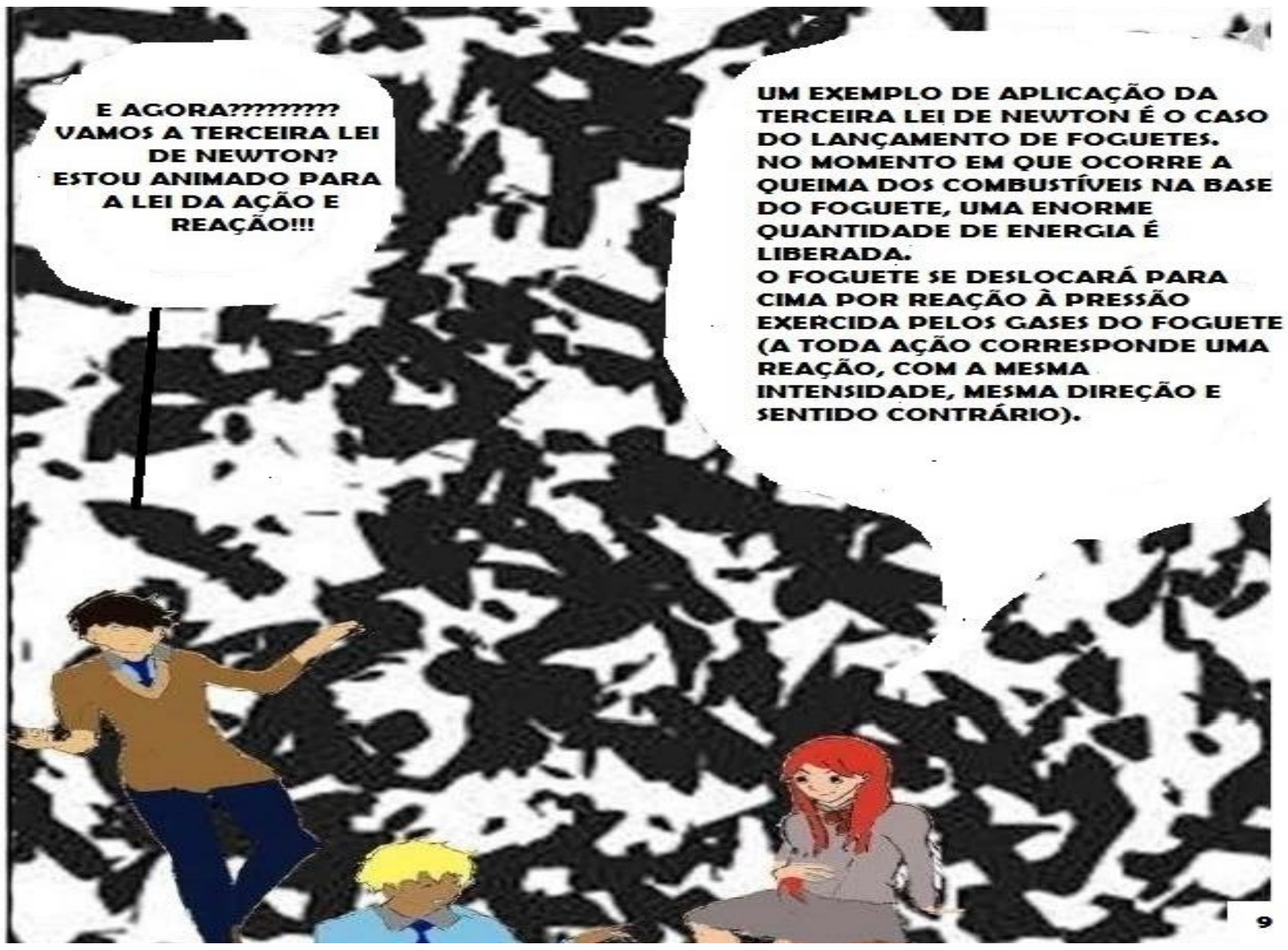

Fonte: Elaborado pela autora (2018)

Figura 54 – Acionamento do AR-Physics página 12: força, 7º Gif "Capitão América"

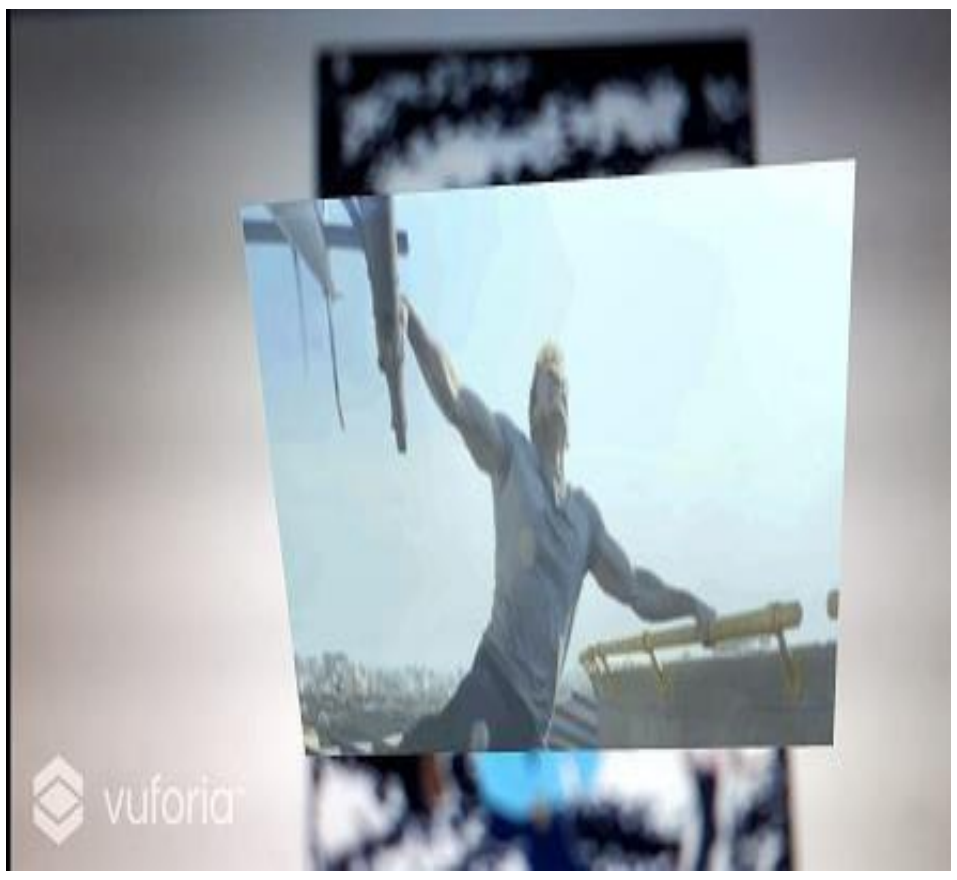

Fonte: BESTCLIPS (2018)

Nas Figuras 53 e 54 retiradas da revista, temos o conceito de força começando a ser explicado. O exercício no qual os alunos deviam interpretar o Gif (Figura 54), observamos avanços no processo de aprendizagem.

A seguir temos a Figura 55 com as respostas do grupo 1. Apesar de existir na resposta uma mistura referente ao conceito de força, ela define a força como conceito vetorial, mostrando avanço no conteúdo.

Figura 55 – Resposta dos alunos grupo 1

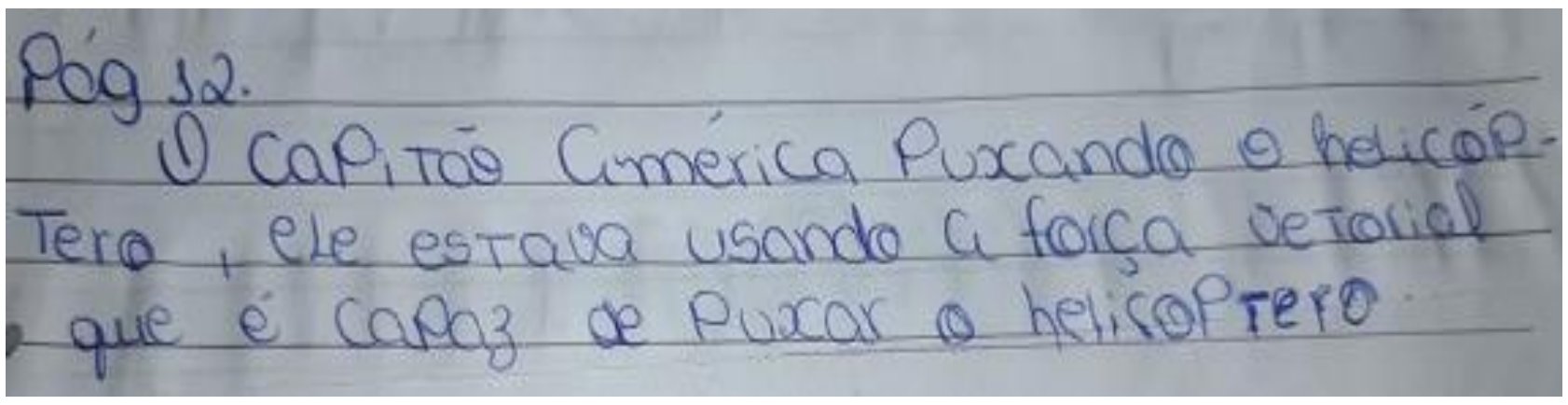
Pág 12.
O capitão América puxando o helicóp-
tero, ele estava usando a força vetorial
que é capaz de puxar o helicoptero.

Fonte: Elaborado pela autora (2018)

O fato de as forças serem grandezas vetoriais, com direção e sentido, e sua identificação na resposta demonstra que o aluno percebeu que, para que o Capitão América conseguisse puxar o helicóptero, essa força teria que ser definida com intensidade e direção. A resposta do grupo 1 (Figura 55) também demonstra que o Capitão América está usando a força vetorial, ou seja, essa intensidade pode ser representada pela força que está sendo empregada pelo Capitão América, o que apresenta ganho de aprendizagem. Podemos chegar a essa conclusão se compararmos com as respostas dadas ao questionário inicial, onde foi perguntado sobre conceito de força. Tivemos respostas como força de vontade e força espiritual, demonstrando pouco ou nenhum sentido científico.

### 8.8.1 Força resultante

Na exposição do conceito força resultante, os alunos deveriam entender que a força resultante é a soma vetorial de todas as forças presentes em um corpo. Ela pode ser igual a zero (vetor nulo – um ponto no sistema cartesiano) ou diferente de zero (um vetor força resultante). O Gif (Figura 57) ilustra a força resultante devido aos vetores força peso, força normal, força de atrito e força exercida pelo homem.

Figura 56 – Página 13 da revista

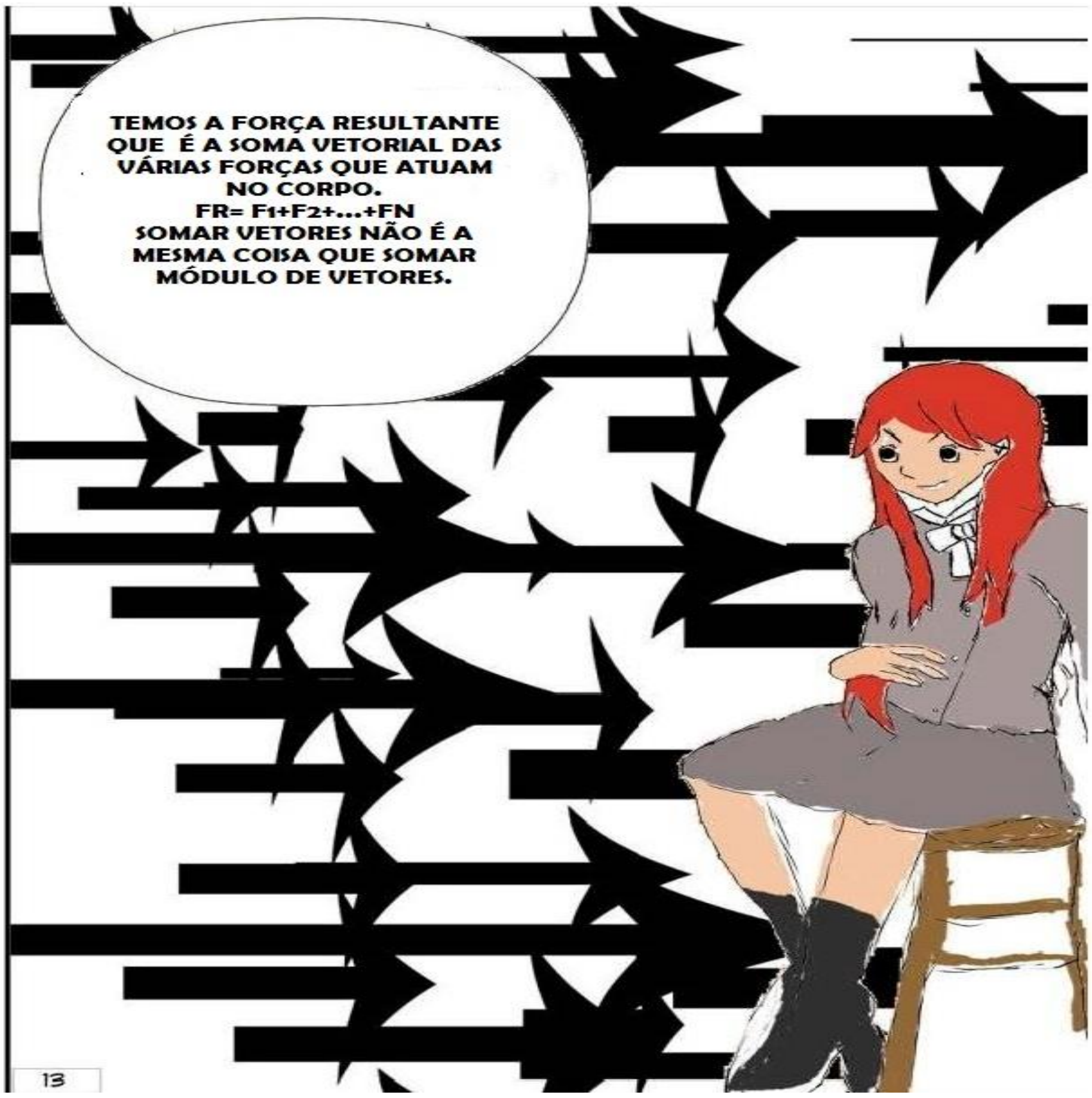

Fonte: Elaborado pela autora (2018)

Figura 57 – Gif acionado pelo aplicativo AR-Physics, página 13, 8º Gif “homem caminhando”

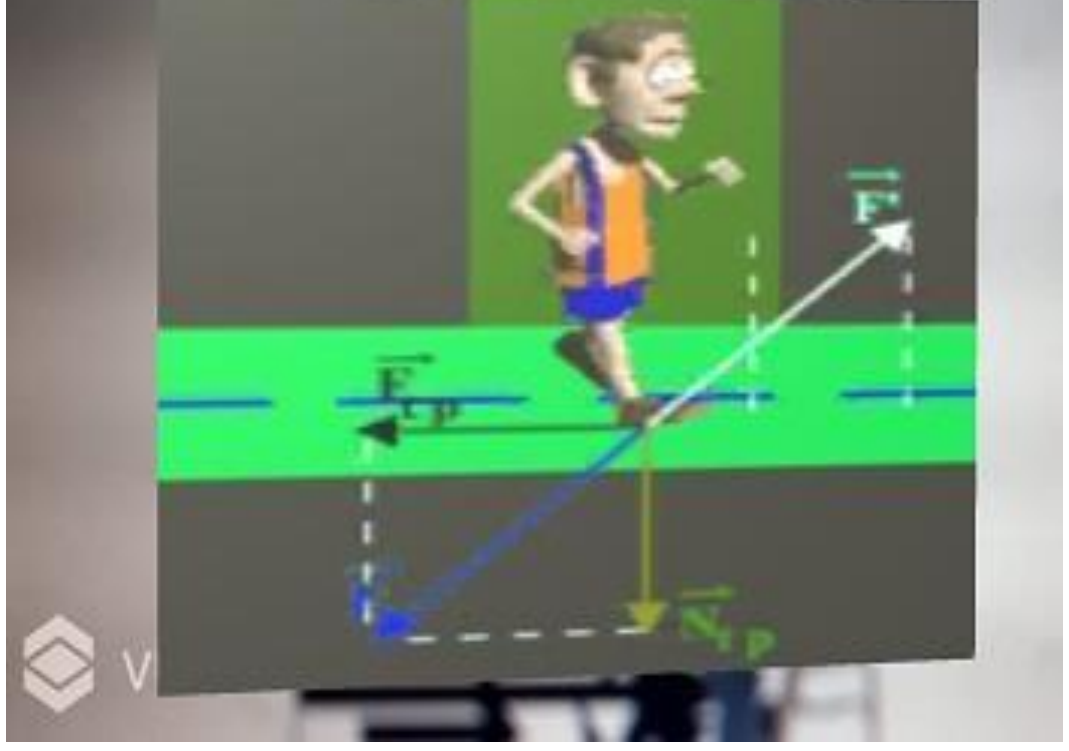

Fonte: GIFSDEFÍSICA (2018)

Os alunos deveriam interpretar o Gif demonstrando o conceito estudado e identificar todas as forças e, consequentemente, a força resultante. Após a aplicação foi percebido que havia um erro na posição dos vetores que não tinha sido analisado anteriormente. Esse erro se deve à análise vetorial feita através da superfície tocada pelo pé do homem e não pelo homem (objeto de estudo), caminhando. Por esse motivo essa análise foi invalidada pela pesquisadora.

### 8.8.2 Força peso (10ª aplicação)

Nesta seção, quisemos apresentar aos alunos o conceito de campo gravitacional no processo de queda livre. Eles deveriam chegar ao entendimento de que a força de atração gravitacional que age sobre o corpo no momento em que ele é abandonado próximo da superfície da Terra (força peso) é responsável pela aceleração que ele adquire durante a queda, que chamamos de aceleração da gravidade (Figura 58).

No 9º Gif representado na Figura 59 ilustramos o peso de um peixe, através do vetor peso, apontado para baixo. Os alunos devem observar essa indicação e descrever a força peso.

Figura 58 – Página 14 da revista

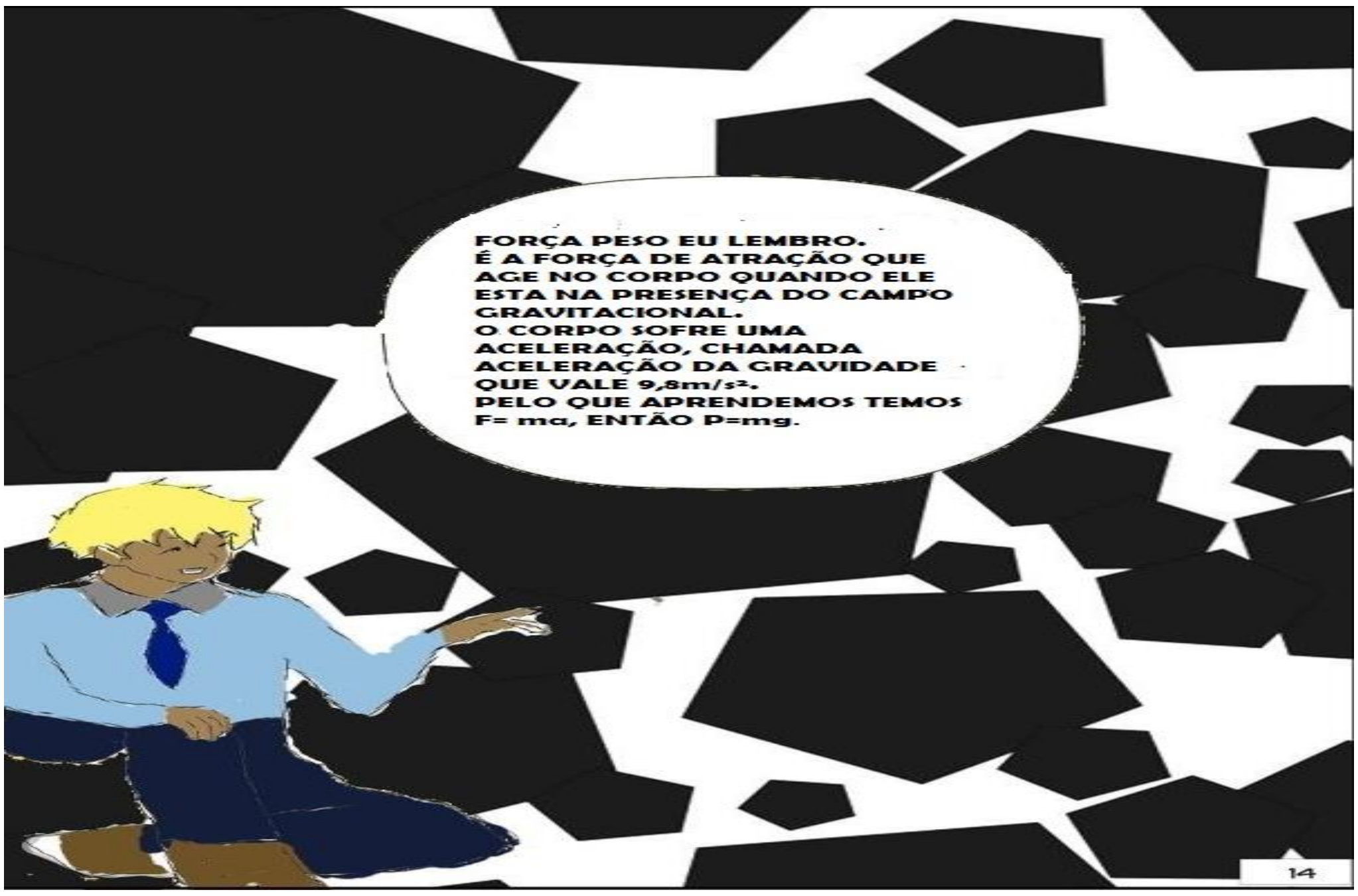

Fonte: Elaborado pela autora (2018)

Figura 59 – Gif acionado pelo aplicativo AR-Physics página 14: 9º Gif “peixe sendo pesado”

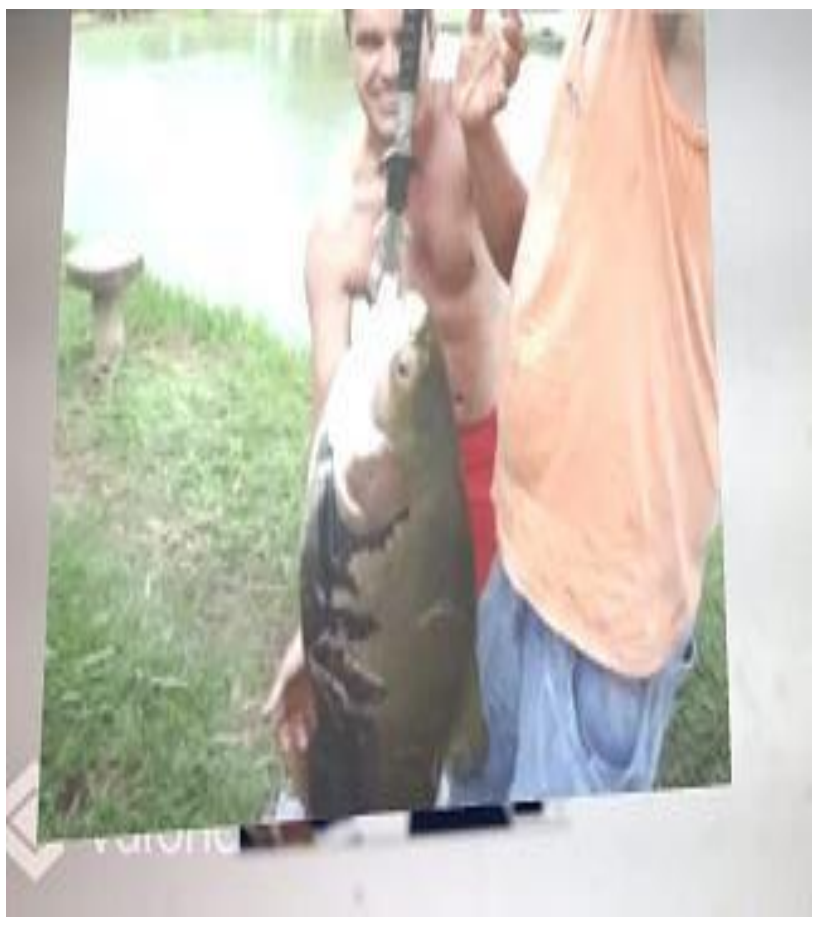

Fonte: OLIVEIRA (2018)

Em seguida, apresentamos a interpretação do grupo 7 sobre esse Gif (Figura 60). O grupo identificou a medida da força peso pelo dinamômetro, porém eles não mencionam o vetor peso.

Figura 60 – Resposta dos alunos grupo 7

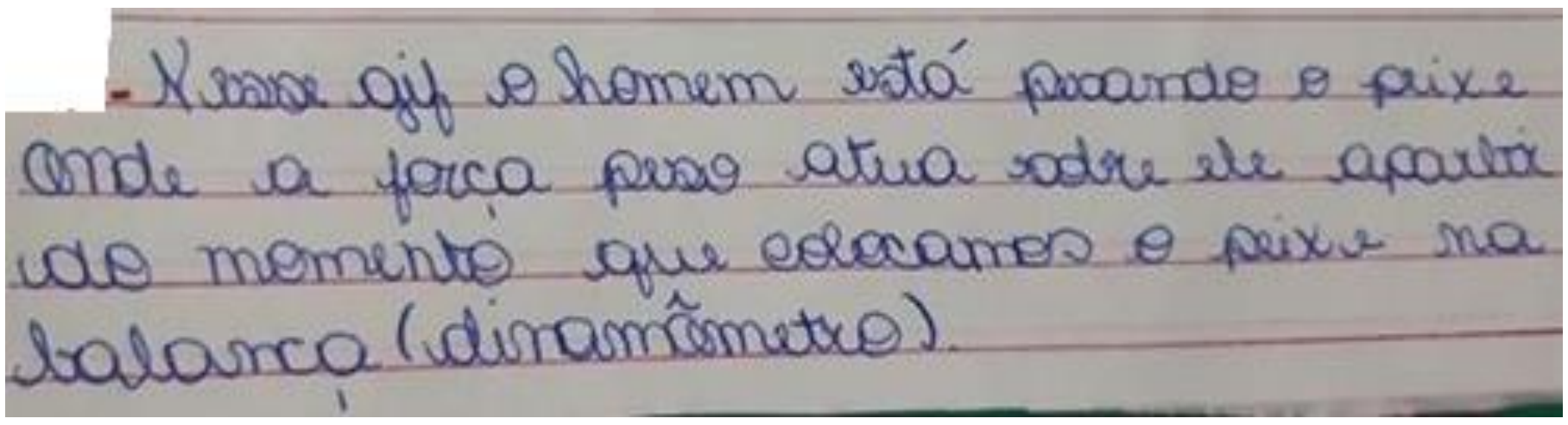

Transcrição: “Nesse Gif o homem está pesando o peixe onde a força peso atua sobre ele apartir do momento que colocamos o peixe na balança (dinamômetro)”.

Fonte: Elaborado pela autora (2018)

A descrição leva ao entendimento de que a força peso só atuaria no momento em que colocamos o peixe na balança e, por esse equívoco, a resposta foi redigida parcialmente correta.

Na Figura 61, do grupo 2, observamos que os alunos conseguem compreender o conceito de força peso, mas ainda não conseguem identificar sua origem, bem como sua característica vetorial. A descrição do grupo “a força de atração que age sobre um corpo quando ele é abandonado” apresenta uma informação incompleta, pois o peixe é atraído pela Terra mesmo que não esteja em queda livre.

Figura 61 – Resposta grupo 2: 9º Gif (Figura 59), força peso

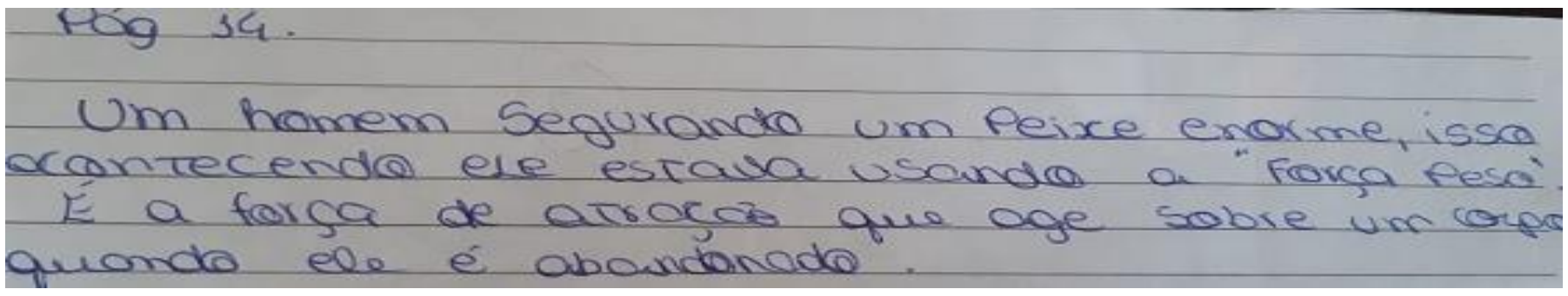
Pág 14.
Um homem Segurando um Peixe enorme, isso acontecendo ele estava usando a "Força Peso". É a força de atração que age sobre um corpo quando ele é abandonado.

Transcrição: “Um homem segurando um peixe enorme, isso acontecendo ele estava usando a “força peso”. É a força de atração que age sobre um corpo quando ele é abandonado”.

Fonte: Elaborado pela autora (2018)

### 8.8.3 Apresentação das forças de interação à distância e atrito (11ª aplicação)

Essa seção foi dividida na abordagem de dois conceitos diferentes. Na primeira parte (parte superior da página 15 da revista, Figura 62), temos uma figura 3D acionada pelo aplicativo AR Physis que apresenta a interação à distância da força gravitacional, na qual os alunos deveriam identificar essa força por meio de campos, sem que houvesse contato entre os corpos. Além da força gravitacional, foi explicado, também, que existem outras forças como, por exemplo, as interações magnéticas e as elétricas. Os alunos fizeram a leitura da seção da revista, observaram a Figura 3D e discutiram sobre o assunto por meio da entrega das interpretações.

Ao final das atividades, a pedido dos alunos, apresentamos a conclusão a que deveriam chegar, explicando sobre a órbita que a Lua faz em torno da Terra com velocidade escalar de aproximadamente 3,70 km/h, descrevendo uma trajetória elíptica estável devido à força gravitacional. Ao término da exposição, muitos alunos estavam satisfeitos por apresentarem resenhas bem próximas dos conceitos expostos pela professora. Uma aluna relatou: “eu sabia!! Liguei essa figura à aula que fizemos na sala de vídeo com o joguinho dos planetas. A Lua fica presa à Terra porque é atraída, tenta fugir, mas a força da gravidade não deixa”. Isso é uma evidência de que os alunos tiveram avanços conceituais durante a aplicação da sequência didática.

A segunda metade da página 15 apresenta a força de interação por atrito, que ocorre por meio do contato direto entre os corpos envolvidos. Foi usado um carro de corrida, onde as faíscas decorrentes do contato do assoalho do carro e o piso indicam a presença da força de atrito (Figura 64).

Figura 62 – Página 15 da revista

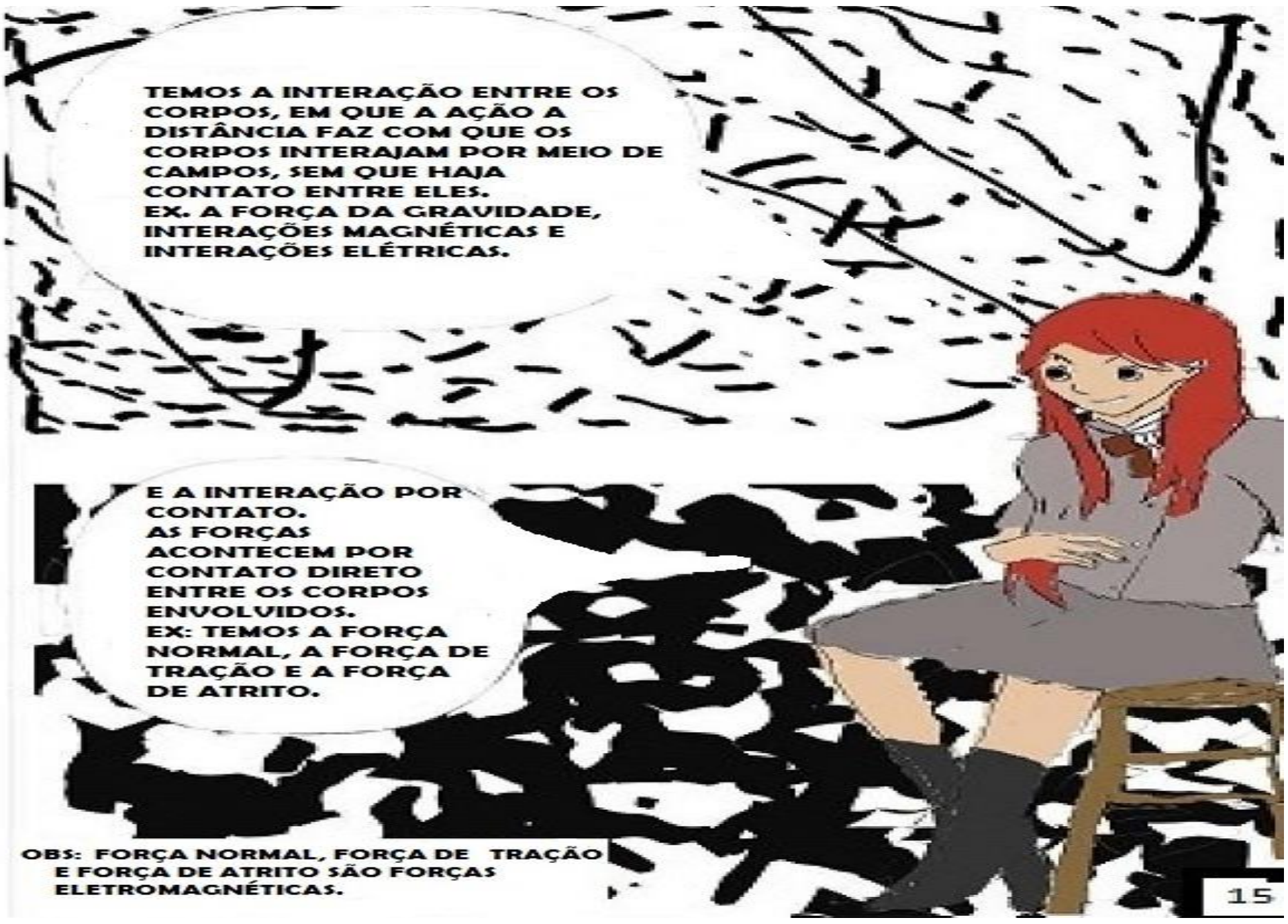

Fonte: Elaborado pela autora (2018)

Figura 63 – Acionamento pelo AR Physics na página 15 da revista: Terra e Lua em 3 D, parte superior da página

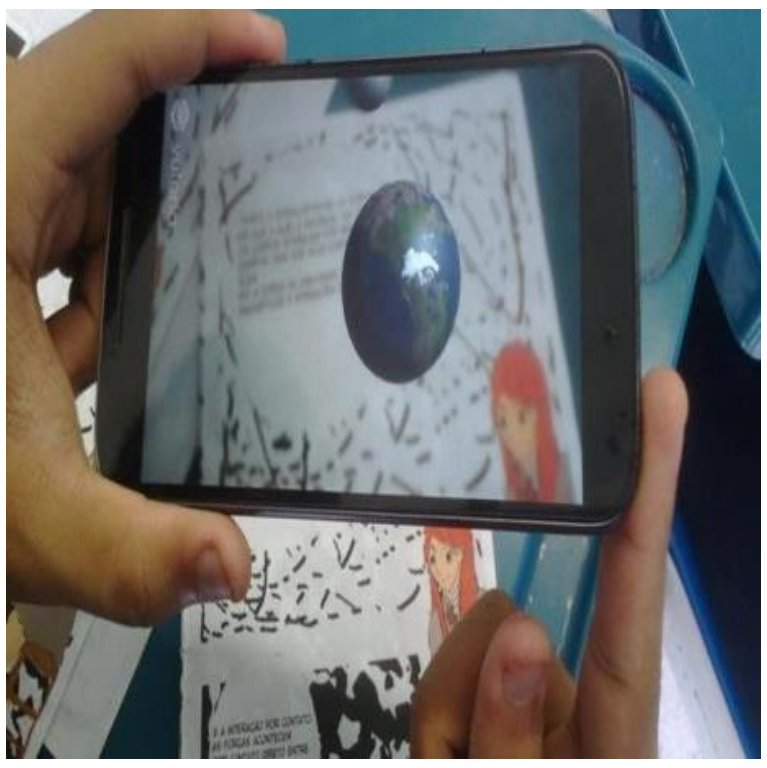

Fonte: Elaborado pela autora (2018)

Figura 64 – Acionamento pelo AR-Physics página 15 da revista: força de atrito, 10º Gif, parte inferior da página

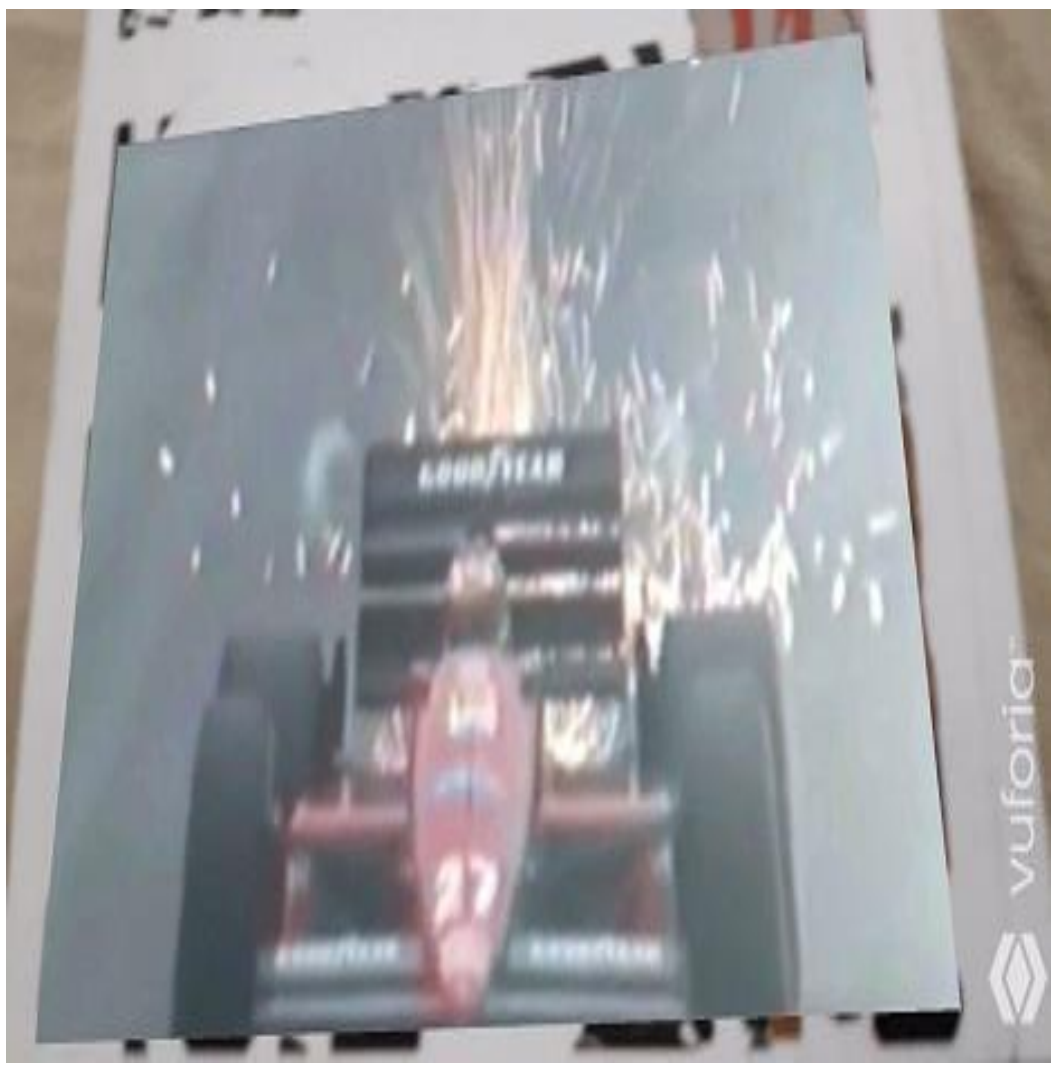

Fonte: FACEBOOK AYRTON SENNA (s.d.)

Nessa atividade, observamos uma continuidade nas respostas cientificas. As respostas mais completas decorreram da interpretação da Figura 63 que tinha uma projeção 3D de interação Terra-Lua.

Na Figura 65 temos a resposta do grupo 5, que apresenta uma explicação simples, mas que demonstra ganho de conhecimento.

Figura 65 – Resposta dos alunos do grupo 5

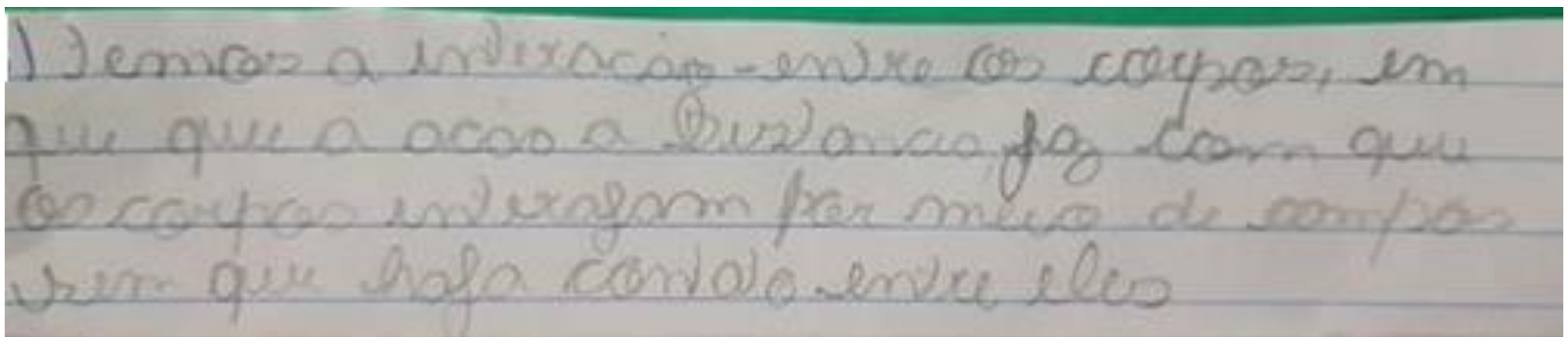

Transcrição: “temos a interação-entre os corpos, em que que a ação a distância faz com que os corpos interajam por meio de campos sem que haja contato entre eles.”

Fonte: Elaborado pela autora (2018)

Podemos observar na Figura 67 que nas perguntas sorteadas também tivemos respostas mais científicas.

Figura 66 – Resposta da pergunta sobre força de atrito: grupo 7

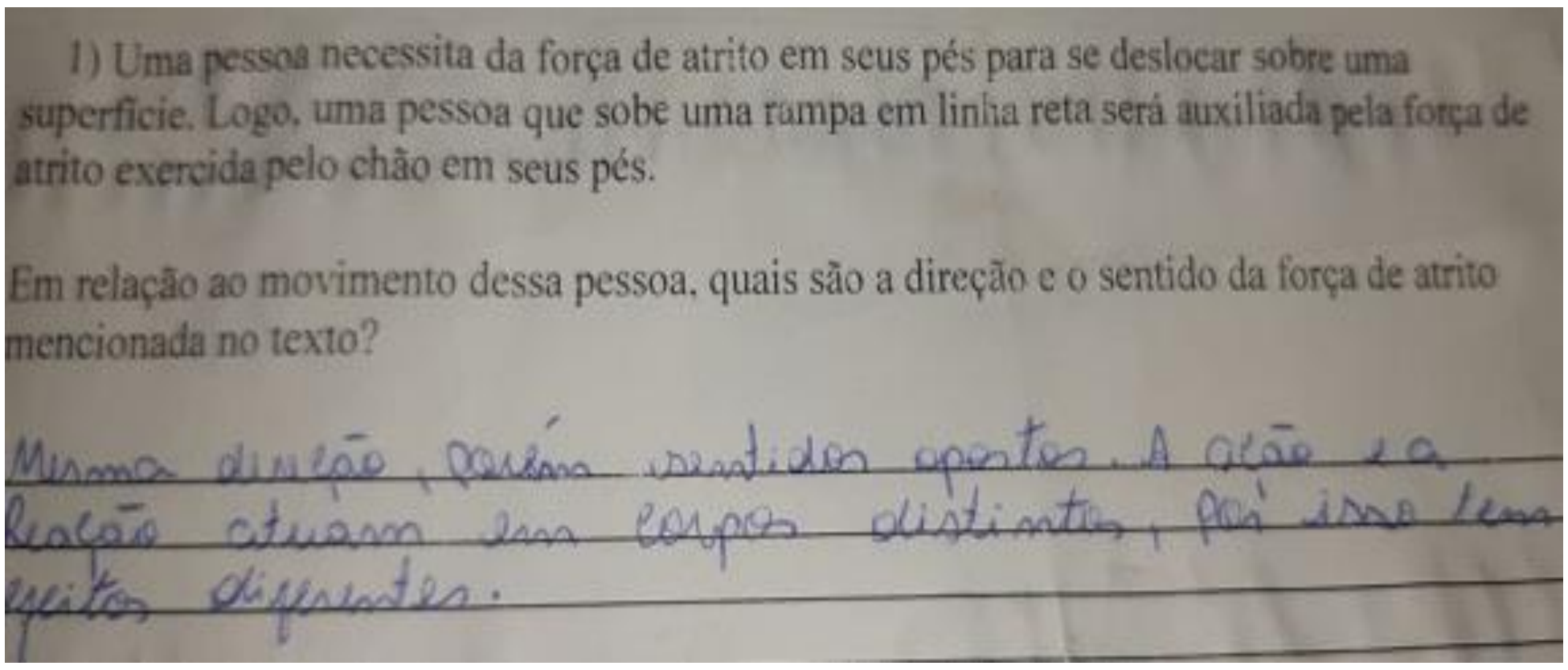

1) Uma pessoa necessita da força de atrito em seus pés para se deslocar sobre uma superfície. Logo, uma pessoa que sobe uma rampa em linha reta será auxiliada pela força de atrito exercida pelo chão em seus pés.

Em relação ao movimento dessa pessoa, quais são a direção e o sentido da força de atrito mencionada no texto?

Mesma direção, porém sentidos opostos. A ação e a reação atuam em corpos distintos, por isso tem efeitos diferentes.

Transcrição: “Mesma direção, porém sentidos opostos. A ação e a reação atuam em corpos distintos, por isso tem efeitos diferentes”.

Fonte: Elaborado pela autora (2018)

O grupo 7 (Figura 66) conseguiu identificar a ligação entre a força de atrito e a terceira lei de Newton, o que demonstra crescimento do conhecimento científico. Essa questão se encontra no Apêndice 5, item 3.

### 8.8.4 Força normal (12ª Aplicação)

Na página 16 (Figura 67), temos a apresentação da força normal. Os alunos deveriam analisar o 11º Gif (Figura 68), o qual contém uma representação gráfica com vetores, através de reação normal a qualquer força de contato na superfície dos corpos quando eles são encontrados. A representação dos vetores apresentada mostra que essa força terá sempre a direção perpendicular à superfície de contato, formando um ângulo de 90º com a superfície tocada. Os vetores também demonstram que a força é contrária à localização do toque. Foram organizados 3 tipos de posições para melhor entendimento dos alunos.

Figura 67 – Página 16 da revista

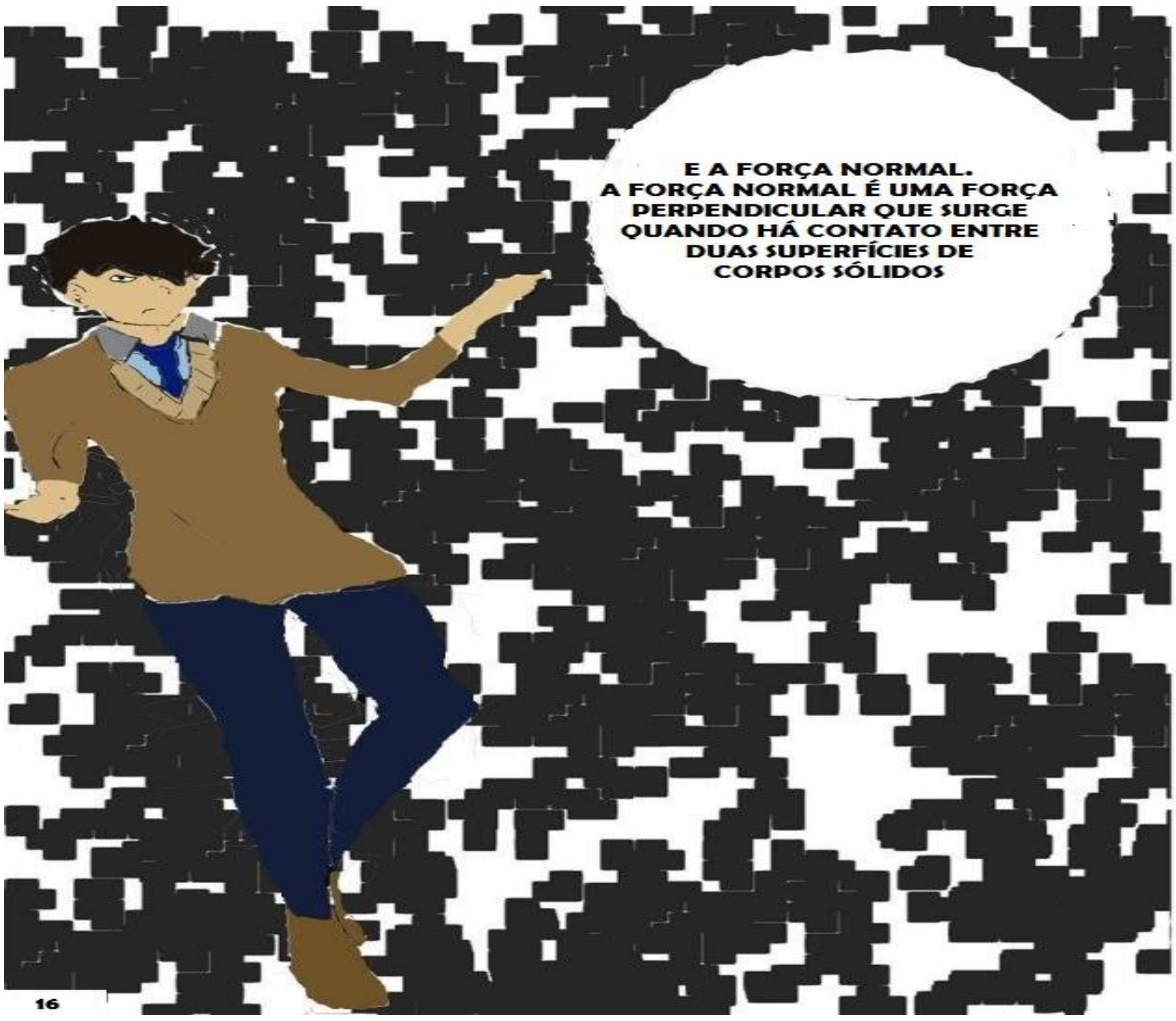

Fonte: Elaborado pela autora (2018)

Figura 68 – Gif acionado pelo aplicativo AR-Physics: 11º Gif, Força Normal

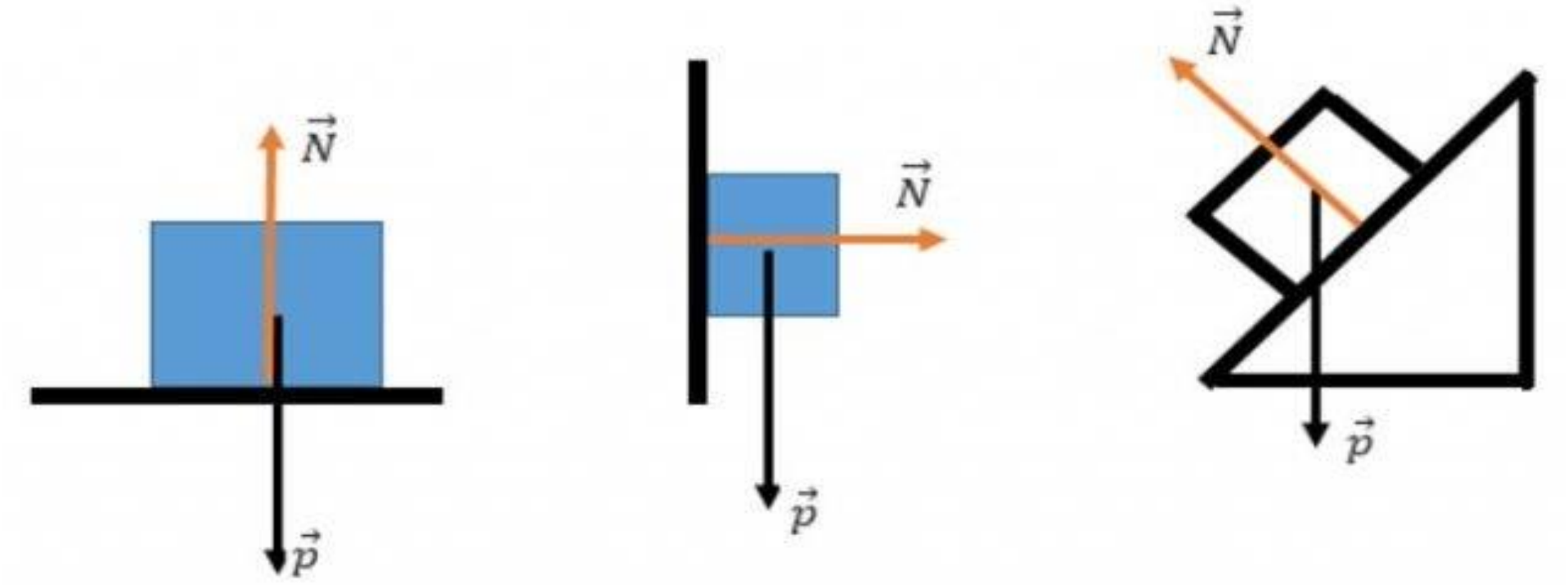

Fonte: PACHANI (2018)

Temos a resposta do grupo 4 (Figura 69) que apresenta a interpretação do Gif (Figura 68). O grupo 4 apresenta uma conclusão satisfatória sobre a força normal o que indica ganho de conhecimento.

Figura 69 – Resposta dos alunos: grupo 4

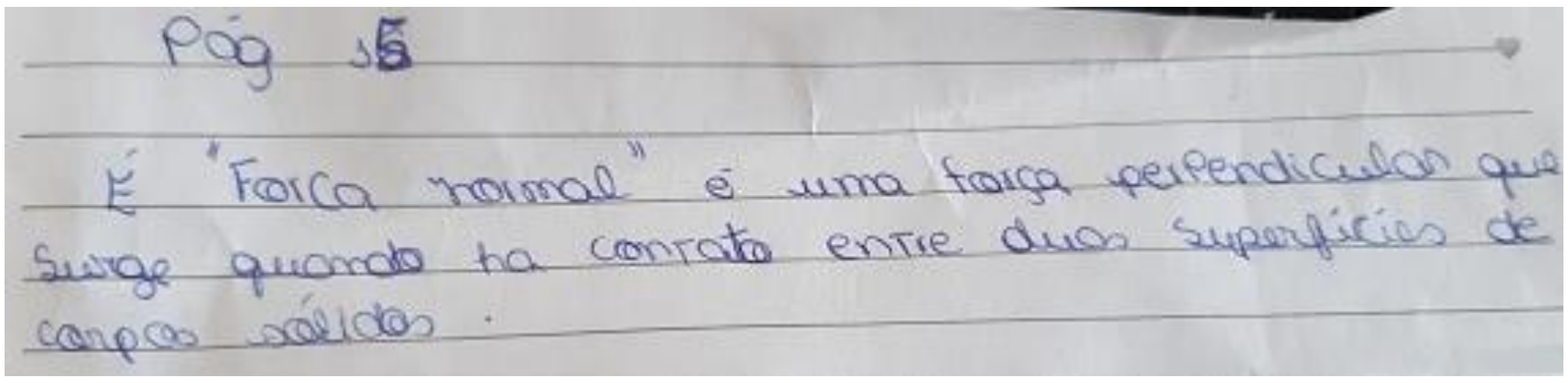
Pág 15

É "Força normal" é uma força perpendicular que surge quando ha contato entre duas superfícies de corpos sólidos.

Transcrição: "É "força normal" é uma força perpendicular que surge quando há contato entre duas superfícies de corpos sólidos".

Fonte: Elaborado pela autora (2018)

### 8.8.5 Desafio

Na página final da revista (Figura 70), temos como demonstração o 12º Gif (Figura 71). Foi explanado como finalização aos alunos um Gif com apresentação de uma experiência chamada mola maluca, que apresenta a queda de uma mola.

Figura 70 – Última página da revista: 17

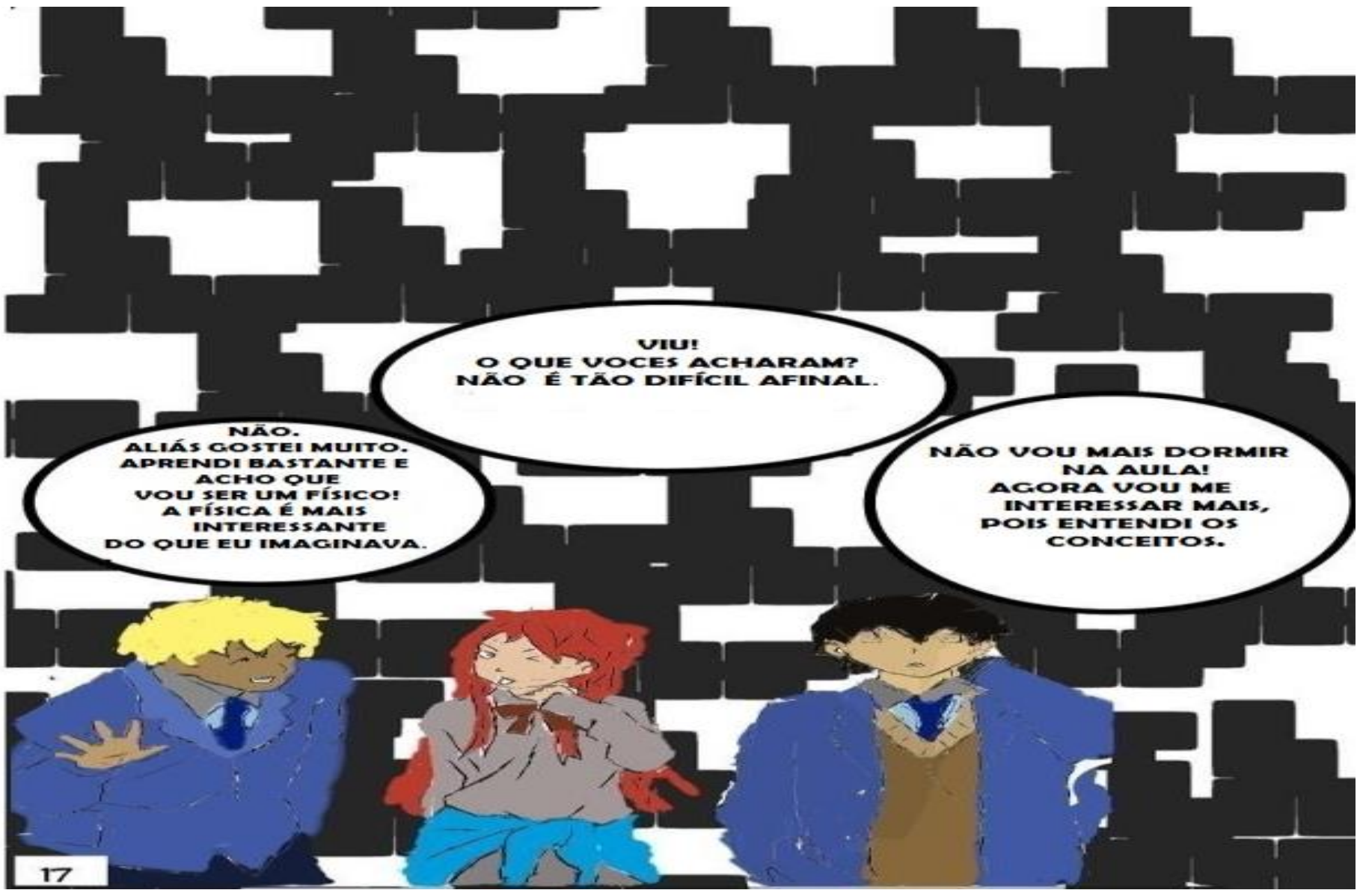

Fonte: Elaborado pela autora (2018)

Figura 71 – Acionamento pelo aplicativo AR-Physics página 17: 12º Gif “mola maluca”

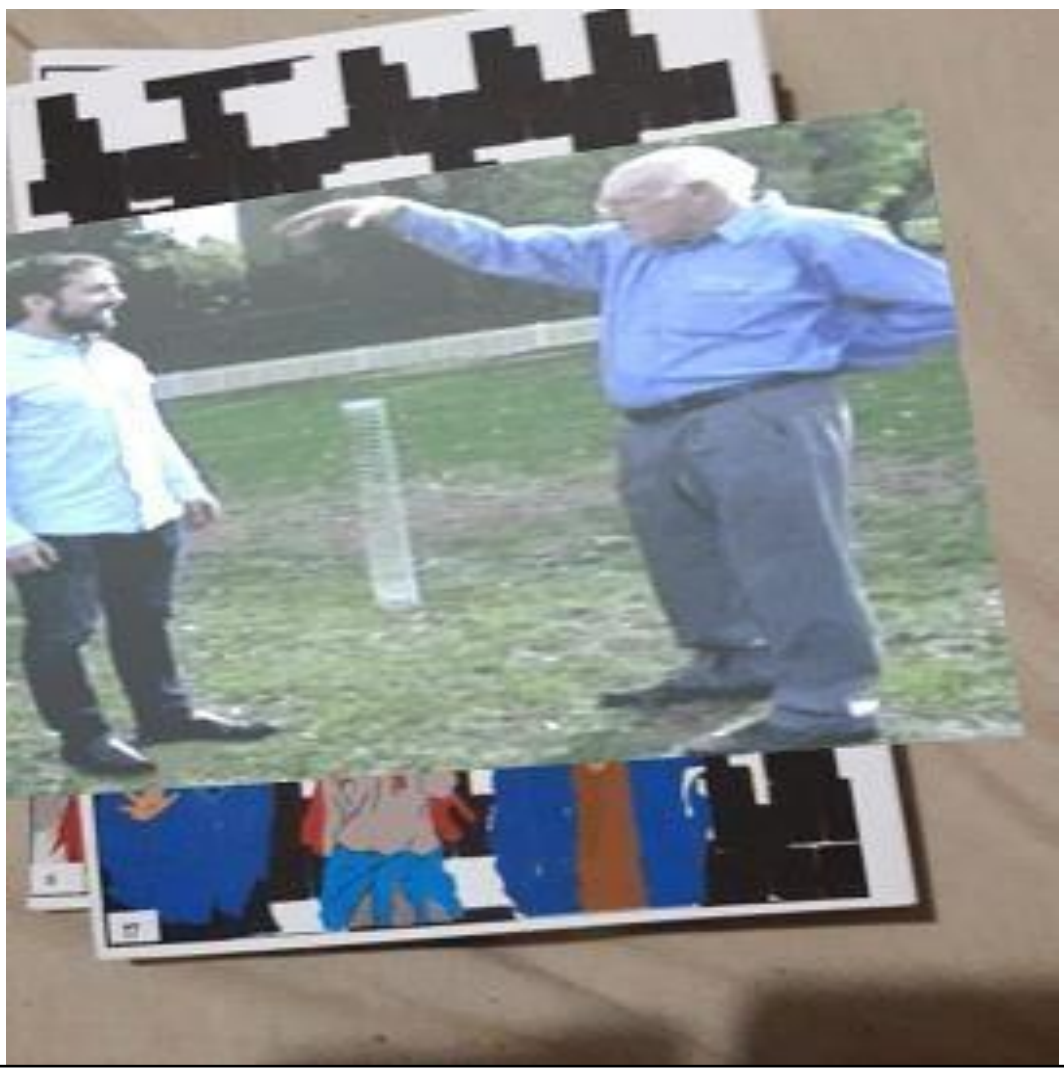

Fonte: SEDENTARISMO HIPERATIVO (2011)

De acordo com as leis de Newton se um corpo está sujeito a um campo gravitacional externo uniforme tal como os campos existentes nas proximidades da superfície da Terra e se ele tiver livre de outras ações externas, o seu centro de massa é acelerado verticalmente para baixo à razão de 9,8 m por segundo aproximadamente. O vídeo apresentado na última aplicação como finalização do processo mostra uma mola em queda livre. Ao analisarmos o Gif percebemos que uma das extremidades permanece em repouso durante a queda enquanto a outra extremidade desce com acelerações que parecem ser maiores que a aceleração gravitacional. O interessante é que tendo já iniciado a queda livre do centro de massa da mola, a parte inferior permanece em repouso durante algum tempo e, só depois, começa a descer com aceleração crescente. Enquanto isso a extremidade superior da mola está descendo a uma aceleração inicial maior que a aceleração da gravidade.

Nessa atividade, os alunos observaram o Gif da mola maluca e a pesquisadora distribuiu uma mola maluca para cada grupo para que eles estudassem o que estava acontecendo com a queda da mola. Eles deveriam observar o Gif, estudar o movimento e reproduzir a experiência, filmar com o celular e explicar no vídeo o que entenderam. Essa atividade foi avaliada por meio das apresentações dos vídeos enviados para a professora.

## 8.9 APRESENTAÇÃO DOS VÍDEOS PRODUZIDOS PELOS ALUNOS (13ª APLICAÇÃO)

Outra forma de análise e avaliação do aprendizado foi desafiá-los a produzir Gifs que apresentem os conceitos adquiridos por meio da sequência didática. O objetivo era demonstrar o conhecimento aplicado a eventos do cotidiano. Os Gifs produzidos foram apresentados na 13ª aplicação, momento em que os alunos fizeram exposição de seus Gifs demonstrando como as leis se encaixavam a cada vídeo. Alguns foram substituídos, no produto final, como resultado de aprendizagem, com autorização dos alunos. Ao todo foram produzidos 12 Gifs.

Figura 72 – Demonstração pelo grupo 5 da 1ª lei de Newton

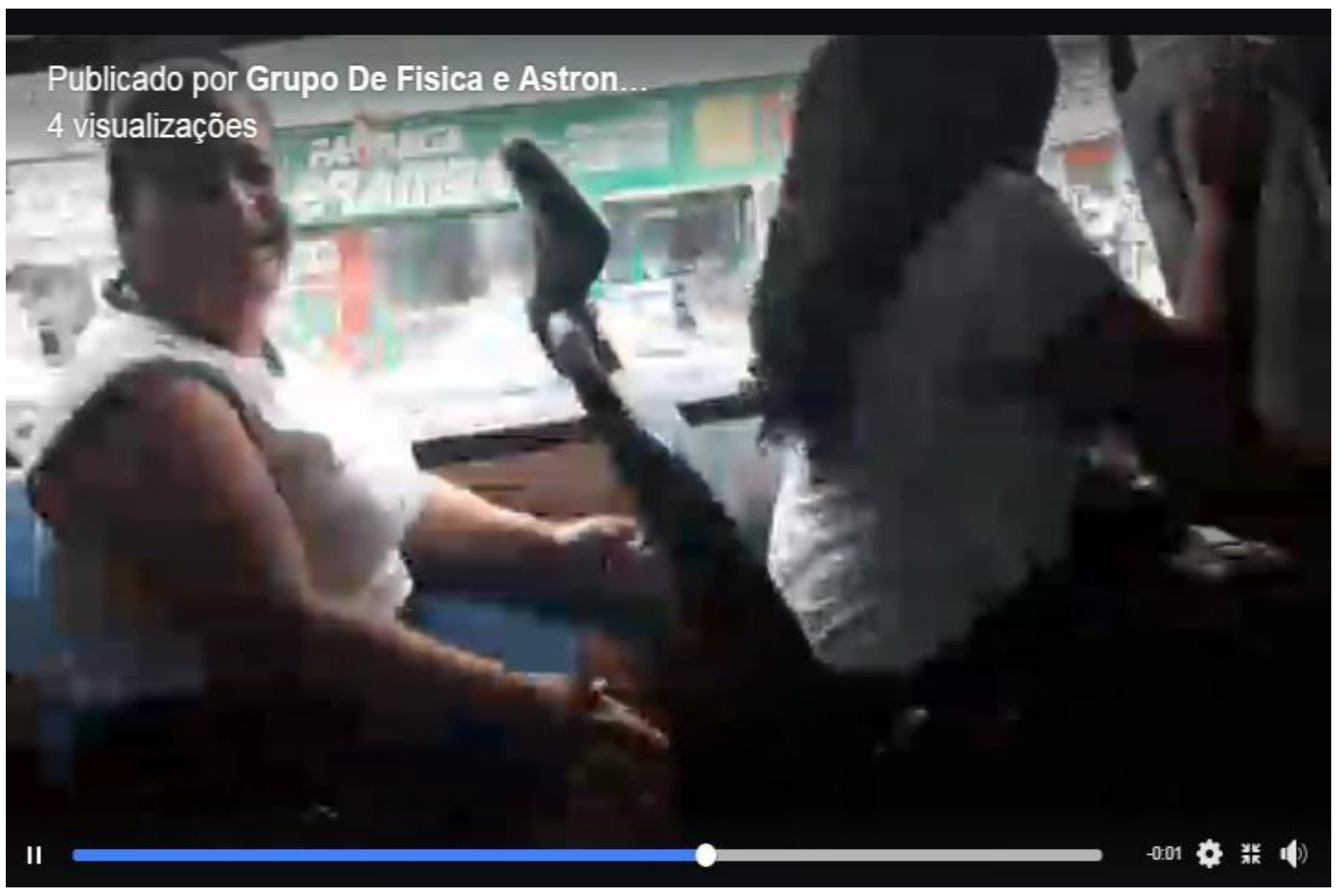

Fonte: GRUPO DE FÍSICA E ASTRONOMIA GUARAPARI (2018)

No Gif da Figura 72 os alunos do grupo 5 representaram a primeira lei de Newton dentro de um ônibus em movimento. A intenção era responder à pergunta: Porque somos jogados para frente do ônibus quando ele freia rapidamente? Os alunos explicitaram a primeira lei de Newton criando um Gif demonstrativo. Eles apresentaram a primeira Lei de Newton pelo princípio da inércia, onde um corpo tende a manter seu estado de movimento se não houver nenhuma força resultante que nele atue. Desta forma, quando o ônibus para, o corpo tende a manter o estado do movimento em que o ônibus estava anteriormente, por isso as pessoas foram arremessadas para frente. Fica claro nesse vídeo a identificação da inércia pelos alunos, demonstrando a resistência que a matéria oferece à aceleração.

Figura 73 – Demonstração pelo grupo 1 da 3ª lei de Newton

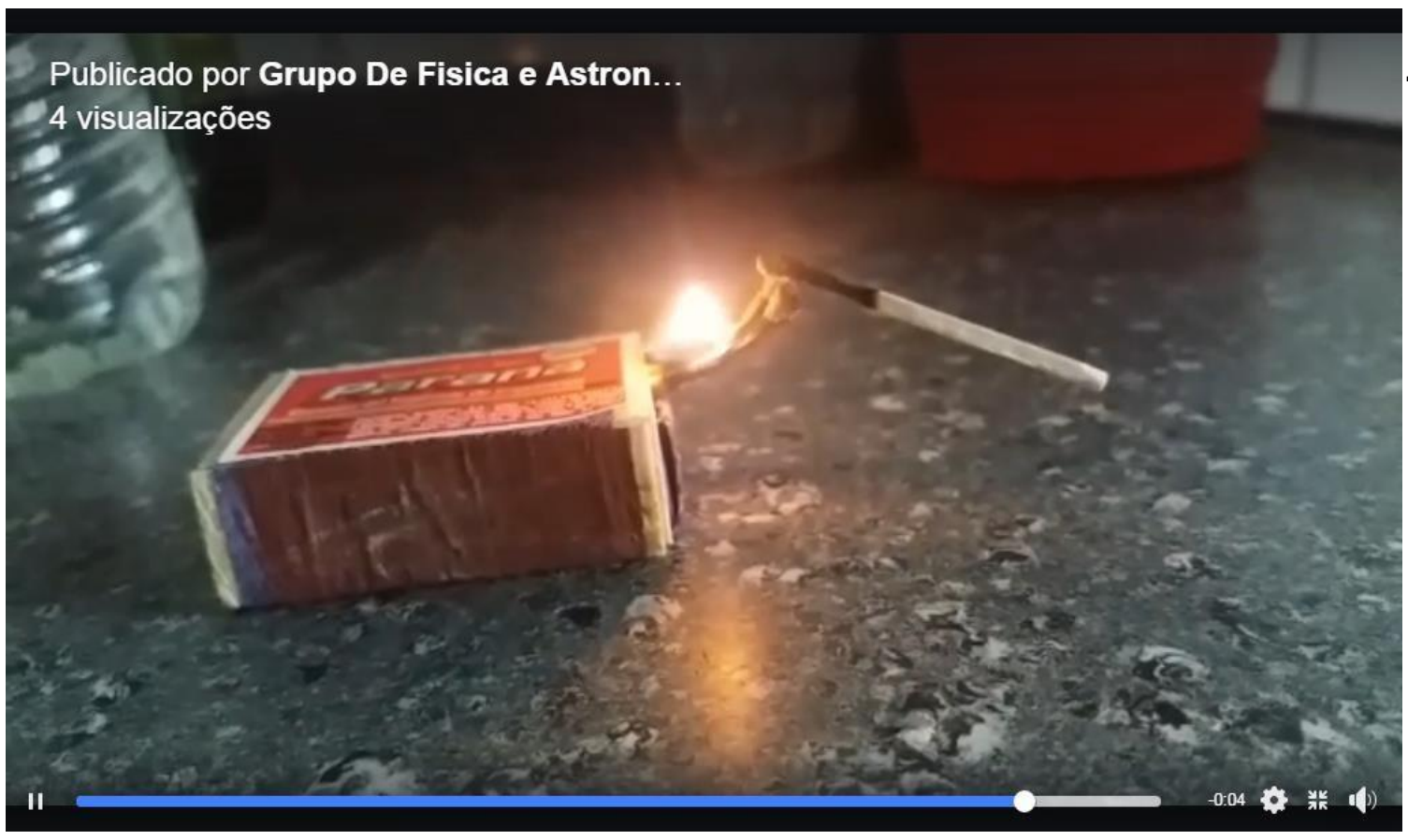

Fonte: GRUPO DE FÍSICA E ASTRONOMIA GUARAPARI (2018)

Uma aluna do grupo 1 descreveu o Gif como “Ação e reação - Os fósforos presos na caixa fazem efeito de ação sobre o que está apenas encostado, sendo assim o que tá apoiado sofre reação deve ser levantado”.

O Gif do grupo 1 demonstra a segunda lei, na qual, o produto da massa do corpo por sua aceleração é igual à força resultante que atua sobre ele; o grupo fez a demonstração com fósforos queimados, evidenciando seu aprendizado.

Foi solicitado que os grupos fizessem pequenos Gifs apresentando as leis de Newton. Alguns alunos fizeram Gifs separados, além do que tinha sido pedido aos sete grupos, mas, no total, foram produzidos 12 Gifs. Alguns grupos tiveram dificuldades para produzir os vídeos ligados aos conceitos. A maioria dos Gifs é bem simples, pois lançaram mão de materiais disponíveis em sala de aula, apesar de eles utilizarem o período de 3 dias para confeccionar os vídeos.

Alguns grupos apresentaram mais de um Gif por escolha própria, o que demonstra interesse pela tarefa e aprendizado. Todos os Gifs feitos pelos alunos, após aplicação da sequência didática, foram disponibilizados na página de física e astronomia de Guarapari no endereço: *https://www.facebook.com/GFAGuarapari/.*

# 9 APLICAÇÃO DE QUESTIONÁRIO PARA VERIFICAÇÃO DE APRENDIZADO: FINALIZANDO A SEQUÊNCIA DIDÁTICA (14ª APLICAÇÃO)

Nos resultados finais, observamos um ganho de conceitos científicos. Constatamos que, nos acertos do questionário final, de 386 perguntas respondidas, tivemos 217 respostas corretas, 38 respostas incompletas, 120 respostas erradas e 11 respostas em branco. No primeiro questionário (Apêndice 1), os alunos deixaram muitas questões em branco o que diminuiu muito no questionário final (Apêndice 6). Em geral, percebemos ganho de conceitos científicos, o que ficou evidenciado na linguagem apresentada nas interpretações. A época escolhida para aplicação ocorreu quando os alunos já estavam cansados e, praticamente, se preparando para as férias de fim de ano, isso enriquece o resultado, pois geralmente não se consegue atrair muito a atenção do aluno nesse período, mas a atividade fez com que eles adquirissem conhecimento, fossem participativos e colaborativos. A sequência didática aplicada tinha o objetivo de construir conhecimento de forma lúdica, o que apresentou um bom resultado, se observados os índices do questionário final.

O índice apresentado no Gráfico 3 da avaliação final apresenta ganho de conhecimento tanto nas respostas consideradas corretas, pois, no Gráfico 1 (questionário de conhecimentos prévios), a maioria das respostas (86%), não revelava conhecimentos por parte dos alunos.

Gráfico 3 – Verificação de conhecimentos questionário final (Apêndice 7)

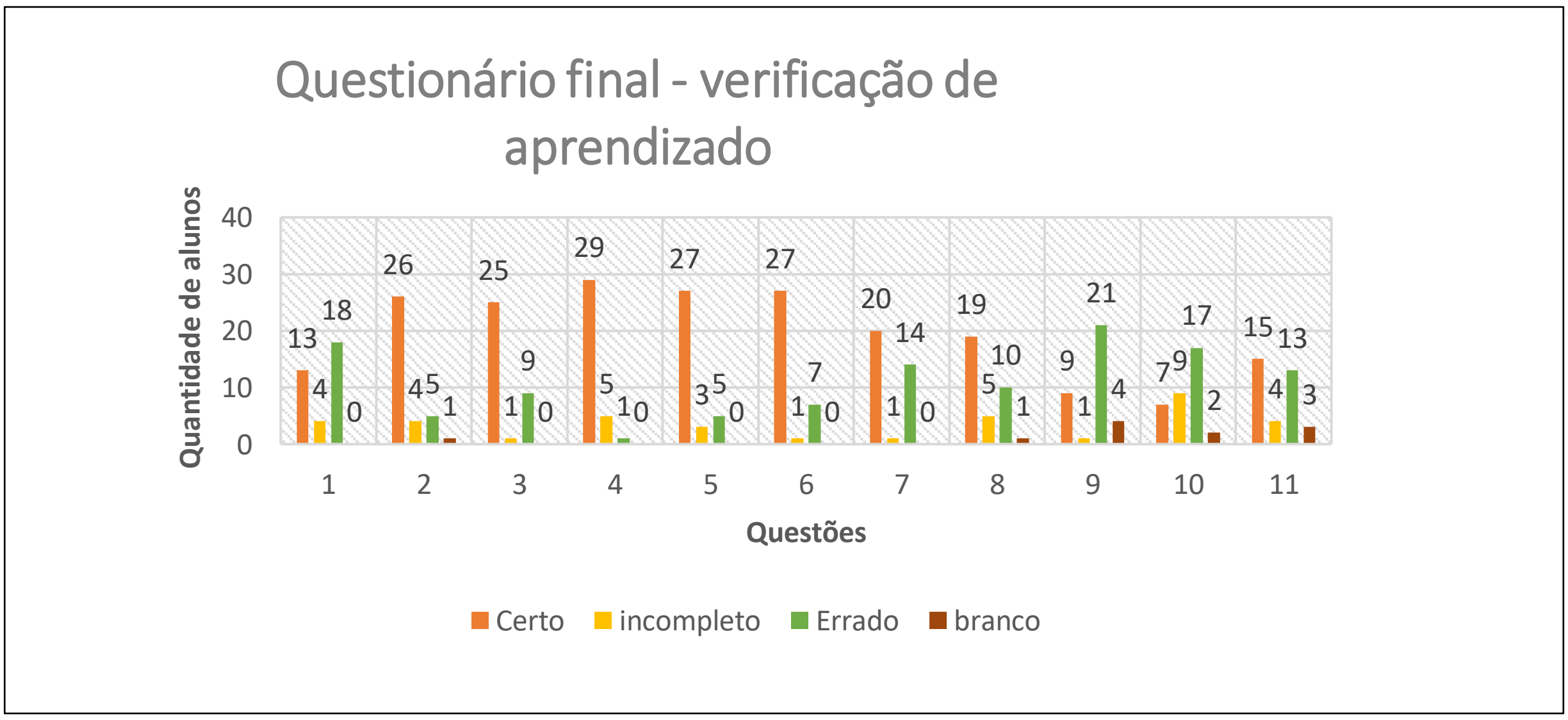

Fonte: Elaborado pela autora (2018)

Observando o Gráfico 3, temos um aumento considerável de acertos e de respostas erradas e os casos de respostas em branco praticamente desapareceram no gráfico de avaliação final (Gráfico 3), com apenas 5 respostas em branco. O índice de incompletos também diminuiu.

O Gráfico 4 apresenta a comparação entre as respostas apresentadas no questionário inicial (Gráfico 1) e as apresentadas no questionário final (Gráfico 3).

## 9.1 COMPARAÇÃO DE RESULTADOS QUESTIONÁRIO INICIAL X QUESTIONÁRIO FINAL

Gráfico 4 – Gráfico de comparação inicial x final

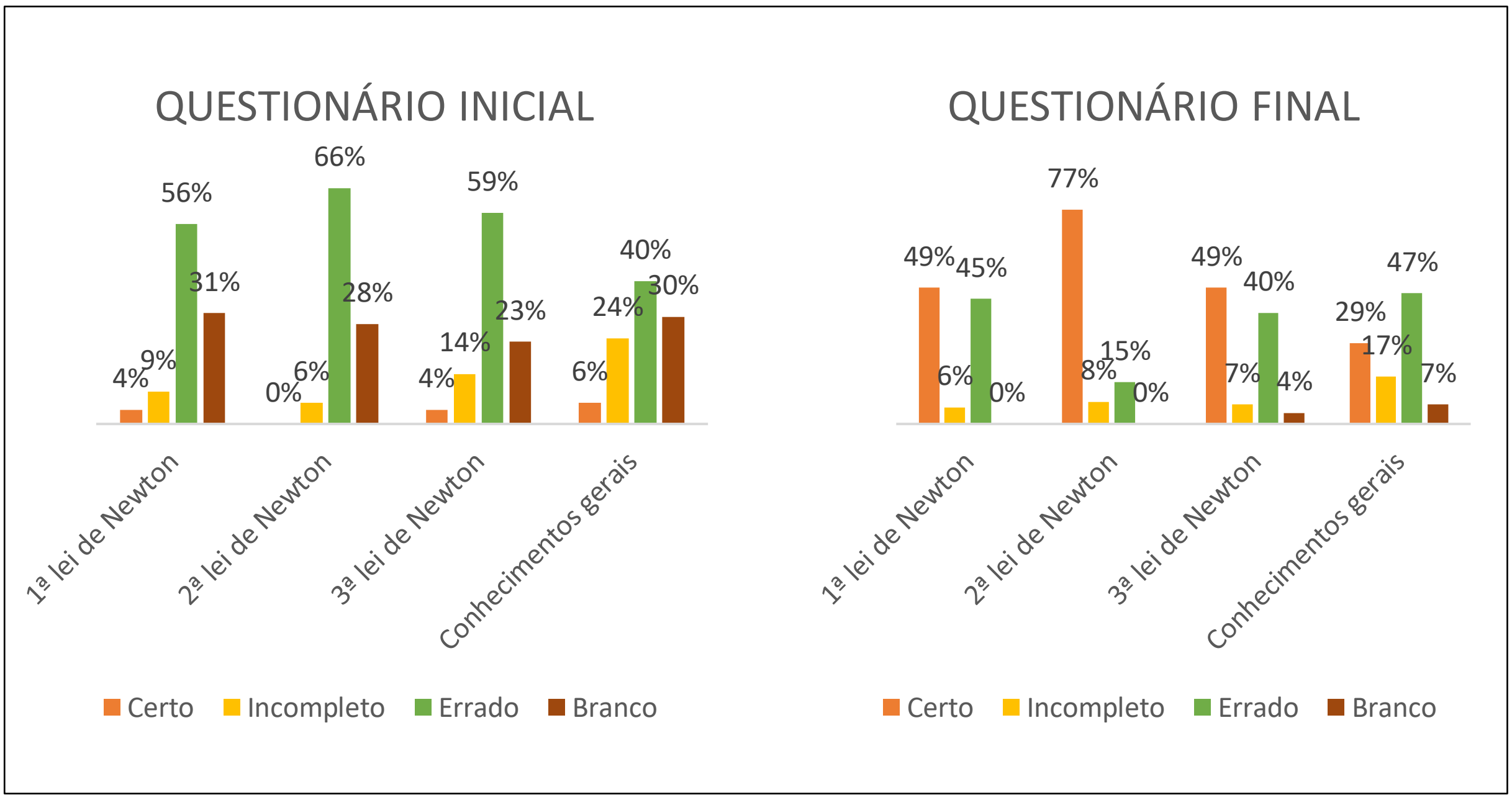

Fonte: Elaborado pela autora (2018)

Podemos observar que no primeiro questionário, no resultado da coluna referente às perguntas que continham conteúdo da primeira lei de Newton, tivemos 56% de perguntas respondidas de forma errada e 31% em branco, ou seja, se contarmos a soma das duas colunas teríamos 87% das questões respondidas sem embasamento cientifico. Tivemos 13% perguntas, apenas, respondidas de forma a articular a resposta ao conceito científico.

Concluímos, então, que, no início da aplicação, a maioria dos alunos não tinha conhecimento científico sobre a primeira Lei de Newton. Se compararmos com a coluna das respostas sobre a primeira Lei de Newton, no questionário final, temos 0% de respostas em branco e 45% erradas, o que apresenta um índice muito menor se comparado com as questões respondidas no primeiro questionário.

Analisando o índice de conhecimento cientifico no questionário final, temos, na coluna da primeira Lei de Newton, 49% de respostas corretas e 6% incompletas, somando o total de 55%, indicando avanços no conhecimento. Isso demonstra um índice bem maior nas respostas científicas revelando ganho no conhecimento sobre a primeira lei de Newton.

Na segunda coluna temos as respostas referentes à segunda lei de Newton. No questionário de conhecimentos prévios (inicial), esta é a barra que apresenta maior número de questões (66%) erradas. O índice de respostas em branco também é grande, temos 28% respostas em branco. Se compararmos com o questionário final, temos, também, a maior barra da aplicação com 77% de questões, só que, no segundo questionário, essas respostas estão corretas. Além de as respostas corretas aumentarem, temos as respostas em branco, que também ficaram em 0%, as incompletas foram 8% e as erradas diminuíram muito, passando a apenas 15%. Esse resultado também nos leva a concluir que os alunos tiveram ganho conceitual da segunda Lei de Newton, apresentando o melhor resultado entre os 4 módulos analisados.

Com relação à terceira Lei de Newton, o primeiro gráfico (Gráfico 1) apresenta um índice alto de respostas erradas e em branco, com um total de 59% de erradas e 23% em branco. Já no questionário final, temos 40% erradas e 4% em branco, os dois índices diminuíram no questionário final e os índices de acerto aumentaram de 4%, no primeiro questionário, para 49% no último questionário. O índice de respostas incompletas também diminuiu de 14%, no primeiro questionário, para 7%, no segundo questionário. Essa análise comparativa nos permite concluir que houve ganho de conhecimento na terceira lei de Newton.

Nas perguntas que apresentavam conhecimentos gerais ligados a conceitos aprendidos na sequência didática temos, também, melhora. No gráfico inicial, podemos verificar 6% de respostas certas, já, no gráfico final, temos 29% respostas certas. As respostas em branco também mostraram uma melhora de 30% no questionário inicial, para 7% no questionário final. Somente não tivemos melhoria de resultado no questionário final das respostas erradas. No questionário inicial, tivemos 40% e, no final, tivemos 47%. Dados assim nos ajudam a compreender que o aumento de respostas erradas, nesse último item, está relacionado com a migração de respostas em branco para respostas erradas, já que os alunos que não tinham conhecimento nem para tentar responder no questionário inicial e poderiam ter respondido de forma errada no questionário final, já que o índice de respostas em branco reduz mais da metade se compararmos o questionário inicial com o questionário final.

Observando os gráficos e os avanços alcançados pelos alunos inferimos que a aprendizagem do conteúdo foi gradual. Apesar de os resultados terem sido muito significativos, uma análise mais apurada deve ser feita para confirmação. A sequência pareceu satisfatória, demandando alguns ajustes nas perguntas apresentadas e nos Gifs escolhidos. Houve grande interesse dos alunos pelas atividades da sequência didática, mesmo sendo final de trimestre.

Fazendo uma comparação dos resultados do questionário inicial para o questionário final, podemos observar algumas diferenças significativas. Podemos comparar os acertos do primeiro questionário compreendendo 4% do total de questões, no questionário final, temos 51% de acertos do total das questões apresentadas: o índice de respostas incompletas diminuiu de 13% no primeiro questionário para 10% no último questionário. No total de respostas erradas, temos 55% no primeiro questionário e 37% no último. E, finalmente, nas respostas em branco tivemos uma diferença expressiva de 28% no primeiro questionário para 3% no último.

Os melhores resultados apontam para as respostas certas, que tiveram um crescimento de 50% e, nas respostas em branco, as quais, no primeiro questionário, representaram 29% e no último, 3%. As respostas erradas, apesar de terem diminuído, são consideradas altas. As melhorias que devem ser feitas para futuras aplicações devem levar esse índice em conta para adaptação e aperfeiçoamento do material.

Outra evidência que o Gráfico 4 indica é que os alunos que antes não tinham interesse em responder ao questionário, ou não souberam dar as respostas, passaram a participar ativamente, se observarmos os índices do segundo gráfico, mesmo que no índice de erros, mostrando, de uma forma ou outra, interesse na apresentação da revista.

# 10 APLICAÇÃO DO QUESTIONÁRIO DE INTERESSE E SUGESTÃO (15ªAPLICAÇÃO)

No questionário complementar (Apêndice 7), foram feitas 7 perguntas sobre a metodologia aplicada referente aos conceitos aprendidos na revista de realidade aumentada. Foram distribuídos 35 questionários aos alunos que participaram ativamente da aplicação da sequência didática. Apresentamos a seguir uma breve análise das respostas e em alguns casos destacamos respostas aleatórias que os alunos registraram.

Na primeira questão, ao perguntarmos se o conteúdo restou compreendido de maneira clara e objetiva, 85% responderam que sim e 15% responderam ter ficado com dúvida em algum dos conteúdos.

Ao serem questionados se enfrentaram dificuldades com a metodologia, 94% avaliaram a metodologia fácil de entender e gostaram, mas 6% julgaram a metodologia com nota mediana (Figura 74).

Figura 74 – Pergunta 1-b: resposta de aluno

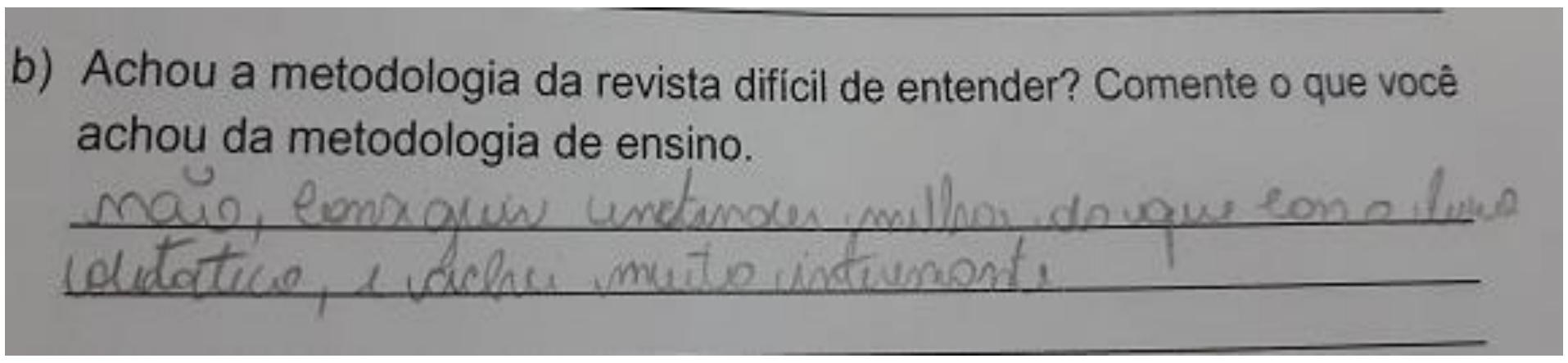

b) Achou a metodologia da revista difícil de entender? Comente o que você achou da metodologia de ensino.

Transcrição: “não, consegui entender melhor do que com o livro didático, e achei muito interessante”.

Fonte: Elaborado pela autora (2018)

A terceira pergunta buscava saber se gostariam de mais atividades como essas. Nesse caso, 100% dos alunos responderam que sim.

Figura 75 – Pergunta 1-c: resposta de aluno

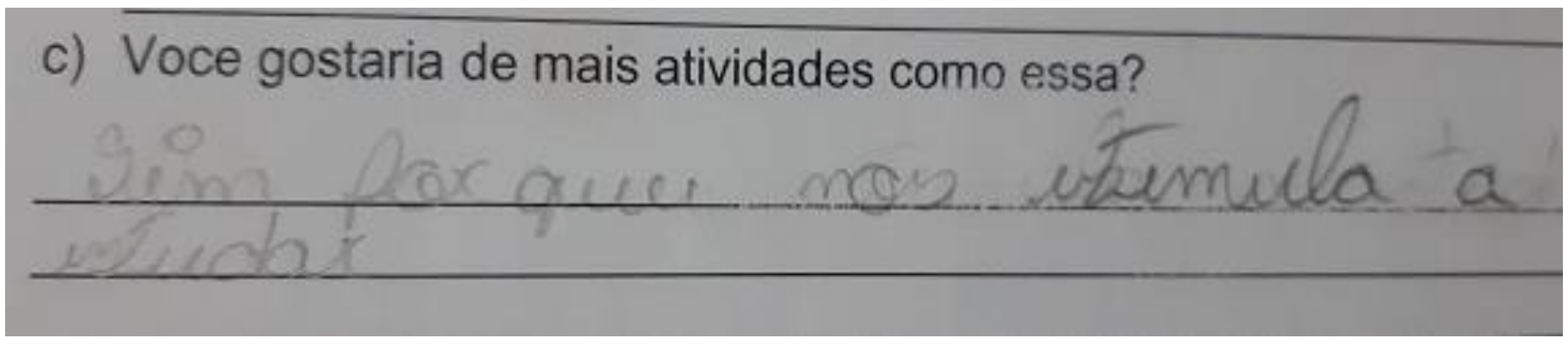

Transcrição: “sim porque nos estimula a estudar”

Fonte: Elaborado pela autora (2018)

A quarta pergunta indagava se na opinião dos alunos, com a revista, eles poderiam aprender mais e também nessa resposta 100% dos alunos disseram que sim.

A quinta pergunta tratou de entender qual conceito e Gif foram melhor explicados e 42% responderam que gostaram do Gif que explicou a primeira Lei de Newton, 11% gostaram dos Gifs que explicaram a segunda lei, 28% acharam que os Gifs que representaram a terceira lei foram mais explicativos e os outros 19% dividiram-se entre as demais Gifs da revista.

A sexta pergunta buscava saber qual dos Gifs exigiu maior esforço de interpretação e, nesta questão, 6% dos alunos tiveram dificuldades de entendimento no Gif do cãozinho pug (segunda Lei de Newton), 6% no Gif que explode a bala na parede (terceira Lei de Newton), 14% no Gif do lançamento do foguete, 26% no Gif da queda da maçã (primeira Lei de Newton), 6% no Gif onde um senhor empurra o carro (segunda Lei de Newton), 6% tiveram dificuldade na interpretação do Gif do Capitão América, 6% no Gif final da demonstração da ação de uma mola para introdução da lei de Hooke. Dez por cento dos alunos responderam não ter tido nenhum tipo de dificuldades para a interpretação das demonstrações apresentadas.

Na sétima pergunta, a pesquisadora pediu sugestões para aprimoramento da revista ou da atividade. Entre os alunos consultados, 85% disseram não haver sugestões ou mudanças, vez que aprenderam muito com a revista, pois se tratava de uma atividade divertida.

Houve algumas sugestões, nos outros 15% das respostas como: melhorar os gatilhos ou *targets* de acionamento dos Gifs; fazer essa atividade fora da sala de aula; uso do aplicativo utilizado em todos os celulares da sala, não somente no celular do professor e usar a metodologia nas provas. Foi sugerido, também, que se expandisse a metodologia para outros conteúdos, principalmente nos conteúdos mais difíceis, pois poderia facilitar o entendimento.

# 11 CONCLUSÃO

Nesse trabalho propusemos uma sequência didática mediada sobre as Leis de Newton utilizando uma revista que apresentava *targets* que direcionavam o acionamento de Gifs através do uso de um celular.

Em análise do primeiro questionário de verificação de conhecimentos prévios do assunto abordado, percebemos que os alunos, ao ingressarem no ensino médio, não trazem praticamente nenhum conhecimento sobre elementos básicos que servem de conhecimentos prévios para o aprendizado das leis de Newton. Se o conteúdo foi visto, o resultado demonstra que muito pouco foi aprendido.

A aplicação da revista ajudou a melhorar os conhecimentos sobre as três leis de Newton, mostrando ser uma boa aliada na prática pedagógica do professor. Essa estratégia pode ser melhor trabalhada com uso constante em sala de aula, adaptando-se ao contexto onde será feita a aplicação.

A pesquisadora conclui que a sequência didática mediada, aplicada com a ajuda da revista, teve uma boa aceitação entre os adolescentes e por esse motivo pôde ajudar no aprendizado. As sugestões dos alunos podem melhorar de forma a aumentar a identificação do adolescente com o material, o qual, por ser um material único, pode ser testado em outros conteúdos, acompanhando as mudanças com a opinião dos usuários, no caso alunos.

Desde a aplicação em 2018, já utilizamos o material em 7 turmas em 2019, sendo 2 turmas de EJA (Educação de Jovens e Adultos). Por perceber a grande aceitação do material, pretendemos utilizar em mais 7 turmas no ano de 2020 e ir procedendo às adaptações necessárias, de acordo com as observações em cada aplicação.

# REFERÊNCIAS

# APÊNDICE 1 – Questionário inicial de avaliação

| |
|---|
| Mestrado profissional em ensino de Física – SBF |
| Instituto Federal do Espirito Santo |
| Campus Cariacica |
| Dissertação: |
| Material didático para ensino das leis de newton com uso da realidade aumentada |
| Lucia Helena Horta Oliveira |
| EEEM Dr. Silva Mello |
| Questionário Inicial de Avaliação |

01) Como você percebe se um corpo está parado ou em movimento? Justifique sua resposta.

____________________________________________________________________

____________________________________________________________________

02 ) Você sabe o que é inércia? Se sim, explique o que é inércia.

____________________________________________________________________

____________________________________________________________________

03) O que provoca a mudança de velocidade em um corpo? Justifique sua resposta.

____________________________________________________________________

____________________________________________________________________

04) Um automóvel trafegando a uma certa velocidade e leva 1 segundo para ser parar numa freada de emergência.

a) O que aconteceu com a velocidade do carro?

____________________________________________________________________

____________________________________________________________________

b) Existiu aceleração? Explique.

____________________________________________________________________

____________________________________________________________________

05) Um corpo de massa m está sujeito à ação de uma força F que o desloca segundo um eixo vertical em sentido contrário ao da gravidade. Porque esse corpo se move com velocidade constante?

______________________________________________________________________

______________________________________________________________________

______________________________________________________________________

06) Um corpo de massa igual a 15 kg move-se com aceleração de modulo igual a $3m/s^2$. Qual o módulo da força resultante que atua no corpo?

______________________________________________________________________

______________________________________________________________________

______________________________________________________________________

07) O que é força?

______________________________________________________________________

______________________________________________________________________

______________________________________________________________________

08) Escreva quais são as leis de Newton, e quais você conhece.

______________________________________________________________________

______________________________________________________________________

______________________________________________________________________

09) Qual lei de Newton utilizo para explicar porque você é arremessado para frente enquanto seu carro para bruscamente? Explique como ela se aplica a situação.

______________________________________________________________________

______________________________________________________________________

______________________________________________________________________

10) O que você acha que faz com que você seja arremessado para frente quando estamos dentro de um ônibus e o motorista faz uma parada brusca?

______________________________________________________________________

______________________________________________________________________

______________________________________________________________________

11) Escreva as principais forças vetoriais que você conhece?

______________________________________________________________________

______________________________________________________________________

12) Em um dia chuvoso e frio, João tentou ligar seu carro e percebeu que a bateria não funcionava. Para tentar resolver o problema, permaneceu sentado confortavelmente e empurrou o painel de controle para frente.

a) O carro se movimentou? Explique.

______________________________________________________________________

______________________________________________________________________

13) O que é força de atrito?

______________________________________________________________________

______________________________________________________________________

APÊNDICE 2 – Apresentação de Slides

Slide 1

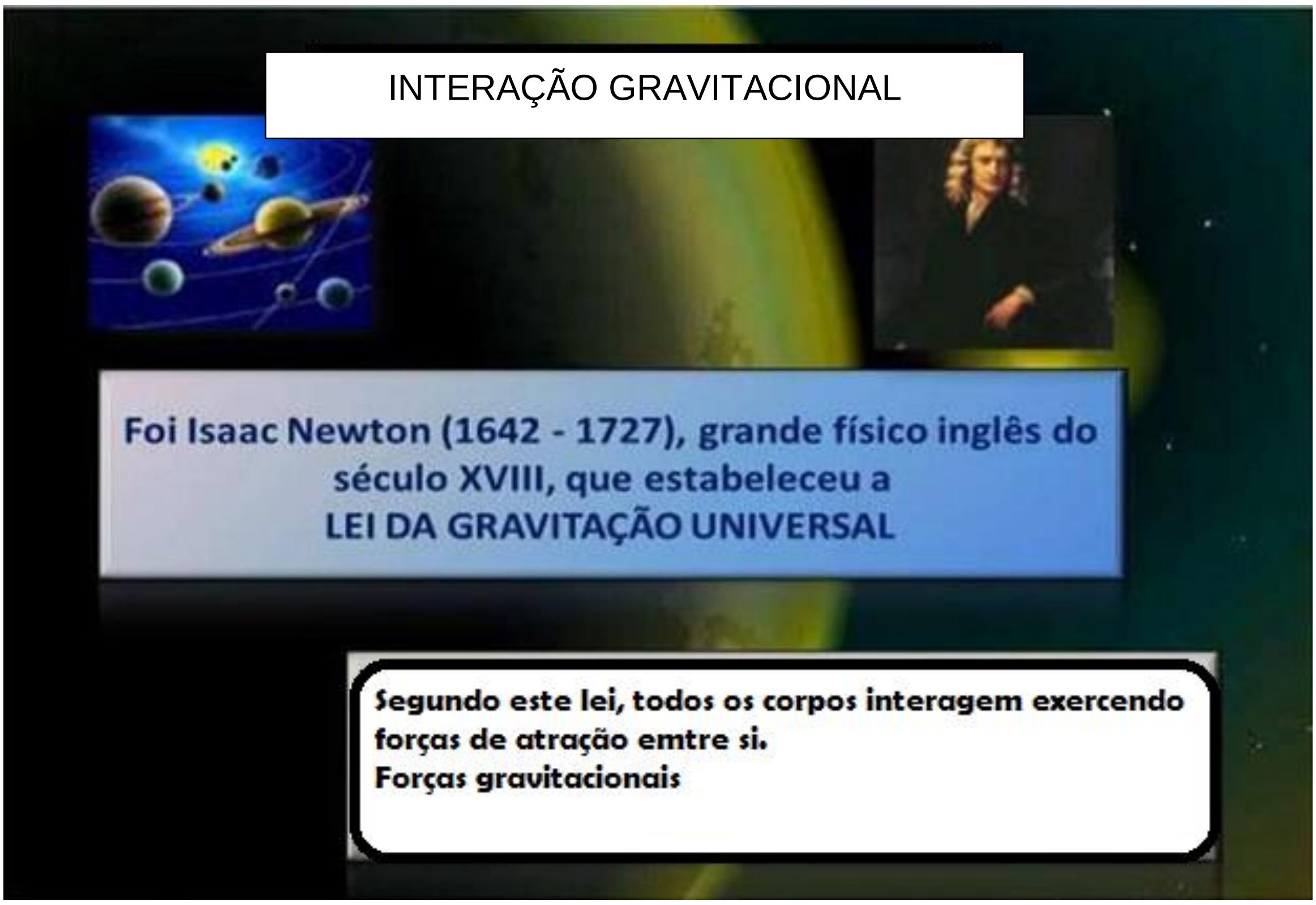

Slide 2

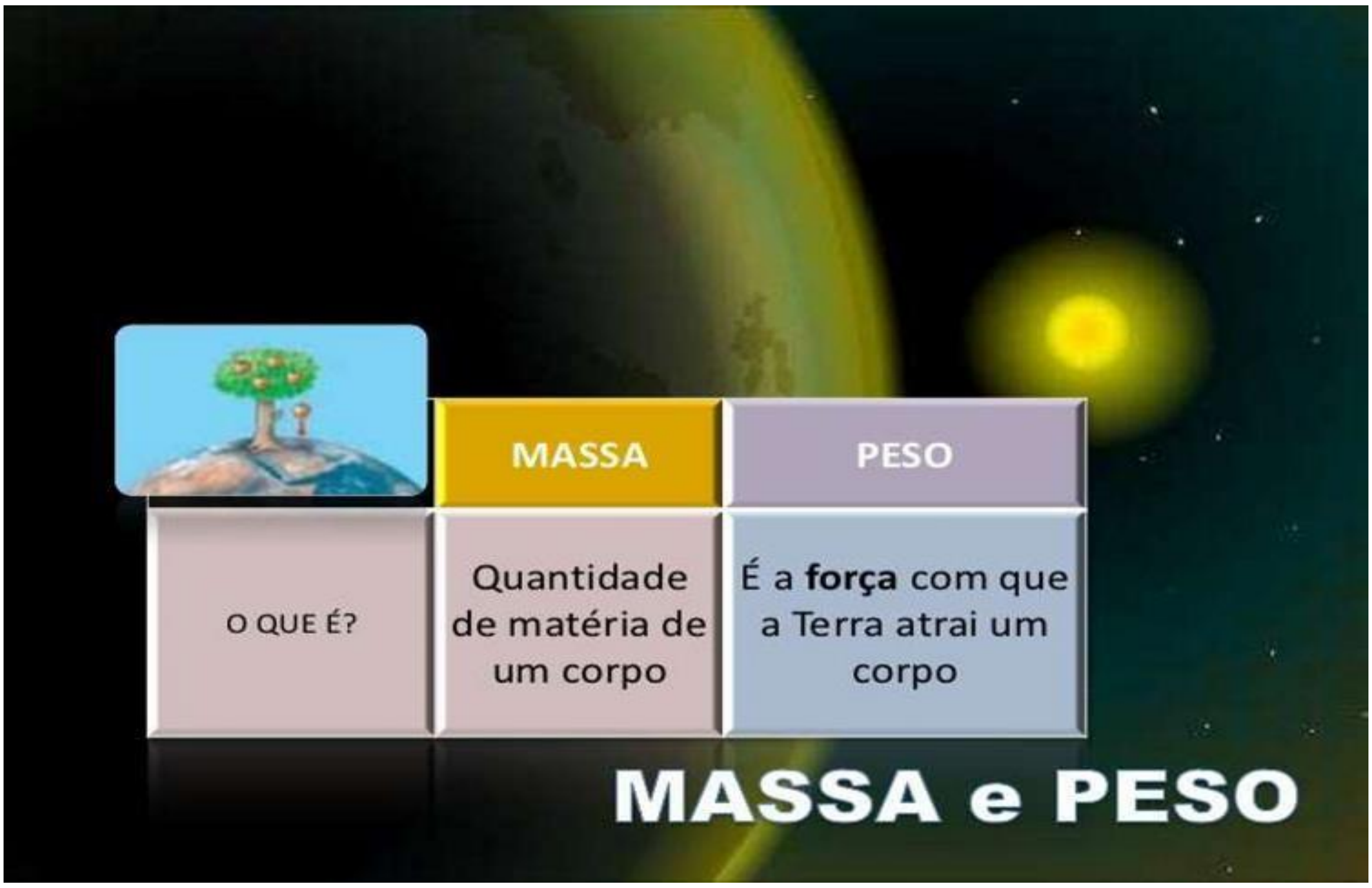

Slide 3

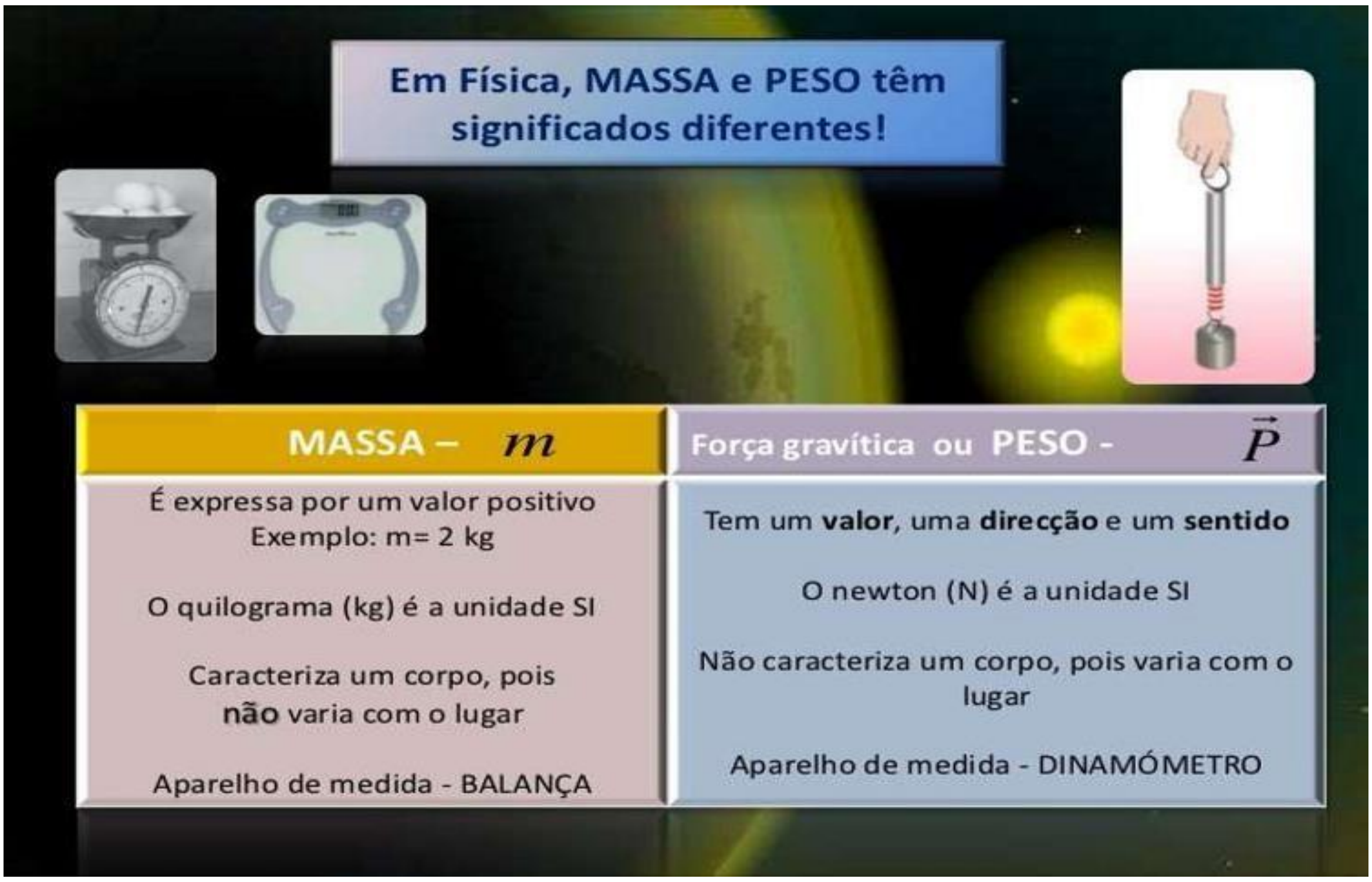

Slide 4

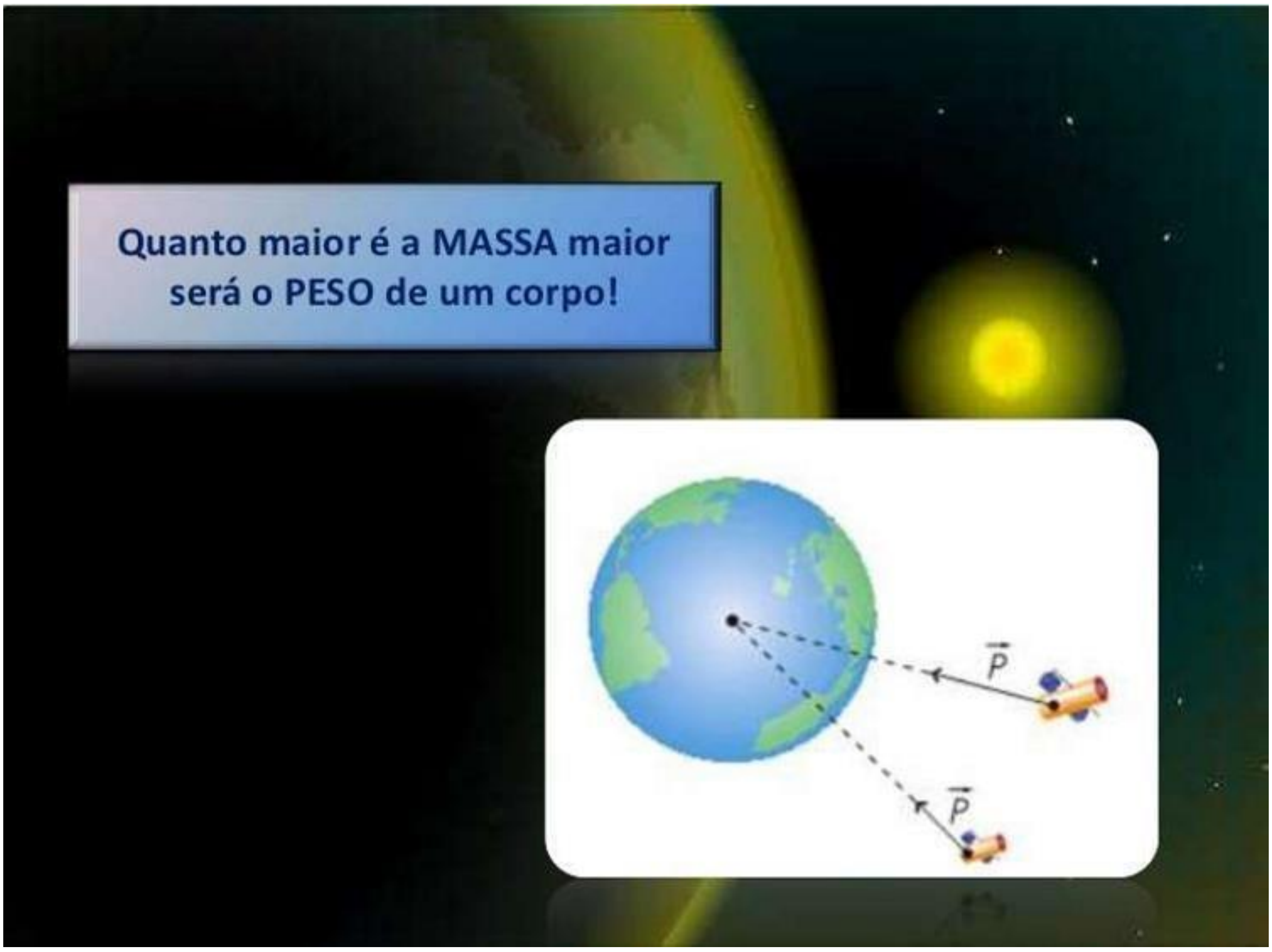

Slide 5

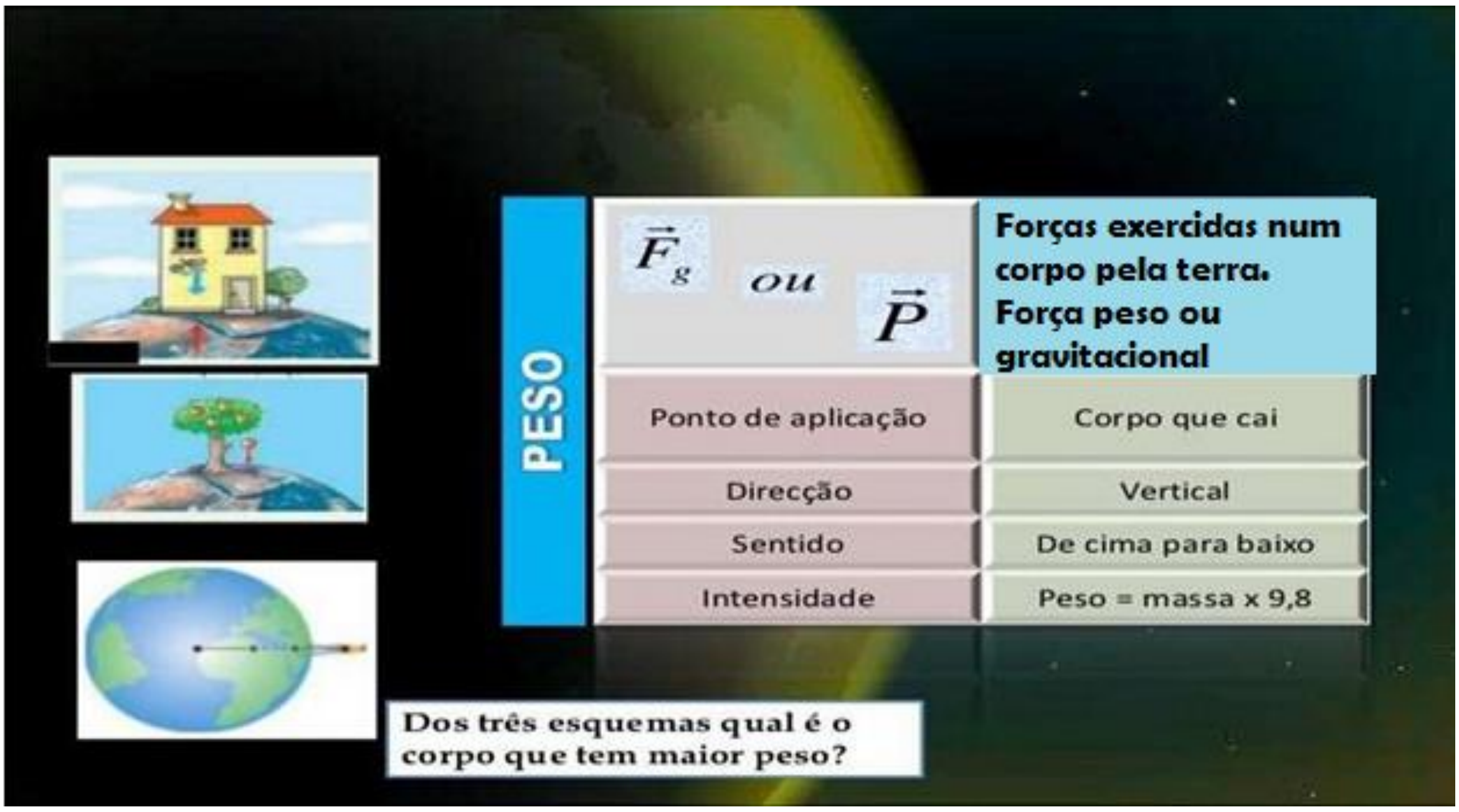

Slide 6

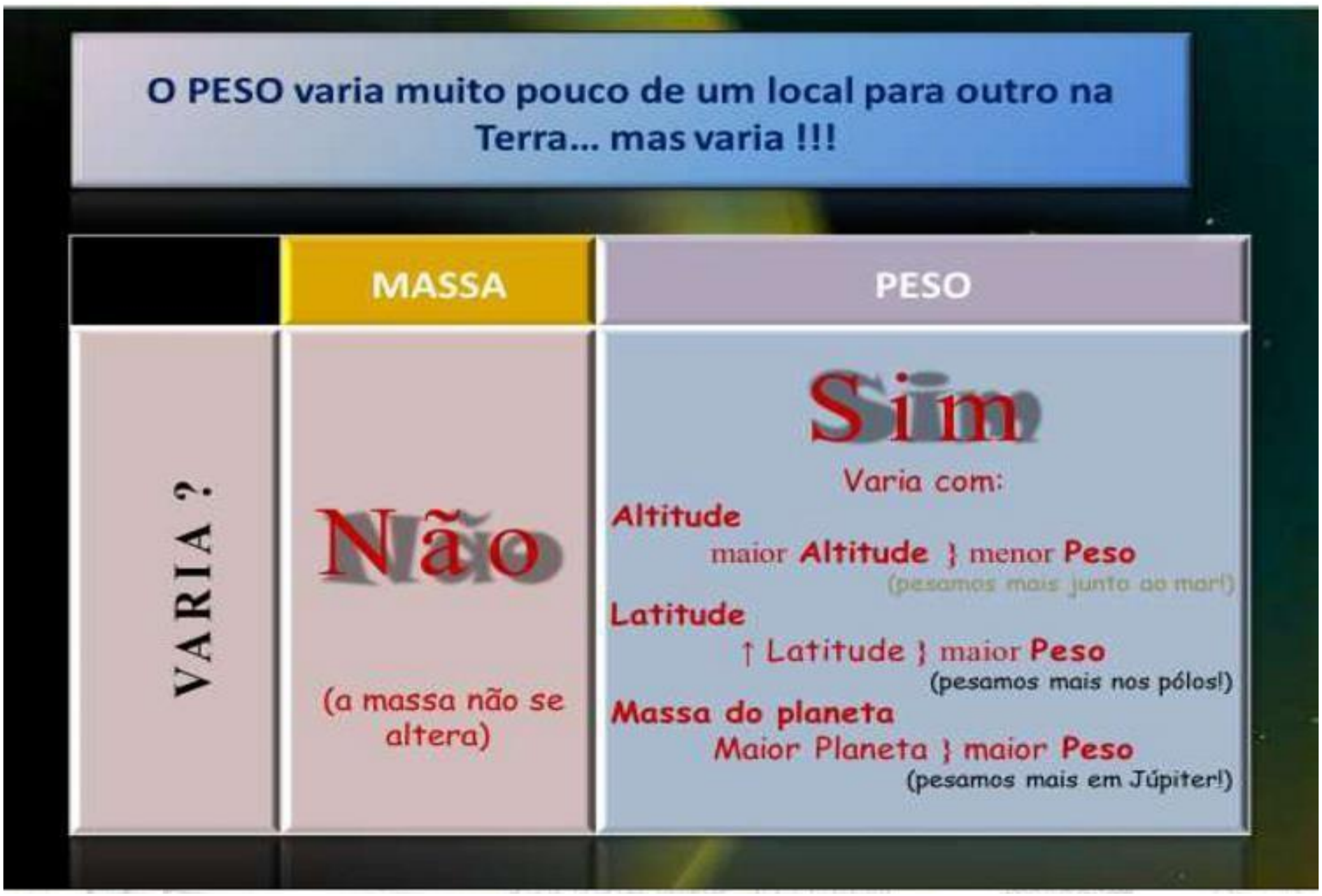

Slide 7

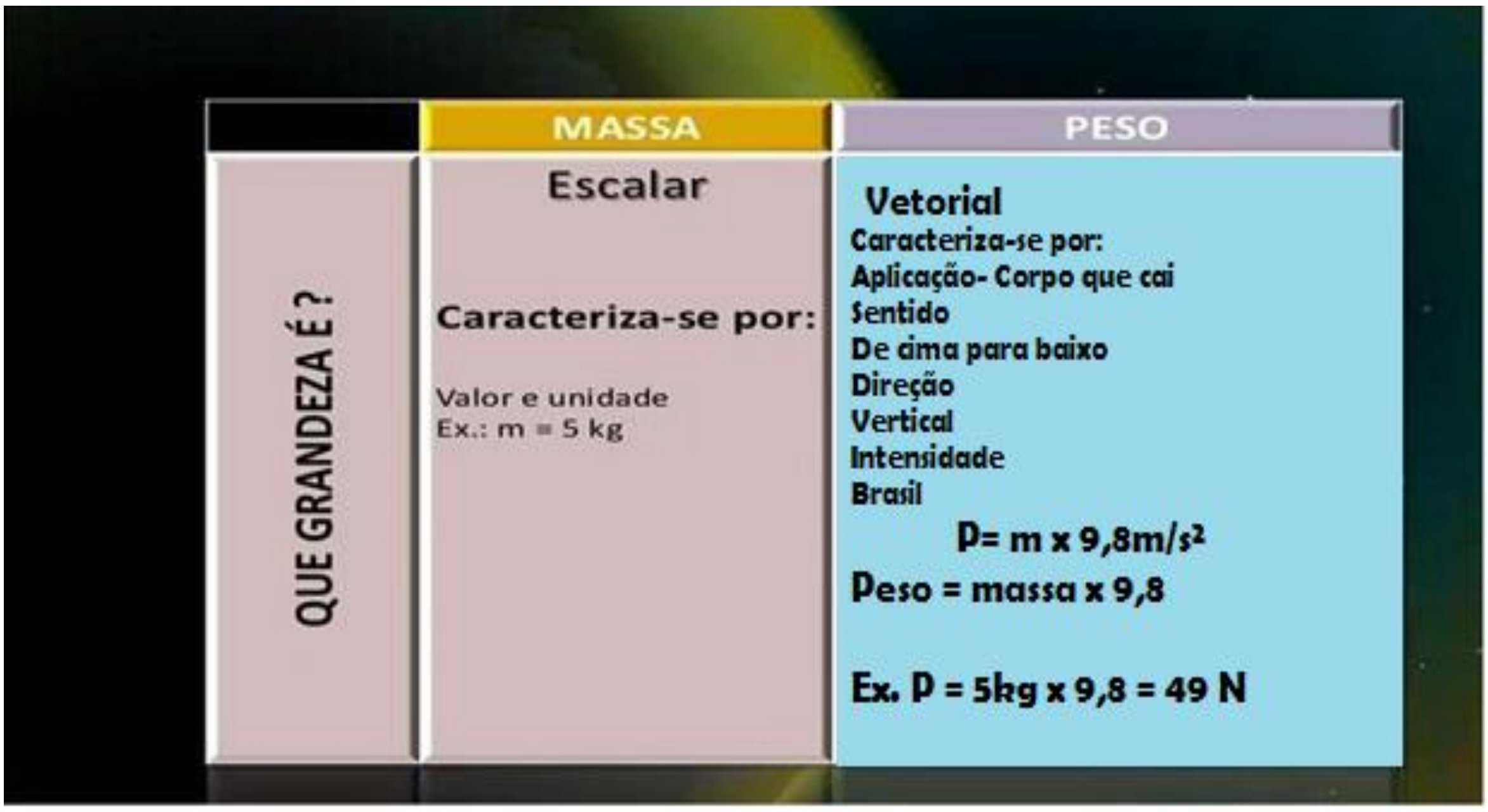

Slide 8

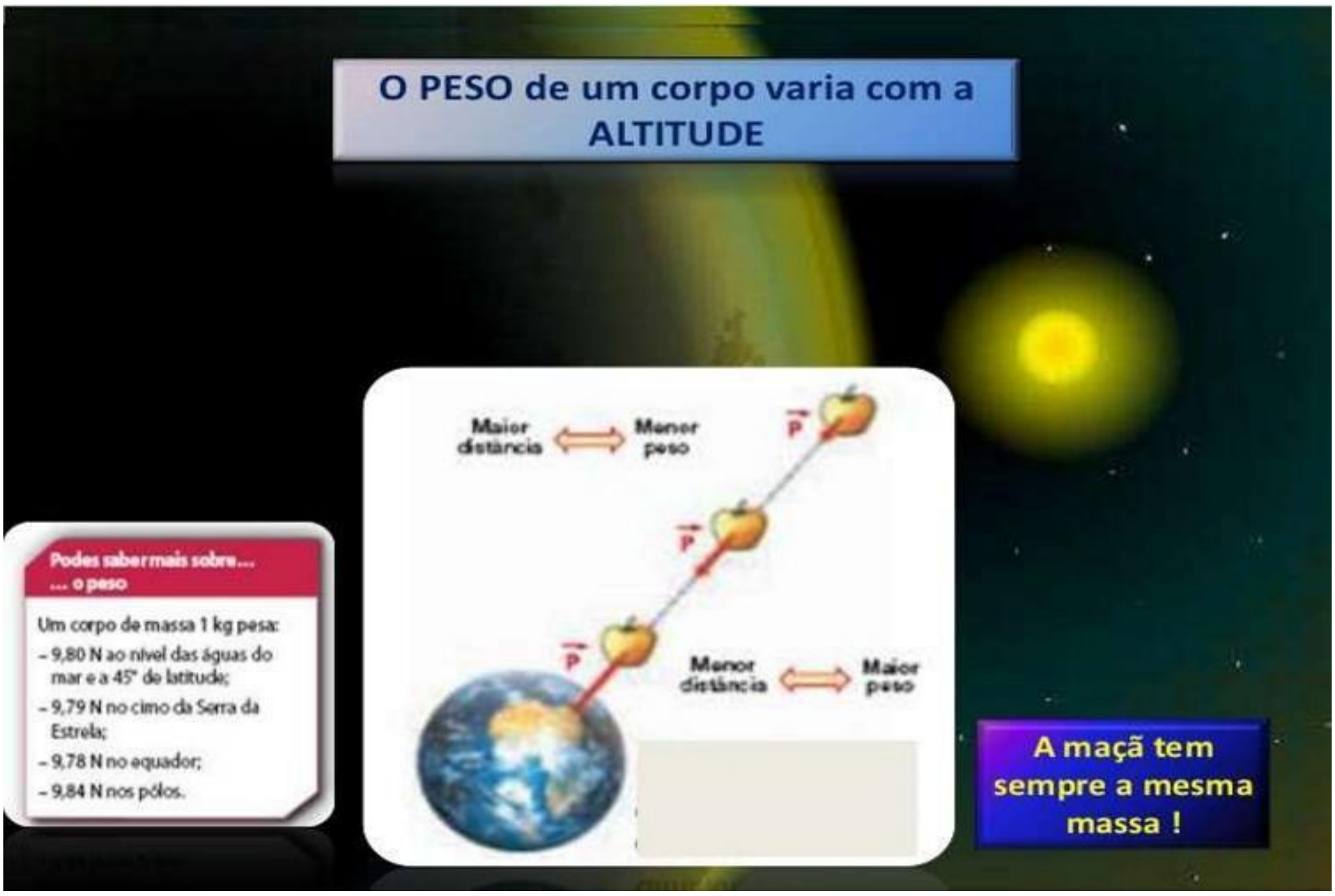

Slide 9

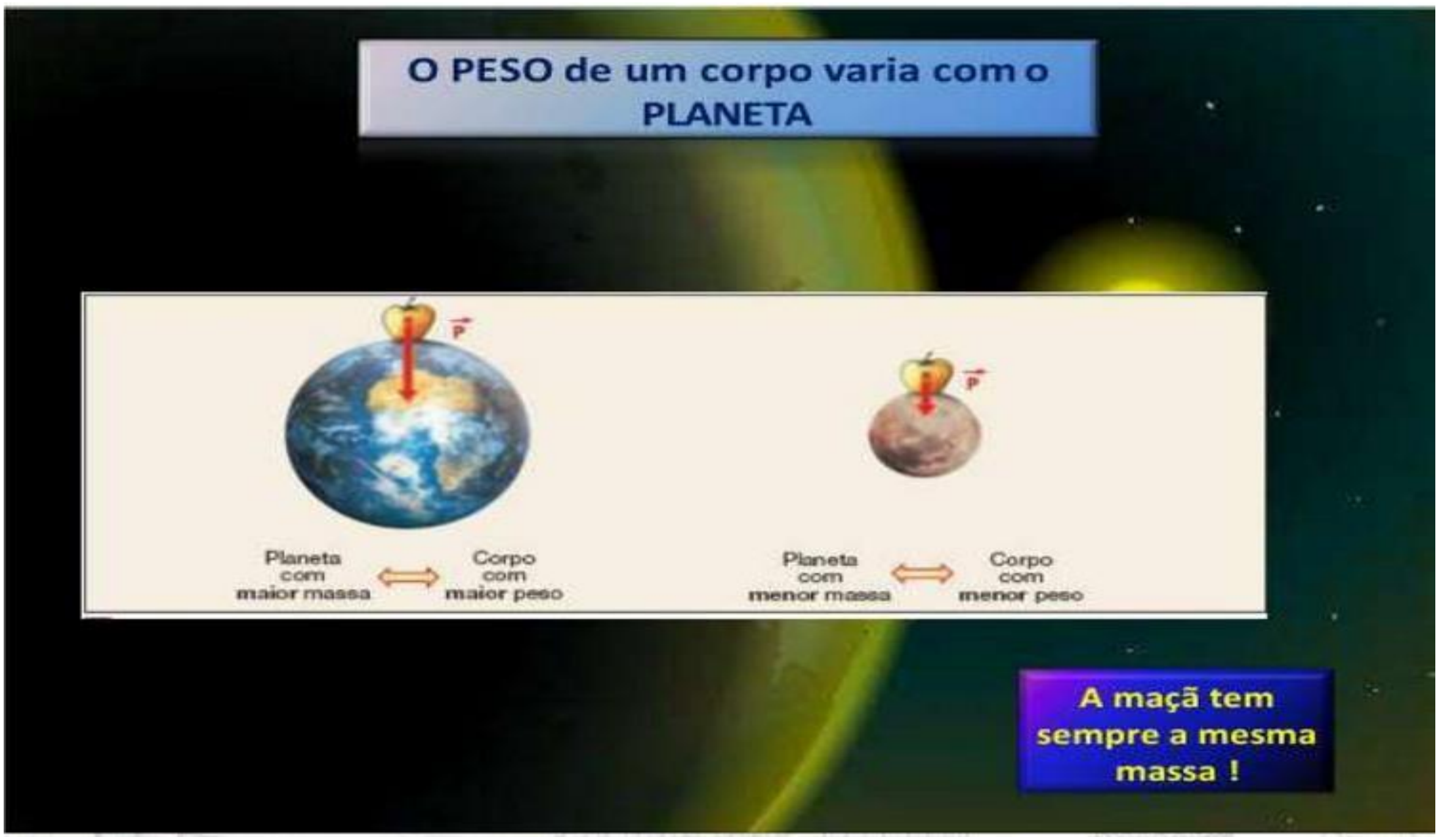

Slide 10

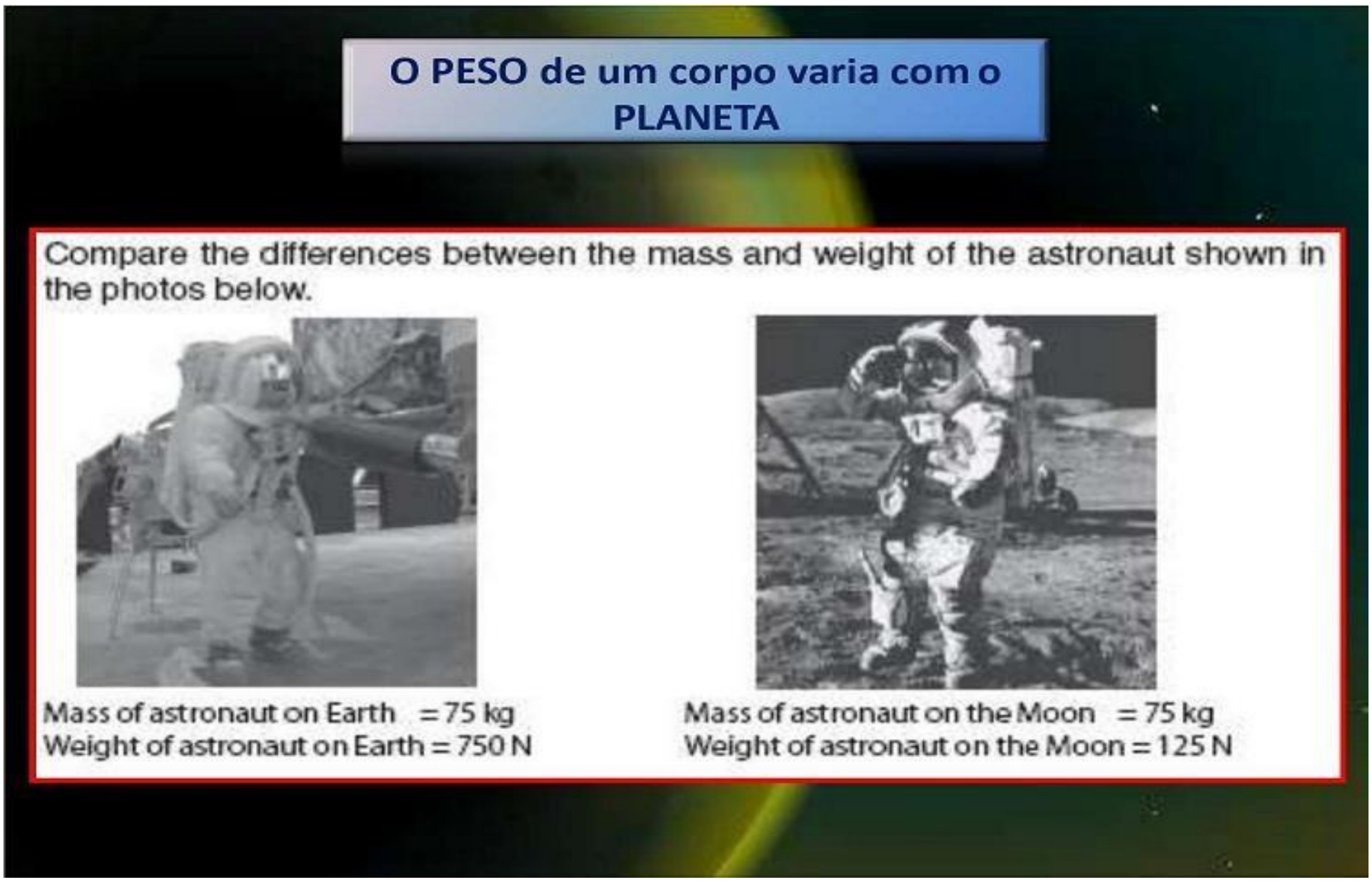

Slide 11

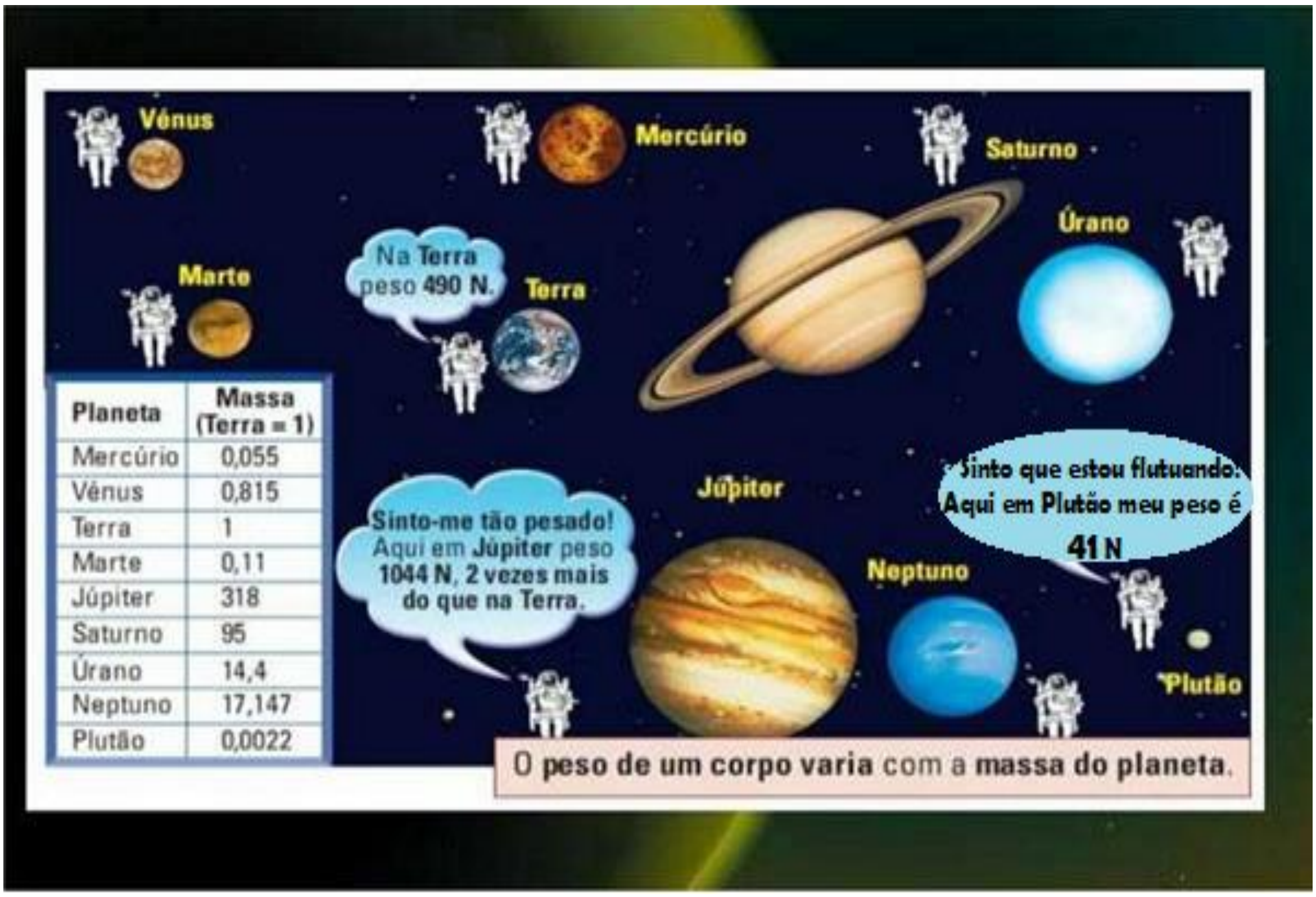

Slide 12

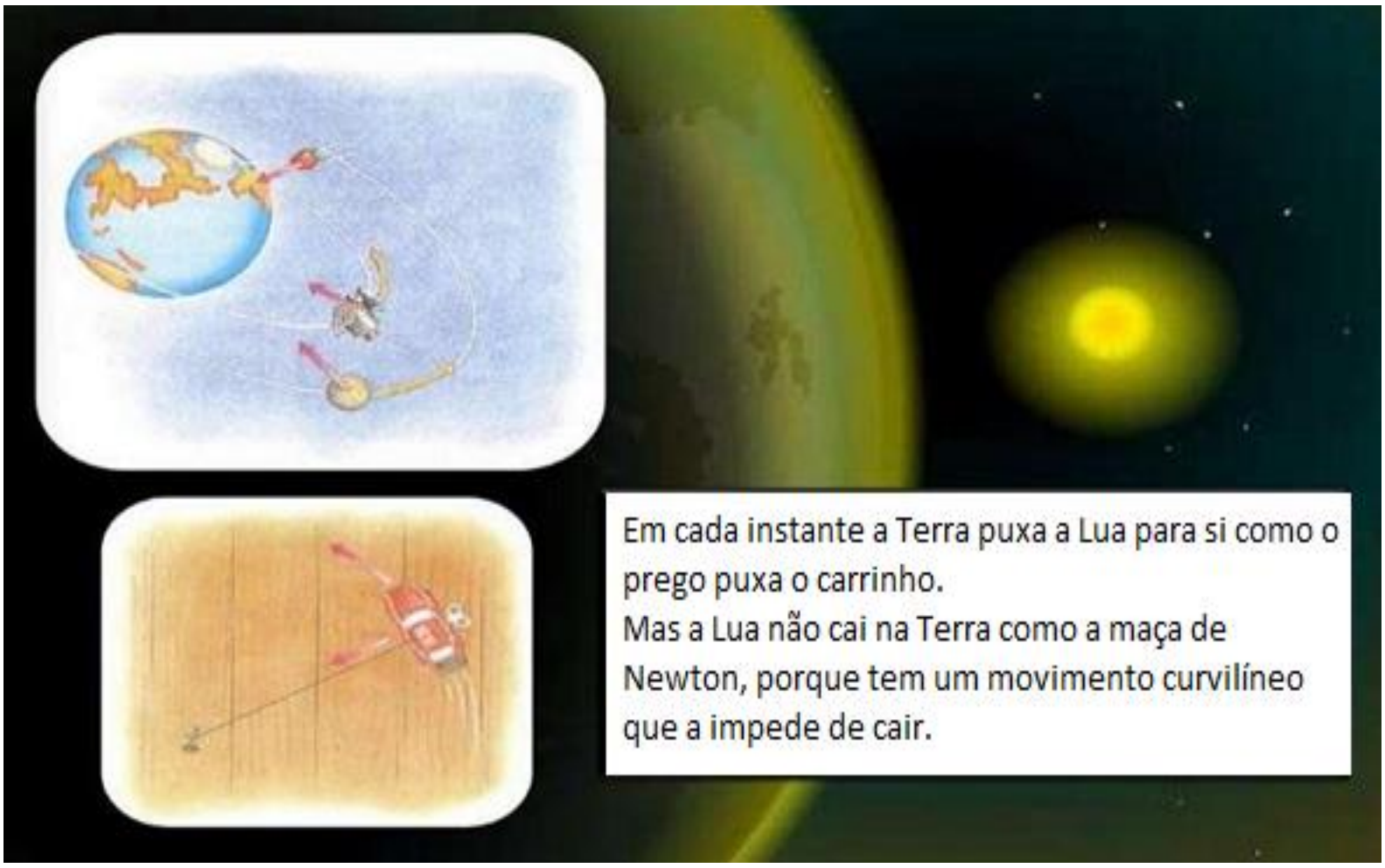

Slide 13

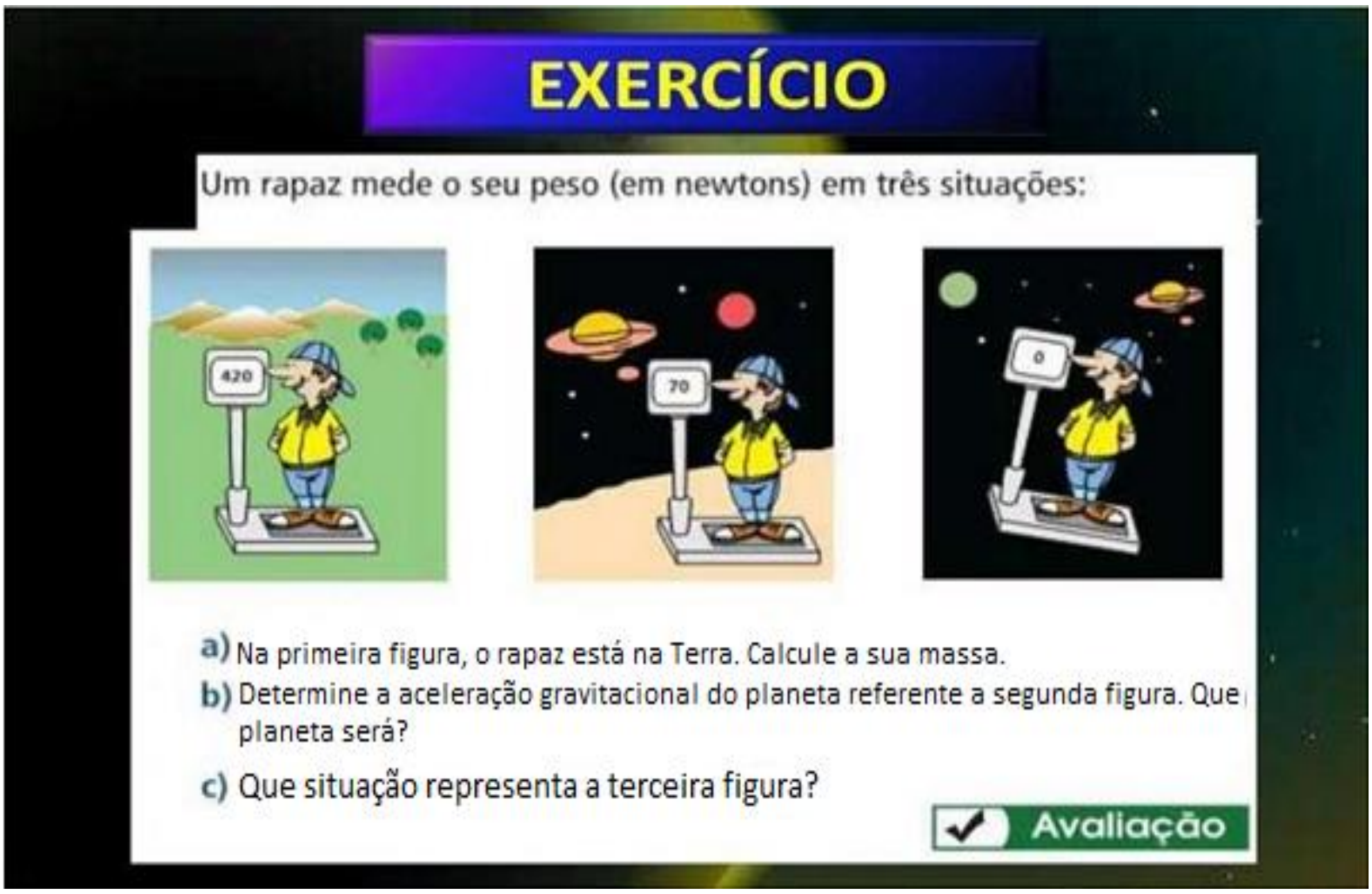

Slide 14

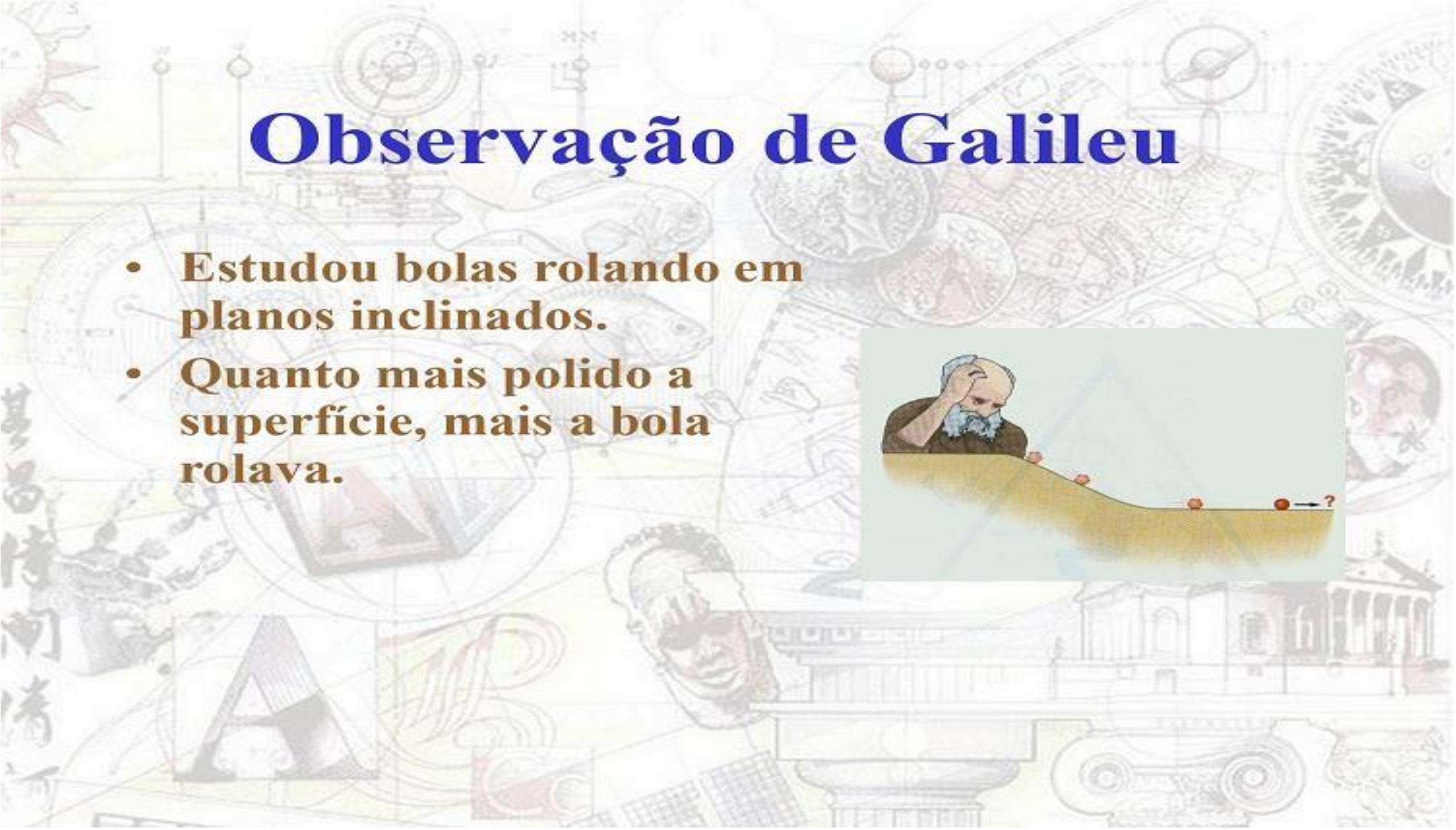

Slide 15

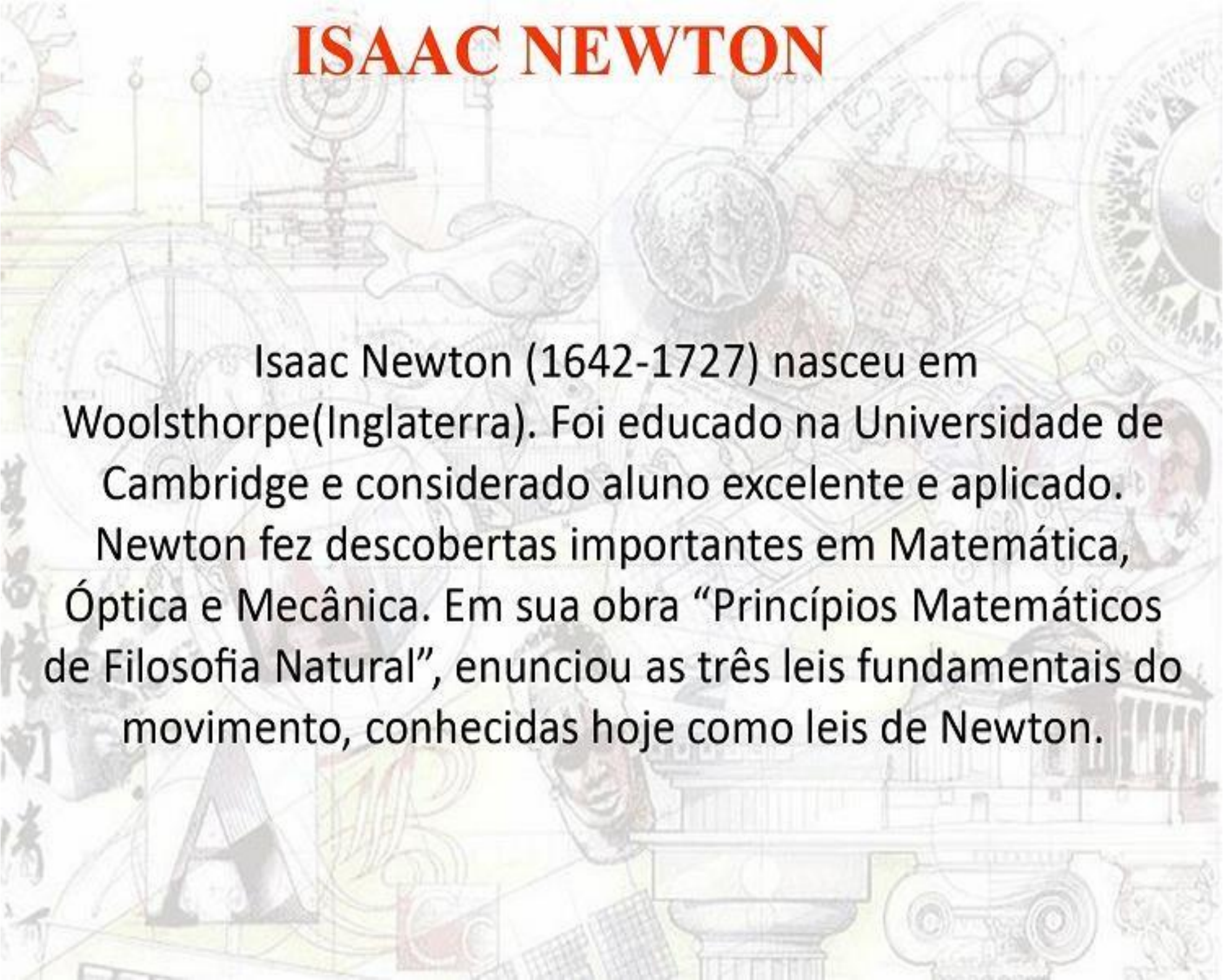
ISAAC NEWTON
Isaac Newton (1642-1727) nasceu em Woolsthorpe(Inglaterra). Foi educado na Universidade de Cambridge e considerado aluno excelente e aplicado. Newton fez descobertas importantes em Matemática, Óptica e Mecânica. Em sua obra "Princípios Matemáticos de Filosofia Natural", enunciou as três leis fundamentais do movimento, conhecidas hoje como leis de Newton.

APÊNDICE 3 – Exercícios entregues aos grupos durante a sequência didática – Primeira Lei de Newton

## 1ª lei de Newton

Grupo 1 –

a) Um veículo trafega por uma estrada retilínea com velocidade constante de 90 km/h. É correto afirmar que a resultante das forças que atuam no veículo é nula?

______________________________________________________________________

______________________________________________________________________

______________________________________________________________________

______________________________________________________________________

b) Você empurra um carrinho e ele se move. Soltando-o você nota que em poucos segundos ele para. Esse fato contradiz a 1ª lei de Newton? Justifique a resposta e explique para a sala.

______________________________________________________________________

______________________________________________________________________

______________________________________________________________________

______________________________________________________________________

______________________________________________________________________

Grupo 2 –

Em nosso cotidiano, podemos observar o princípio da inércia em várias situações. Uma delas é quando andamos de ônibus ou trem. Pensando nessa situação responda:

a) Por que quando o ônibus breca os passageiros vão para frente?

______________________________________________________________________

______________________________________________________________________

______________________________________________________________________

______________________________________________________________________

______________________________________________________________________

b) Quando o trem entra em movimento acelerado e uma pessoa, em seu interior, inclina o corpo para trás, é porque existe uma força puxando essa pessoa?
Justifique as respostas acima e apresente para a sala

______________________________________________________________________

______________________________________________________________________

______________________________________________________________________

______________________________________________________________________

______________________________________________________________________

Grupo 3 –

a) O princípio da inércia é válido quando se aplica a um corpo uma única força?

______________________________________________________________________

______________________________________________________________________

______________________________________________________________________

______________________________________________________________________

______________________________________________________________________

b) Na Física Assim como em outras disciplinas, é necessário organizar os estudos que buscam soluções para nossas indagações. Por exemplo, na mecânica concentramos os estudos relativos à ideia de movimento, que por sua vez, podem ter objetivos ligados à cinemática ou dinâmica. Descreva uma situação que justifique a diferença entre os objetivos da cinemática e os da Dinâmica, ao analisar o movimento de um ponto material.

______________________________________________________________________

______________________________________________________________________

______________________________________________________________________

______________________________________________________________________

Grupo 4 –

a) A primeira Lei de Newton afirma que, se a soma de todas as forças atuando sobre o corpo for nulo o que acontecerá?

______________________________________________________________________

______________________________________________________________________

______________________________________________________________________

______________________________________________________________________

______________________________________________________________________

______________________________________________________________________

b) Um ponto material sob ação de um sistema de forças realiza MRU. O que se pode afirmar a respeito da resultante das forças que agem sobre o ponto material?

______________________________________________________________________

______________________________________________________________________

______________________________________________________________________

______________________________________________________________________

______________________________________________________________________

APÊNDICE 4 – Exercícios da Segunda Lei de Newton

## 2ª Lei de Newton

1) Você arremessa uma bola no ar diretamente para cima. Imediatamente após ter sido lançada a bola, que forças são exercidas sobre a bola? Demonstre cada força exercida nesse lançamento em um desenho:

______________________________________________________________________
______________________________________________________________________
______________________________________________________________________
______________________________________________________________________
__________________________________

a) Descida se ela é uma força de contato ou ação a distância.

______________________________________________________________________
______________________________________________________________________
______________________________________________________________________
______________________________________________________________________
__________________________________

b) Identifique o agente correspondente

______________________________________________________________________
______________________________________________________________________
______________________________________________________________________
______________________________________________________________________
__________________________________

2) Uma força constante exercida sobre A o fez acelerar a 5m/s². Exercida sobre B, a menor força o fez acelerar a 3m/s². Aplicando a C, ela faz acelerar a 8/ms².

a) Qual dos objetos possui a maior massa? Explique.

______________________________________________________________________
______________________________________________________________________
______________________________________________________________________

b) Qual possui a menor massa? Explique.

c) A segunda Lei de Newton é dada por $\vec{F}_{res} = m\vec{a}$, logo $m\vec{a}$ é uma força? Explique.

2.1) Um objeto experimenta uma força constante que o acelera a $10 m/s^2$ . Qual será a aceleração do objeto se:

a) A força for duplicada?

b) A massa for duplicada?

c) A força e a massa forem duplicadas?

2.2) Um objeto experimenta uma força constante e acelera a 8m/s². Qual será a aceleração desse objeto se:

a) A força for reduzida à metade?

______________________________________________________________________

b) A massa for reduzida à metade?

______________________________________________________________________

c) A força e a massa forem ambas, reduzidas à metade?

______________________________________________________________________

3.1) Redija um texto de um parágrafo sobre força e movimento. Explique com palavras próprias a ligação entre a força e o movimento. Onde for possível cite evidências que sustentam suas afirmações.

______________________________________________________________________

3.2) A Figura mostra um gráfico da aceleração versus força para um objeto de 200g. que valores de força completam corretamente as lacunas referentes à escala horizontal?

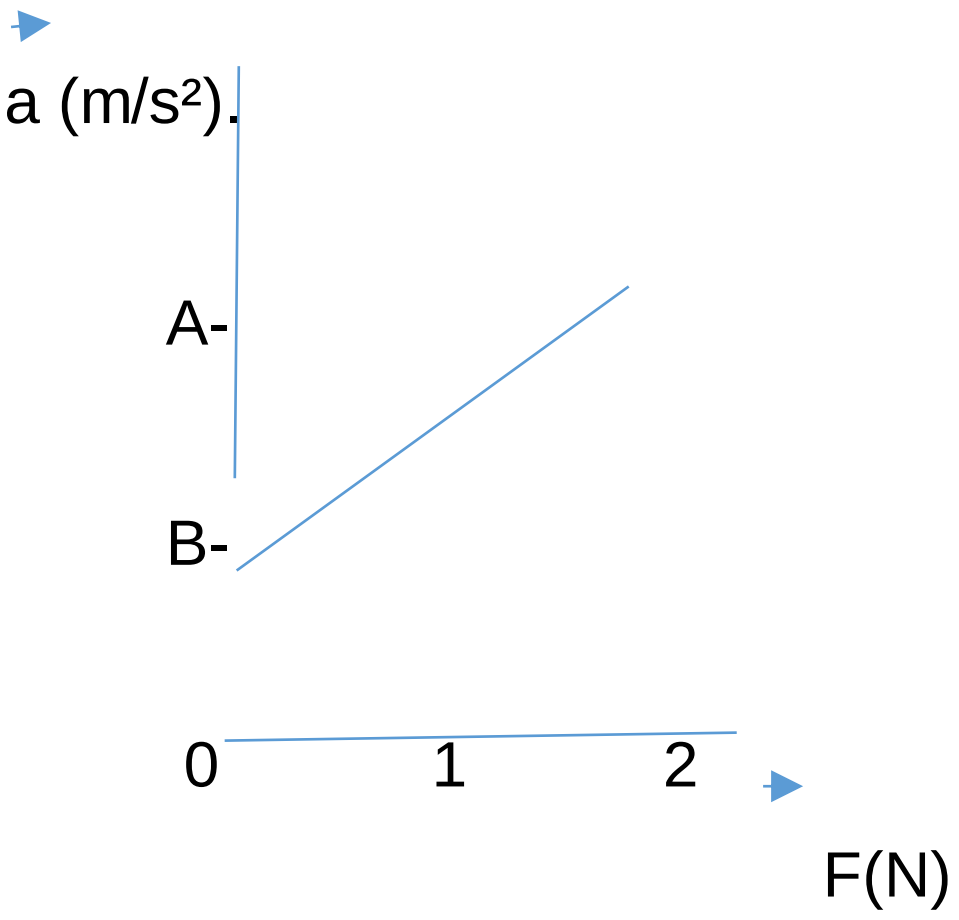

De acordo com a mesma Figura mostrando o gráfico aceleração versos força para um abjeto de 500g. que valores de aceleração completam corretamente as lacunas no gráfico abaixo:

______________________________________________________________________

______________________________________________________________________

______________________________________________________________________

______________________________________________________________________

__________________________________

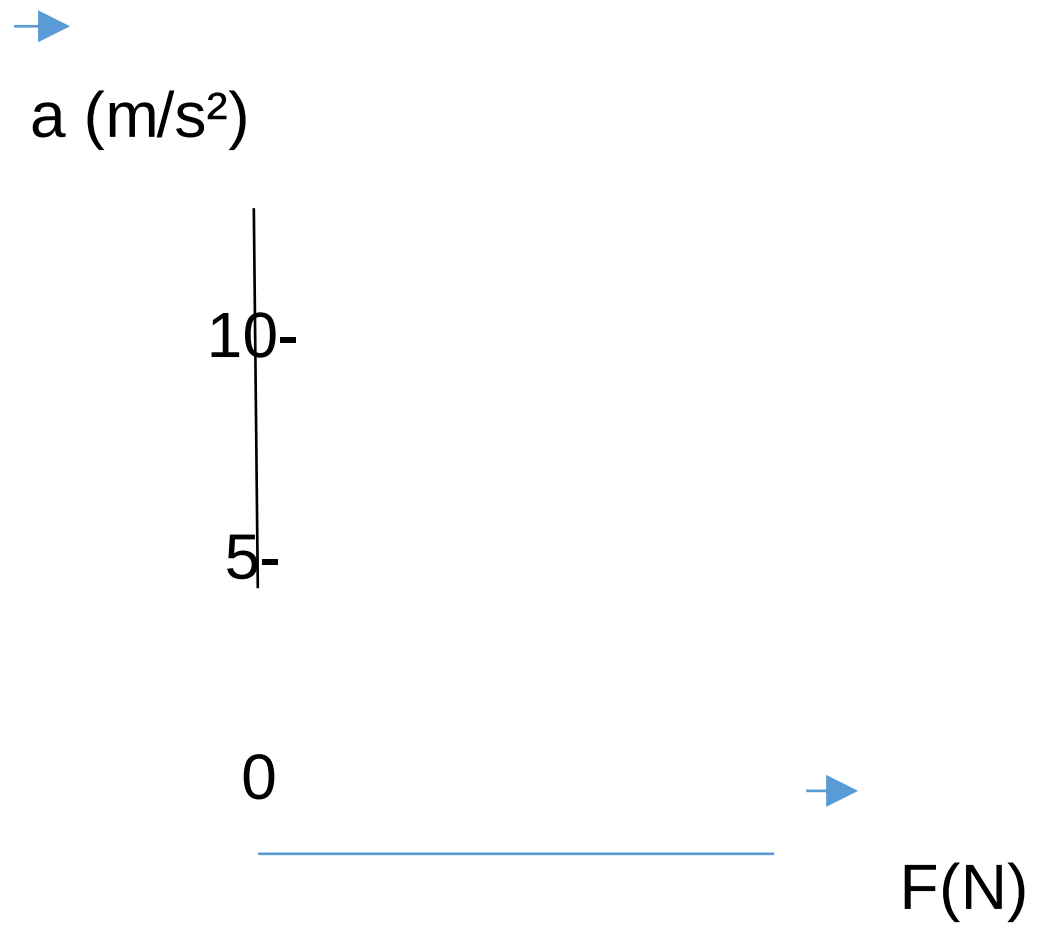

A ________ B_________

4.1) Um objeto está se movendo para o norte e a sua velocidade está aumentando. De posse dessa informação o que você conclui? Justifique sua resposta.

______________________________________________________________________
______________________________________________________________________
______________________________________________________________________
______________________________________________________________________
__________________________________

4.2) Um tijolo de ferro com 2 kg tem 2 vezes mais massa do que um tijolo de ferro de 1 kg ? Tem duas vezes mais volume? Seria mais fácil sustentar um caminhão de cimento sobre a terra ou sobre a lua?

______________________________________________________________________
______________________________________________________________________
______________________________________________________________________
______________________________________________________________________
__________________________________

# APÊNDICE 5 – Exercícios da Terceira Lei de Newton

1) Uma garrafa é empurrada sobre uma mesa e escorrega para fora da extremidade da mesa. Não desprezando a resistência do ar.

a) Quais forças atuam sobre a garrafa enquanto ela cai da mesa até o chão?

b) Quais são as reações dessas forças; ou seja, sobre quais corpos e por quais corpos as reações são exercidas?

2) O piso de um elevador exerce uma força normal de 620N de baixo para cima sobre o passageiro que pesa 650N.

a) Quais são as reações dessas duas forças?

b) O passageiro está sendo acelerado?

2) Uma pessoa necessita da força de atrito em seus pés para se deslocar sobre uma superfície. Logo, uma pessoa que sobe uma rampa em linha reta será auxiliada pela força de atrito exercida pelo chão em seus pés.

Em relação ao movimento dessa pessoa, quais são a direção e o sentido da força de atrito mencionada no texto?

___________________________________________________________________

___________________________________________________________________

___________________________________________________________________

3) Uma bala de um rifle 22, deslocando-se a 350 m/s, atinge o tronco de uma arvore grande, no qual ela penetra até a profundidade de 0,130m. A massa da bala é de 1,80g. Supunha uma força retardadora constante.

a) Qual o tempo necessário para a bala parar?

___________________________________________________________________

___________________________________________________________________

___________________________________________________________________

___________________________________________________________________

b) Qual a força em Newtons, que o tronco da arvore exerce sobre a bala

___________________________________________________________________

___________________________________________________________________

___________________________________________________________________

3.1) Sabemos que a terra puxa a lua, isso significa que a lua também puxa a terra?

___________________________________________________________________

___________________________________________________________________

___________________________________________________________________

___________________________________________________________________

___________________________________________________________________

4) Você pode identificar as forças de ação e reação no caso de um objeto em queda no vácuo?

________________________________________________________________________

________________________________________________________________________

________________________________________________________________________

4.1) Um carro acelera em uma rodovia. Identifique a força que move o carro.

________________________________________________________________________

________________________________________________________________________

________________________________________________________________________

________________________________________________________________________

________________________________________________________________________

4.2) Quando você deixa uma bola de borracha cair no chão, ela repica até quase a altura original. O que causa o repique da bola?

________________________________________________________________________

________________________________________________________________________

________________________________________________________________________

# APÊNDICE 6 – Avaliação final

| |
|---|
| Mestrado profissional em ensino de Física – SBF<br>Instituto Federal do Espirito Santo<br>Campus Cariacica<br>Dissertação:<br>Material didático para ensino das leis de Newton com uso da realidade aumentada<br>Lucia Helena Horta Oliveira<br>EEEM Dr. Silva Mello<br>Questionário Final de Avaliação |

1) Comente a seguinte afirmação: " Inércia é um tipo de força que mantém os corpos parados ou em movimento retilíneo uniforme".

______________________________________________________________________

______________________________________________________________________

______________________________________________________________________

______________________

2) É possível um corpo exercer força sobre si mesmo?

______________________________________________________________________

______________________________________________________________________

______________________________________________________________________

______________________

3) Você empurra um carrinho e ele se move. Soltando-o você nota que em poucos segundos ele para. Esse fato contradiz a 1ª lei de Newton? Se sim, pela primeira lei como deveria ser o movimento do carrinho? Se não justifique porquê.

______________________________________________________________________

______________________________________________________________________

______________________________________________________________________

______________________________________________________________________

______________________________________________________________________

______________________________________________________________________

_____________________

4) Pode-se assistir pela tv os astronautas saltando na superfície da lua quando chegaram ao satélite natural da Terra. É possível observar que os saltos deles são mais demorados do que se estivessem na Terra (eles alcançam uma altura maior e descem mais lentamente). Essa diferença se deve ao fato de a aceleração da gravidade terrestre valer perto de 10m/s² e a aceleração gravitacional lunar valer aproximadamente 1,6 m/s². Um astronauta de massa 80 Kg apresenta pesos diferentes na terra e na lua. Determine os valores desses pesos.

________________________________________________________________________

________________________________________________________________________

________________________________________________________________________

________________________________________________________________________

________________________________________________________________________

______________________________________

5) Um corpo de massa igual a 4 Kg é submetido à ação simultaneamente e exclusiva de duas forças constantes de intensidades iguais a 4N e 6N, respectivamente. O maior valor possível para a aceleração desse corpo é de:

________________________________________________________________________

________________________________________________________________________

________________________________________________________________________

________________________________________________________________________

________________________________________________________________________

______________________________________

6) Deseja-se pesar uma girafa. Devido ao seu tamanho, usam-se duas balanças idênticas, do seguinte modo: Colocam-se as patas dianteiras sobre a balança 1 e as traseiras sobre a balança 2, como mostra a Figura. Como a girafa em repouso, m a balança 1 indica 400 Kgf e a balança 2 indica 300 Kgf. Qual é o peso da girafa?

________________________________________________________________________

________________________________________________________________________

________________________________________________________________________

________________________________________________________________________

______________________________

7) O burro puxa a carroça, exercendo sobre ela uma força horizontal para frente. Pelo princípio da ação e reação, a carroça aplicará no burro uma força de igual intensidade e direção, porém para trás. Como são opostas e de igual valor. Essas forças devem se cancelar, assim nem o burro nem a carroça deverão sair do lugar.

a) De acordo com a lei da ação e reação o paradoxo do burro contém um erro de conceito. Identifique esse erro.

b) Como se justifica o fato de o burro conseguir se mover para frente, se a força aplicada sobre ele, pela carroça é para trás?

8) Quando você deixa uma bola de borracha cair no chão, ela repica até quase a altura original. O que causa o repique da bola?

9) Sabemos que a Terra puxa a Lua, isso significa que a lua também puxa a terra? Justifique sua resposta.

______________________________________________________________________
______________________________________________________________________
______________________________________________________________________
______________________________________________________________________
_______________________________

10 ) Uma pessoa necessita de força de atrito em seus pés para se deslocar sobre uma superfície. Logo, uma pessoa que sobe uma rampa em linha reta auxiliada pela força de atrito exercida pelo chão em seus pés.

Em relação ao movimento dessa pessoa, quais são a direção e o sentido da força de atrito mencionado no texto? Desenhe indicando essas forças através de vetores.

______________________________________________________________________
______________________________________________________________________
______________________________________________________________________
______________________________________________________________________
_______________________________

# APÊNCICE 7 – Questionário extra

| |
|---|
| Mestrado profissional em ensino de Física – SBF<br>Instituto Federal do Espirito Santo<br>Campus Cariacica<br>Dissertação:<br>Material didático para ensino das leis de Newton com uso da realidade aumentada<br>Lucia Helena Horta Oliveira<br>EEEM Dr. Silva Mello<br>Questionário para avaliação da qualidade do produto |

1 – Referente aos conceitos que você aprendeu com a revista de realidade aumentada responda:

a) O conteúdo ficou entendido de maneira clara e objetiva?

_______________________________________________________________________

_______________________________________________________________________

_______________________________________________________________________

_______________________________________________________________________

______________________________

b) Você achou a metodologia da revista difícil de entender? Comente o que você achou da metodologia de ensino.

_______________________________________________________________________

_______________________________________________________________________

_______________________________________________________________________

_______________________________________________________________________

______________________________

c) Você gostaria de mais atividades como essa?

_______________________________________________________________________

_______________________________________________________________________

_______________________________________________________________________

_______________________________________________________________________

______________________________

d) Você acha que com atividades como essa você consegue aprender mais ou não?

_____________________________________________________________________________________________________________________________________________________________________________________

e) Qual conceito ficou melhor explicado por qual Gif da revista na sua opinião?

_____________________________________________________________________________________________________________________________________________________________________________________

f) Qual Gif você teve maior dificuldade de interpretar?

_____________________________________________________________________________________________________________________________________________________________________________________

g) Você tem alguma sugestão para melhorar ou mudar a revista para uma próxima atividade?

_____________________________________________________________________________________________________________________________________________________________________________________

Obrigada! Sua opinião faz a diferença para o nosso melhoramento contínuo.

APÊNDICE 8 – Revista para impressão (Aplicação)

(As páginas estão em sequência de impressão frente e verso, mas deve-se imprimir a capa em apenas um lado).

## Orientações ao professor.

O material foi desenvolvido para ensino das Leis de Newton, mas também conta com demonstrações de algumas forças mais utilizadas no processo ensino aprendizado das Leis de Newton.

Para melhor utilização da revista, é interessante que o aluno tenha uma base que inclui grandezas escalares e vetoriais, conceito de movimento, espaço percorrido, velocidade escalar, ponto material, referencial, trajetória, movimento retilíneo uniforme, movimento retilíneo uniformemente variado, movimentos sob a ação da gravidade, lançamento vertical e lançamento obliquo de projéteis.

Com esse conteúdo a interpretação dos alunos tende a ficar mais científica. Caso o aluno não tenha uma base dos assuntos abordados mesmo que superficialmente as interpretações podem ficar apenas na observação não sendo observada a parte científica. No caso de ensino de jovens e adultos (EJA), a sugestão é uma base mais simples como a utilizada na sequência da dissertação de apresentação da revista.

Para instalar o aplicativo acesse: https://wordpress.com/view/physics7.school.blog

Ao final da página temos o aplicativo para androide na versão inicial e final. Abra com o celular a página. Clique no link e baixe o aplicativo. O aplicativo pode ser usado sem internet, mas para instalar o usuário precisa de internet.
Depois de instalado abra o aplicativo. Você verá uma câmera aberta posicione a câmera nas páginas da revista. A realidade aumentada aparece nas páginas: capa - frente e trás, 06, 07, 08, 09, 10, 11, 12, 13, 14, 15 (temos 2 Figuras – posicione acima e abaixo na página), 16 e 17.

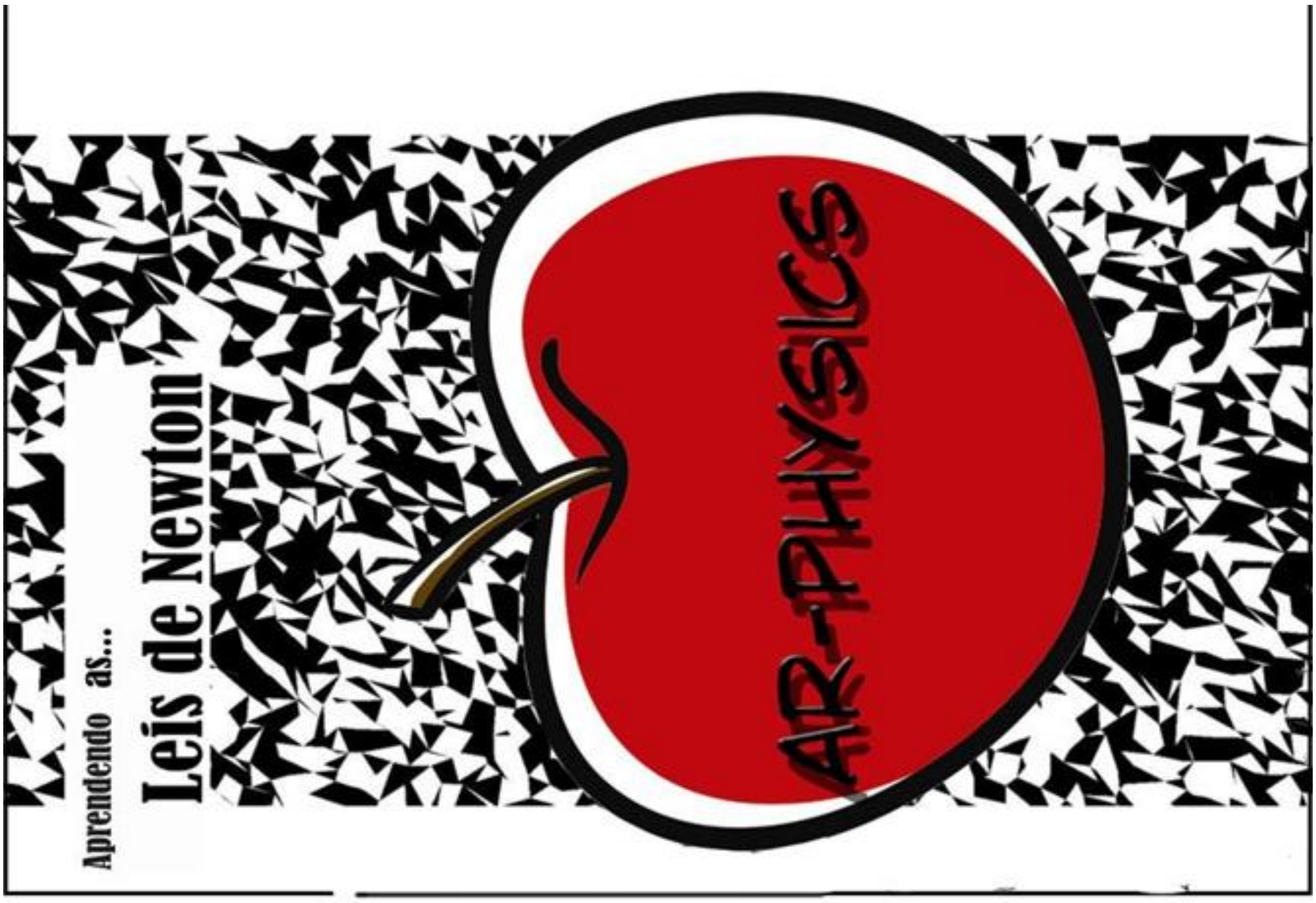
Aprendendo as...
Leis de Newton
AR-PHYSICS

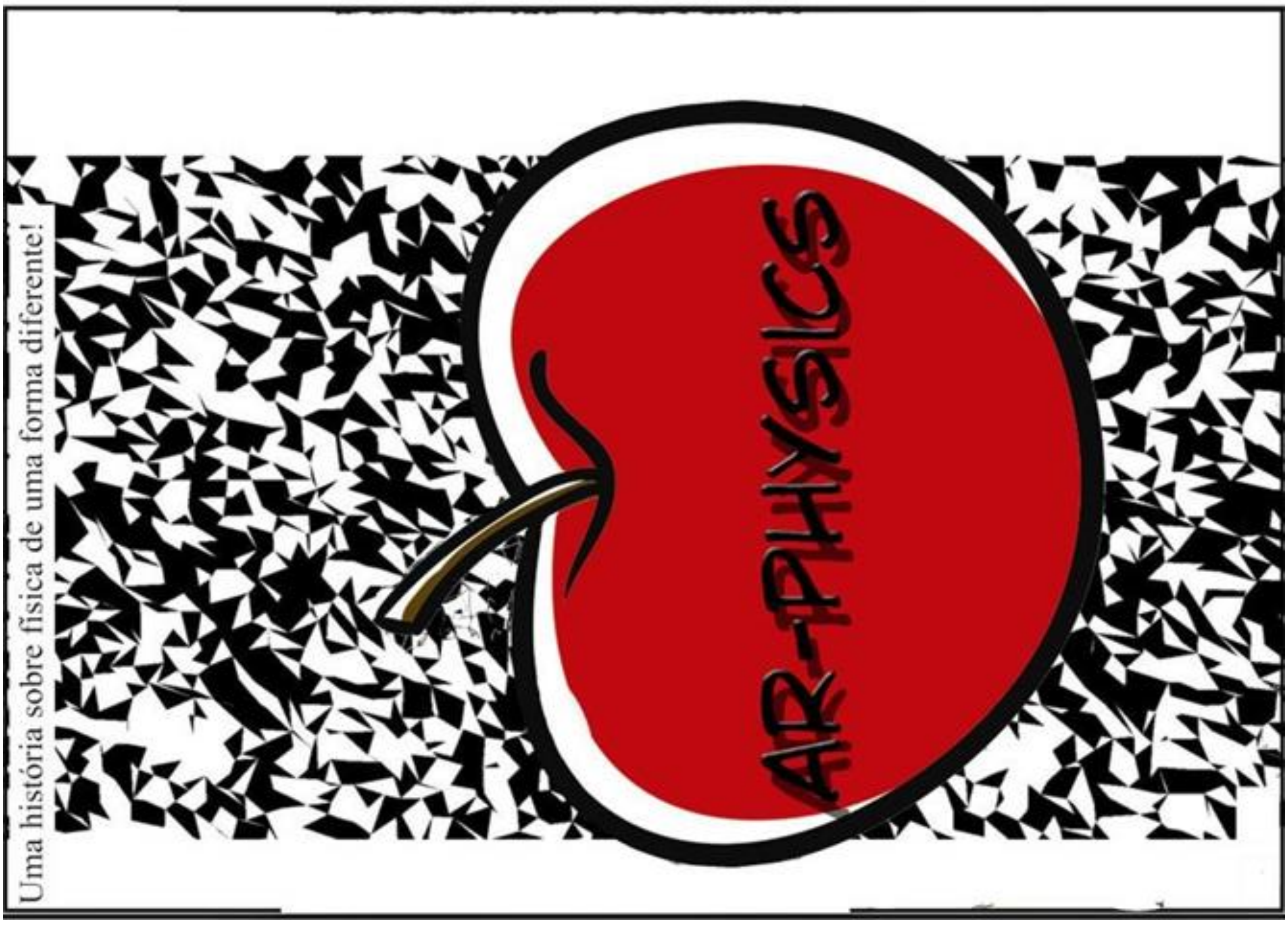
Uma história sobre física de uma forma diferente!
AR-PHYSICS

**Orientações ao professor.**

**O material foi desenvolvido para ensino das Leis de Newton, mas também conta com demonstrações de algumas forças mais utilizadas no processo ensino aprendizado das Leis de Newton. Para melhor utilização da revista, é interessante que o aluno tenha uma base que inclui grandezas escalares e vetoriais, conceito de movimento, espaço percorrido, velocidade escalar, ponto material, referencial, trajetória, movimento retilíneo uniforme, movimento retilíneo uniformemente variado, movimentos sob a ação da gravidade, lançamento vertical e lançamento obliquo de projéteis.**

**Com esse conteúdo a interpretação dos alunos tende a ficar mais científica. Caso o aluno não tenha uma base dos assuntos abordados mesmo que superficialmente as interpretações podem ficar apenas na observação não sendo observada a parte científica.**

**No caso de ensino de jovens e adultos (EJA), a sugestão é uma base mais simples como a utilizada na sequência da dissertação de apresentação da revista.**

**Para instalar o aplicativo acesse:**

**https://wordpress.com/view/physics7.school.blog**

**Ao final da página temos o aplicativo para androide na versão inicial e final. Abra com o celular a página. Clique no link e baixe o aplicativo. O aplicativo pode ser usado sem internet mas para instalar o usuário precisa de internet.**

**Depois de instalado abra o aplicativo. Você verá uma câmera aberta posicione a câmera na s paginas da revista.**

**A realidade aumentada aparece nas páginas: capa - frente e trás, 06, 07, 08, 09, 10, 11 , 12, 13, 14, 15 (temos 2 figuras – posicione a cima e a baixo da página), 16 e 17 .**

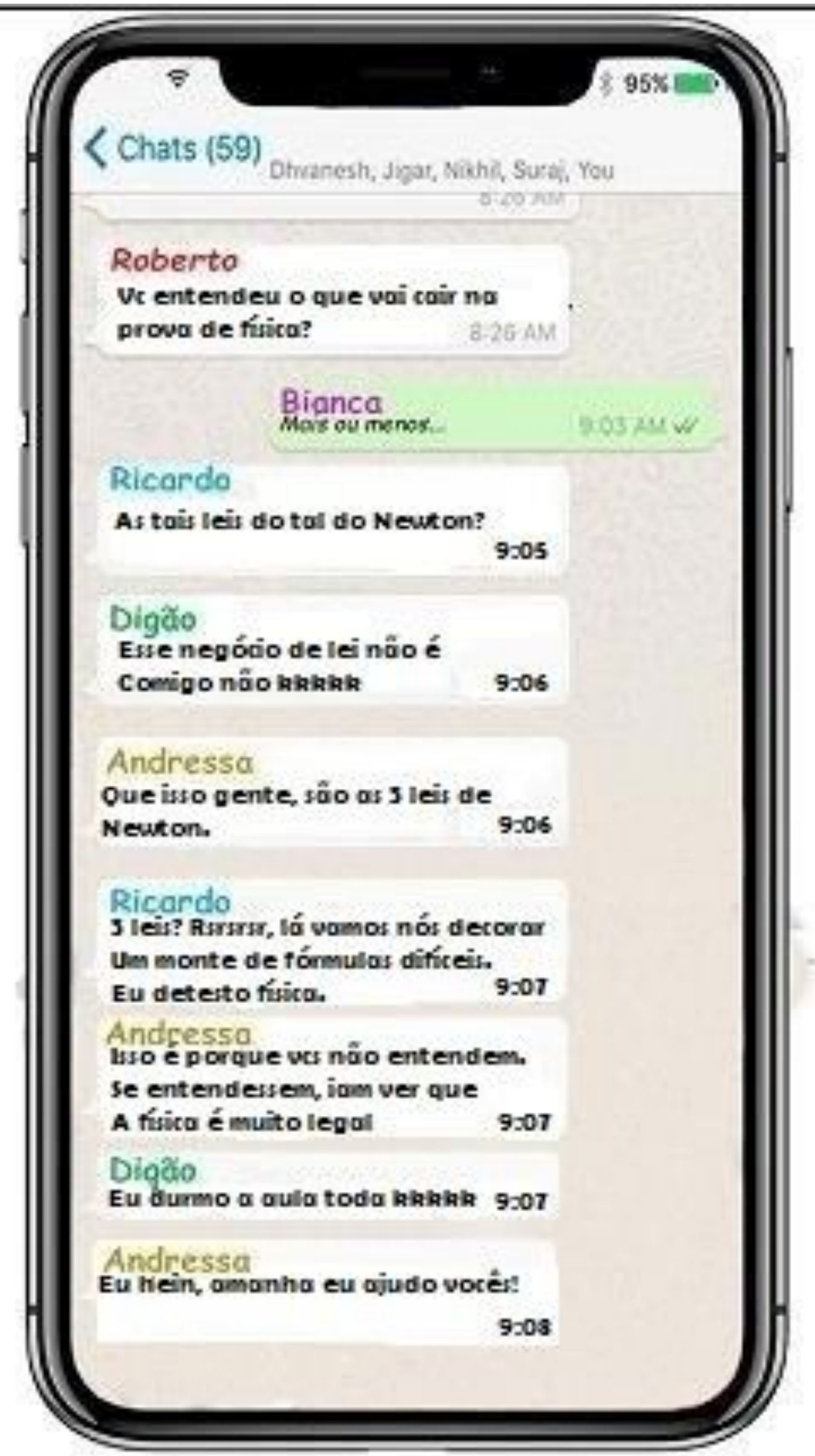

1

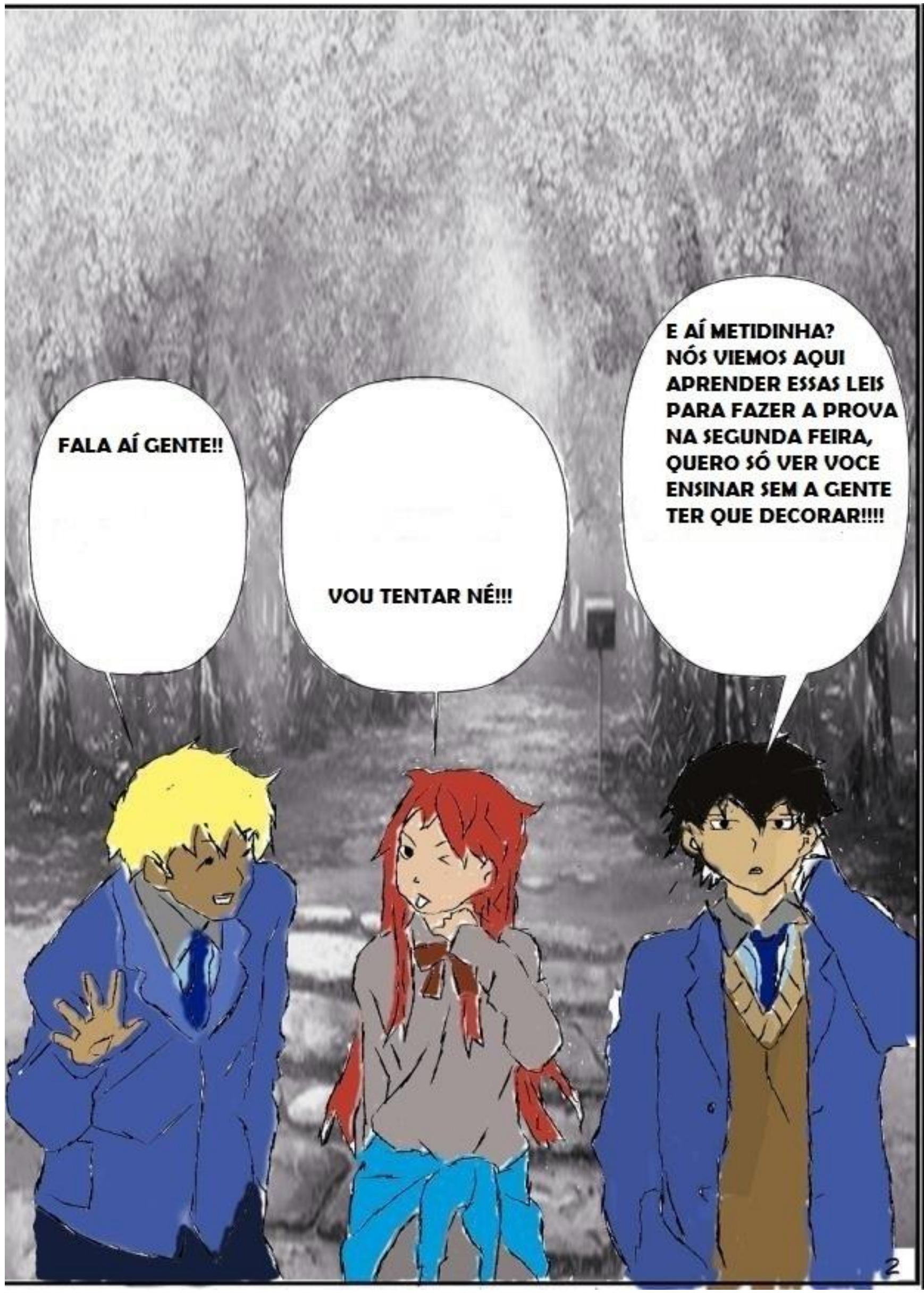

**PRODUTO USADO COMO COMPLEMENTO NA APLICAÇÃO DA DISSERTAÇÃO:**
**PROPOSTA DE SEQUÊNCIA DIDÁTICA PARA ENSINO DAS LEIS DE NEWTON,**
**UTILIZANDO GIFs E VÍDEOS.**
**DISSERTAÇÃO PARA O MESTRADO PROFISSIONAL DE ENSINO DE FÍSICA.**

SBF(SOCIEDADE BRASILEIRA DE FÍSICA).
MNPF (MESTRADO NACIONAL PROFISSIONAL EM ENSINO DE FÍSICA).
POLO IFES DE CARIACICA - E.S.

**AUTOR: LUCIA HELENA HORTA OLIVEIRA**
**ORIENTADOR: DR. SAMIR LACERDA DA SILVA**
**DESIGNER E PROGRAMAÇÃO: AYLER FRANCISCO HORTA OLIVEIRA.**

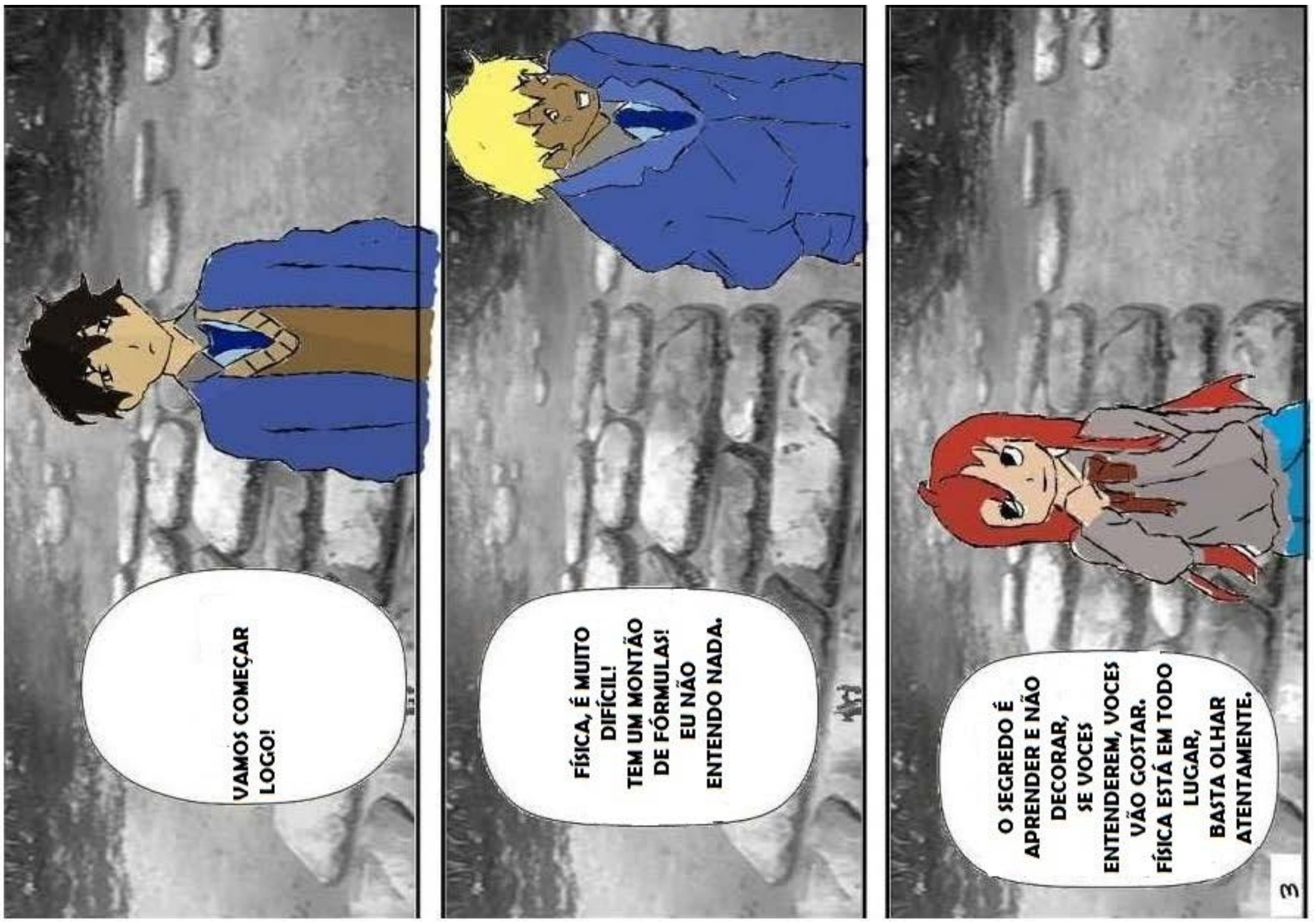
VAMOS COMEÇAR LOGO!
FÍSICA, É MUITO DIFÍCIL! TEM UM MONTÃO DE FÓRMULAS! EU NÃO ENTENDO NADA.
O SEGREDO É APRENDER E NÃO DECORAR, SE VOCES ENTENDEREM, VOCES VÃO GOSTAR. FÍSICA ESTÁ EM TODO LUGAR, BASTA OLHAR ATENTAMENTE.
3

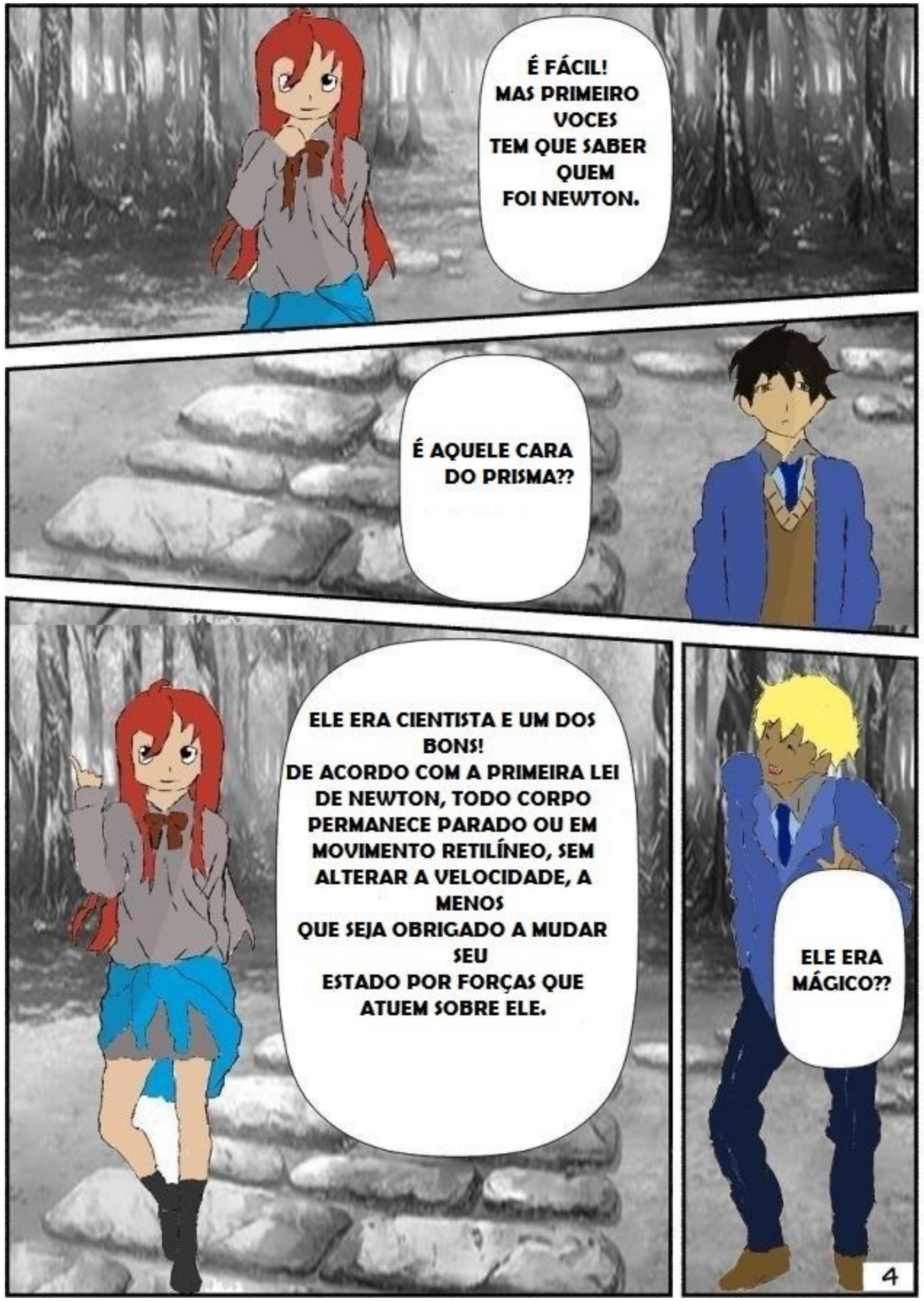
É FÁCIL!
MAS PRIMEIRO
VOCES
TEM QUE SABER
QUEM
FOI NEWTON.
É AQUELE CARA
DO PRISMA??
ELE ERA CIENTISTA E UM DOS
BONS!
DE ACORDO COM A PRIMEIRA LEI
DE NEWTON, TODO CORPO
PERMANECE PARADO OU EM
MOVIMENTO RETILÍNEO, SEM
ALTERAR A VELOCIDADE, A
MENOS
QUE SEJA OBRIGADO A MUDAR
SEU
ESTADO POR FORÇAS QUE
ATUEM SOBRE ELE.
ELE ERA
MÁGICO??
4

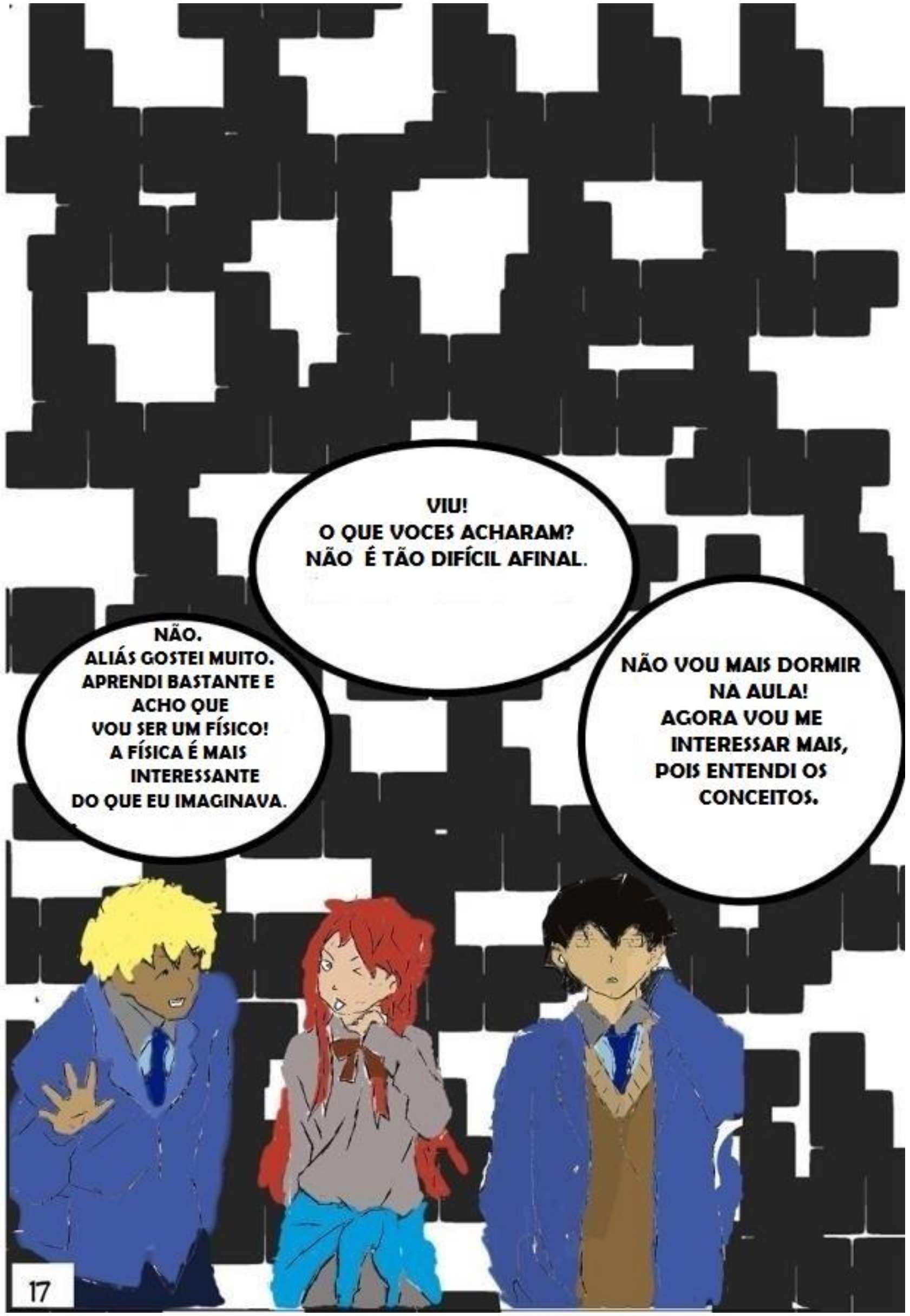
VIU!
O QUE VOCES ACHARAM?
NÃO É TÃO DIFÍCIL AFINAL.
NÃO.
ALIÁS GOSTEI MUITO.
APRENDI BASTANTE E
ACHO QUE
VOU SER UM FÍSICO!
A FÍSICA É MAIS
INTERESSANTE
DO QUE EU IMAGINAVA.
NÃO VOU MAIS DORMIR
NA AULA!
AGORA VOU ME
INTERESSAR MAIS,
POIS ENTENDI OS
CONCEITOS.
17

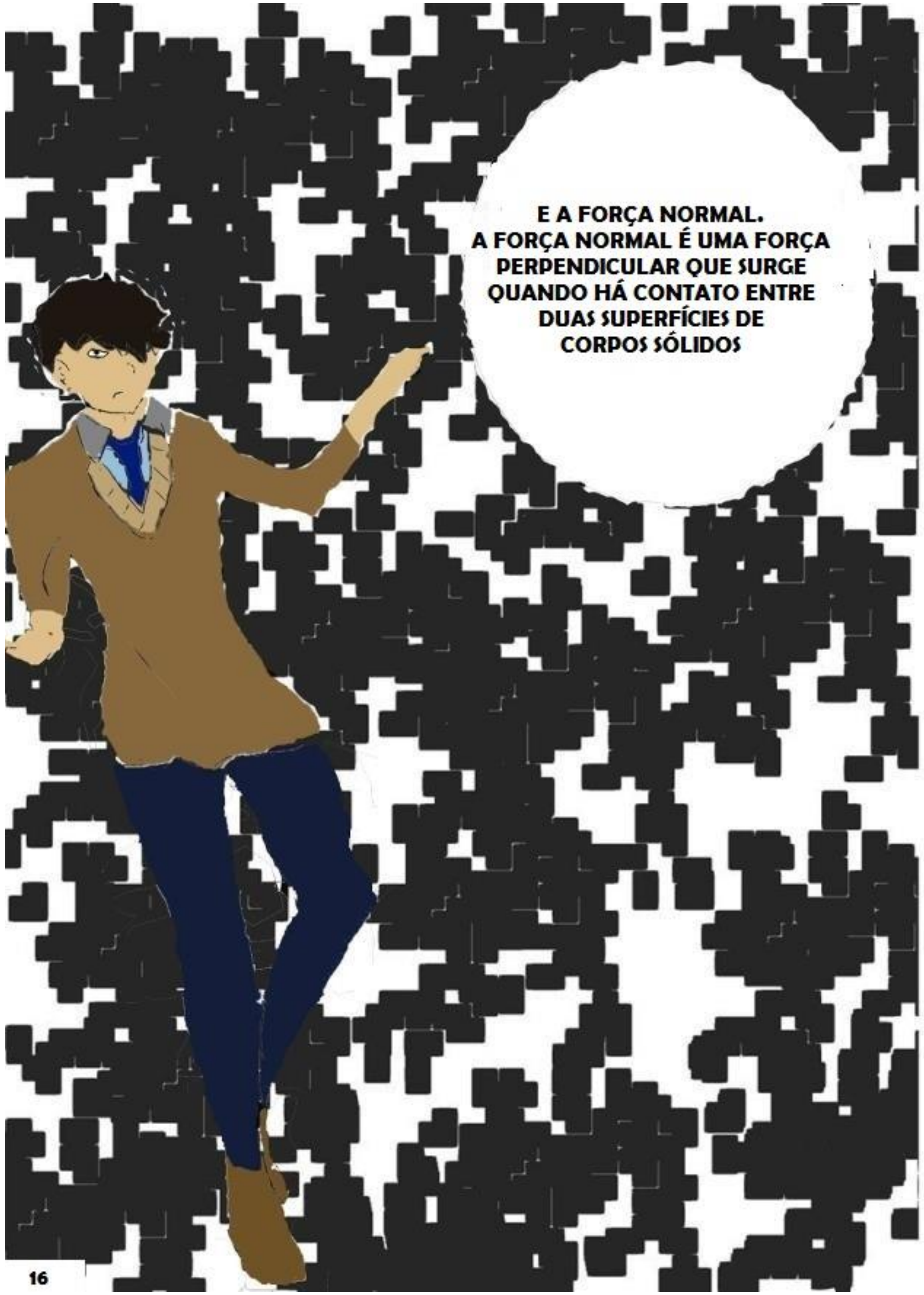
E A FORÇA NORMAL.
A FORÇA NORMAL É UMA FORÇA PERPENDICULAR QUE SURGE QUANDO HÁ CONTATO ENTRE DUAS SUPERFÍCIES DE CORPOS SÓLIDOS
16

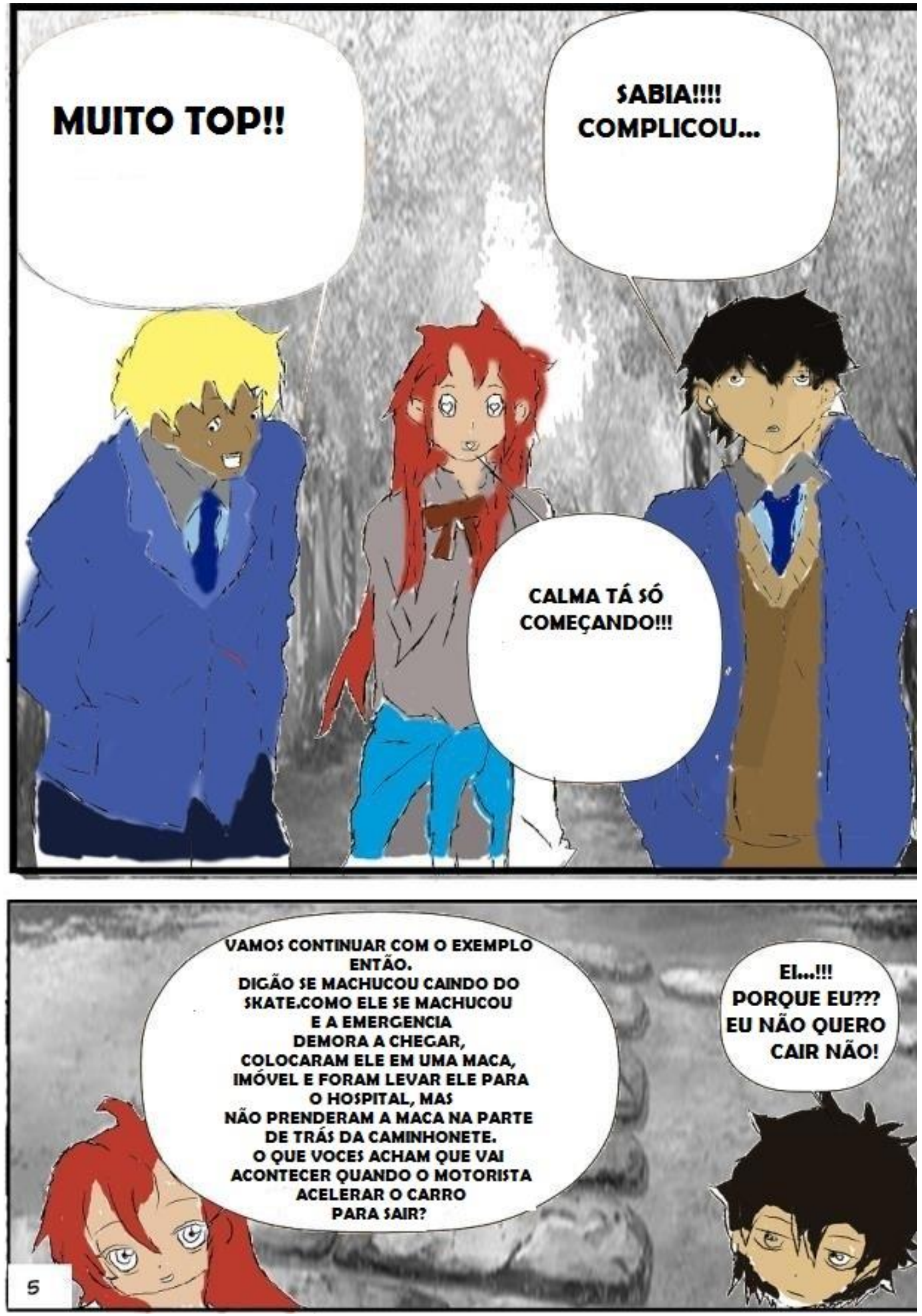
MUITO TOP!!
SABIA!!!! COMPLICOU...
CALMA TÁ SÓ COMEÇANDO!!!
VAMOS CONTINUAR COM O EXEMPLO ENTÃO. DIGÃO SE MACHUCOU CAINDO DO SKATE.COMO ELE SE MACHUCOU E A EMERGENCIA DEMORA A CHEGAR, COLOCARAM ELE EM UMA MACA, IMÓVEL E FORAM LEVAR ELE PARA O HOSPITAL, MAS NÃO PRENDERAM A MACA NA PARTE DE TRÁS DA CAMINHONETE. O QUE VOCES ACHAM QUE VAI ACONTECER QUANDO O MOTORISTA ACELERAR O CARRO PARA SAIR?
EI...!!! PORQUE EU??? EU NÃO QUERO CAIR NÃO!
5

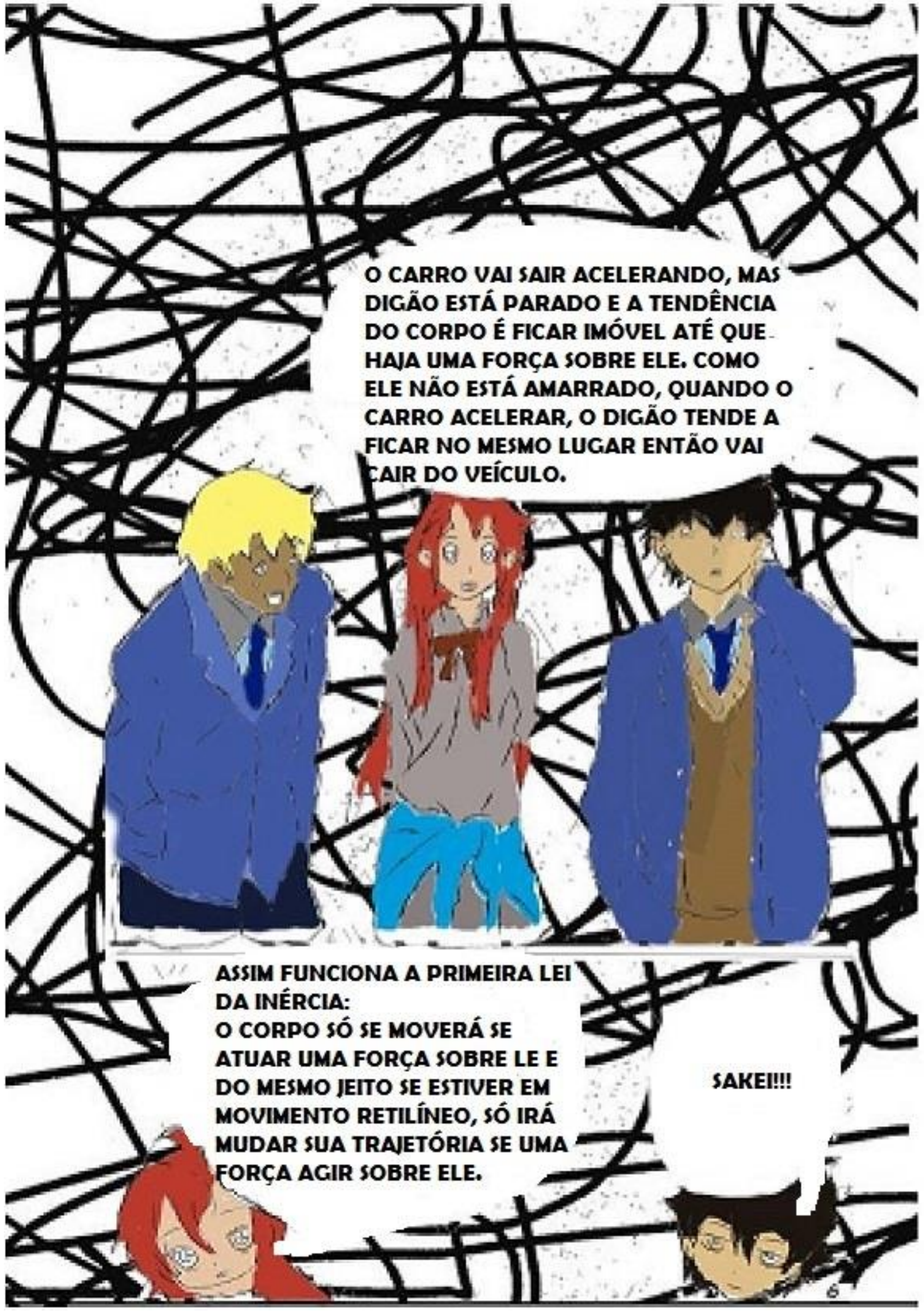
O CARRO VAI SAIR ACELERANDO, MAS DIGÃO ESTÁ PARADO E A TENDÊNCIA DO CORPO É FICAR IMÓVEL ATÉ QUE HAJA UMA FORÇA SOBRE ELE. COMO ELE NÃO ESTÁ AMARRADO, QUANDO O CARRO ACELERAR, O DIGÃO TENDE A FICAR NO MESMO LUGAR ENTÃO VAI CAIR DO VEÍCULO.
ASSIM FUNCIONA A PRIMEIRA LEI DA INÉRCIA: O CORPO SÓ SE MOVERÁ SE ATUAR UMA FORÇA SOBRE LE E DO MESMO JEITO SE ESTIVER EM MOVIMENTO RETILÍNEO, SÓ IRÁ MUDAR SUA TRAJETÓRIA SE UMA FORÇA AGIR SOBRE ELE.
SAKEI!!!

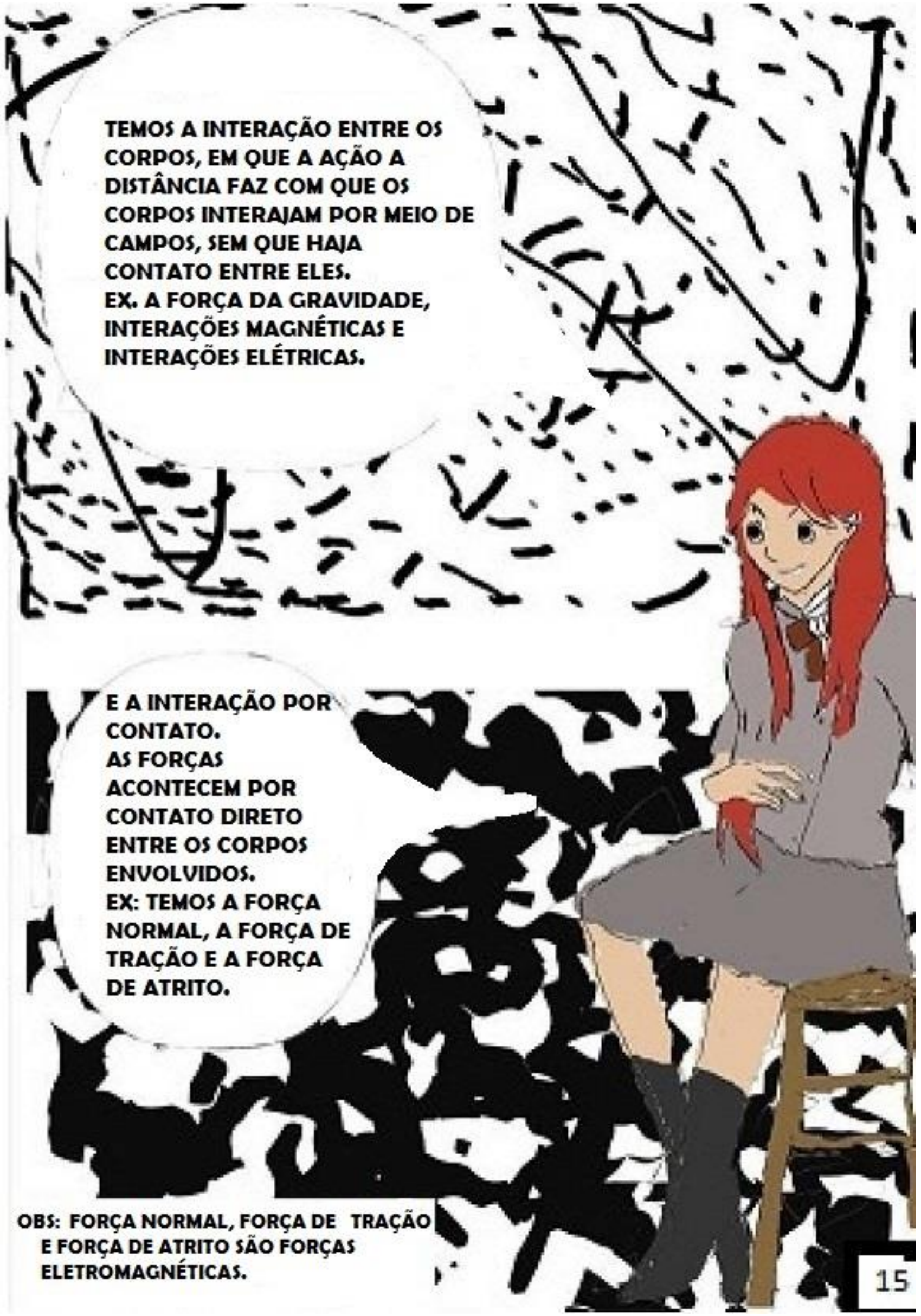
TEMOS A INTERAÇÃO ENTRE OS CORPOS, EM QUE A AÇÃO A DISTÂNCIA FAZ COM QUE OS CORPOS INTERAJAM POR MEIO DE CAMPOS, SEM QUE HAJA CONTATO ENTRE ELES. EX. A FORÇA DA GRAVIDADE, INTERAÇÕES MAGNÉTICAS E INTERAÇÕES ELÉTRICAS.
E A INTERAÇÃO POR CONTATO. AS FORÇAS ACONTECEM POR CONTATO DIRETO ENTRE OS CORPOS ENVOLVIDOS. EX: TEMOS A FORÇA NORMAL, A FORÇA DE TRAÇÃO E A FORÇA DE ATRITO.
OBS: FORÇA NORMAL, FORÇA DE TRAÇÃO E FORÇA DE ATRITO SÃO FORÇAS ELETROMAGNÉTICAS.
15

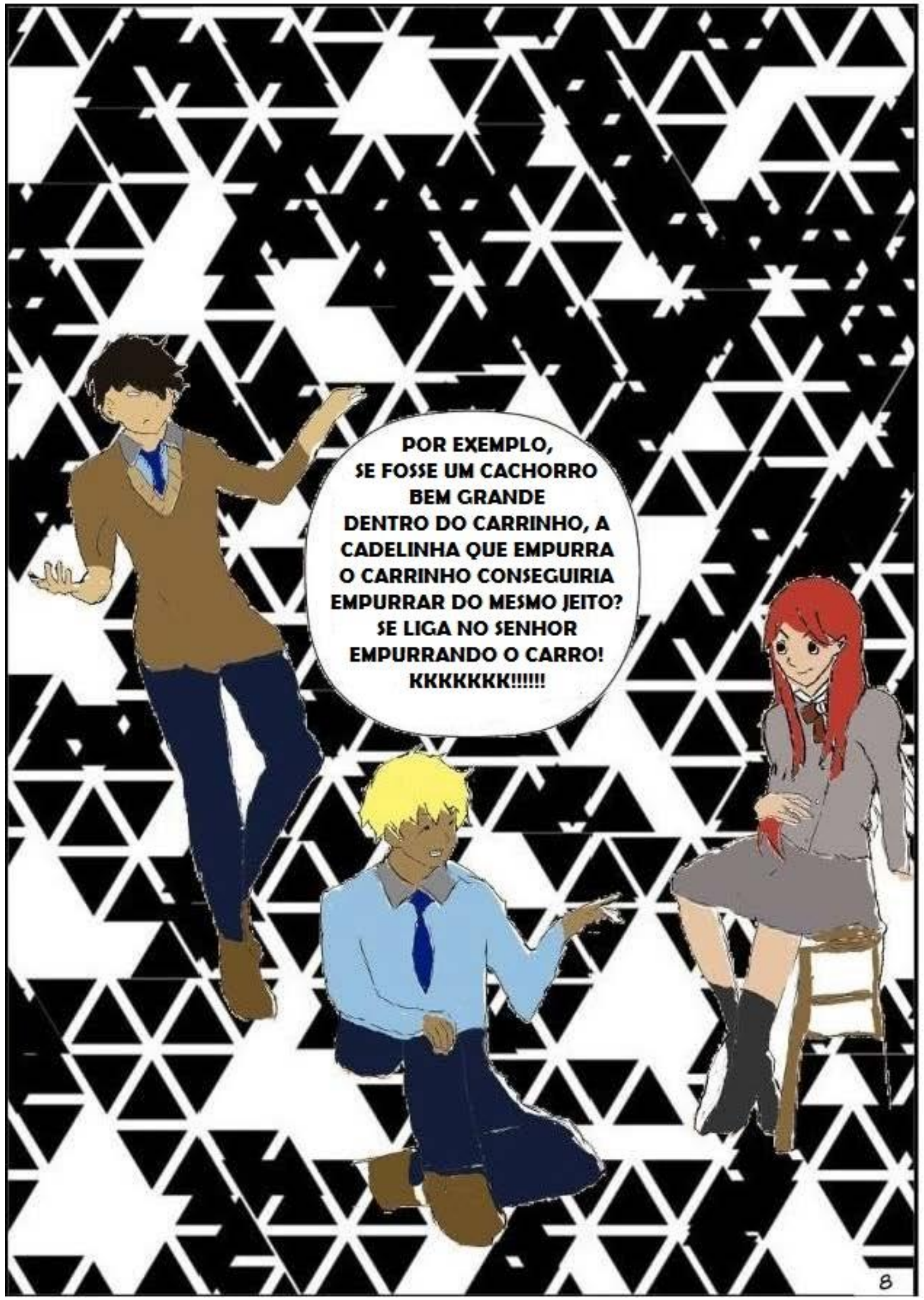
POR EXEMPLO,
SE FOSSE UM CACHORRO
BEM GRANDE
DENTRO DO CARRINHO, A
CADELINHA QUE EMPURRA
O CARRINHO CONSEGUIRIA
EMPURRAR DO MESMO JEITO?
SE LIGA NO SENHOR
EMPURRANDO O CARRO!
KKKKKKK!!!!!!
8

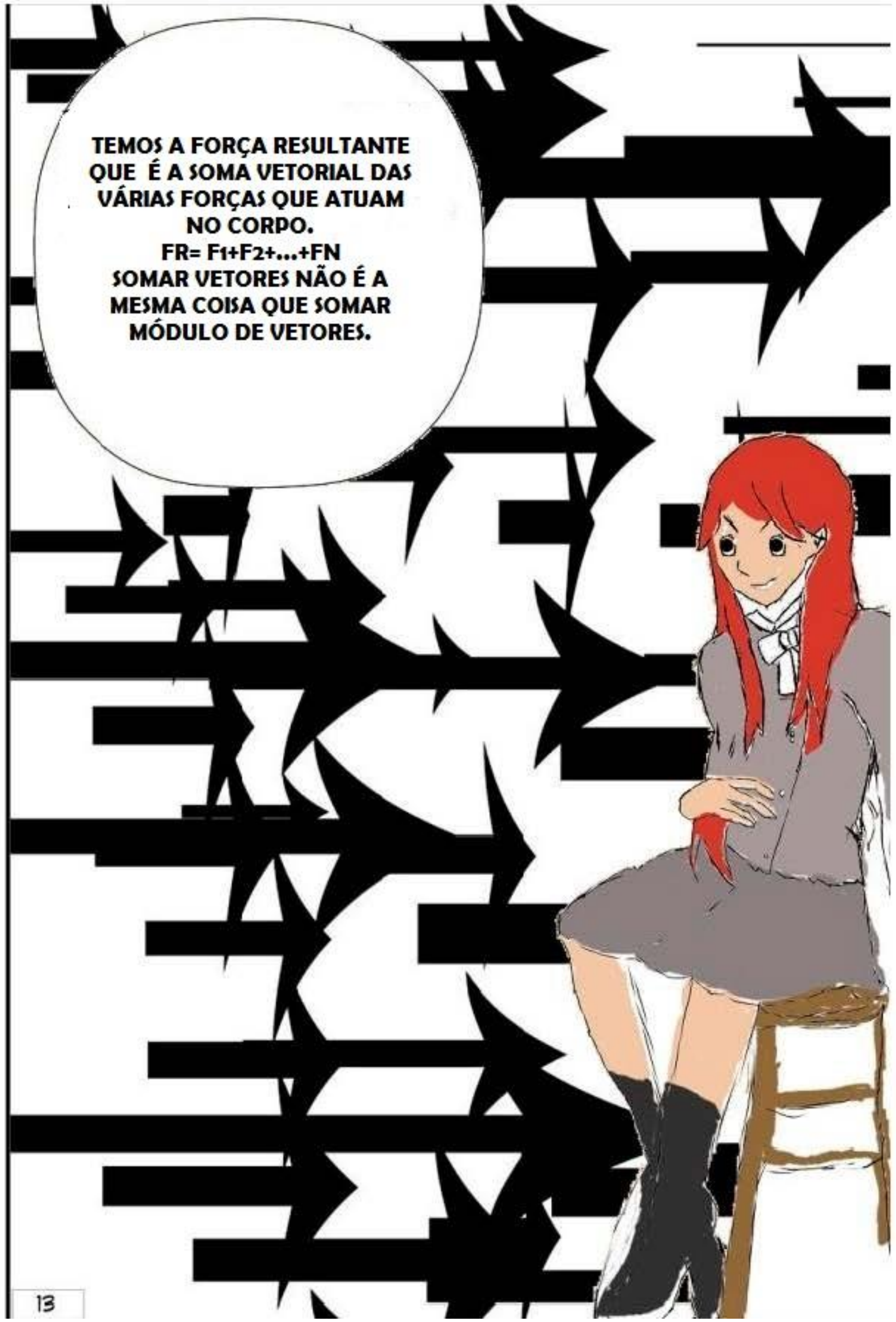
TEMOS A FORÇA RESULTANTE
QUE É A SOMA VETORIAL DAS
VÁRIAS FORÇAS QUE ATUAM
NO CORPO.
FR= F1+F2+...+FN
SOMAR VETORES NÃO É A
MESMA COISA QUE SOMAR
MÓDULO DE VETORES.
13

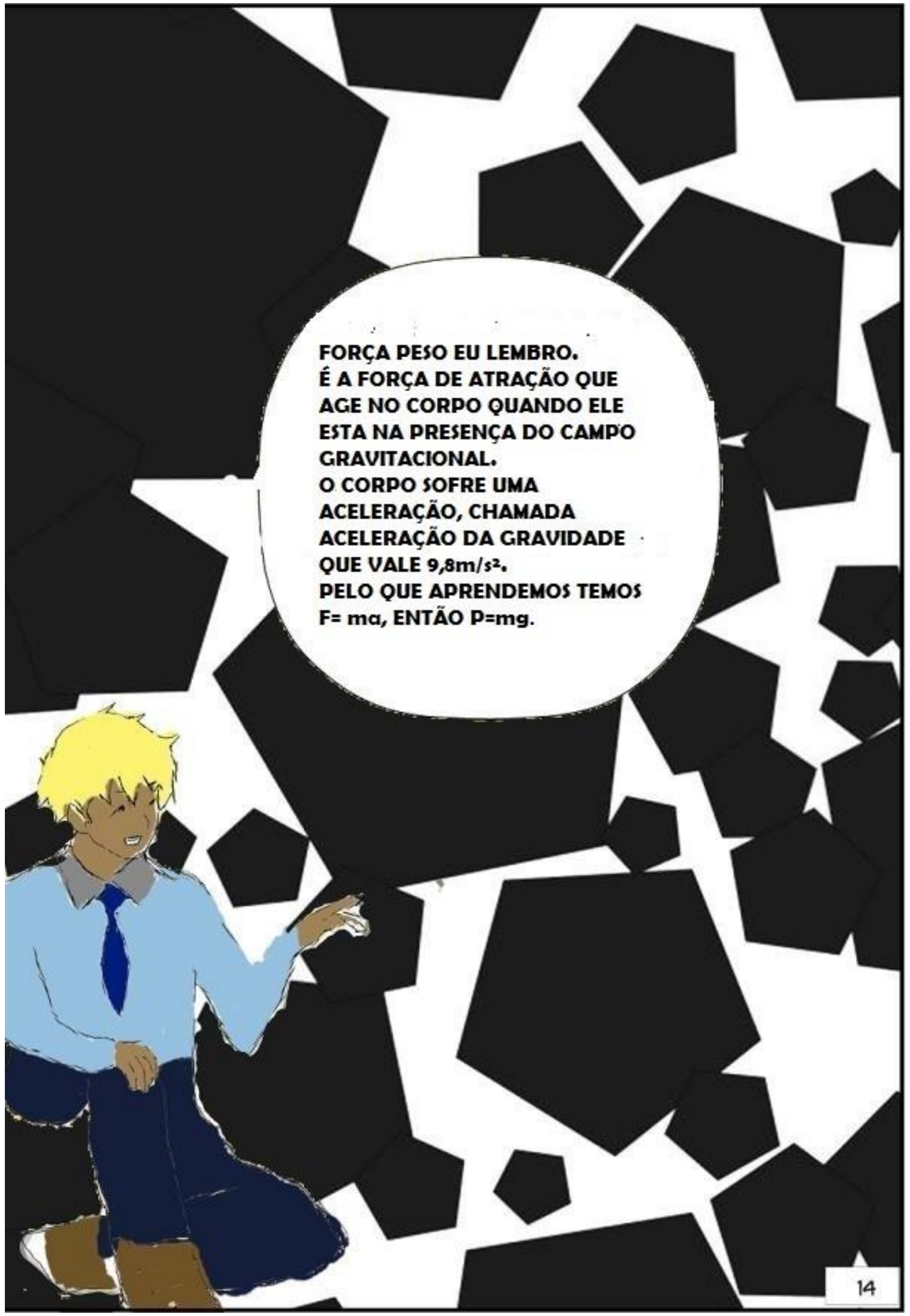
FORÇA PESO EU LEMBRO.
É A FORÇA DE ATRAÇÃO QUE
AGE NO CORPO QUANDO ELE
ESTA NA PRESENÇA DO CAMPO
GRAVITACIONAL.
O CORPO SOFRE UMA
ACELERAÇÃO, CHAMADA
ACELERAÇÃO DA GRAVIDADE
QUE VALE 9,8m/s².
PELO QUE APRENDEMOS TEMOS
F= ma, ENTÃO P=mg.
14

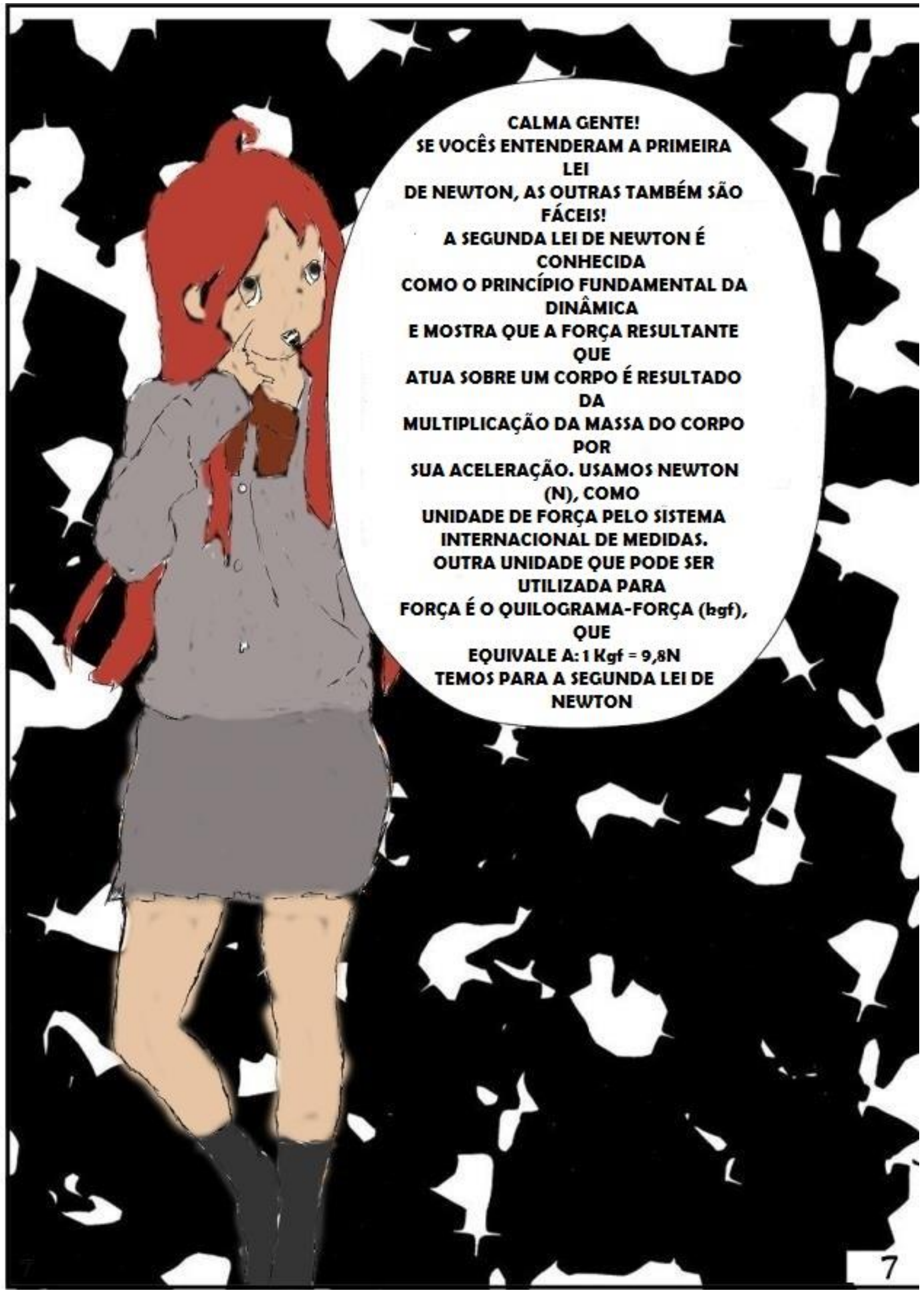
CALMA GENTE!
SE VOCÊS ENTENDERAM A PRIMEIRA
LEI
DE NEWTON, AS OUTRAS TAMBÉM SÃO
FÁCEIS!
A SEGUNDA LEI DE NEWTON É
CONHECIDA
COMO O PRINCÍPIO FUNDAMENTAL DA
DINÂMICA
E MOSTRA QUE A FORÇA RESULTANTE
QUE
ATUA SOBRE UM CORPO É RESULTADO
DA
MULTIPLICAÇÃO DA MASSA DO CORPO
POR
SUA ACELERAÇÃO. USAMOS NEWTON
(N), COMO
UNIDADE DE FORÇA PELO SISTEMA
INTERNACIONAL DE MEDIDAS.
OUTRA UNIDADE QUE PODE SER
UTILIZADA PARA
FORÇA É O QUILOGRAMA-FORÇA (kgf),
QUE
EQUIVALE A: 1 Kgf = 9,8N
TEMOS PARA A SEGUNDA LEI DE
NEWTON
7

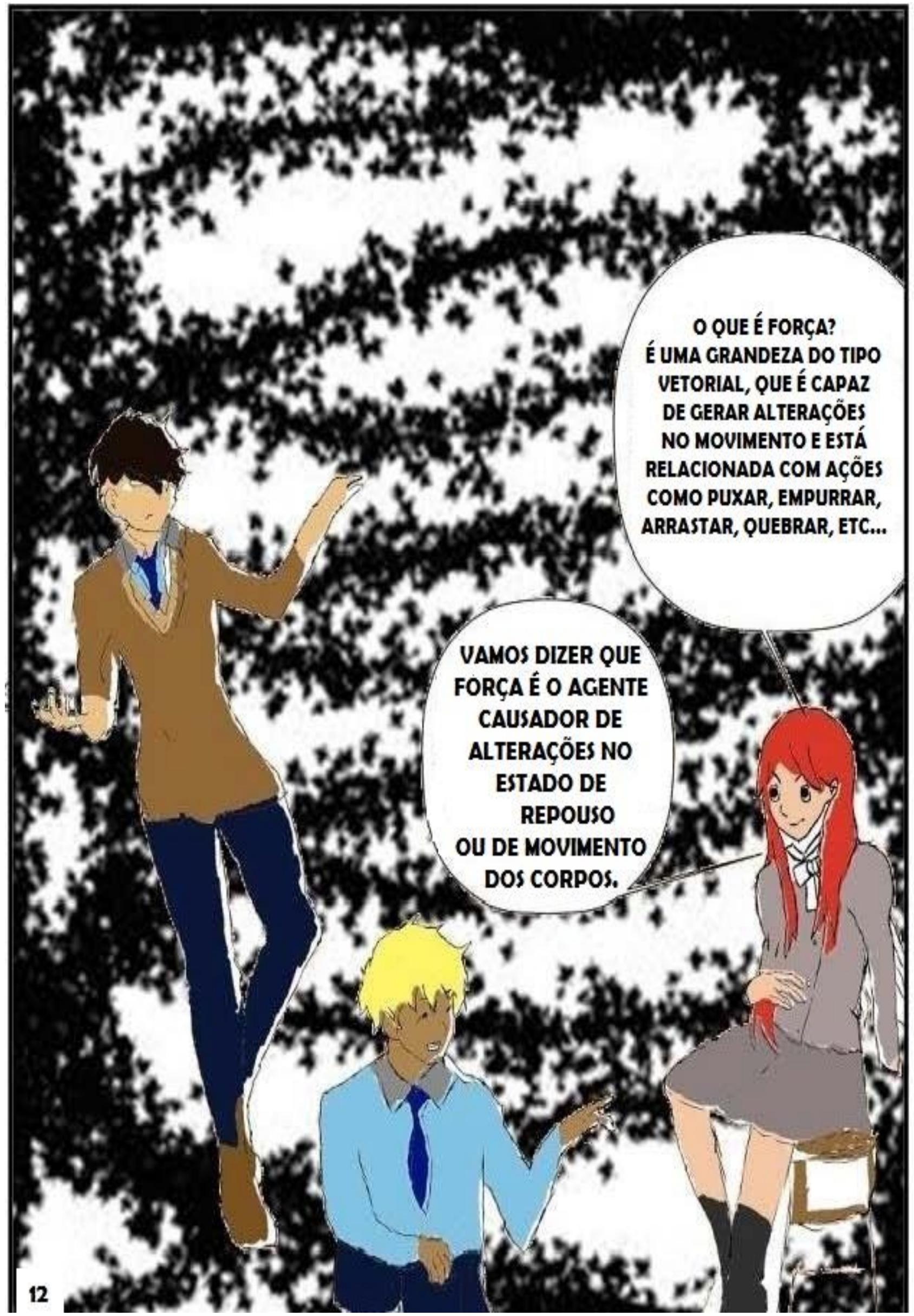
O QUE É FORÇA?
É UMA GRANDEZA DO TIPO VETORIAL, QUE É CAPAZ DE GERAR ALTERAÇÕES NO MOVIMENTO E ESTÁ RELACIONADA COM AÇÕES COMO PUXAR, EMPURRAR, ARRASTAR, QUEBRAR, ETC...
VAMOS DIZER QUE FORÇA É O AGENTE CAUSADOR DE ALTERAÇÕES NO ESTADO DE REPOUSO OU DE MOVIMENTO DOS CORPOS.
12

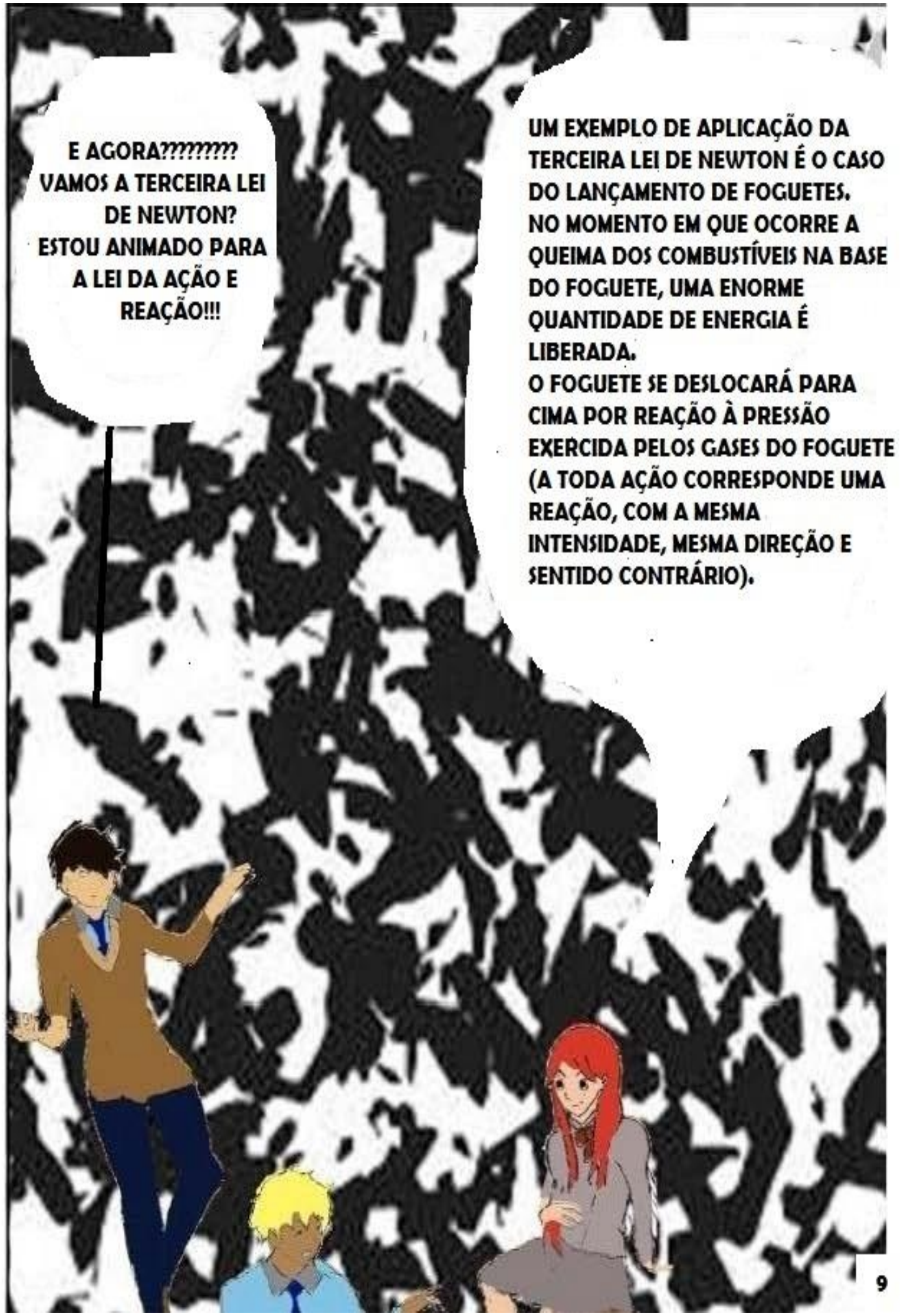
E AGORA?????????
VAMOS A TERCEIRA LEI DE NEWTON?
ESTOU ANIMADO PARA A LEI DA AÇÃO E REAÇÃO!!!
UM EXEMPLO DE APLICAÇÃO DA TERCEIRA LEI DE NEWTON É O CASO DO LANÇAMENTO DE FOGUETES.
NO MOMENTO EM QUE OCORRE A QUEIMA DOS COMBUSTÍVEIS NA BASE DO FOGUETE, UMA ENORME QUANTIDADE DE ENERGIA É LIBERADA.
O FOGUETE SE DESLOCARÁ PARA CIMA POR REAÇÃO À PRESSÃO EXERCIDA PELOS GASES DO FOGUETE (A TODA AÇÃO CORRESPONDE UMA REAÇÃO, COM A MESMA INTENSIDADE, MESMA DIREÇÃO E SENTIDO CONTRÁRIO).
9

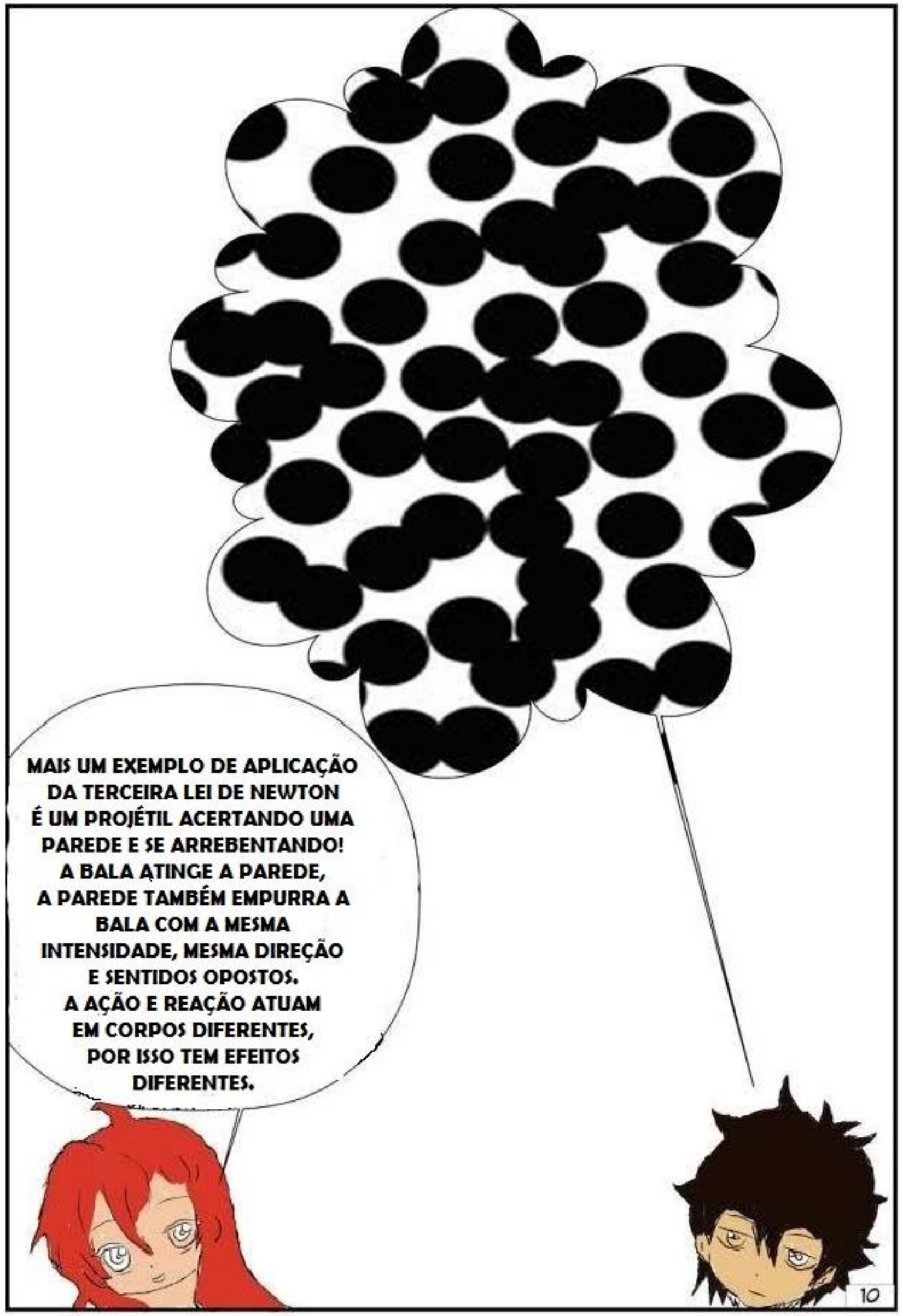
MAIS UM EXEMPLO DE APLICAÇÃO DA TERCEIRA LEI DE NEWTON É UM PROJÉTIL ACERTANDO UMA PAREDE E SE ARREBENTANDO! A BALA ATINGE A PAREDE, A PAREDE TAMBÉM EMPURRA A BALA COM A MESMA INTENSIDADE, MESMA DIREÇÃO E SENTIDOS OPOSTOS. A AÇÃO E REAÇÃO ATUAM EM CORPOS DIFERENTES, POR ISSO TEM EFEITOS DIFERENTES.
10

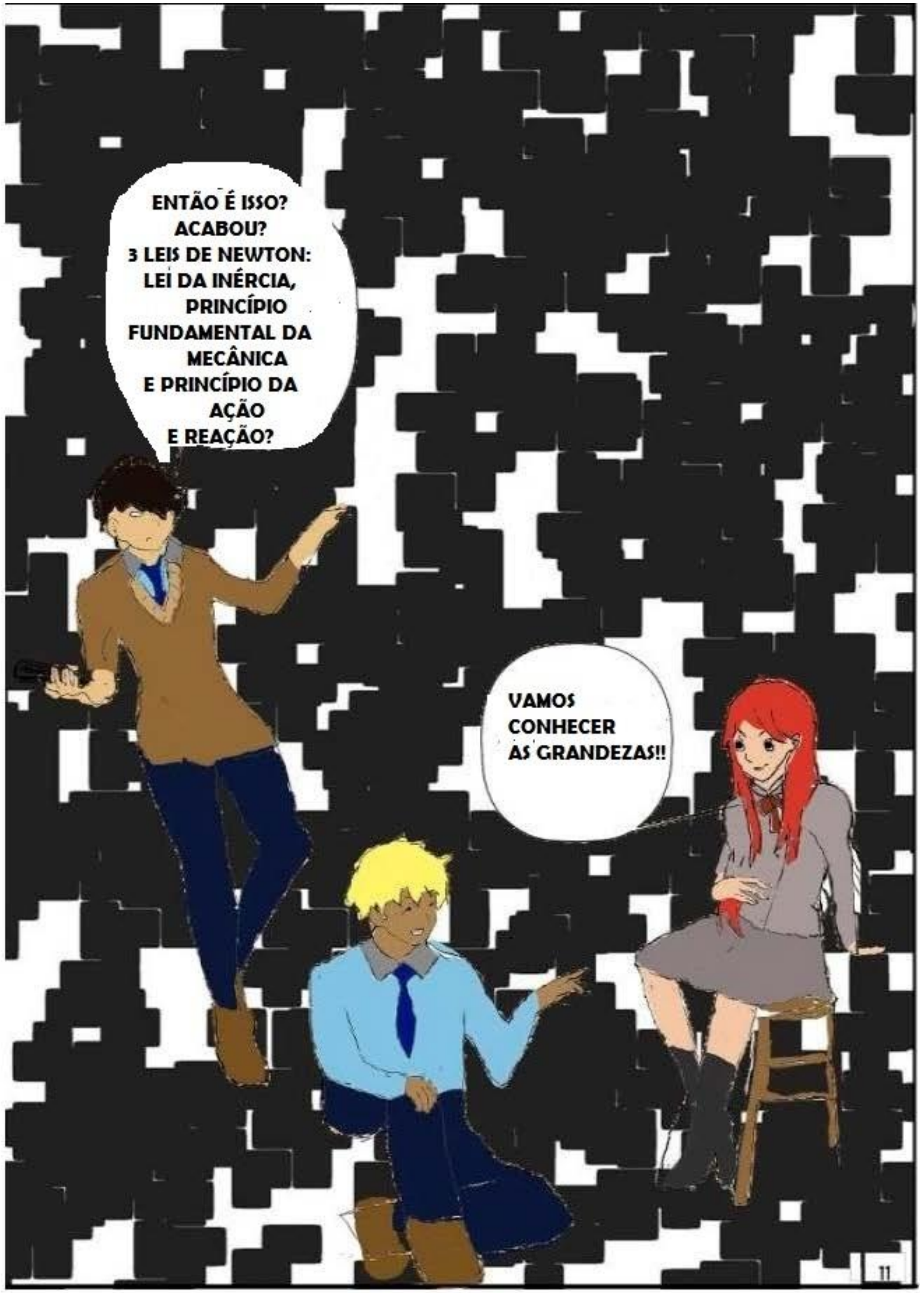
ENTÃO É ISSO? ACABOU? 3 LEIS DE NEWTON: LEI DA INÉRCIA, PRINCÍPIO FUNDAMENTAL DA MECÂNICA E PRINCÍPIO DA AÇÃO E REAÇÃO?
VAMOS CONHECER AS GRANDEZAS!!
11

# APÊNDICE 9 – Produto Final (Apresentação)

O produto final é constituído da apresentação da sequência didática conforme aconteceu a aplicação do produto com seus devidos passos.

Por ele o professor pode acompanhar como foi a realização da aplicação da sequência didática da dissertação **PROPOSTA DE SEQUÊNCIA DIDÁTICA PARA ENSINO DAS LEIS DE NEWTON, UTILIZANDO GIFs E VÍDEOS.**

A qualquer apêndice ou anexo que se deseje imprimir favor buscar na dissertação original já que a apresentação do produto final é apenas a demonstração de como ele foi aplicado.

INSTITUTO FEDERAL ESPÍRITO SANTO

SBF SOCIEDADE BRASILEIRA DE FÍSICA

CAPES

MNPEF Mestrado Nacional Profissional em Ensino de Física

# PROPOSTA DE MATERIAL DIDÁTICO PARA ENSINO DAS LEIS DE NEWTON COM USO DE VÍDEOS E GIFS

Autora: Lucia Helena Horta Oliveira
Orientador: Samir Lacerda da Silva

# INTRODUÇÃO

- Escola tradicional x construtivismo;
- Utilização de tecnologias norteadas pela mediação (Vygotsky).
- Tecnologias de Informação e Comunicação (TICs) ;
- Estudo das leis de Newton através de revista em quadrinhos com uso de vídeos e Gifs;
- Aplicativo próprio para utilização do R.A;

# MOTIVAÇÃO

- Integrar a tecnologia ao ensino tradicional ;
- Expansão da realidade tecnológica e digital;
- Mudança na forma de apresentação do conteúdo.

## JUSTIFICATIVA

- Mediação como ponto central das relações de aprendizagem do aluno.
- Contexto a sua volta envolvido no processo. Professor como mediador do conhecimento.
- Utilizar os recursos digitais e tecnológicos - PCNs
- Nova opção de recurso didático
- Uso do aplicativo- android 5.0 com câmera ou superior
- Não precisa de internet ;
- App gratuito

## PERGUNTA NORTEADORA

- **O uso de vídeos e Gifs como símbolo na mediação de Vygoksky aplicada ao ensino das leis de Newton, pode auxiliar na transmissão e compreensão do conteúdo?**

## OBJETIVO - GERAL

- Sequência didática baseada na mediação (Vygotsky);
- Revista em quadrinhos com tecnologia;
- Leis de Newton;

## REFERENCIAL TEÓRICO

- Abordagem da primeira , segunda e terceira leis de Newton;

- Mediação de Vygotsky ;

## Primeira lei de Newton

- Primeira lei de Newton – Lei da inércia

Repouso ou MRV;

Estado natural;

Forças externas.

Resistencia;

Depende da quantidade de matéria envolvida – inercia.

## FÍSICA TEÓRICA ENVOLVIDA NA APLICAÇÃO DA SEQUÊNCIA DIDÁTICA

Alterações no estado de movimento

- Primeira lei de Newton

- Segunda lei de Newton

- Terceira lei de Newton

## Segunda lei de Newton

- Lei da força
- Relaciona a quantidade de movimento de um corpo e sua variação com a ação de uma força externa;
- Modificação do movimento é proporcional à força atuante;

- $\vec{F} = \frac{d\vec{p}}{dt}$

m = Constante $\vec{F} = m.\frac{d\vec{v}}{dt}$

Geral $\vec{F} = \frac{dm\vec{v}}{dt} = \frac{md\vec{v}}{dt} + \frac{\vec{v}dm}{dt}$

## Terceira lei de Newton

Ação e reação;

Iguais e opostas

Não se cancelam

Corpos diferente

# METODOLOGIA

- Uso de novas tecnologias;
- Signo - revista ;
- Os símbolos em Realidade aumentada, Gifs e vídeos representados por Target
- Tema – As leis de Newton.
- 20 páginas;
- Exercicios ao final de cada lei.
- Dois questionários.
- Aplicação 15 aulas.

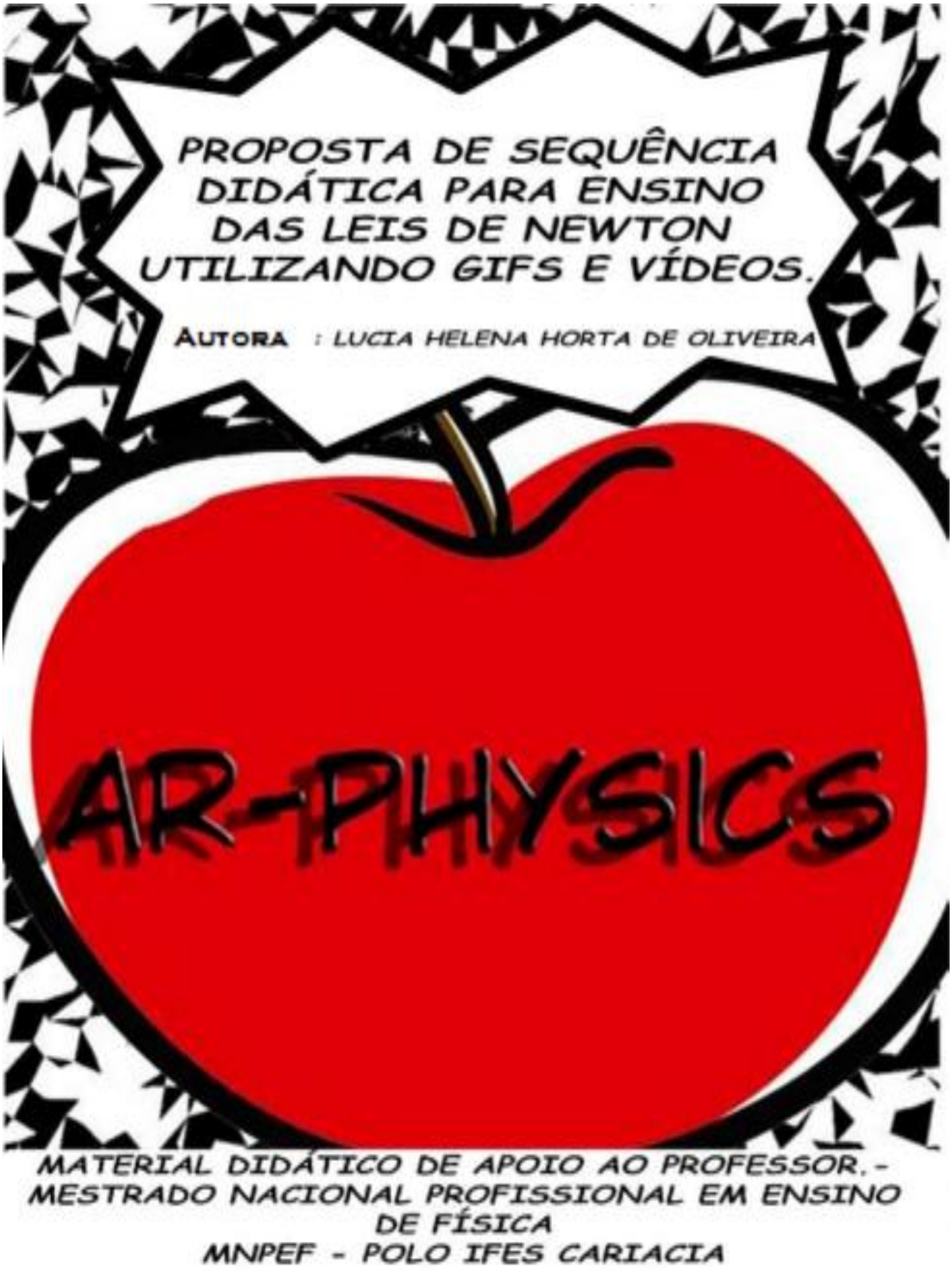
PROPOSTA DE SEQUÊNCIA DIDÁTICA PARA ENSINO DAS LEIS DE NEWTON UTILIZANDO GIFS E VÍDEOS.
AUTORA : LUCIA HELENA HORTA DE OLIVEIRA
AR-PHYSICS
MATERIAL DIDÁTICO DE APOIO AO PROFESSOR. - MESTRADO NACIONAL PROFISSIONAL EM ENSINO DE FÍSICA
MNPEF - POLO IFES CARIACIA

# PROPOSTA DE SEQUÊNCIA DIDÁDICA PARA ENSINO DAS LEIS DE NEWTON UTILIZANDO GIFS E VÍDEOS

MATERIAL DIDÁTICO DE APOIO AO PROFESSOR

AUTORA: LUCIA HELENA HORTA OLIVEIRA
ORIENTADOR : SAMIR LACERDA DA SILVA
MESTRADO NACIONAL PROFISSIONAL EM ENSINO DE FÍSICA
POLO IFES CARIACICA

ORIENTAÇÕES

- A aplicação  composta de 15 momentos.
- Material de apoio 3 questionários .
- 12 slides.
- Data show
- 7 revistas coloridas.
- 1 celular .
- O produto foi aplicado com o aplicativo de testes que retirou da internet a maioria  dos vídeos que foram transformados em Gifs.

1ºMOMENTO: Questionário de conhecimentos prévios aplicado individualmente.
Contando com 13 perguntas .

## Intervenção

- Análise do primeiro questionário
- Falta de conhecimentos prévios
- Necessário apresentação de conhecimentos básicos
- Massa
- Peso
- Referencial;
- Movimento;
- Bases que deveriam ser vistas no 9º ano ensino fundamental na disciplina de ciências.

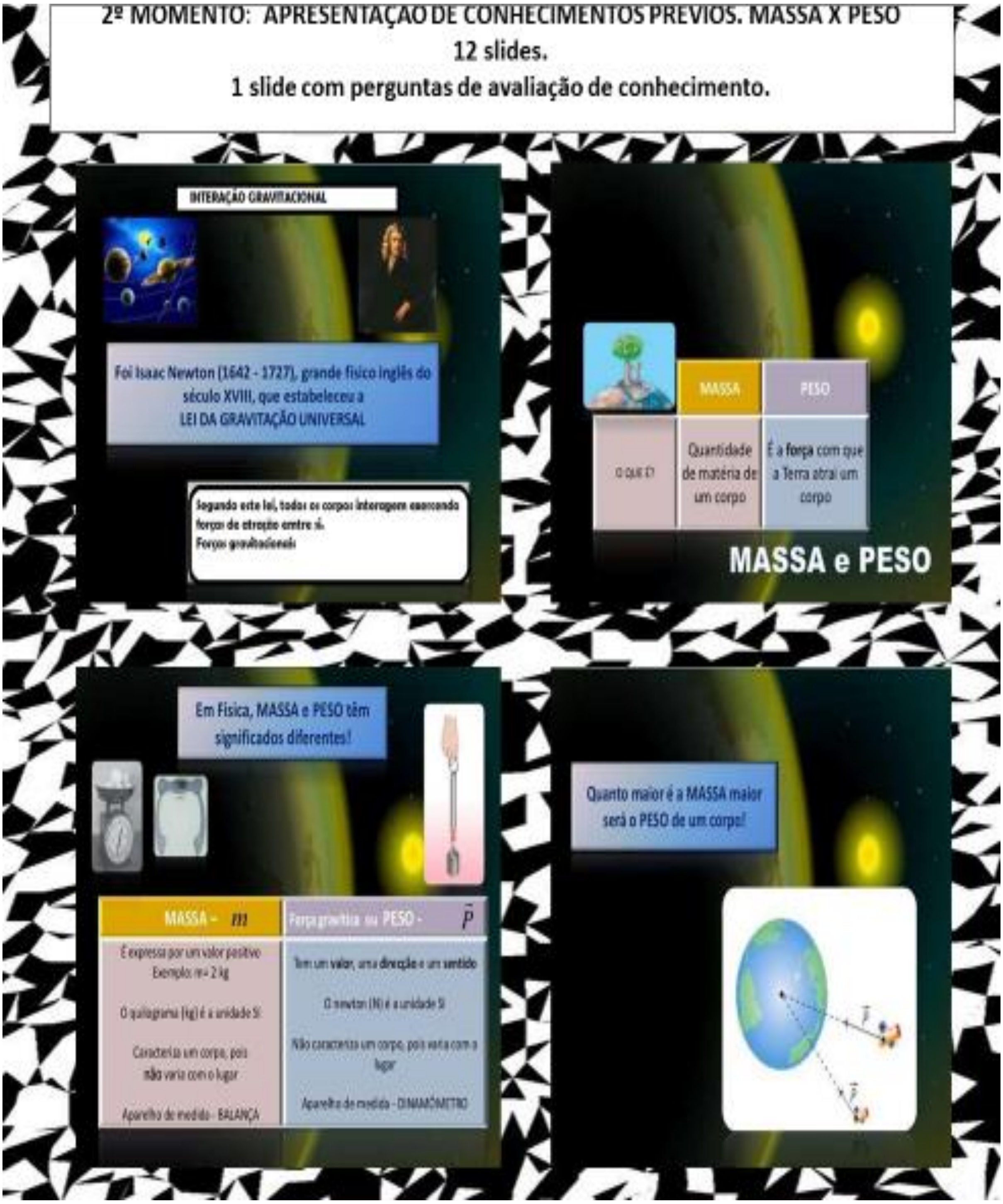
2º MOMENTO: APRESENTAÇÃO DE CONHECIMENTOS PRÉVIOS. MASSA X PESO
12 slides.
1 slide com perguntas de avaliação de conhecimento.
INTERAÇÃO GRAVITACIONAL
Foi Isaac Newton (1642 - 1727), grande físico inglês do século XVIII, que estabeleceu a LEI DA GRAVITAÇÃO UNIVERSAL
Segundo esta lei, todos os corpos interagem exercendo forças de atração entre si.
Forças gravitacionais
MASSA
PESO
O QUE É?
Quantidade de matéria de um corpo
É a força com que a Terra atrai um corpo
MASSA e PESO
Em Física, MASSA e PESO têm significados diferentes!
MASSA – m
Força gravítica ou PESO – P
É expressa por um valor positivo
Exemplo: m= 2 kg
O quilograma (kg) é a unidade SI
Caracteriza um corpo, pois não varia com o lugar
Aparelho de medida - BALANÇA
Tem um valor, uma direção e um sentido
O newton (N) é a unidade SI
Não caracteriza um corpo, pois varia com o lugar
Aparelho de medida - DINAMÔMETRO
Quanto maior é a MASSA maior será o PESO de um corpo!

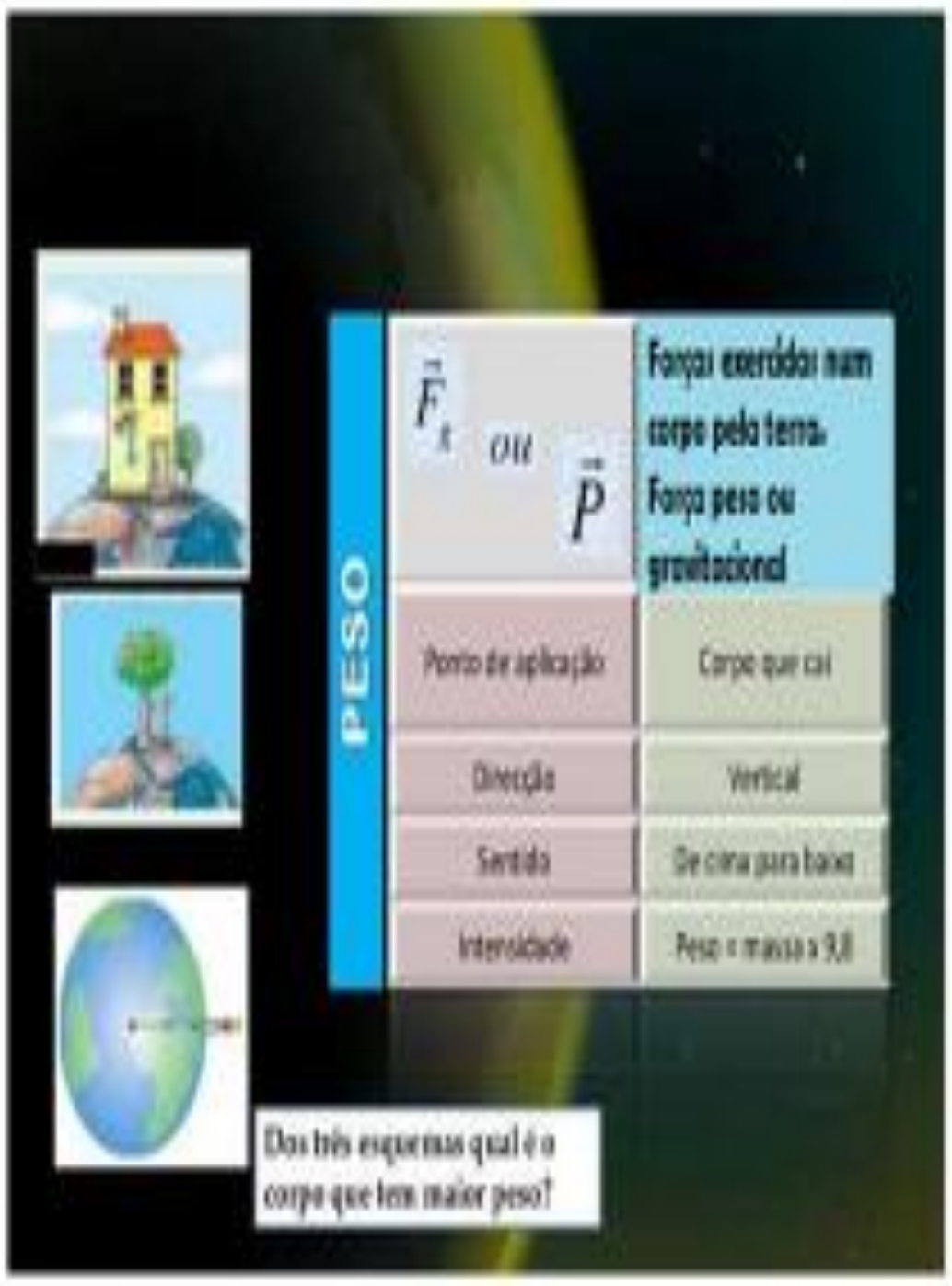
PESO
Forças exercidas num corpo pela terra.
Força peso ou gravitacional
Ponto de aplicação
Corpo que cai
Direcção
Vertical
Sentido
De cima para baixo
Intensidade
Peso = massa x 9,8
Dos três esquemas qual é o corpo que tem maior peso?

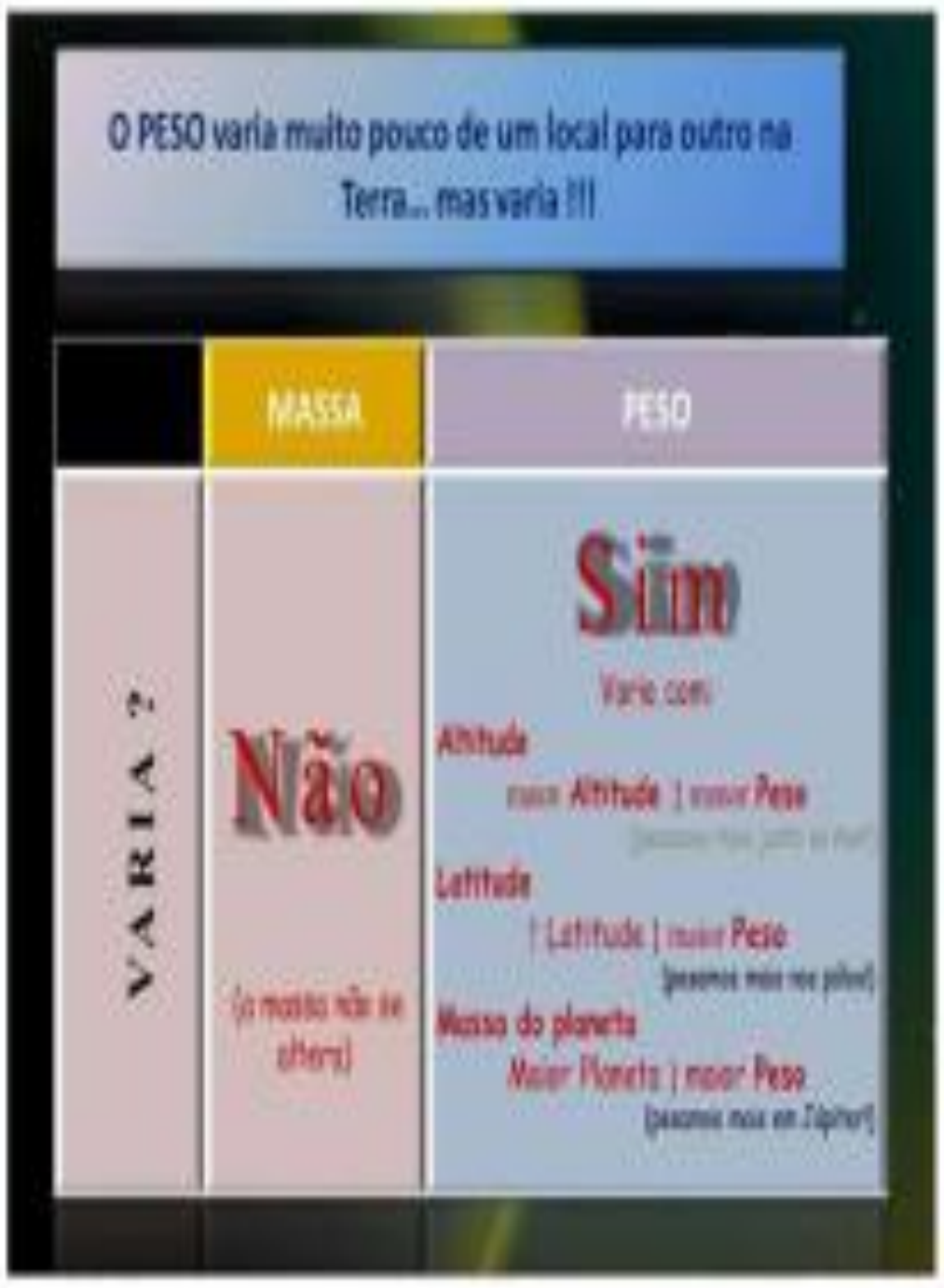
O PESO varia muito pouco de um local para outro na Terra... mas varia !!!
MASSA
PESO
VARIA ?
Não
Sim
Varia com:
Altitude
Latitude
Massa do planeta

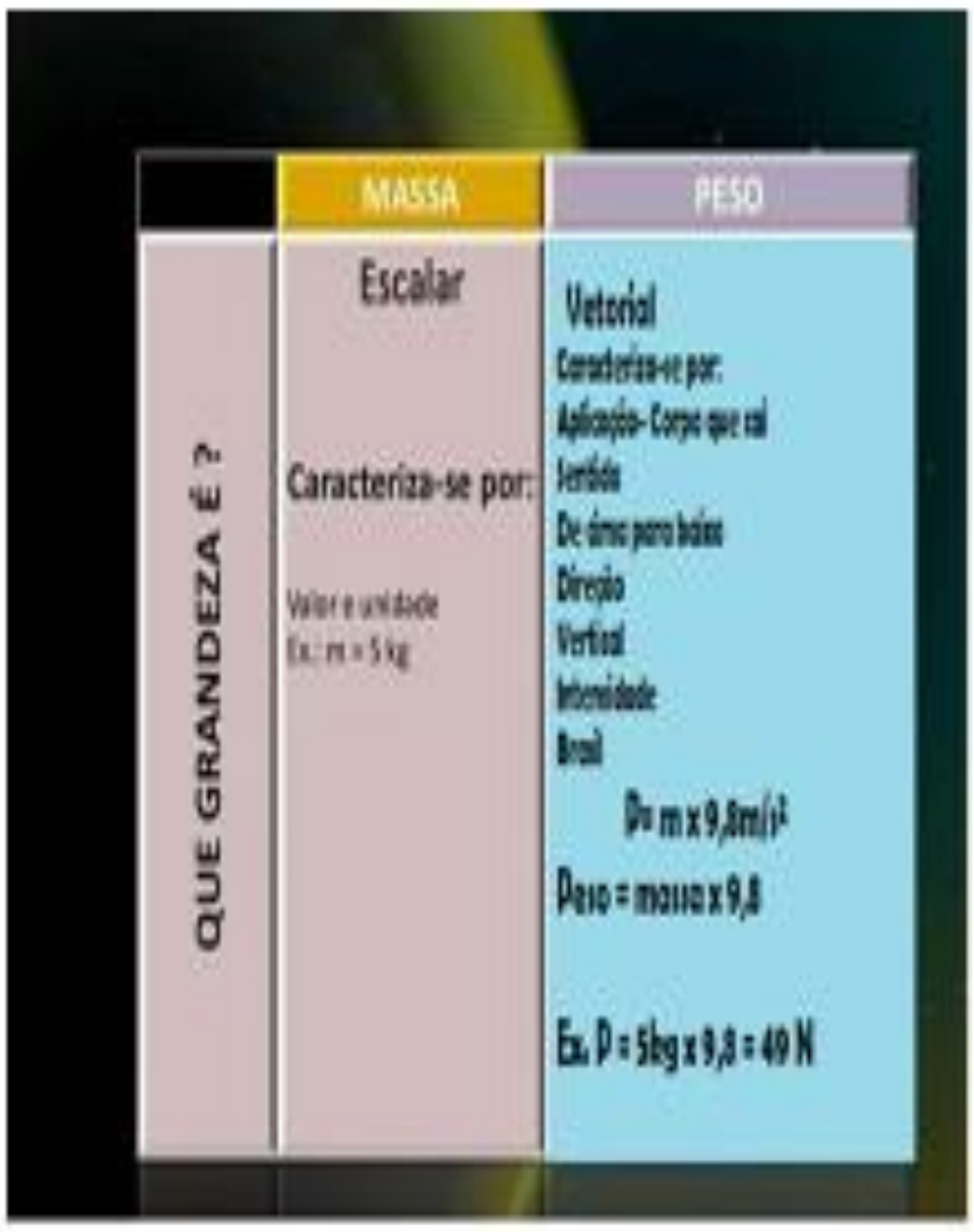
MASSA
PESO
QUE GRANDEZA É ?
Escalar
Caracteriza-se por:
Valor e unidade
Vetorial
Peso = massa x 9,8

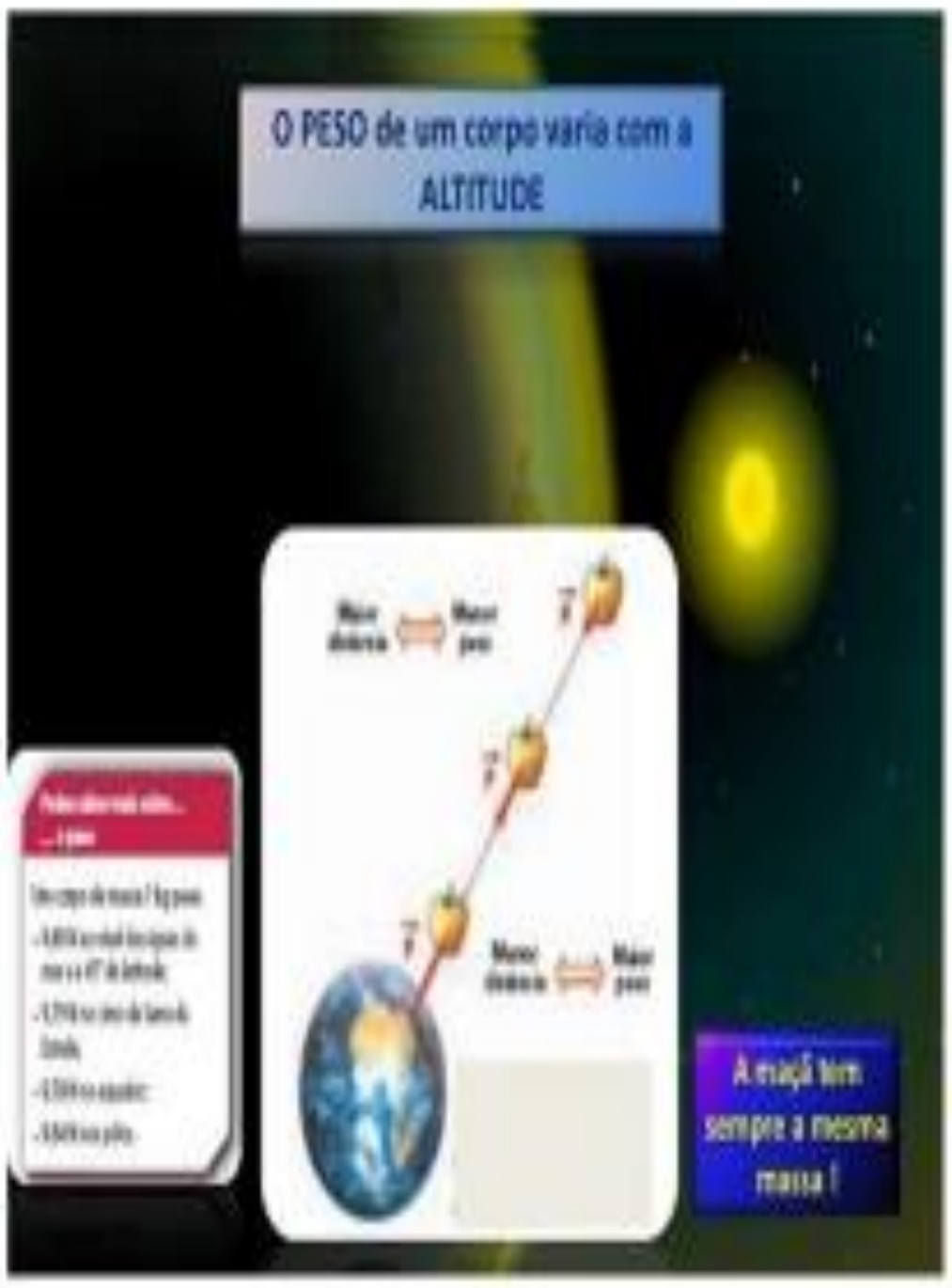
O PESO de um corpo varia com a ALTITUDE
A maçã tem sempre a mesma massa !

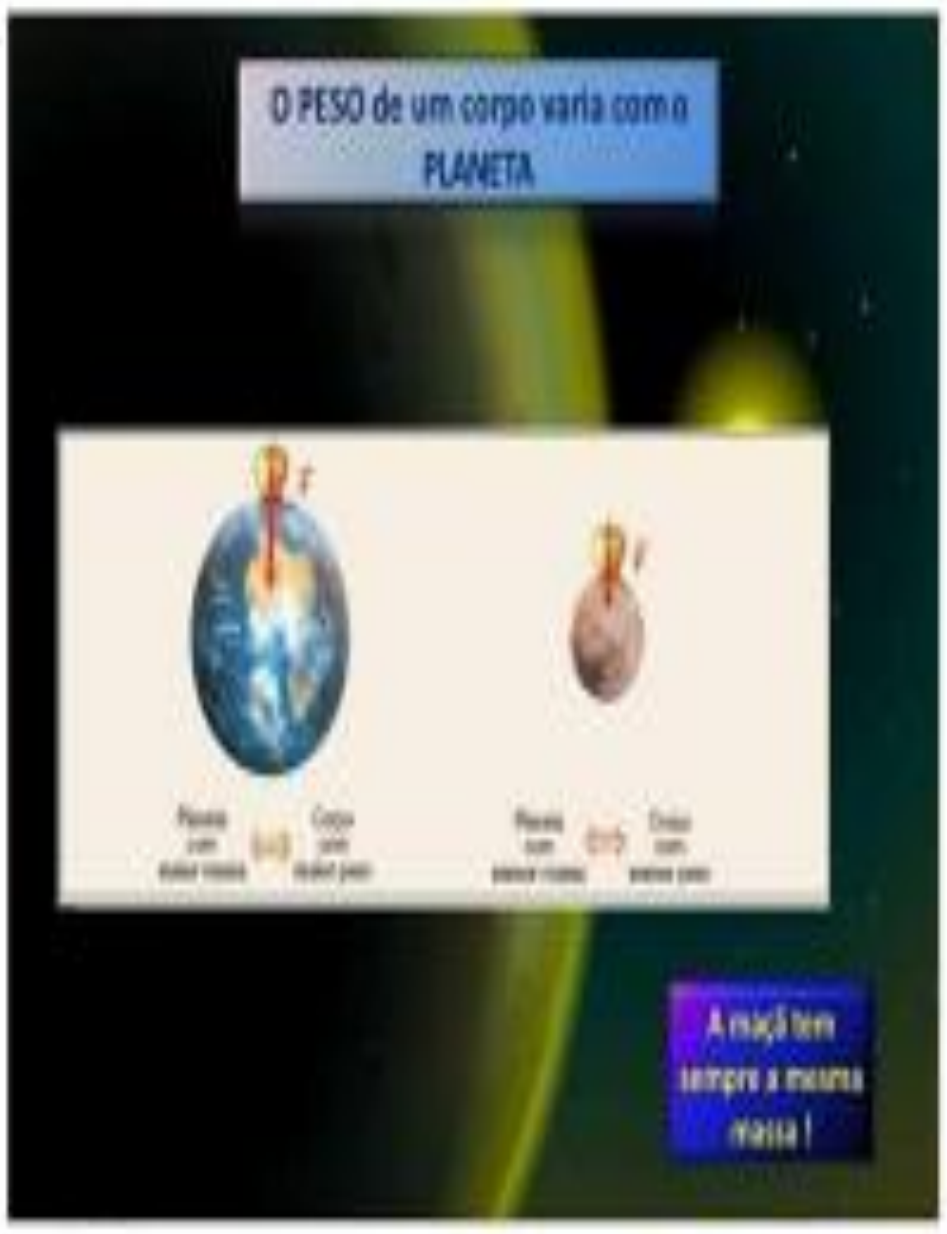
O PESO de um corpo varia com o PLANETA
A maçã tem sempre a mesma massa !

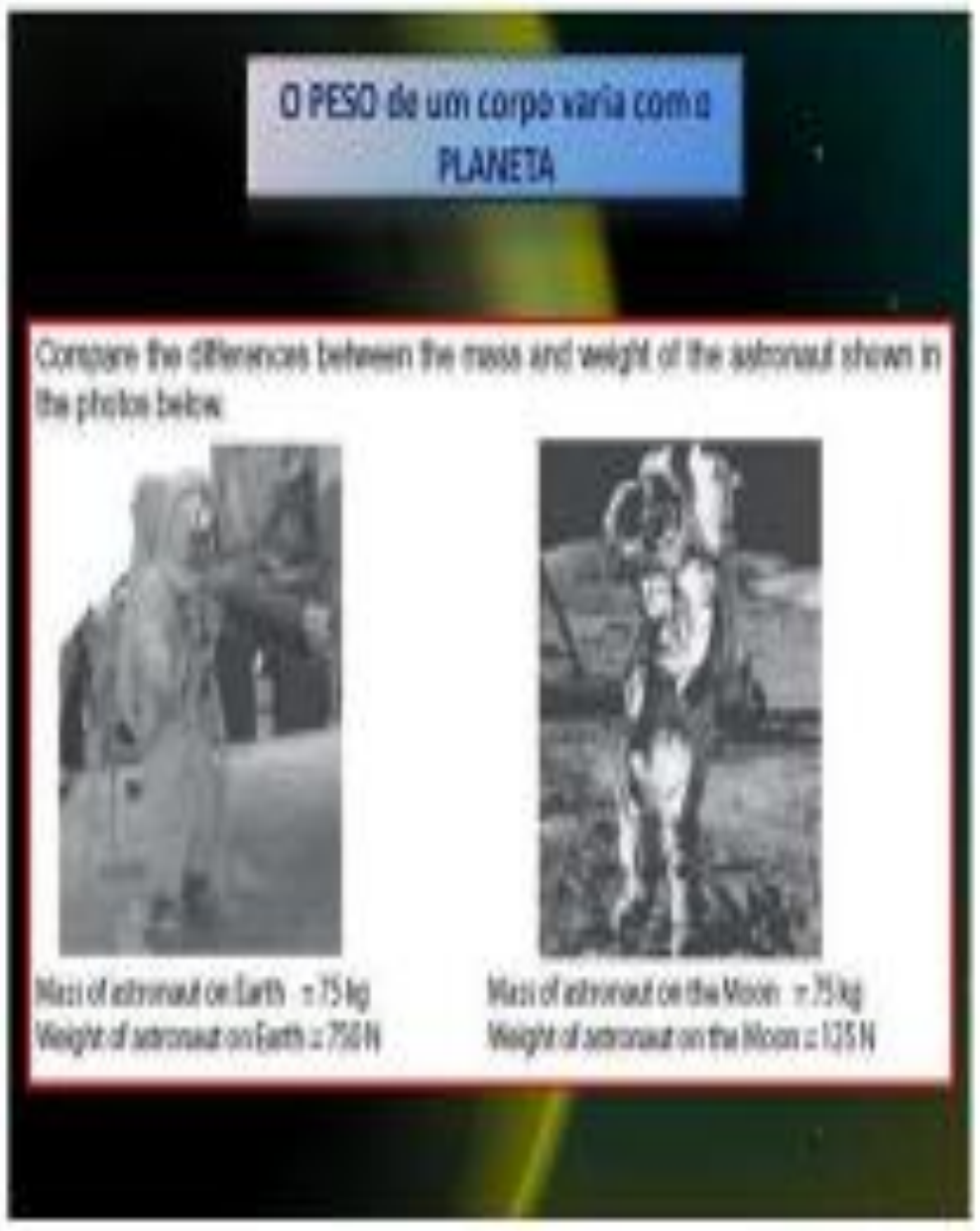
O PESO de um corpo varia com o PLANETA
Compare the differences between the mass and weight of the astronaut shown in the photos below.
Mass of astronaut on Earth = 75 kg
Weight of astronaut on Earth ≈ 750 N
Mass of astronaut on the Moon = 75 kg
Weight of astronaut on the Moon ≈ 125 N

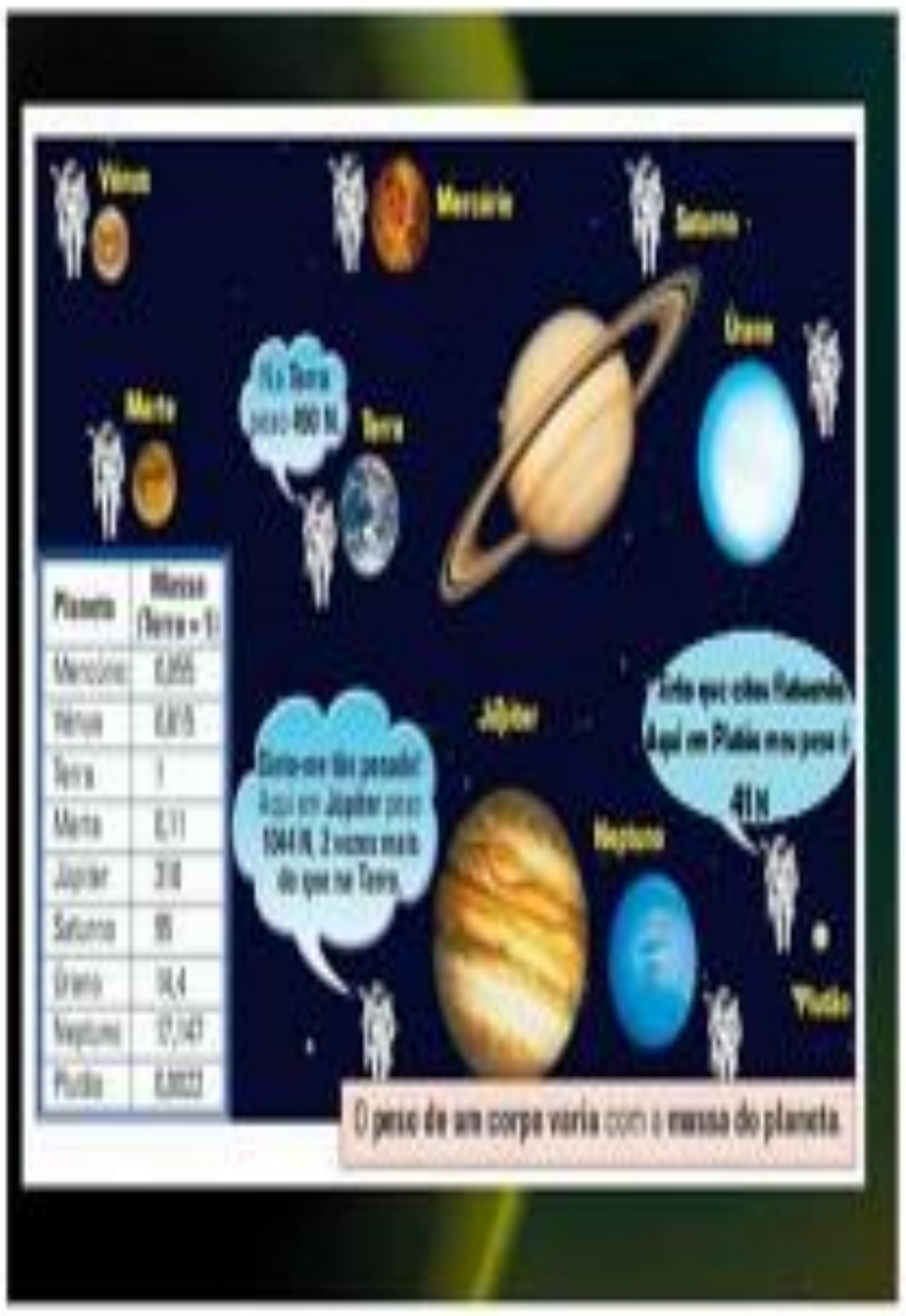
Vénus
Mercúrio
Saturno
Urano
Marte
Terra
Júpiter
Neptuno
Plutão
Planeta
Massa (Terra = 1)
Mercúrio 0,055
Vénus 0,815
Terra 1
Marte 0,11
Júpiter 318
Saturno 95
Urano 14,4
Neptuno 17,147
Plutão 0,0022
41 N
O peso de um corpo varia com a massa do planeta

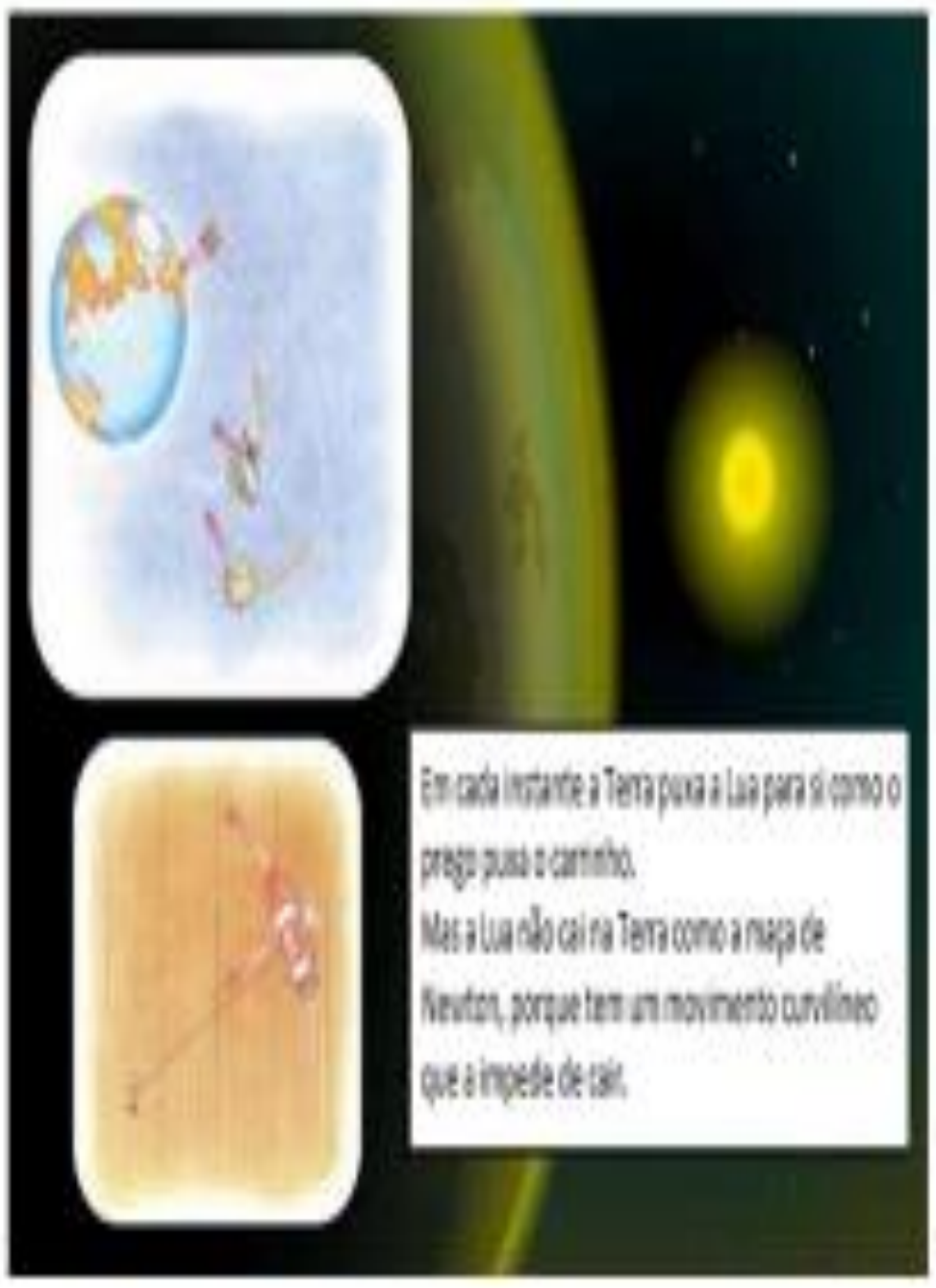
Em cada instante a Terra puxa a Lua para si como o prego puxa o caminho.
Mas a Lua não cai na Terra como a maça de Newton, porque tem um movimento curvilíneo que a impede de cair.

**Perguntas de finalização da aula expositiva**

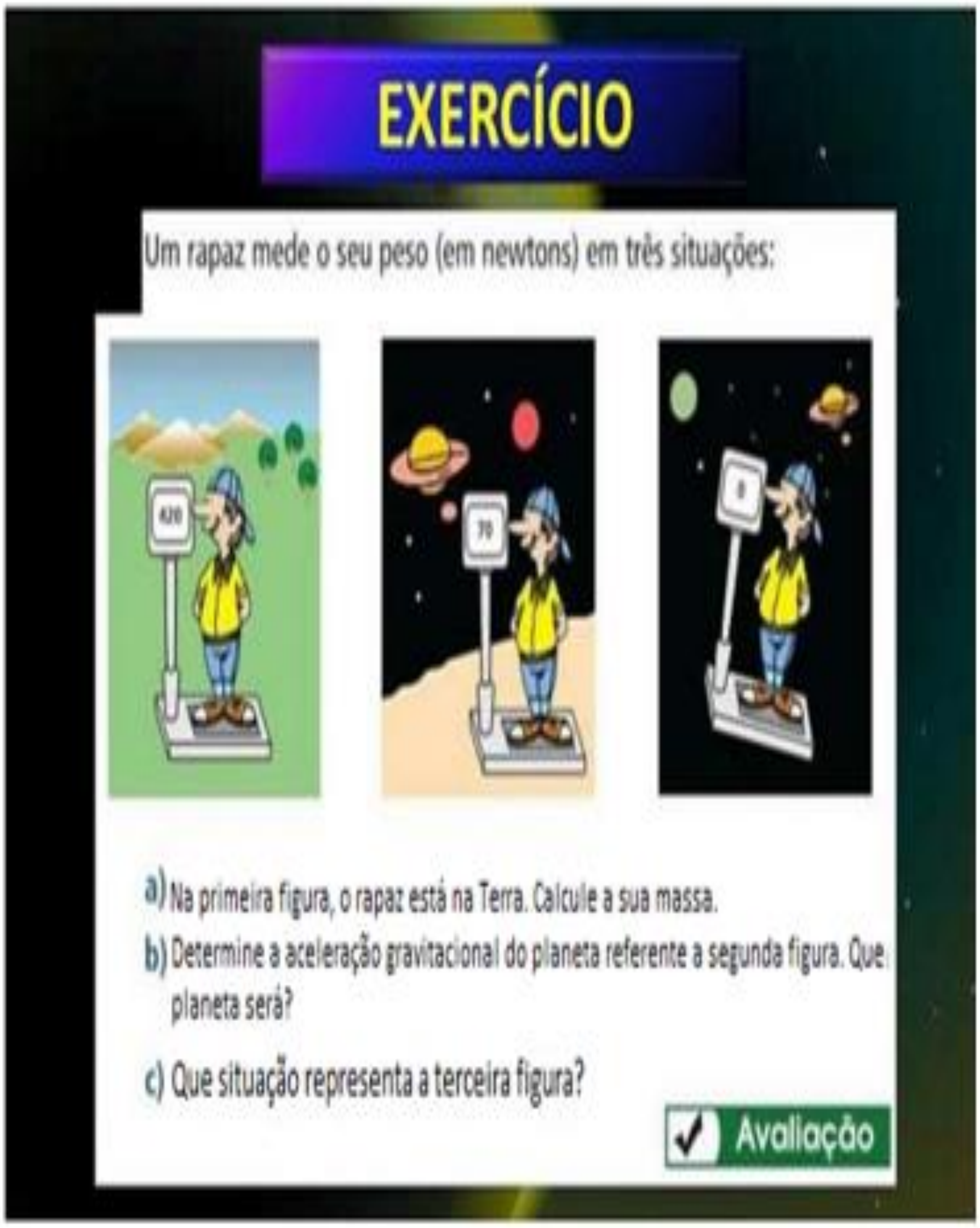

3º MOMENTO: **AULA MEDIADA COM RECURSOS TECNOLÓGICOS - SIMULADOR PHET**

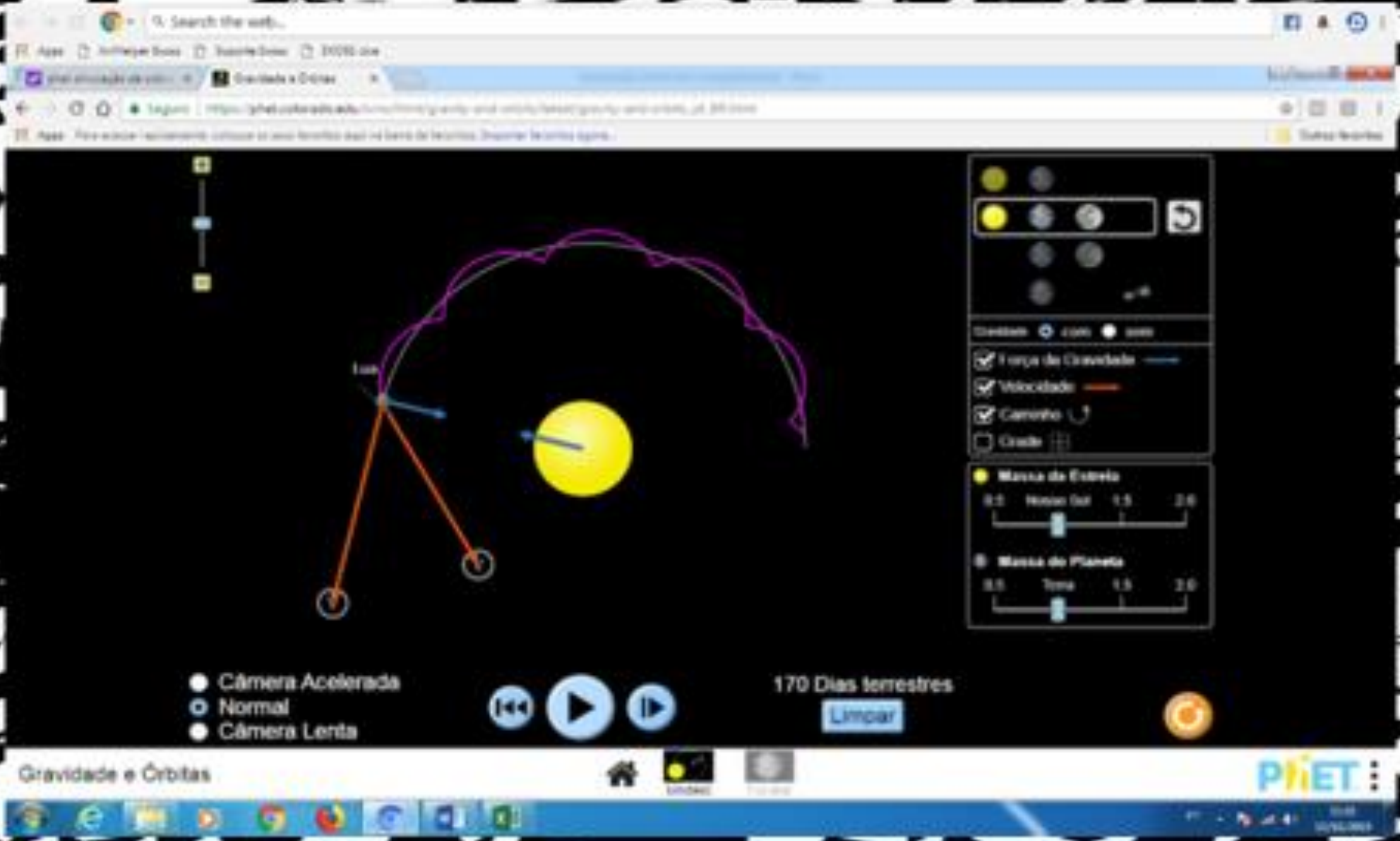

< https://phet.colorado.edu/pt_BR/simulation/gravity-and-orbits > Acesso em: 23 out. 2018-

**4º MOMENTO: UM BREVE RELATO COM HISTÓRIA DA FÍSICA – INTRODUÇÃO A APRESENTAÇÃO DA REVISTA**

## Observação de Galileu

- Estudou bolas rolando em planos inclinados.
- Quanto mais polido a superfície, mais a bola rolava.

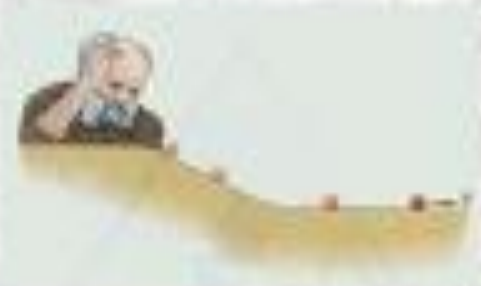

## ISAAC NEWTON

Isaac Newton (1642-1727) nasceu em Woolsthorpe(Inglaterra). Foi educado na Universidade de Cambridge e considerado aluno excelente e aplicado. Newton fez descobertas importantes em Matemática, Óptica e Mecânica. Em sua obra "Princípios Matemáticos de Filosofia Natural", enunciou as três leis fundamentais do movimento, conhecidas hoje como leis de Newton.

4º Momento: Continuação- Introduzindo a primeira lei de Newton com a

# Revista:

O 4º momento, iniciou com a leitura da revista pelos alunos da páginas 01, 02, a 05. Na página 06 foi acionado o primeiro Gif com o aplicativo AR PHYSICS para aprendizagem da primeira Lei de Newton

Página 01

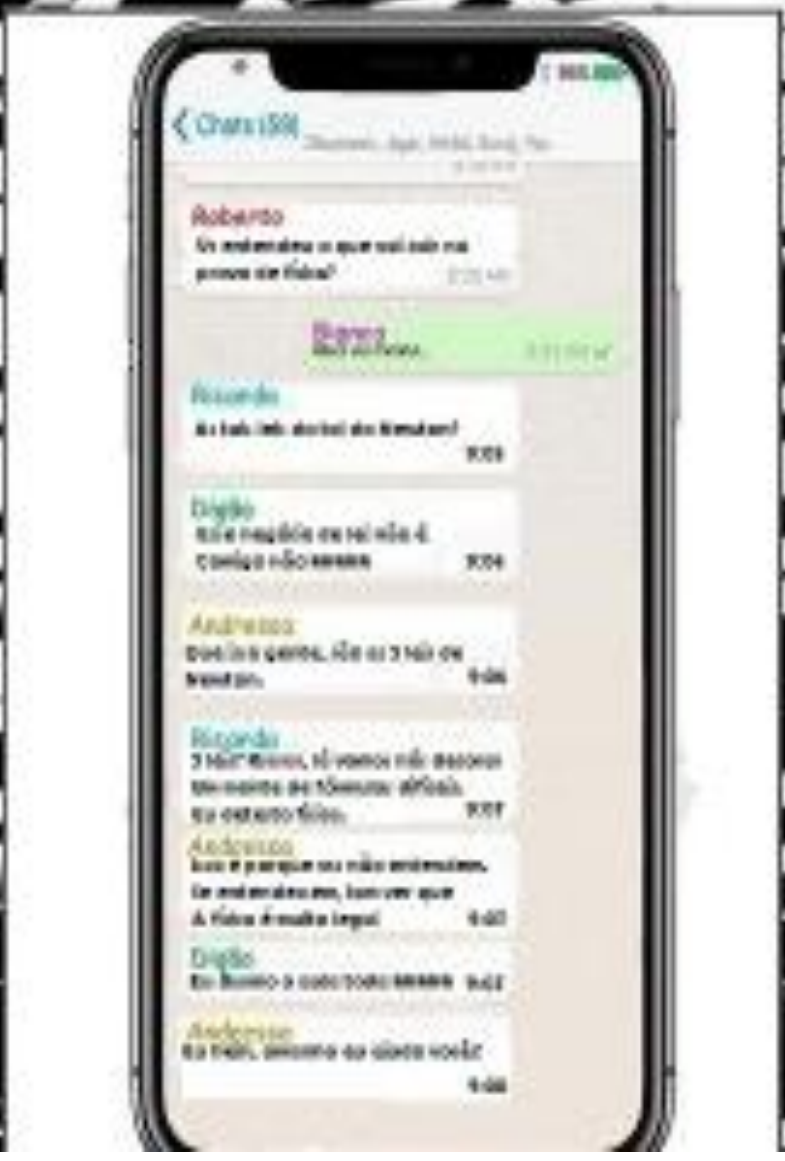

Página 02

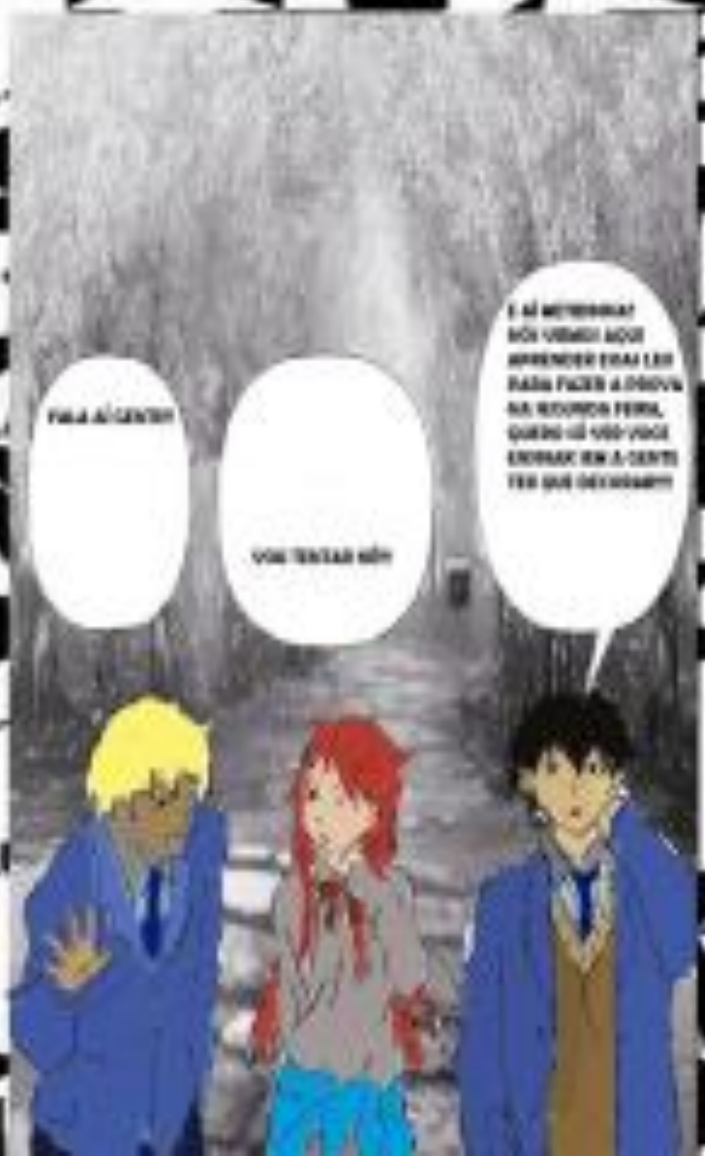

Página 03

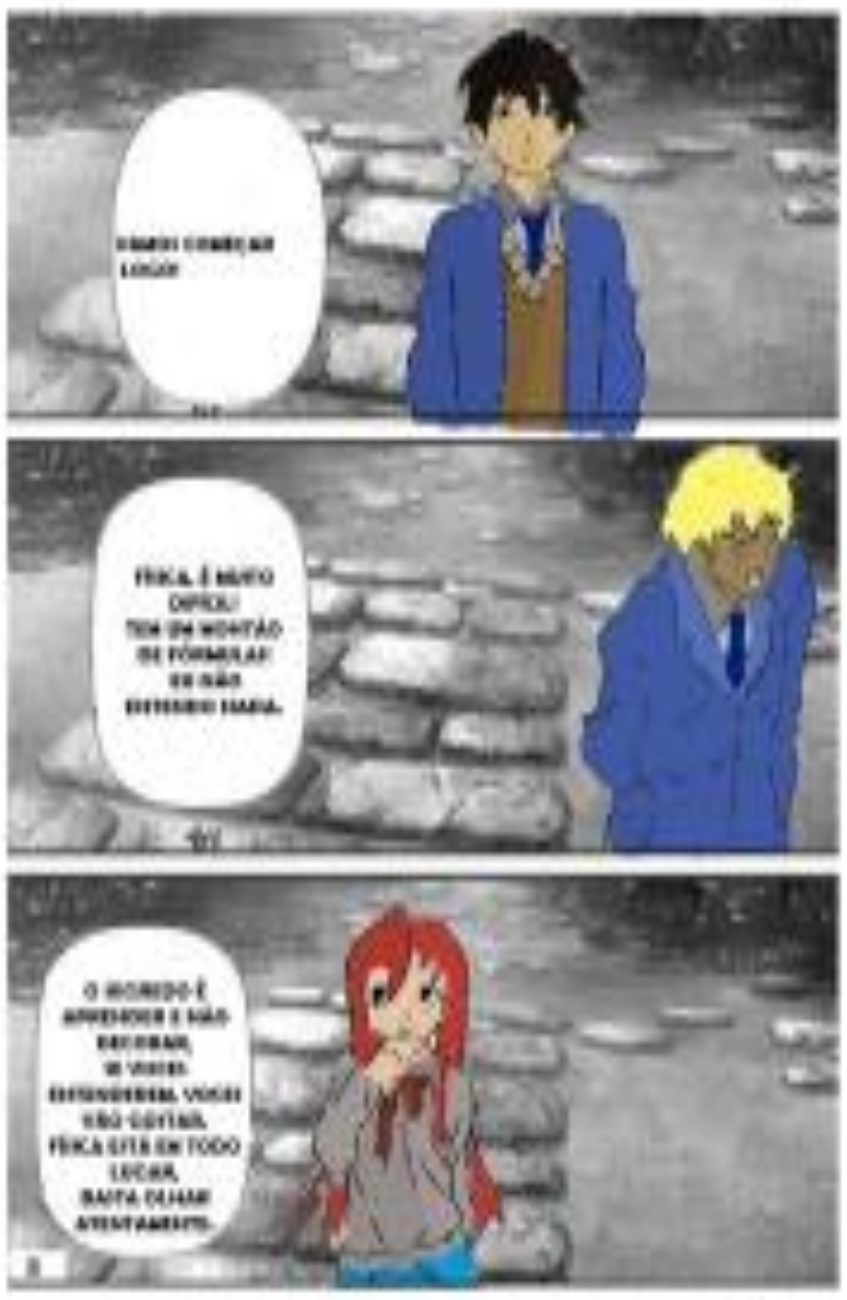

Página 04

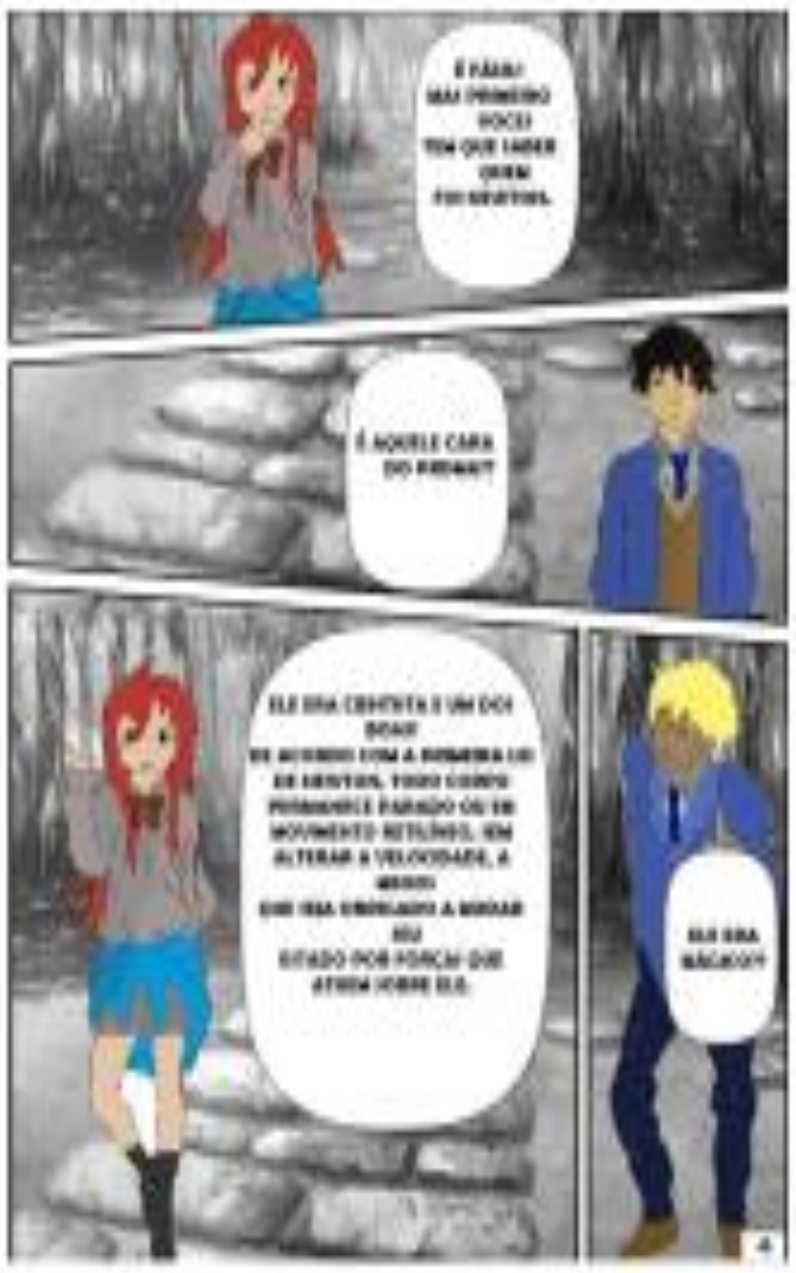

Página 05

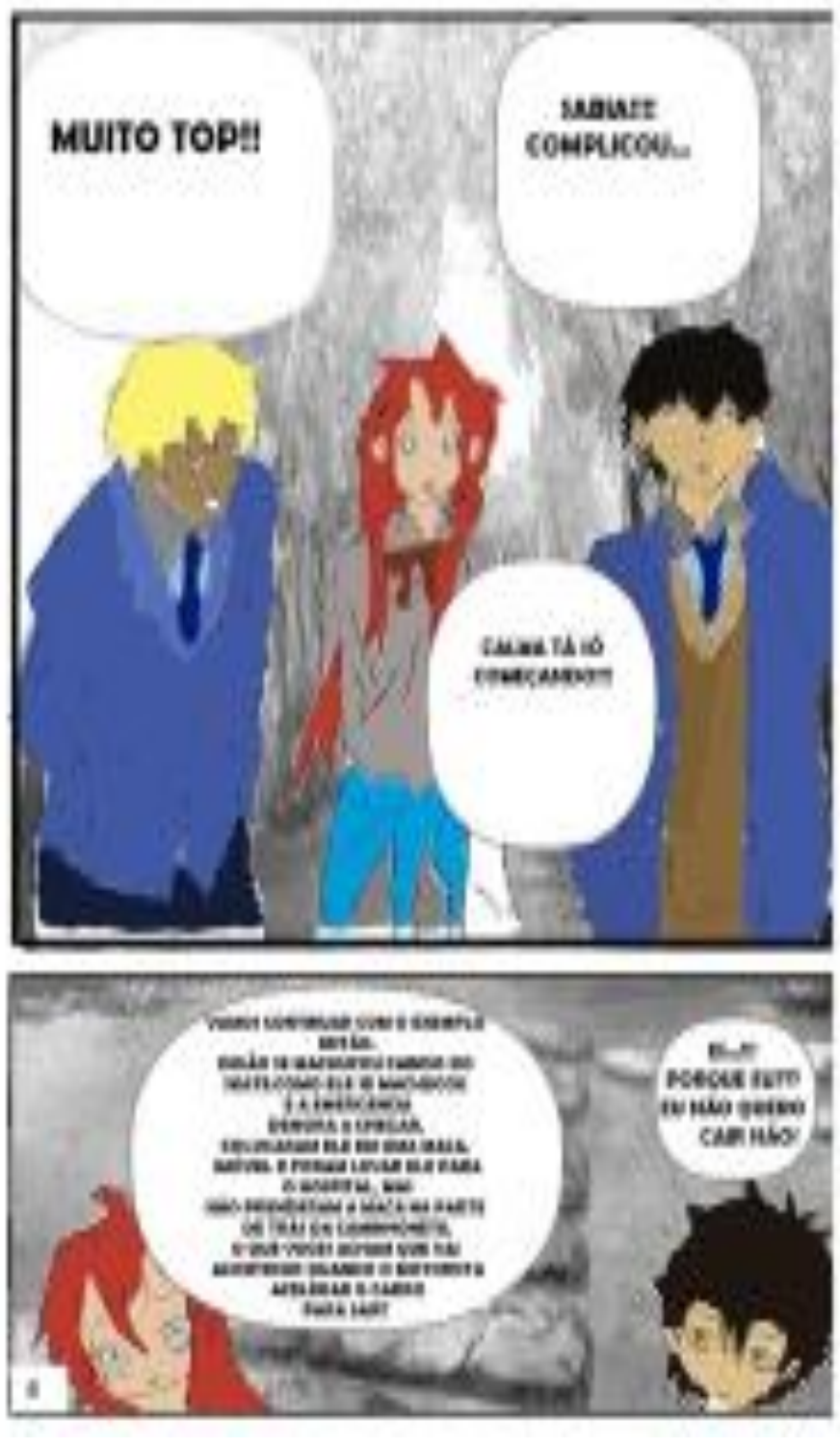

Página 06

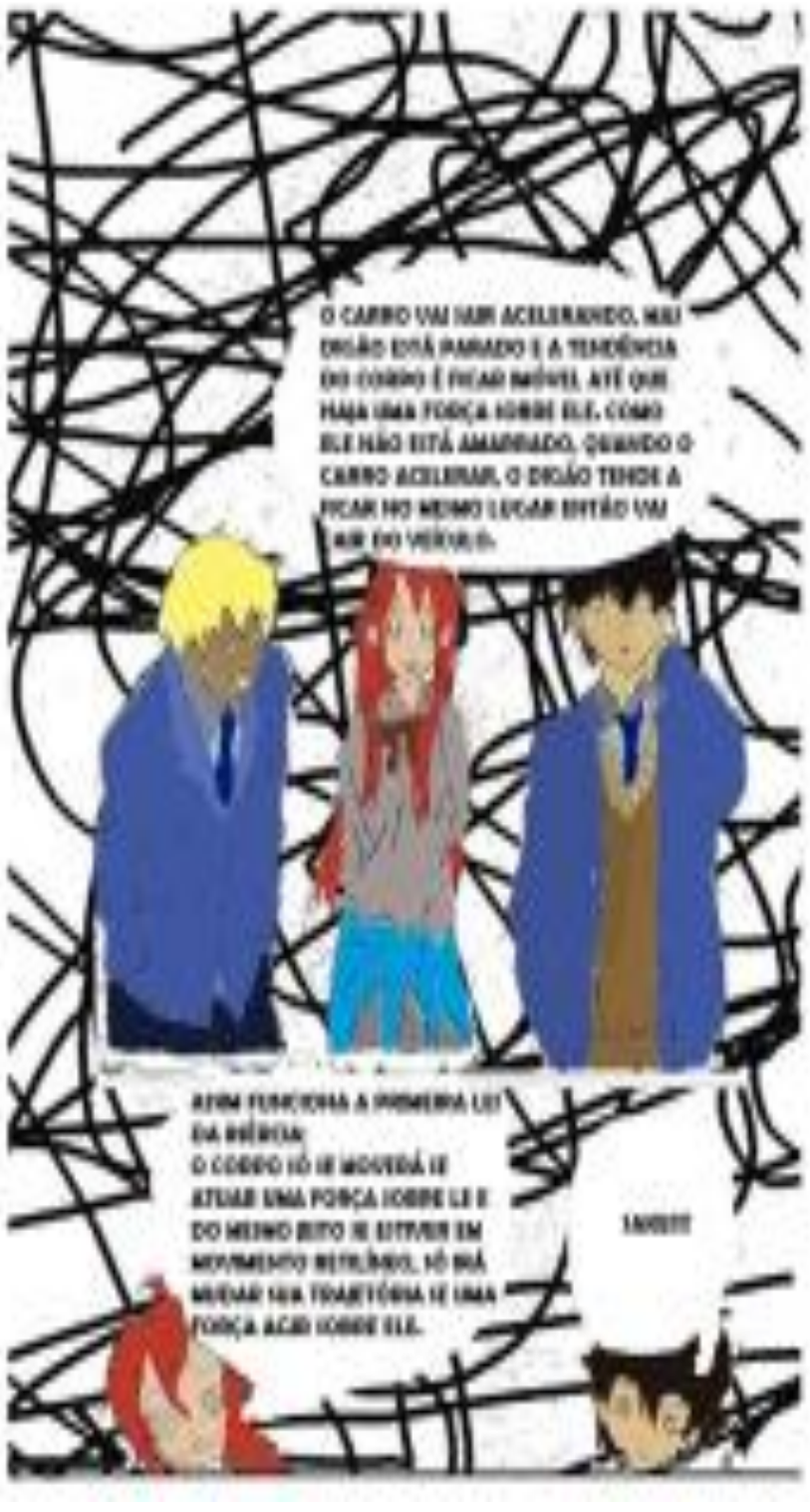

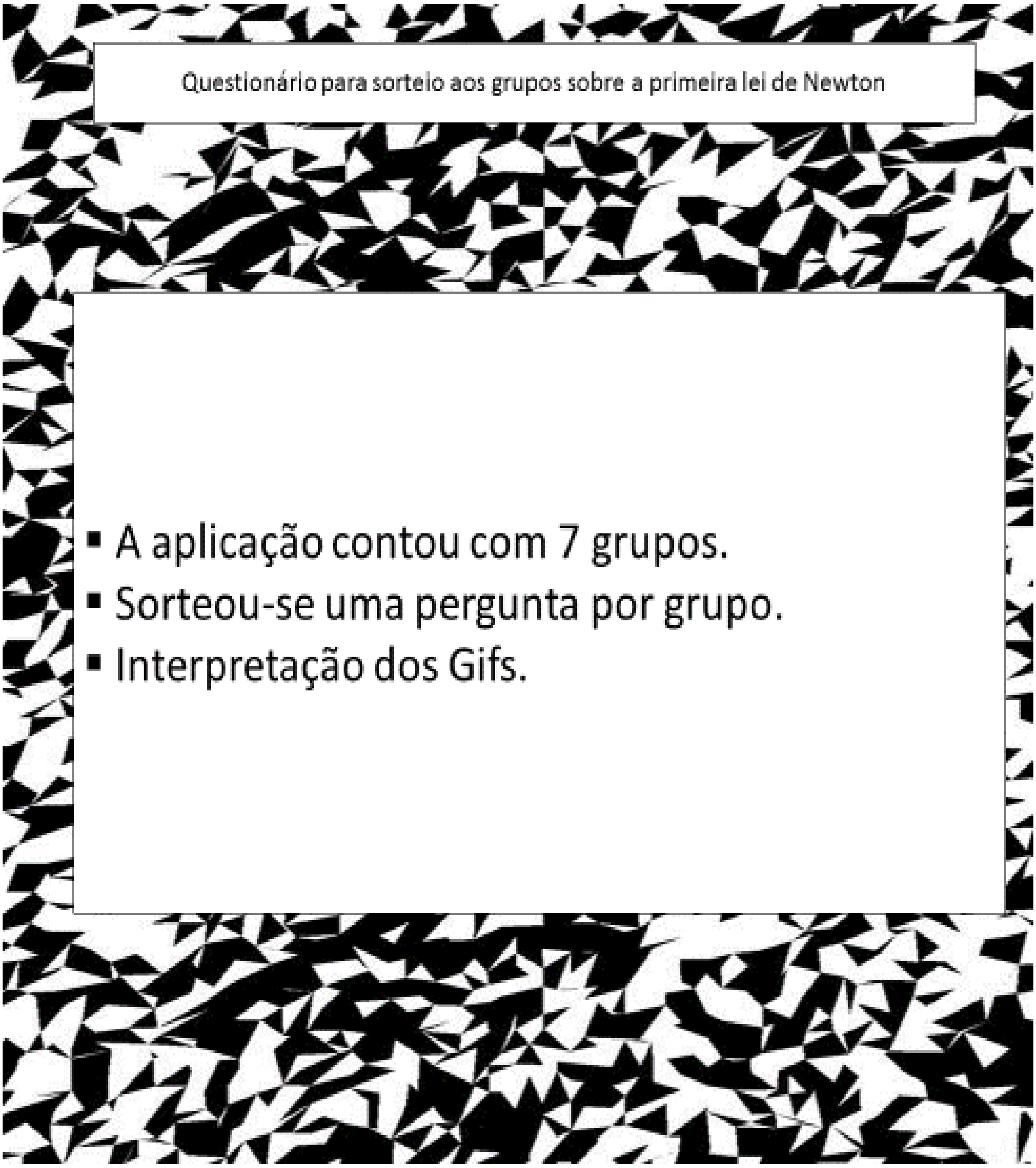
Questionário para sorteio aos grupos sobre a primeira lei de Newton
▪ A aplicação contou com 7 grupos.
▪ Sorteou-se uma pergunta por grupo.
▪ Interpretação dos Gifs.

**5ª momento: Ação mediada da segunda Lei de Newton**

Na página 07 inicia-se a 2ª Lei de Newton.
Na segunda Lei os alunos analisaram 2 Gifs diferentes.
Foi analisado 1 Gif em cada aula.
Sorteou-se perguntas sobre a segunda lei aos grupo.
Os grupos interpretaram os Gifs .

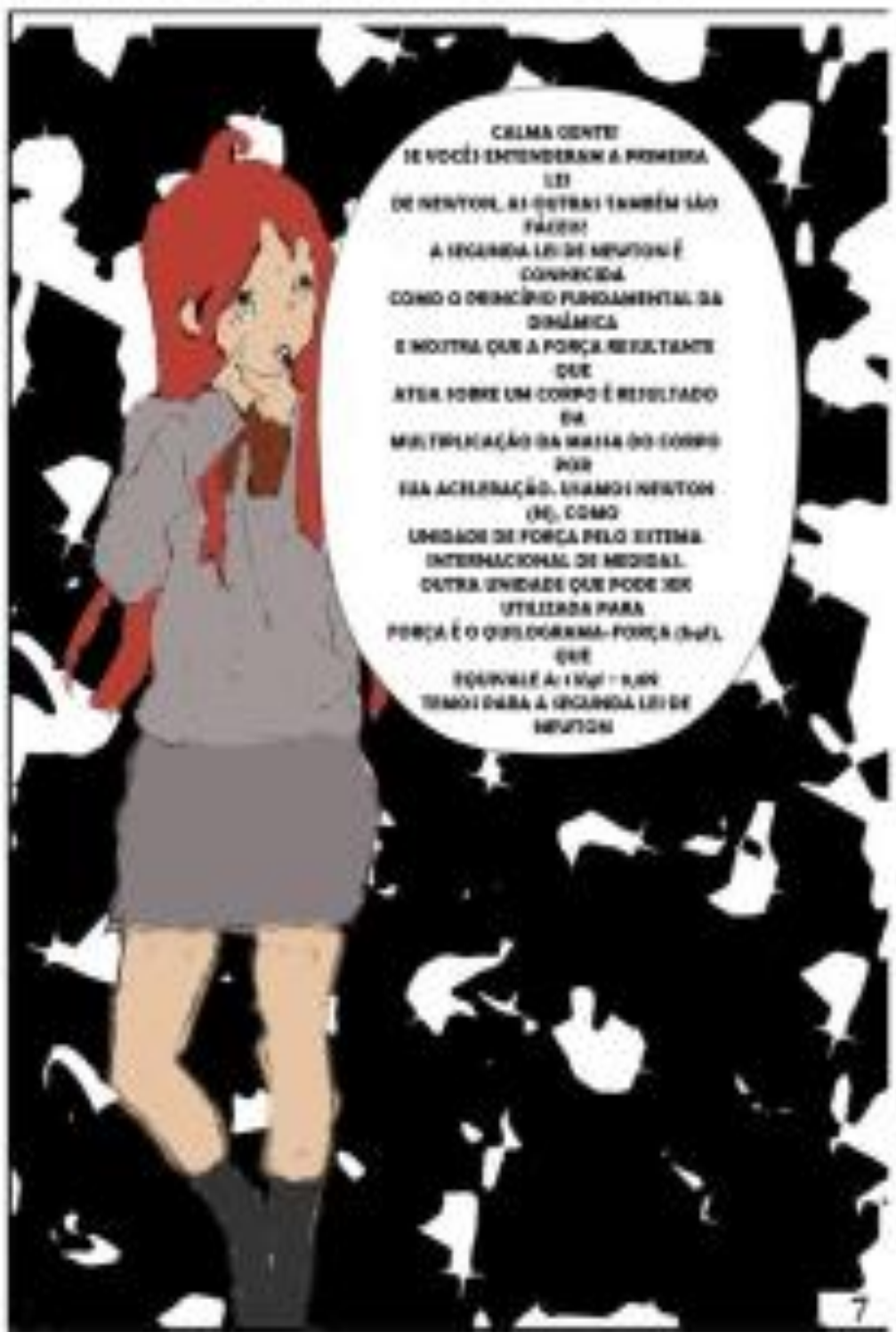

## 6ª aplicação:

Continuação da segunda lei de Newton com a página 08.

Nessa aplicação eles analisaram o Gif e fizeram pequenos resumos explicando a lei aplicada aos dois gifs apresentados no 5ª e 6ª momentos.

Página 08

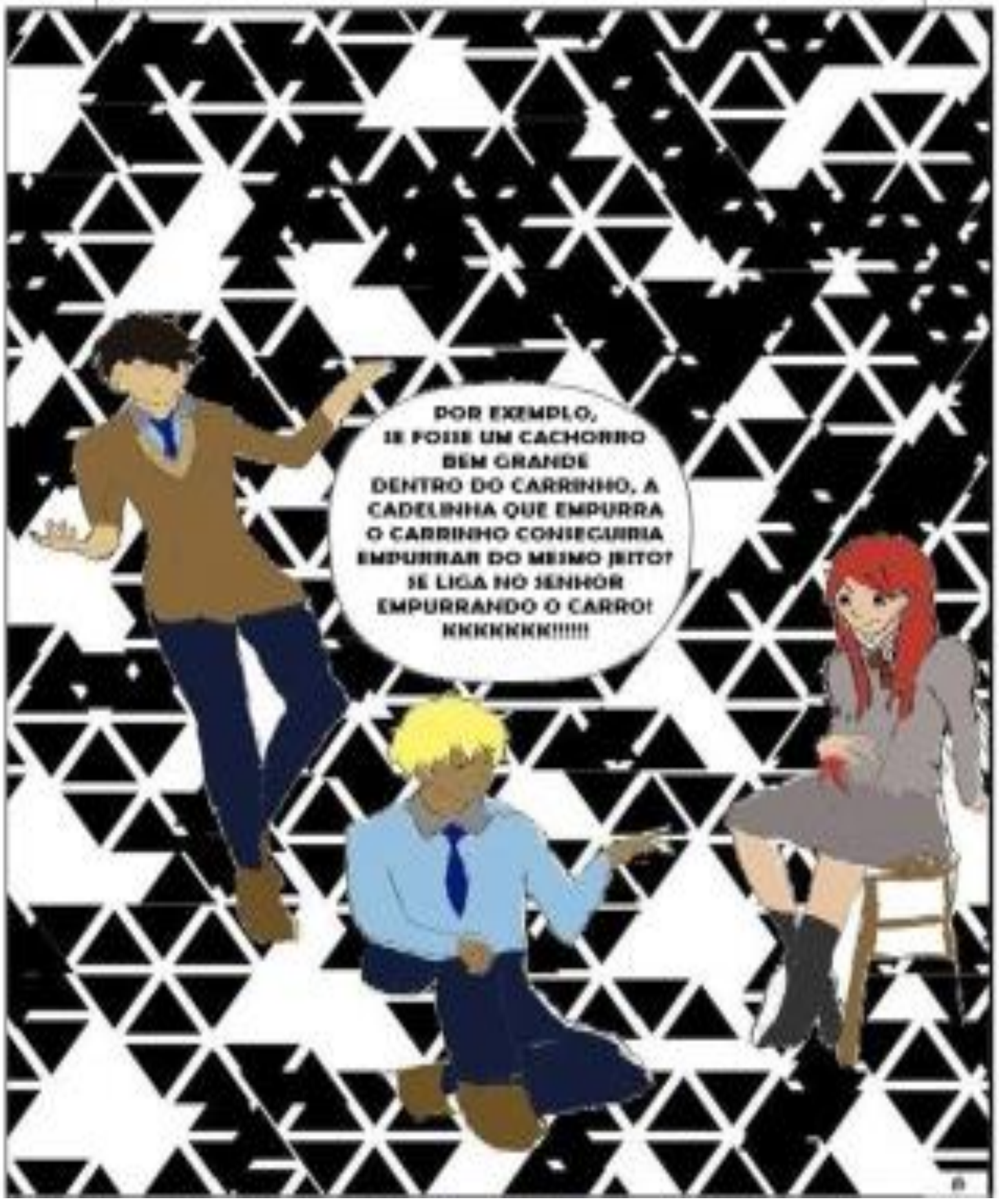

# 7º Momento: Estudo da terceira lei de Newton -

Na 3ª Lei de Newton tivemos 3 Gifs para observação e análise. Primeiro a página 09 lida e observada. Logo após o acionamento do Gif da página 09 foram entregues as perguntas referentes a terceira Lei de Newton.

Página 09

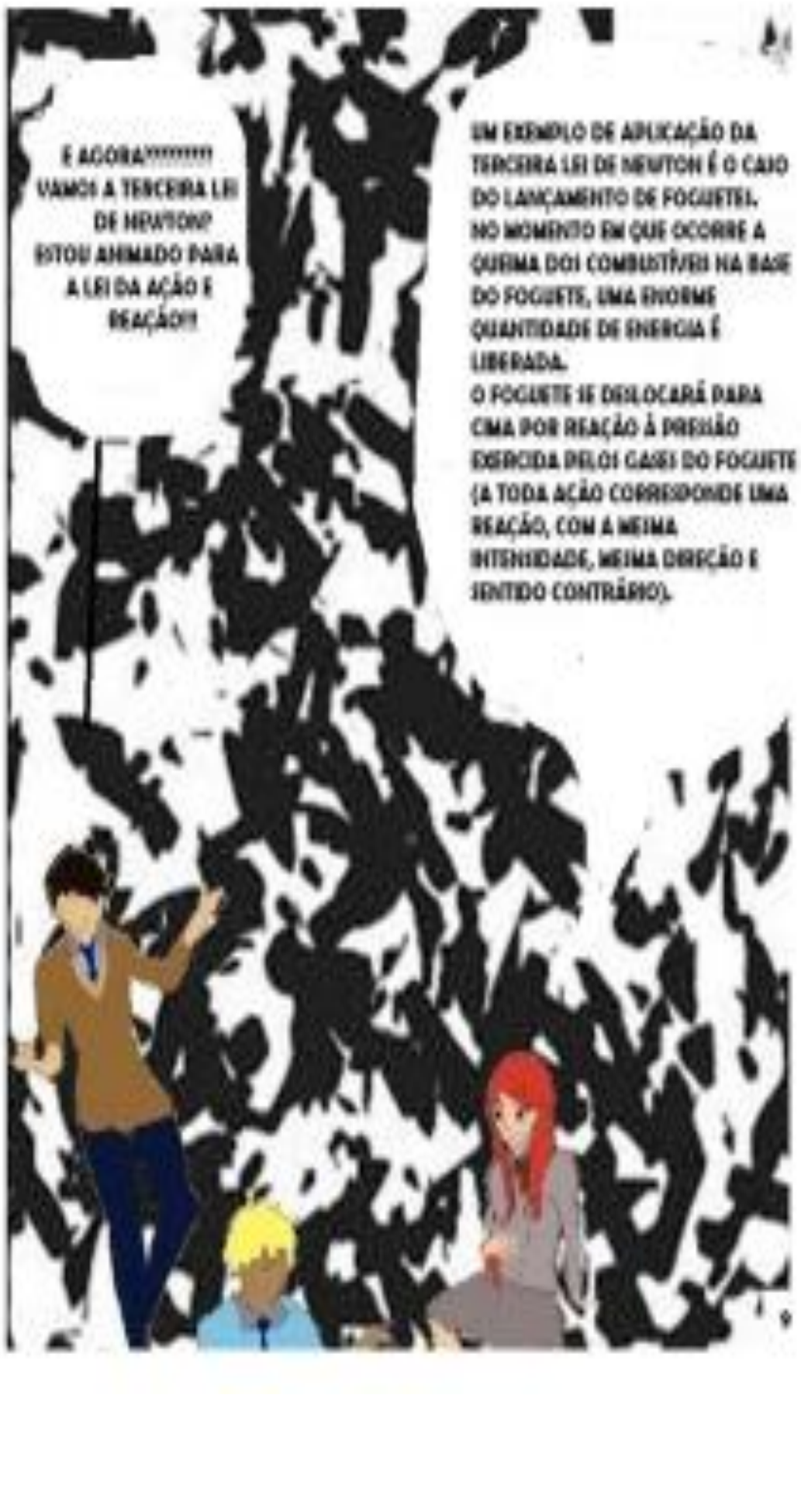

# 8º momento: 3ª Lei de Newton, análise dos gifs

No 8º momento os alunos acionam os dois Gifs das páginas 10 e 11 e analisam a terceira lei em cada um deles.
Os grupos escreveram pequenos textos sobre os 3 Gifs, em seguida um grupo foi sorteado e expôs para toda turma sua interpretação do Gif escolhido.

Página 10

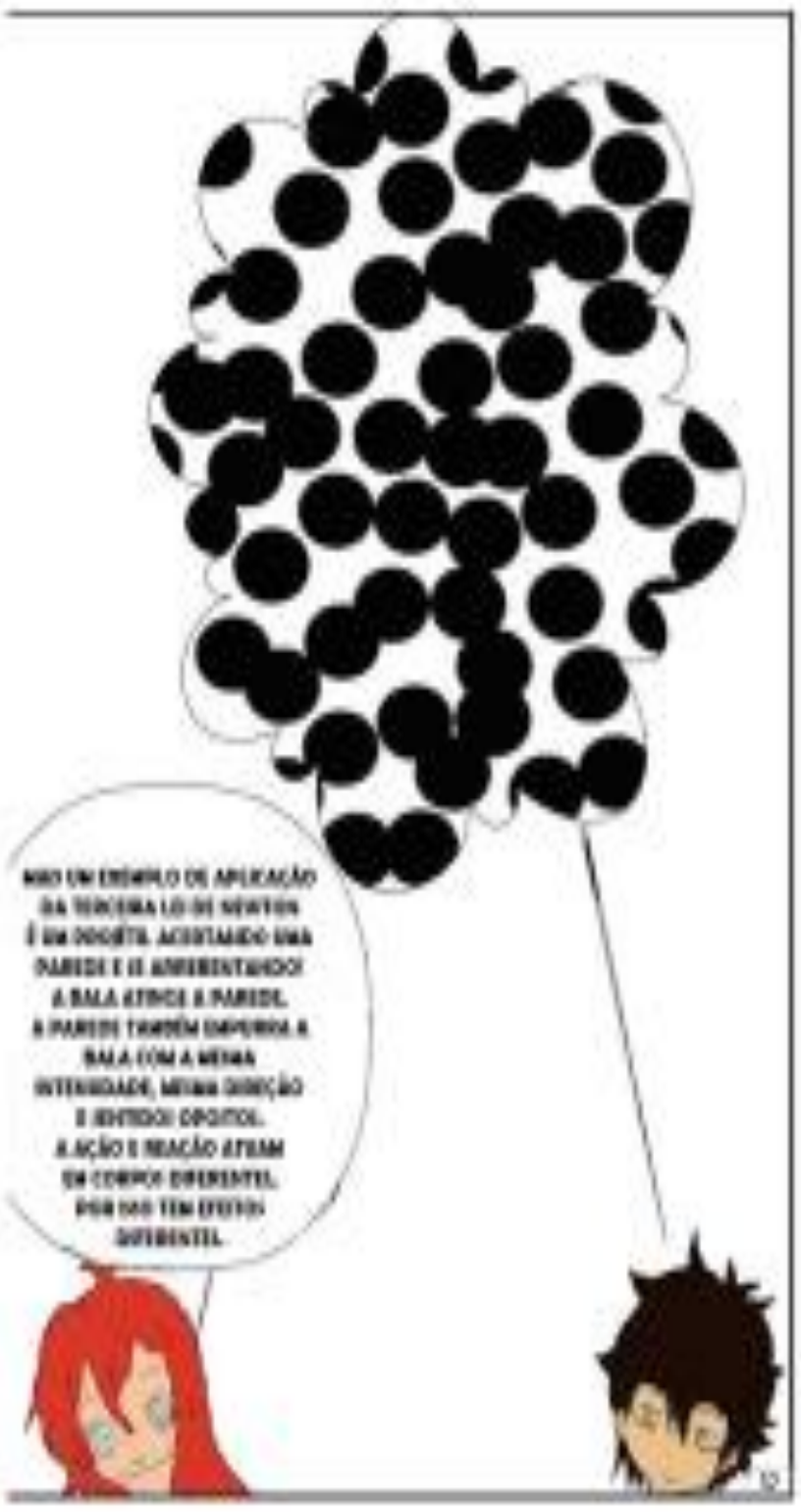

Página 11

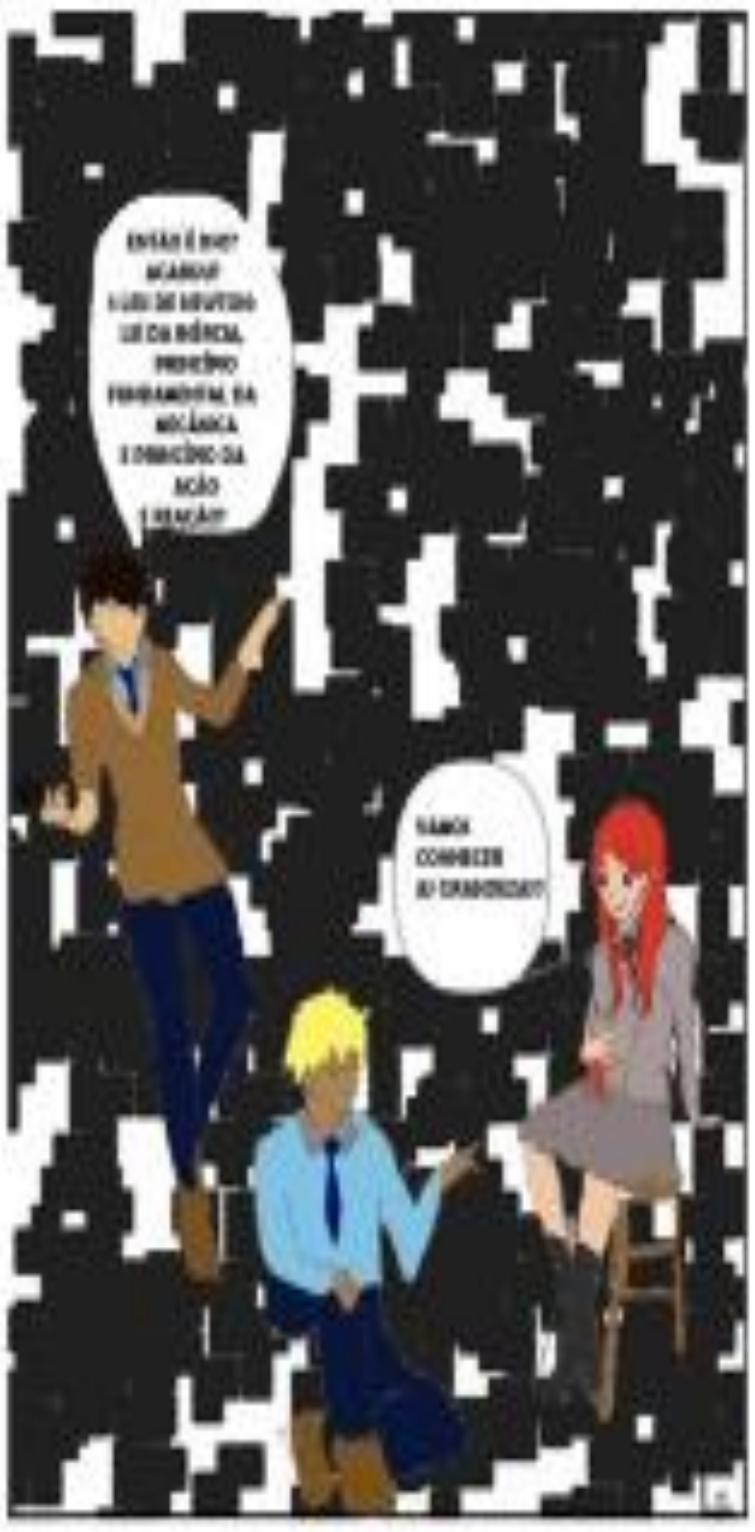

# 9º Momento: Conhecendo as forças

Página 12

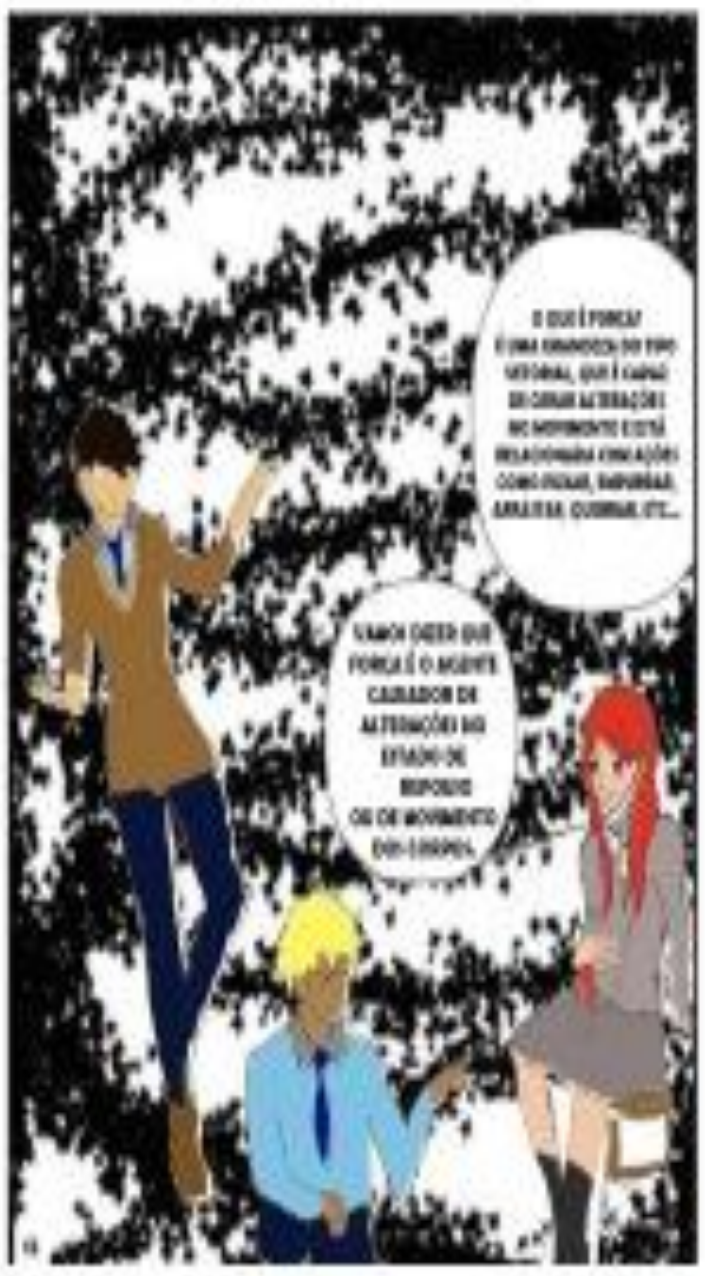

Página 13

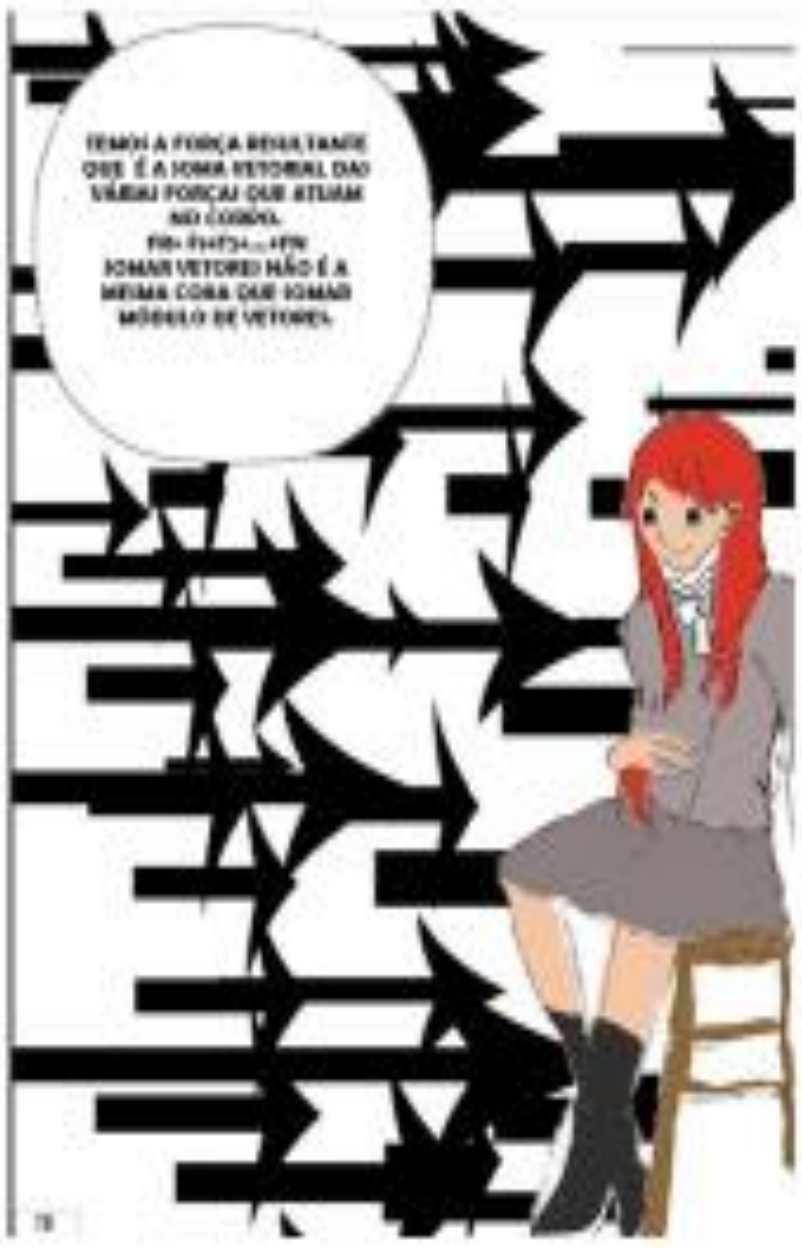

No 9º momento os alunos fizeram a leitura das páginas, acionam o Gif da página 12 e 13 interpretaram o conceito de força e força resultante. Sorteou-se 2 grupos para falar sobre as conclusões que chegaram cada um sobre um dos Gifs diferentes.

Observação: O Gif da página 13 no aplicativo de teste tem um erro conceitual e não deve ser reproduzido e sim o da versão final

10º Momento: **Força Peso**

Os alunos leram a revista, acionaram o Gif e analisaram, descrevendo a força peso.
Ao final um grupo foi sorteado para falar sobre a conclusão que chegaram expondo seu texto para toda a turma.

14ª Página

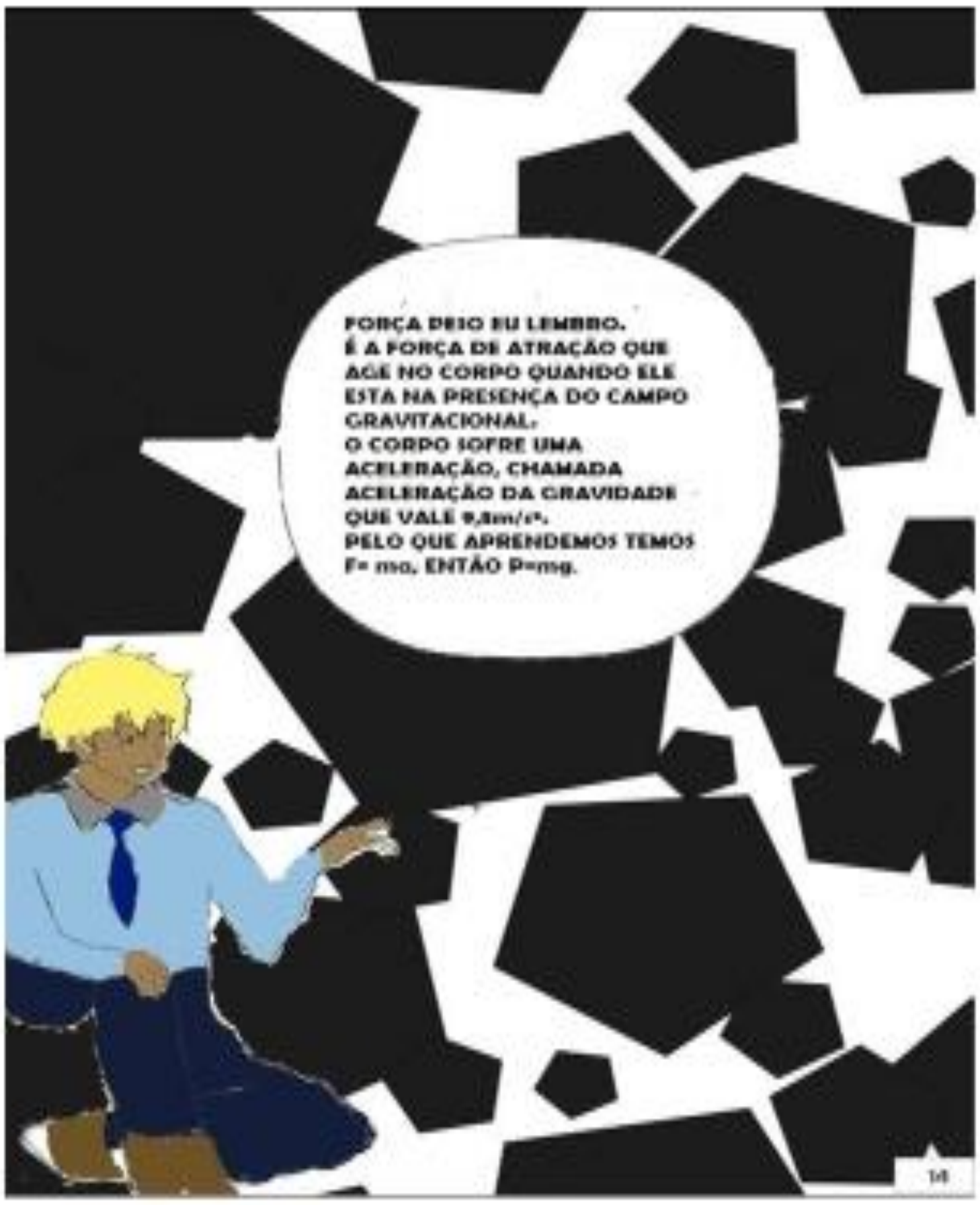

## 11º Momento: Força de interação e de atrito

Na página 15 temos uma figura 3D para estudo da interação entre os corpos a distância deve-se posicionar o aplicativo na parte de cima da revista para observar .

Na parte de baixo temos um Gif representando a força de atrito. Nesse momento os alunos leram a revista, analisaram a figura 3D e o Gif, escreveram sua análise e dois grupos foram escolhidos para expor suas anotações.

Página 15

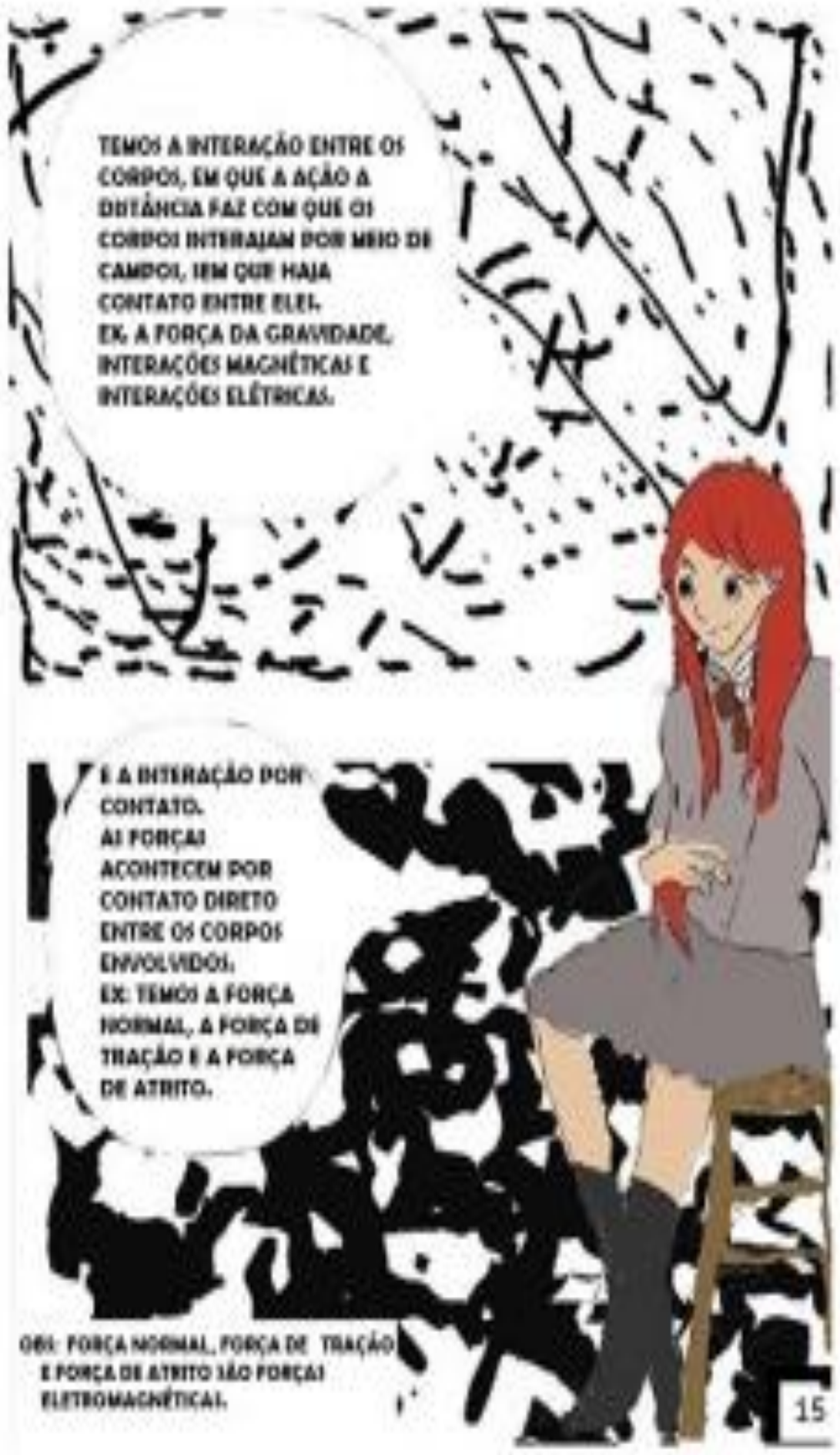

# 12º Momento: Força Normal

Na página 16 temos a representação da força normal.
Os alunos leram a revista e observaram o Gif, analisando junto com seus grupos.
Um grupo foi sorteado para expor sua resposta para toda a turma.

Página 16

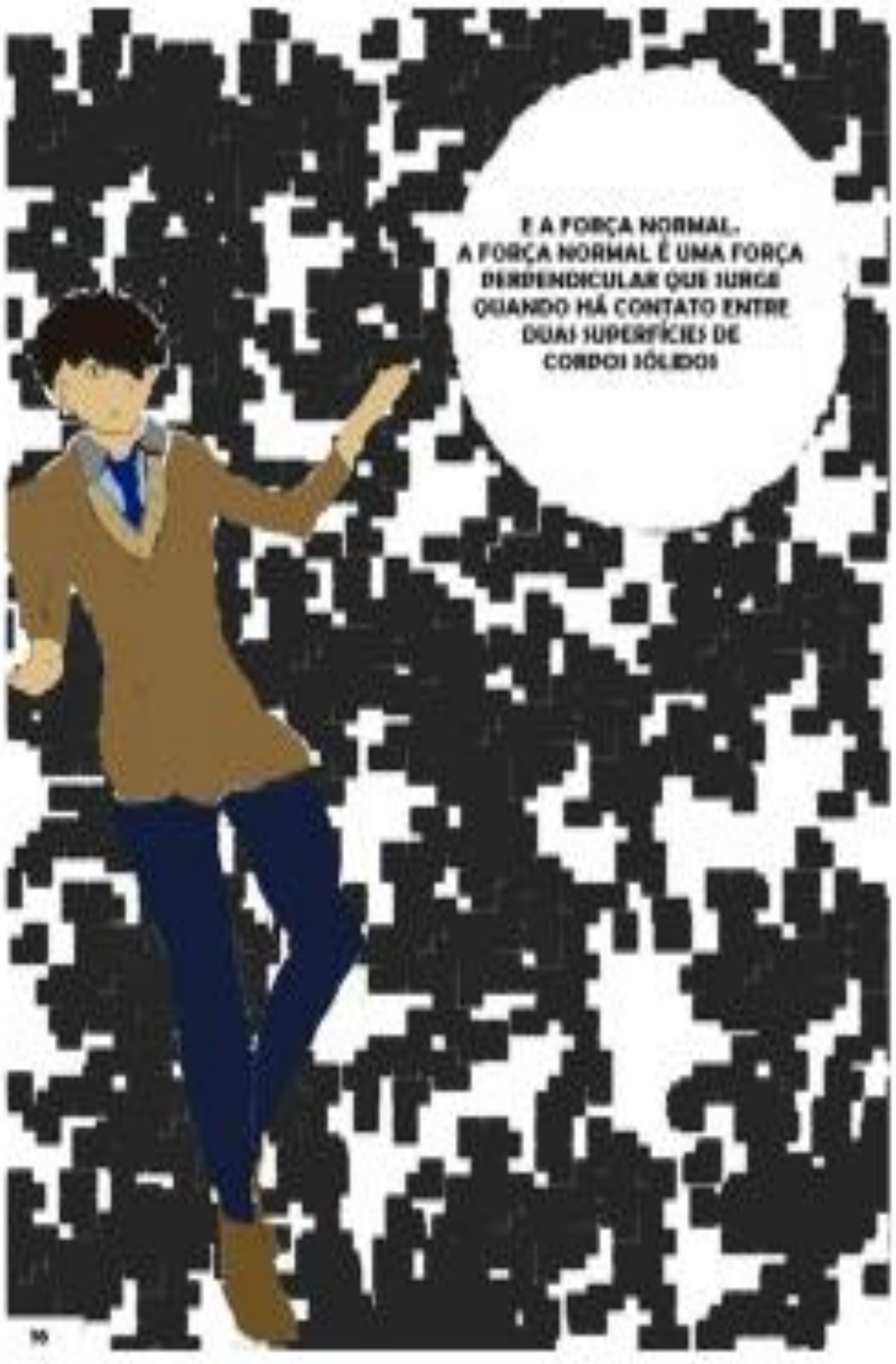

# Desafio

Na página 17 foi proposto um desafio para análise da queda de uma mola maluca.

Os alunos observaram a mola maluca e reproduziram em sala de aula o movimento mostrado no Gif.
Após reprodução desse movimento os grupos escreveram suas conclusões e compartilharam com toda turma.

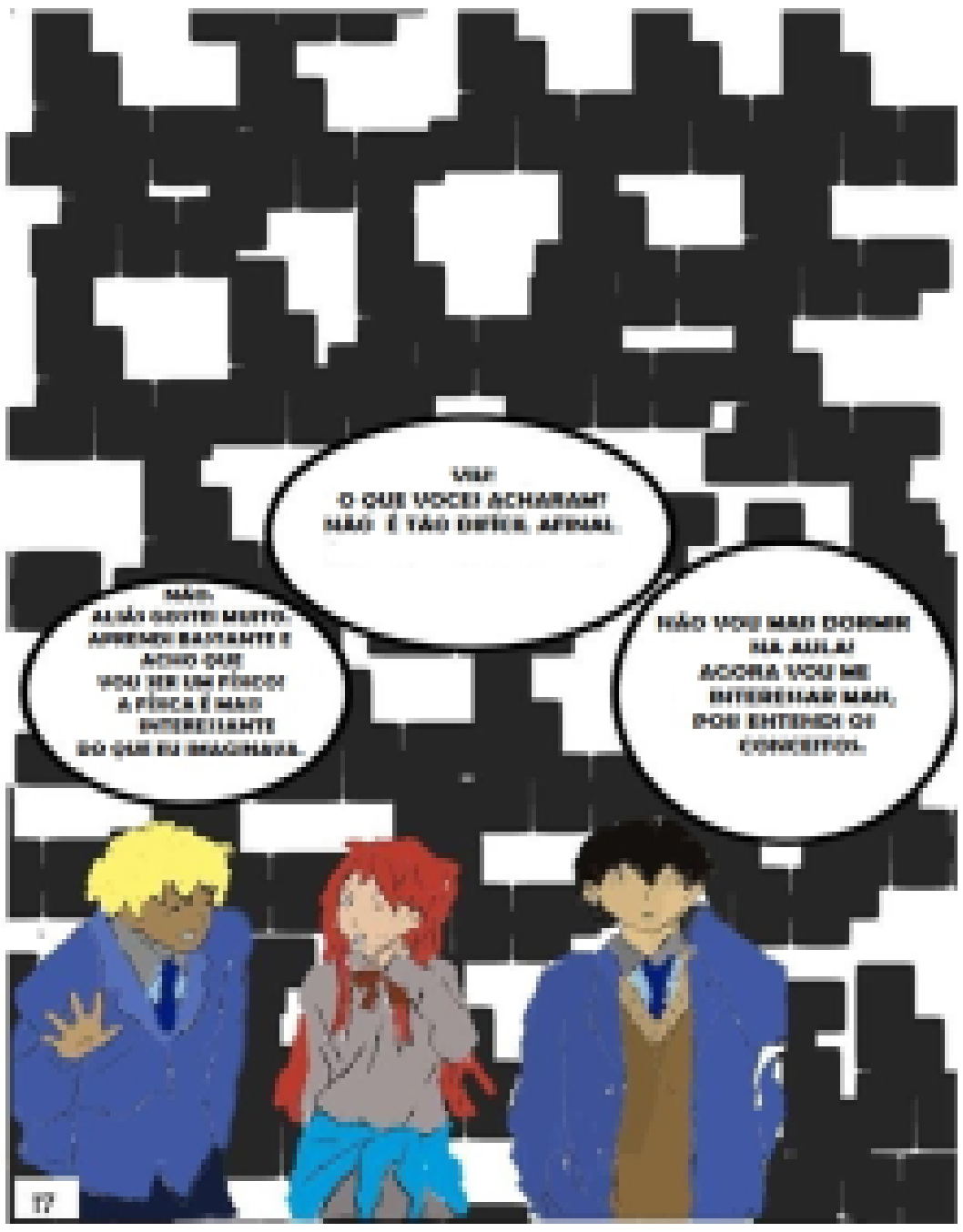

# 13º Momento:

Os grupos devem receber a tarefa de reproduzirem vídeos dos temas estudados e apresentar na sala de aula em forma de Gifs.

# 14º Momento: Aplicação do questionário final para verificação de aprendizagem

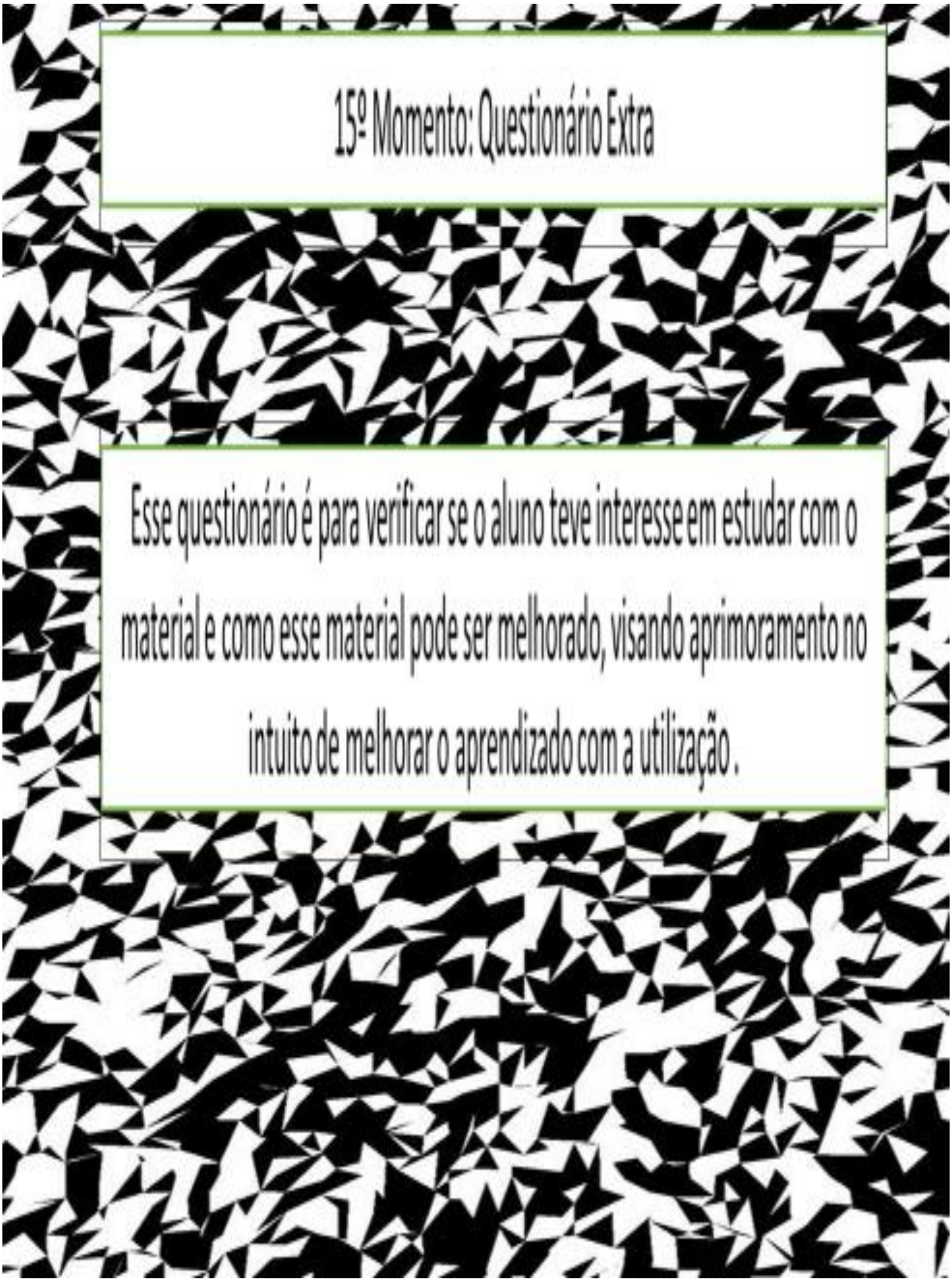
15º Momento: Questionário Extra
Esse questionário é para verificar se o aluno teve interesse em estudar com o material e como esse material pode ser melhorado, visando aprimoramento no intuito de melhorar o aprendizado com a utilização .

# REFERÊNCIAS

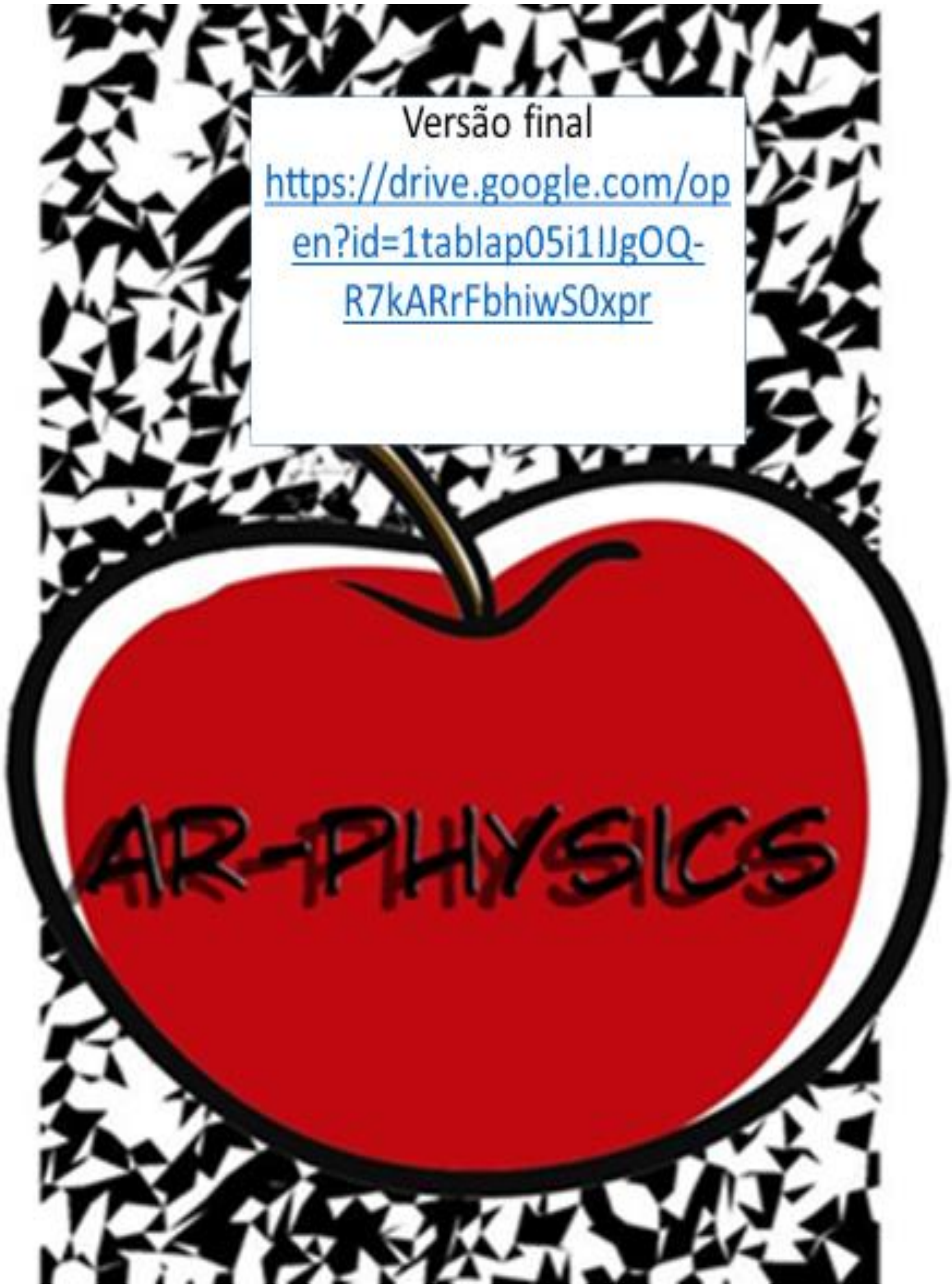
Versão final
https://drive.google.com/open?id=1tablap05i1IJgOQ-R7kARrFbhiwS0xpr
AR-PHYSICS